\documentclass[10pt,twoside]{IEEEtran}  

\usepackage{caption, subcaption, graphicx,epstopdf,algorithm,amsmath,url,amssymb,algorithmic,epsfig, color}

\newcommand{\betahat}{\hat{\beta}}

\newcommand{\matr}[2]{ \left[\begin{array}{cc}
     #1 \\
     #2
   \end{array}
  \right]
  }

\newcommand{\beq}{\begin{equation}}
\newcommand{\eeq}{\end{equation}}
\newcommand{\bdm}{\begin{displaymath}}
\newcommand{\edm}{\end{displaymath}}

\newcommand{\se}{&=&}

\newcommand{\sle}{& \le &}

\newtheorem{theorem}{Theorem}[section]
\newtheorem{lem}[theorem]{Lemma}

\newtheorem{definition}[theorem]{Definition}
\newtheorem{remark}[theorem]{Remark}

\newcommand{\bd}{\begin{definition}}
\newcommand{\ed}{\end{definition}}

\newcommand{\bv}{\begin{vugraph}}
\newcommand{\ev}{\end{vugraph}}
\newcommand{\bi}{\begin{itemize}}
\newcommand{\ei}{\end{itemize}}
\newcommand{\ben}{\begin{enumerate}}
\newcommand{\een}{\end{enumerate}}

\newcommand{\bean}{\begin{eqnarray*} }
\newcommand{\eean}{\end{eqnarray*} }
\newcommand{\bea}{\begin{eqnarray} }
\newcommand{\eea}{\end{eqnarray} }
\newcommand{\nn}{\nonumber}
\newcommand{\ba}{\begin{array} }
\newcommand{\ea}{\end{array} }

\newcommand{\Thresh}{\text{Thresh}}
\newcommand{\Prune}{\text{Prune}}
\newcommand{\LS}{\text{LS}}

\newcommand{\Span}{\text{range}}
\newcommand{\train}{\text{train}}

\newcommand{\cs}{\text{cs}}
\newcommand{\new}{\text{new}}
\newcommand{\old}{\text{old}}

\newcommand{\Shat}{\hat{S}}
\newcommand{\That}{\hat{T}}
\newcommand{\Lhat}{\hat{L}}

\newcommand{\Phat}{\hat{P}}
\newcommand{\Qhat}{\hat{Q}}
\newcommand{\rank}{\text{rank}}
\newcommand{\basis}{\text{basis}}
\newcommand{\abasis}{\text{approx-basis}}

\newcommand{\rhat}{\hat{r}}
\newcommand{\SE}{\text{SE}}

\begin{document}
\title{An Online Algorithm for Separating Sparse and Low-dimensional Signal Sequences from their Sum}
\author{Han Guo, Chenlu Qiu, Namrata Vaswani, \\
Dept of Electrical and Computer Engineering, Iowa State University, Ames, IA \\
\{hanguo,chenlu,namrata\}@iastate.edu
\thanks{A part of this work was presented at Allerton 2010, Allerton 2011 and ICASSP 2014 \cite{rrpcp_allerton,rrpcp_allerton11,rrpcp_icassp14}. This research was partially supported by NSF grants CCF-0917015, CCF-1117125 and IIS-1117509. 
}
}
\maketitle

\begin{abstract}
This paper designs and extensively evaluates an online algorithm, called practical recursive projected compressive sensing (Prac-ReProCS), for recovering a time sequence of sparse vectors $S_t$ and a time sequence of dense vectors $L_t$ from their sum, $M_t:= S_t + L_t$, when the $L_t$'s lie in a slowly changing low-dimensional subspace of the full space. A key application where this problem occurs is in real-time video layering where the goal is to separate a video sequence into a slowly changing background sequence and a sparse foreground sequence that consists of one or more moving regions/objects on-the-fly. Prac-ReProCS is a practical modification of its theoretical counterpart which was analyzed in our recent work. Experimental comparisons demonstrating the advantage of the approach for both simulated and real videos, over existing batch and recursive methods, are shown.  Extension to the undersampled case is also developed.
\end{abstract}

\section{Introduction} 
This paper designs and evaluates a practical algorithm for recovering a time sequence of sparse vectors $S_t$ and a time sequence of dense vectors $L_t$ from their sum, $M_t:= S_t + L_t$, when the $L_t$'s lie in a slowly changing low-dimensional subspace of $\mathbf{R}^n$. The magnitude of the entries of $L_t$ could be larger, roughly equal or smaller than that of the nonzero entries of $S_t$. The extension to the undersampled case,  $M_t:= A S_t + BL_t$, is also developed.
The above problem can be interpreted as one of online/recursive sparse recovery from potentially large but structured noise. In this case, $S_t$ is the quantity of interest and $L_t$ is the potentially large but structured low-dimensional noise. Alternatively it can be posed as a  recursive / online robust principal components analysis (PCA) problem. In this case $L_t$, or in fact, the subspace in which it lies, is the quantity of interest while $S_t$ is the  outlier.

A key application where the above problem occurs is in video layering where the goal is to separate a slowly changing background from moving foreground objects/regions \cite{Torre03aframework,rpca}. The foreground layer, e.g. moving people/objects, is of interest in applications such as automatic video surveillance, tracking moving objects, or video conferencing. The background sequence is of interest in applications such as background editing (video editing applications). In most static camera videos, the background images do not change much over time and hence the background image sequence is well modeled as lying in a fixed or slowly-changing low-dimensional subspace of $\mathbf{R}^n$ \cite{error_correction_PCP_l1,rpca}. Moreover the changes are typically global, e.g. due to lighting variations, and hence modeling it as a dense image sequence is valid too \cite{rpca}. The foreground layer usually consists of one or more moving objects/persons/regions that move in a correlated fashion, i.e. it is a sparse image sequence that often changes in a correlated fashion over time. Other applications where the above problem occurs include solving the video layering problem from compressive video measurements, e.g. those acquired using a single-pixel camera; online detection of brain activation patterns from full or undersampled functional MRI (fMRI) sequences (the ``active" part of the brain forms the sparse image, while the rest of the brain which does not change much over time forms the low-dimensional part); or  sensor networks based detection and tracking of abnormal events such as forest fires or oil spills. 
The single pixel imaging and undersampled fMRI applications are examples of the compressive case, $M_t=A S_t+ B L_t$ with $B=A$.

{\em Related Work. }
Most high dimensional data often approximately lie in a lower dimensional subspace. Principal components' analysis (PCA) is a widely used dimension reduction technique that finds a small number of orthogonal basis vectors (principal components), along which most of the variability of the dataset lies. For a given dimension, $r$, PCA finds the $r$-dimensional subspace that minimizes the mean squared error between data vectors and their projections into this subspace \cite{PCA}.  It is well known that PCA is very sensitive to outliers. Computing the PCs in the presence of outliers is called robust PCA. Solving the robust PCA problem recursively as more data comes in is referred to as online or recursive robust PCA. ``Outlier" is a loosely defined term that usually refers to any corruption that is not small compared to the true signal (or data vector) and that occurs only occasionally.
As suggested in \cite{dense_error_correct}, an outlier can be nicely modeled as a sparse vector.



In the last few decades, there has been a large amount of work on robust PCA, e.g. \cite{Torre03aframework,Roweis98emalgorithms,novel_m_estimator,outlier_pursuit, mccoy_tropp11}, and recursive robust PCA e.g. \cite{sequentialSVD,ipca_weightedand,Li03anintegrated}. In most of these works, either the locations of the missing/corruped data points are assumed known \cite{sequentialSVD} (not a practical assumption); or they first detect the corrupted data points and then replace their values using nearby values \cite{ipca_weightedand}; or weight each data point in proportion to its reliability (thus soft-detecting and down-weighting the likely outliers) \cite{Torre03aframework,Li03anintegrated}; or just remove the entire outlier vector \cite{outlier_pursuit, mccoy_tropp11}. 
Detecting or soft-detecting outliers ($S_t$) as in \cite{ipca_weightedand,Torre03aframework,Li03anintegrated} is easy when the outlier magnitude is large, but not when it is of the same order or smaller than that of the $L_t$'s.

In a series of recent works \cite{rpca,rpca2}, a new and elegant solution to robust PCA called Principal Components' Pursuit (PCP)  has been proposed, that does not require a two step outlier location detection/correction process and also does not throw out the entire vector. It redefines batch robust PCA as a problem of separating a low rank matrix, ${\cal L}_t := [L_1,\dots,L_t]$, from a sparse matrix, ${\cal S}_t := [S_1,\dots,S_t]$, using the measurement matrix, ${\cal M}_t := [M_1,\dots,M_t] = {\cal L}_t+ {\cal S}_t$. Other recent works that also study batch algorithms for recovering a sparse ${\cal S}_t$ and a low-rank ${\cal L}_t$ from ${\cal M}_t := {\cal L}_t+ {\cal S}_t$ or from undersampled measurements include \cite{rpca_tropp, linear_inverse_prob, rpca_hu,SpaRCS,rpca_vayatis,rpca_zhang, rpca_Giannakis,compressivePCP,rpca_reduced,noisy_undersampled_yuan}.
It was shown in \cite{rpca} that by solving PCP: 
\beq
\underset{{\cal L},{\cal S}} {\min}\|{\cal L}\|_* + \lambda\|{\cal S}\|_1 \ \text{subject to}  \ \ {\cal L} + {\cal S} = {\cal M}_t
\label{PCP}
\vspace{-2mm}
\eeq
one can recover ${\cal L}_t$ and ${\cal S}_t$ exactly, provided that (a) ${\cal L}_t$ is ``dense";
(b) any element of the matrix ${\cal S}_t$ is nonzero w.p. $\varrho$, and zero w.p. $1-\varrho$, independent of all others (in particular, this means that the support sets of the different $S_t$'s are independent over time); and (c) the rank of ${\cal L}_t$ and the support size of ${\cal S}_t$ are small enough. Here $\|A\|_*$ is the nuclear norm of a matrix $A$ (sum of singular values of $A$) while $\|A\|_1$ is the $\ell_1$ norm of $A$ seen as a long vector.

Notice that most applications described above require an online solution. A batch solution would need a long delay; and would also be much slower and more memory-intensive than a recursive solution.
Moreover, the assumption that the foreground support is independent over time is not usually valid.
To address these issues, in the conference versions of this work \cite{rrpcp_allerton,rrpcp_allerton11}, we introduced a novel recursive solution called Recursive Projected Compressive Sensing (ReProCS). In recent work \cite{rrpcp_perf,rrpcp_perf2,rrpcp_correct}, we have obtained performance  guarantees for ReProCS. Under  mild assumptions (denseness, slow enough subspace change of $L_t$ and ``some" support change at least every $h$ frames of $S_t$), we showed that, with high probability (w.h.p.), ReProCS can exactly recover the support set of $S_t$ at all times; and the reconstruction errors of both $S_t$ and $L_t$ are upper bounded by a time invariant and small value. The work of \cite{rrpcp_perf,rrpcp_perf2} contains a partial result while \cite{rrpcp_correct} is a complete correctness result.
{\em Contributions. }
The contributions of this work are as follows.
(1) We design a practically usable modification of the ReProCS algorithm studied in \cite{rrpcp_perf,rrpcp_perf2,rrpcp_correct}. By ``practically usable", we mean that (a) it requires fewer parameters and we develop simple heuristics to set these parameters without any model knowledge; (b) it exploits practically motivated assumptions and we demonstrate that these assumptions are valid for real video data. While denseness and gradual support change are also used in earlier works - \cite{rpca,rpca2} and \cite{modcsjournal} respectively - {\em slow subspace change} is the key new (and valid) assumption introduced in ReProCS.
(2) We show via extensive simulation and real video experiments that practical-ReProCS is more robust to correlated support change of $S_t$ than PCP and other existing work. Also, it is also able to recover small magnitude sparse vectors significantly better than other existing recursive as well as batch algorithms.
(3) We also develop a compressive practical-ReProCS algorithm that can recover $S_t$ from $M_t:= A S_t + B L_t$. In this case $A$ and $B$ can be fat, square or tall.





{\em More Related Work. }
{
Other very recent work on recursive / online robust PCA includes \cite{grass_undersampled,xu_nips2013_1,xu_nips2013_2,mateos_anomaly2,mateos_anomaly}.
}

{ 
Some other related work includes work that uses structured sparsity models, e.g. \cite{jia_et_al_face}. For our problem, if it is known that the sparse vector consists of one or a few connected regions, these ideas could be incorporated into our algorithm as well. On the other hand, the advantage of the current approach that only uses sparsity is that it works both for the case of a few connected regions as well as for the case of multiple small sized moving objects, e.g. see the airport video results at \url{http://www.ece.iastate.edu/~chenlu/ReProCS/Video_ReProCS.htm}.
}


{\em Paper Organization. }
We give the precise problem definition and assumptions in Sec \ref{probdef}. The practical ReProCS algorithm is developed in Sec \ref{pracrep}. The algorithm for the compressive measurements' case is developed in Sec \ref{compressive}. In Sec \ref{modelverify}, we demonstrate using real videos that the key assumptions used by our algorithm are true in practice. Experimental comparisons on simulated and real data are shown in Sec \ref{expts}. Conclusions and future work are discussed in Sec \ref{conc}.

\subsection{Notation}

For a set $T \subseteq \{1,2,\cdots n\}$, we use $|T|$ to denote its cardinality; and we use $T^c$ to denote its complement, i.e. $T^c:= \{i \in \{1,2,\dots n\}: i \notin T \}$. The symbols $\cup,\cap, \setminus$ denote set union set intersection and set difference respectively (recall $T_1 \setminus T_2:=T_1 \cap T_2^c$).
%
For a vector $v$, $v_i$ denotes the $i$th entry of $v$ and $v_T$ denotes a vector consisting of the entries of $v$ indexed by $T$. We use $\|v\|_p$ to denote the $\ell_p$ norm of $v$. The support of $v$, $\text{supp}(v)$, is the set of indices at which $v$ is nonzero, $\text{supp}(v) := \{i : v_i\neq 0\}$. We say that $v$ is s-sparse if $|\text{supp}(v)| \leq s$.

%
For a matrix $B$, $B'$ denotes its transpose, and $B^{\dag}$ denotes its pseudo-inverse. For a matrix with linearly independent columns, $B^{\dag} = (B'B)^{-1}B'$. The notation $[.]$ denotes an empty matrix. We use $I$ to denote an identity matrix. For an $m \times n$ matrix $B$ and an index set $T \subseteq \{1,2,\dots n\}$, $B_T$ is the sub-matrix of $B$ containing columns with indices in the set $T$. Notice that $B_T= B I_T$. We use $B \setminus B_{T}$ to denote $B_{T^c}$.
Given another matrix $B_2$ of size $m \times n_2$, $[B \ B_2]$ constructs a new matrix by concatenating matrices $B$ and $B_2$ in horizontal direction. Thus, $[(B\setminus B_{T}) \  B_2] = [B_{T^c} \  B_2]$.
%
%
We use the notation $B \overset{SVD}{=} U \Sigma V'$ to denote the singular value decomposition (SVD) of $B$ with the diagonal entries of $\Sigma$ being arranged in non-decreasing order.

The interval notation $[t_1, t_2]:=\{t_1, t_1+1, \cdots, t_2\}$ and similarly the matrix $[L_{t_1}, \dots L_{t_2}]:=[L_{t_1}, L_{t_1+1}, \cdots, L_{t_2}]$ 

\begin{definition}\label{defn_delta}
The {\em $s$-restricted isometry constant (RIC)} \cite{decodinglp}, $\delta_s$, for an $n \times m$ matrix $\Psi$ is the smallest real number satisfying $(1-\delta_s) \|x\|_2^2 \leq \|\Psi_T x\|_2^2 \leq (1+\delta_s) \|x\|_2^2$
for all sets $T$ with $|T| \leq s$ and all real vectors $x$ of length $|T|$.%
\end{definition} 

\begin{definition} For a matrix $M$,
\bi
\item $\Span(M)$ denotes the subspace spanned by the columns of $M$.
\item  $M$ is a {\em basis matrix} if $M'M = I$.
\item  The {\em notation $Q = \basis(\Span(M))$, or $Q = \basis(M)$ for short,} means that $Q$ is a {\em basis matrix} for $\Span(M)$ i.e. $Q$ satisfies $Q'Q=I$ and $\Span(Q) = \Span(M)$.
\ei
\end{definition}

\begin{definition} \
\bi
\item The {\em $b\%$ left singular values' set} of a matrix $M$ is the smallest set of indices of its singular values that contains at least $b\%$ of the total singular values' energy. In other words,  if $M\overset{SVD}{=} U \Sigma V'$, it is the smallest set $T$ such that $\sum_{i \in T} (\Sigma)_{i,i}^2 \ge \frac{b}{100}\sum_{i=1}^n (\Sigma)_{i,i}^2$.

\item The corresponding matrix of left singular vectors, $U_T$, is referred to as the {\em $b\%$ left singular vectors' matrix.}

\item The notation {\em $[Q,\Sigma]= \abasis(M,b\%)$} means that $Q$ is the $b\%$ left singular vectors' matrix for $M$ and $\Sigma$ is the diagonal matrix with diagonal entries equal to the b\% left singular values' set.

\item The notation {\em $Q= \abasis(M,r)$} means that $Q$ contains the left singular vectors of $M$ corresponding to its $r$ largest singular values. This also sometimes referred to as: {\em $Q$ contains the $r$ top singular vectors of $M$.}

\ei
\end{definition}



%

\section{Problem Definition and Assumptions}\label{probdef}
The measurement vector at time $t$, $M_t$, is an $n$ dimensional vector which can be decomposed as
\beq
M_t := S_t + L_t.
\label{problem_defn}
\eeq
Let $T_t$ denote the support set of $S_t$, i.e., $$T_t :=\text{supp}(S_t)=\{i: (S_t)_i \neq 0\}.$$ 
We assume that $S_t$ and $L_t$ satisfy the assumptions given below in the next three subsections.
Suppose that an initial training sequence which does not contain the sparse components is available, i.e. we are given  $\mathcal{M}_\train = [M_t; 1\leq t \leq t_\train]$ with $M_t = L_t$. This is used to get an initial estimate of the subspace in which the $L_t$'s lie \footnote{If an initial sequence without $S_t$'s is not available, one can use a batch robust PCA algorithm to get the initial subspace estimate as long as the initial sequence satisfies its required assumptions.}.
At each $t >t_\train$, the goal is to recursively estimate $S_t$ and $L_t$ and the subspace in which $L_t$ lies. By ``recursively" we mean: use $\Shat_{t-1}, \Lhat_{t-1}$ and the previous subspace estimate to estimate $S_t$ and $L_t$.

The magnitude of the entries of $L_t$ may be small, of the same order, or large compared to that of the nonzero entries of $S_t$. In applications where $S_t$ is the signal of interest, the case when $\|L_t\|_2$ is of the same order or larger than $\|S_t\|_2$ is the difficult case. 

A key application where the above problem occurs is in separating a video sequence into background and foreground layers. Let $\text{Im}_t$ denote the image at time $t$, $F_t$ denote the foreground image at $t$ and $B_t$ the background image at $t$, all arranged as 1-D vectors. Then, the image sequence satisfies
\bea
(\text{Im}_t)_i = \left\{ \begin{array}{cc}
    (F_t)_i  & \ \text{if} \ i \in \text{supp}(F_t) \\
    (B_t)_i  & \ \text{if} \ i \notin \text{supp}(F_t)
    \end{array}
\right.
\label{overlay}
\eea
In fMRI, $F_t$ is the sparse active region image while $B_t$ is the background brain image.
In both cases, it is fair to assume that an initial background-only training sequence is available. For video this means there are no moving objects/regions in the foreground. For fMRI, this means some frames are captured without providing any stimulus to the subject.

Let $\mu$ denote the empirical mean of the training background images. If we let $L_t := B_t - \mu$, $M_t  := \text{Im}_t - \mu $,  $T_t := \text{supp}(F_t)$, and
$$(S_t)_{T_t} := (F_t - B_t)_{T_t}, \ (S_t)_{T_t^c} := 0,$$
then, clearly, $M_t = S_t + L_t$.
Once we get the estimates $\Lhat_t$, $\Shat_t$, we can also recover the foreground and background as
$$\hat{B}_t = \Lhat_t + \mu, \ \That_t = \text{supp}(\Shat_t), \ (\hat{F}_t)_{\That_t} = (\text{Im}_t)_{\That_t}, \ (\hat{F}_t)_{\That_t^c} = 0.$$


\subsection{Slowly changing low-dimensional subspace change} \label{lowdim}

We assume that for $\tau$ large enough, any $\tau$ length subsequence of the $L_t$'s lies in a subspace of $\mathbf{R}^n$ of dimension less than $\min(\tau,n)$, and usually much less than $\min(\tau,n)$. In other words, for $\tau$ large enough, $\max_t \rank([L_{t-\tau+1}, \dots L_t]) \ll \min(\tau,n)$. Also, this subspace is either fixed or changes slowly over time.

{ One way to model this is as follows \cite{rrpcp_perf}. Let $L_t = P_{t} a_t$ where $P_{t}$ is an $n \times r_t$ {\em basis matrix} with $r_t \ll n$ that is piecewise constant with time, i.e. $P_{t} = P_{(j)}$ for all $t \in [t_j, t_{j+1})$  and $P_{(j)}$ changes as
$$P_{(j)} = [(P_{(j-1)} R_j \setminus P_{(j),\old}), P_{(j),\new}]$$
where $P_{(j),\new}$ and $P_{(j),\old}$ are basis matrices of size $n \times c_{j,\new}$ and $n \times c_{j,\old}$ respectively with $P_{(j),\new}'P_{(j-1)}=0$ and $R_j$ is a rotation matrix. Moreover,
(a)  $0 \le \sum_{i=1}^{j} (c_{i,\new} - c_{i,\old}) \le c_{\text{dif}}$; (b) $0 \le c_{j,\new} \le c_{\max} < r_0$; (c) $(t_{j+1}-t_j) \gg r_0 + c_{\text{dif}}$; and (d) there are a total of $J$ change times with $J \ll (n-r_0 - c_{\text{dif}})/c_{\max}$.

Clearly, (a) implies that $r_t \le r_0 + c_{\text{dif}}:= r_{\max}$ and (d) implies that $r_{\max} + J c_{\max} \ll n$.
This, along with (b) and (c), helps to ensure that for any $\tau > r_{\max}+c_{\max}$, $r^{t,\tau}:=\max_t \rank([L_{t-\tau+1}, \dots L_t]) < \min(\tau,n)$, and for $\tau \gg r_{\max}+c_{\max}$, $r^{t,\tau} \ll \min(\tau,n)$
\footnote{
{ To address a reviewer comment, we explain this in detail here. Notice first that (c) implies that $(t_{j+1}-t_j) \gg r_{\max}$. Also, (b) implies that $\rank([L_{t_j}, \dots L_{t_{j+k}-1}]) \le r_{\max} + (k-1) c_{\max}$. First consider the case when both $t-\tau+1$ and $t$ lie in $[t_j, t_{j+1}-1]$. In this case, $r^{t,\tau} \le r_{\max}$ for any $\tau$. Thus for any $t_{j+1}-t_j > \tau \gg r_{\max}$, $r^{t,\tau} \ll \min(\tau,n)$. Next consider the case when $t-\tau+1 \in [t_j, t_{j+1}-1]$ and $t  \in [t_{j+1}, t_{j+2}-1]$. In this case, $r^{t,\tau} \le r_{\max}+ c_{\max}$.  Thus,  for any $t_{j+2}-t_j > \tau \gg r_{\max}+c_{\max}$, $r^{t,\tau} \ll \min(\tau,n)$. Finally consider the case when $t-\tau+1 \in [t_j, t_{j+1}-1]$ and $t  \in [t_{j+k+1}, t_{j+k+2}-1]$ for a $0 < k < J-1$. In this case, $\tau$ can be rewritten as $\tau = (t_{j+k+1}-t_{j+1}) + \tau_1 + \tau_2$ with $\tau_1:=t_{j+1}-(t-\tau+1)$ and $\tau_2:=t- (t_{j+k+1}-1)$.
Clearly, $r^{t,\tau} \le (r_{\max} + (k-1) c_{\max}) + \min(\tau_1,c_{\max})  + \min(\tau_2,c_{\max}) < k r_{\max} + \min(\tau_1,c_{\max})  + \min(\tau_2,c_{\max}) \ll (t_{j+k+2}-t_{j+1}) + \min(\tau_1,c_{\max})  + \min(\tau_2,c_{\max}) \le  (t_{j+k+2}-t_{j+1}) + \tau_1 + \tau_2 = \tau$. Moreover, $r^{t,\tau} \le  r_{\max} + (k+1) c_{\max} \le r_{\max} + J c_{\max} \ll n$.  Thus, in this case again for any $\tau$, $r^{t,\tau} \ll \min(\tau,n)$.
}
}.
}


By slow subspace change, we mean that: for $t \in [t_j, t_{j+1})$,  $\| (I - P_{(j-1)} P_{(j-1)} ') L_t\|_2$ is initially small and increases gradually. In particular, we assume that, for $t \in [t_j, t_j + \alpha)$, 
$$\| (I - P_{(j-1)} P_{(j-1)} ') L_t\|_2 \le \gamma_\new   \ll \min(\| L_t\|_2, \|S_t\|_2)$$ 
and increases gradually after $t_j + \alpha$. One model for ``increases gradually" is as given in \cite[Sec III-B]{rrpcp_perf}. Nothing in this paper requires the specific model and hence we do not repeat it here.

The above piecewise constant subspace change model is a simplified model for what typically happens in practice. In most cases, $P_{t}$ changes a little at each $t$ in such a way that the low-dimensional assumption approximately holds. If we try to model this, it would result in a nonstationary model that is difficult to precisely define or to verify (it would require multiple video sequences of the same type to verify)
\footnote{With letting $a_t$ be a zero mean random variable with a covariance matrix that is constant for sub-intervals within $[t_j, t_{j+1})$, the above model is a piecewise wide sense stationary approximation to the nonstationary model.}.

Since background images typically change only a little over time (except in case of a camera viewpoint change or a scene change), it is valid to model the mean-subtracted background image sequence as lying in a slowly changing low-dimensional subspace. We verify this assumption in Sec \ref{modelverify}.

\subsection{Denseness} \label{dense_assum}
To state the denseness assumption, we first need to define the denseness coefficient. This is a simplification of the one introduced in our earlier work \cite{rrpcp_perf}.  
\begin{definition}[denseness coefficient]\label{subspace_kappa}
For a matrix or a vector $B$, define
\beq 
\kappa_s(B)=\kappa_s(\Span(B)) : = \max_{|T| \le s} \|{I_T}' \text{basis}(B)\|_2
\eeq
where $\|.\|_2$ is the vector or matrix $2$-norm. Recall that $\text{basis}(B)$ is short for $\text{basis}(\Span(B))$. Similarly $\kappa_s(B)$ is short for $\kappa_s(\Span(B))$. Notice that $\kappa_s(B)$ is a property of the subspace $\Span(B)$. Note also that $\kappa_s(B)$ is a non-decreasing function of $s$ and of $\rank(B)$.
\end{definition}

{ We assume that the subspace spanned by the $L_t$'s is dense, i.e.
$$\kappa_{2s}(P_{(j)}) = \kappa_{2s}([L_{t_j}, \dots L_{t_{j+1}-1}]) \le \kappa_*$$
for a $\kappa_*$ significantly smaller than one. Moreover, a similar assumption holds for $P_{(j),\new}$ with a tighter bound: $\kappa_{2s}(P_{(j),\new}) \le \kappa_\new < \kappa_*$.
This assumption is similar to one of the denseness assumptions used in \cite{Candes_Recht,rpca}.
In \cite{rpca}, a bound is assumed on $\kappa_1(U)$ and $\kappa_1(V)$ where $U$ and $V$ are the matrices containing the left and right singular vectors of the entire matrix, $[L_1, L_2 \dots L_t]$; and a tighter bound is assumed on $\max_{i,j} |(UV')_{i,j}|$. In our notation, $U = [P_{(0)}, P_{(1),\new}, \dots P_{(J),\new}]$.
}

The following lemma, proved in \cite{rrpcp_perf}, relates the RIC of $I-PP'$, when $P$ is a {\em basis matrix}, to the denseness coefficient for $\Span(P)$. Notice that $I-PP'$ is an $n\times n$ matrix that has rank $(n-\rank(P))$ and so it cannot be inverted.
\begin{lem}\label{delta_kappa}
For a {\em basis matrix}, $P$,
$$\delta_s(I-PP') = \kappa_s(P)^2$$
\end{lem}
Thus, the denseness assumption implies that the RIC of the matrix $(I - P_{(j)} P_{(j)}')$ is small. Using any of the RIC based sparse recovery results, e.g. \cite{candes_rip}, this ensures that for $t \in [t_j, t_{j+1})$, $s$-sparse vectors $S_t$ are recoverable from $(I - P_{(j)} P_{(j)}')M_t = (I - P_{(j)} P_{(j)}') S_t$ by $\ell_1$ minimization.

Very often, the background images primarily change due to lighting changes (in case of indoor sequences) or due to moving waters or moving leaves (in case of many outdoor sequences) \cite{rpca, rrpcp_perf}. All of these result in global changes and hence it is valid to assume that the subspace spanned by the background image sequences is dense.


\subsection{Small support size, some support change, small support change assumption on $S_t$} \label{suppchange}
Let the sets of support additions and removals be
$$\Delta_t := T_t \setminus T_{t-1}, \ \Delta_{e,t} := T_{t-1} \setminus T_t.$$

(1) We assume that
$$|T_t| + \min(|T_t|, |\Delta_t|+|\Delta_{e,t}|) \le s + s_\Delta \ \text{where} \ s_\Delta \ll s$$
In particular, this implies that we either need $|T_t| \le s$ and $|\Delta_t|+|\Delta_{e,t}| \le s_\Delta$ ($S_t$ is sparse with support size at most $s$, and its support changes slowly) or, in cases when the change $|\Delta_t|+|\Delta_{e,t}|$ is large, we need $|T_t| \le 0.5(s + s_\Delta)$ (need a tighter bound on the support size).

(2) We also assume that there is {\em some} support change every few frames, i.e. at least once every $h$ frames, $|\Delta_t| > s_{\Delta,\min}$. Practically, this is needed to ensure that at least some of the background behind the foreground is visible so that the changes to the background subspace can be estimated. 

In the video application, foreground images typically consist of one or more moving objects/people/regions and hence are sparse. Also, typically the objects are not static, i.e. there is some support change at least every few frames. 
On the other hand, since the objects usually do not move very fast, slow support change is also valid most of the time. The time when the support change is almost comparable to the support size is usually when the object is entering or leaving the image, but these are the exactly the times when the object's support size is itself small (being smaller than $0.5(s+s_\Delta)$ is a valid).
We show some verification of these assumptions in Sec \ref{modelverify}.

\section{Prac-ReProCS: Practical ReProCS}\label{pracrep}
We first develop a practical algorithm based on the basic ReProCS idea from our earlier work \cite{rrpcp_perf}. Then we discuss how the sparse recovery and support estimation steps can be improved. The complete algorithm is summarized  in Algorithm \ref{PracReProCS}.  Finally we discuss an alternate subspace update procedure in Sec \ref{recpca}. 

\subsection{Basic algorithm} \label{basicalgo}

We use $\Shat_t, \That_t, \Lhat_t$ to denote estimates of $S_t$, its support, $T_t$, and $L_t$ respectively; and we use $\Phat_{t}$ to denote the {\em basis matrix} for the estimated subspace of $L_t$ at time $t$.
Also, let
\bea \label{defphi}
\Phi_{t}:=(I -  \Phat_{t-1} \Phat_{t-1}')
\eea

Given the initial training sequence which does not contain the sparse components,  $\mathcal{M}_\train = [L_1, L_2, \dots L_{t_\train}]$ we compute $\Phat_0$ as an approximate basis for $\mathcal{M}_\train$, i.e.  $\Phat_0 = \abasis(\mathcal{M}_\train, b\%)$. Let $\hat{r}=\rank(\Phat_0)$.
We need to compute an approximate basis because for real data, the $L_t$'s are only approximately low-dimensional. We use $b\%=95\%$ or $b\%=99.99\%$ depending on whether the low-rank part is approximately low-rank or almost exactly low-rank. After this, at each time $t$, ReProCS involves 4 steps: (a) Perpendicular Projection; (b) Sparse Recovery (recover $T_t$ and $S_t$); (c) Recover $L_t$; (d) Subspace Update (update $\Phat_{t}$).

{\bf Perpendicular Projection. } In the first step, at time $t$, we project the measurement vector, $M_t$, into the space orthogonal to $\Span(\Phat_{t-1})$ to get the projected measurement vector,
\bea
y_t:= \Phi_{t} M_t. 
\label{defyt}
\eea

{\bf Sparse Recovery (Recover $T_t$ and $S_t$). }
With the above projection, $y_t$ can be rewritten as
\bea \label{defbeta}
y_t = \Phi_{t} S_t + \beta_t \ \text{where} \ \beta_t:= \Phi_{t} L_t
\eea
{ Because of the slow subspace change assumption, projecting orthogonal to $\Span(\Phat_{t-1})$ nullifies most of the contribution of $L_t$ and hence $\beta_t$ can be interpreted as small ``noise". We explain this in detail in Appendix \ref{betasmall}.
}

Thus,  the problem of recovering $S_t$ from $y_t$ becomes a traditional noisy sparse recovery / CS problem. Notice that, since the $n \times n$ projection matrix, $\Phi_{t}$, has rank $n- \rank(\Phat_{t-1})$, therefore $y_t$ has only this many ``effective" measurements, even though its length is $n$.
To recover $S_t$ from $y_t$, one can use $\ell_1$ minimization \cite{bpdn,candes_rip}, or any of the greedy or iterative thresholding algorithms from literature. In this work we use $\ell_1$ minimization: we solve
\bea
{\min}_x  \|x\|_1 \   \text{s.t.}  \ \|y_t - \Phi_{t} x\|_2 \le \xi
\label{CS}
\eea
and denote its solution by $\Shat_{t,cs}$.
By the denseness assumption, $P_{t-1}$ is dense. Since $\Phat_{t-1}$ approximates it, this is true for $\Phat_{t-1}$ as well \cite[Lemma 6.6]{rrpcp_perf}. Thus, by Lemma \ref{delta_kappa}, the RIC of $\Phi_{t}$ is small enough. Using \cite[Theorem 1]{candes_rip}, this and the fact that $\beta_t$ is small ensures that $S_t$ can be accurately recovered from $y_t$.
The constraint $\xi$ used in the minimization should equal $\|\beta_t\|_2$ or its upper bound. Since $\beta_t$ is unknown we set $\xi = \|\betahat_t\|_2$ where $\betahat_t:= \Phi_{t} \hat{L}_{t-1}$.

By thresholding on $\Shat_{t,cs}$ to get an estimate of its support followed by computing a least squares (LS) estimate of $S_t$ on the estimated support and setting it to zero everywhere else, we can get a more accurate estimate, $\Shat_t$, as suggested in \cite{dantzig}. We discuss better support estimation and its parameter setting in Sec \ref{supp_improve}.

{\bf Recover $L_t$. }  The estimate $\Shat_t$ is used to estimate $L_t$ as $\Lhat_t = M_t-\Shat_t$. Thus, if $S_t$ is recovered accurately, so will $L_t$. 

{\bf Subspace Update (Update $\Phat_{t}$). } Within a short delay after every subspace change time, one needs to update the subspace estimate, $\Phat_{t}$.
To do this in a provably reliable fashion, we introduced the projection PCA (p-PCA) algorithm in \cite{rrpcp_perf}. The algorithm studied there used knowledge of the subspace change times $t_j$ and of the number of new directions $c_{j,\new}$. Let $\Phat_{(j-1)}$ denote the final estimate of a basis for the span of $P_{(j-1)}$. It is assumed that the delay between change times is large enough so that $\Phat_{(j-1)}$ is an accurate estimate. At $t=t_j+\alpha-1$, p-PCA gets the first estimate of the new directions, $\Phat_{(j),\new,1}$, by projecting the last $\alpha$ $\Lhat_t$'s perpendicular to $\Phat_{(j-1)}$ followed by computing the $c_{j,\new}$ top left singular vectors of the projected data matrix. It then updates the subspace estimate as $\Phat_{t} = [\Phat_{(j-1)}, \Phat_{(j),\new,1}]$. The same procedure is repeated at every $t=t_j+ k\alpha-1$ for $k=2,3, \dots K$ and each time we update the subspace as $\Phat_{t} = [\Phat_{(j-1)}, \Phat_{(j),\new,k}]$. Here $K$ is chosen so that the subspace estimation error decays down to a small enough value within $K$ p-PCA steps.

{ In this paper, we design a practical version of p-PCA which does not need knowledge of $t_j$ or $c_{j,\new}$. This is summarized in Algorithm \ref{PracReProCS}. The key idea is as follows. We let $\hat\sigma_{\min}$  be the $\hat{r}^{th}$ largest singular value of the training dataset. This serves as the noise threshold for approximately low rank data. We split projection PCA into two phases: ``detect subspace change" and ``p-PCA". We are in the detect phase when the previous subspace has been accurately estimated. Denote the basis matrix for this subspace by $\Phat_{(j-1)}$. We detect the subspace change as follows. Every $\alpha$ frames, we project the last $\alpha$ $\Lhat_t$'s perpendicular to $\Phat_{(j-1)}$ and compute the SVD of the resulting matrix. If there are any singular values above $\hat\sigma_{\min}$, this means that the subspace has changed. At this point, we enter the ``p-PCA" phase. In this phase, we repeat the $K$ p-PCA steps described above with the following change: we estimate $c_{j,\new}$ as the number of singular values above $\hat\sigma_{\min}$, but clipped at $\lceil \alpha/3 \rceil$ (i.e. if the number is more than $\lceil \alpha/3 \rceil$ then we clip it to $\lceil \alpha/3 \rceil$). We stop either when the stopping criterion given in step \ref{stop_crit} is achieved ($k \ge K_{\min}$ and the projection of $\Lhat_t$ along $\Phat_{\new,k}$ is not too different from that along $\Phat_{\new,k}$) or when $k \ge K_{\max}$.

For the above algorithm, with theoretically motivated choices of algorithm parameters, under the assumptions from Sec \ref{probdef}, it is possible to show that, w.h.p., the support of $S_t$ is exactly recovered, the subspace of $L_t$'s is accurately recovered within a finite delay of the change time. We provide a brief overview of the proof from \cite{rrpcp_perf,rrpcp_correct} in Appendix \ref{betasmall} that helps explain why the above approach works.

\begin{remark}
The p-PCA algorithm only allows addition of new directions. If the goal is to estimate the span of $[L_1, \dots L_t]$, then this is what is needed. If the goal is sparse recovery, then one can get a smaller rank estimate of $\Phat_{t}$ by also including a step to delete the span of the removed directions, $P_{(j),\old}$. This will result in more ``effective" measurements available for the sparse recovery step and hence possibly in improved performance. The simplest way to do this is to do one simple PCA step every some frames. In our experiments, this did not help much though.
A provably accurate solution is described in \cite[Sec VII]{rrpcp_perf}.
\end{remark} 

\begin{remark}
The p-PCA algorithm works on small batches of $\alpha$ frames. This can be made fully recursive if we compute the SVD of $(I - \Phat_{(j-1)} {\Phat_{(j-1)}}')[\Lhat_{t-\alpha+1}, \dots \Lhat_t]$ using the incremental SVD (inc-SVD) procedure summarized in Algorithm \ref{incsvd} \cite{sequentialSVD} for one frame at a time. As explained in  \cite{sequentialSVD} and references therein, we can get the left singular vectors and singular values of any matrix $M=[M_1,M_2,\dots M_\alpha]$ recursively by starting with $\Phat=[.],\hat\Sigma=[.]$ and calling [$\Phat$, $\hat\Sigma$] = inc-SVD($\Phat$, $\hat\Sigma$, $M_i$) for every column $i$ or for short batches of columns of size of $\alpha/k$. Since we use $\alpha=20$ which is a small value, the use of incremental SVD does not speed up the algorithm in practice and hence we do not report results using it.
\end{remark}
}

\begin{algorithm*}
\caption{Practical ReProCS-pPCA                                                                                                                                                                                         }\label{PracReProCS}
\textbf{Input:} $M_t$;  \textbf{Output:} $\That_t$, $\hat{S}_t$, $\hat{L}_t$; \textbf{Parameters:} $q, b, \alpha, K_{\min}, K_{\max}$.
{ We used $\alpha=20, K_{\min}=3,K_{\max}=10$ in all experiments ($\alpha$ needs to only be large compared to $c_{\max}$); we used $b=95$ for approximately low-rank data (all real videos and the lake video with simulated foreground) and used  $b=99.99$ for almost exactly low rank data (simulated data); we used $q=1$ whenever $\|S_t\|_2$ was of the same order or larger than $\|L_t\|_2$ (all real videos and the lake video) and used $q=0.25$ when it was much smaller (simulated data with small magnitude $S_t$).
}

\textbf{Initialization}
\bi
\item $[\Phat_{0},\hat\Sigma_0] \leftarrow \abasis(\frac{1}{\sqrt{t_{\train}}}[M_1, \dots M_{t_\train}],b\%)$.
\item Set $\hat{r} \leftarrow \rank(\Phat_0)$, $\hat\sigma_{\min} \leftarrow ((\hat\Sigma_0)_{\rhat,\rhat})$, $\hat{t}_0 = t_{\train}$, $\text{flag}=\text{detect}$
\item Initialize $\hat{P}_{(t_\train)} \leftarrow \hat{P}_0$ and  $\hat{T}_t \leftarrow [.]$.
\ei

\textbf{For $t > t_{\train}$ do}
\begin{enumerate}
\item Perpendicular Projection: compute $y_t \leftarrow \Phi_{t} M_t $ with $\Phi_{t} \leftarrow I - \hat{P}_{t-1}{\hat{P}_{t-1}}'$

\item Sparse Recovery (Recover $S_t$ and $T_t$)
\begin{enumerate}
	\item If $\frac{|\hat{T}_{t-2}\cap \hat{T}_{t-1}|}{|\hat{T}_{t-2}|}<0.5$
		\begin{enumerate}
			\item Compute $\hat{S}_{t,\cs}$ as the solution of (\ref{CS}) with $\xi= \|\Phi_{t} \hat{L}_{t-1}\|_2$. 
			\item  $\hat{T}_t \leftarrow \Thresh (\hat{S}_{t,\cs},\omega)$ with $\omega= q \sqrt{\|M_t \|^2/n}$. Here $T \leftarrow \Thresh (x,\omega)$ means that $T = \{i : \ |(x)_i| \geq \omega\}$.
		\end{enumerate}
       Else
	\begin{enumerate}

    \item Compute $\hat{S}_{t,\cs}$ as the solution of (\ref{wCS}) with $T = \hat{T}_{t-1}$, $\lambda =  \frac{|\hat{T}_{t-2} \setminus \hat{T}_{t-1}|}{|\hat{T}_{t-1}|}$,  $\xi= \|\Phi_{t} \hat{L}_{t-1}\|_2$.

    \item $\hat{T}_{\text{add}} \leftarrow  \Prune (\hat{S}_{t,\cs}, 1.4| \hat{T}_{t-1} |)$. Here $T \leftarrow \Prune (x,k)$ returns indices of the $k$ largest magnitude elements of $x$.

  \item  $\hat{S}_{t,\text{add}} \leftarrow \LS(y_t,\Phi_{t}, \hat{T}_{\text{add}})$. Here $\hat{x} \leftarrow \text{LS} (y, A, T)$ means that $\hat{x}_T = ({A_T}'A_T)^{-1}{A_T}'y$ and $\hat{x}_{T^c} = 0$.

 \item  $\hat{T}_t \leftarrow \Thresh (\hat{S}_{t,\text{add}},\omega)$ with $\omega=q \sqrt{\|M_t \|^2/n}$.

\end{enumerate}

\item $\hat{S}_t \leftarrow \LS (y_t,\Phi_{t}, \hat{T}_t)$
\end{enumerate}


\item Estimate $L_t$: $\hat{L}_t \leftarrow M_t - \hat{S}_t $

{
\item Update $\hat{P}_{t}$: projection PCA
\ben
\item If $\text{flag} = \text{detect}$ and  $\text{mod}(t - \hat{t}_j +1, \alpha)=0$, (here $\text{mod}(t,\alpha)$ is the remainder when $t$ is divided by $\alpha$)

\ben
\item  compute the SVD of $ \frac{1}{\sqrt{\alpha}}(I - \Phat_{(j-1)} {\Phat_{(j-1)}}')[\Lhat_{t-\alpha+1}, \dots \Lhat_t]$ and check if any singular values are above $\hat\sigma_{\min}$
\item if the above number is more than zero then set $\text{flag} \leftarrow \text{pPCA}$, increment $j \leftarrow j+1$, set $\hat{t}_j \leftarrow t-\alpha+1$, reset $k \leftarrow 1$
\een

Else $\Phat_{t}  \leftarrow \hat{P}_{t-1}$.

\item If $\text{flag} = \text{pPCA}$ and  $\text{mod}(t - \hat{t}_j +1, \alpha)=0$,

\ben
\item compute the SVD of $ \frac{1}{\sqrt{\alpha}}(I - \Phat_{(j-1)} {\Phat_{(j-1)}}')[\Lhat_{t-\alpha+1}, \dots \Lhat_t]$,
\item  let $\hat{P}_{j,\new,k}$ retain all its left singular vectors with singular values above $\hat\sigma_{\min}$ or all $\alpha/3$ top left singular vectors whichever is smaller,
\item update $\hat{P}_{t} \leftarrow [\Phat_{(j-1)} \ \hat{P}_{j,\new,k}]$, increment $k \leftarrow k+1$

\item If $k \geq  K_{\min}$ and $\frac{\|\sum_{t-\alpha+1}^t (\hat{P}_{j,\new,i-1} {\hat{P}_{j,\new,i-1}}' - \hat{P}_{j,\new,i} {\hat{P}_{j,\new,i}}')L_t\|_2}{\|\sum_{t-\alpha+1}^t \hat{P}_{j,\new,i-1} {\hat{P}_{j,\new,i-1}}' L_t\|_2 }< 0.01$ for $i=k-2,k-1,k$; or $k = K_{\max}$,
\\    then $K \leftarrow  k$, $\Phat_{(j)} \leftarrow [\Phat_{(j-1)} \ \hat{P}_{j,\new,K}]$ and reset $\text{flag} \leftarrow \text{detect}$. 
\label{stop_crit}
\een

Else $\Phat_{t}  \leftarrow \hat{P}_{t-1}$.

\een
}

\een
\end{algorithm*}

\subsection{Exploiting slow support change when valid}
In \cite{rrpcp_perf,rrpcp_perf2}, we always used $\ell_1$ minimization followed by thresholding and LS for sparse recovery. However
if slow support change holds, one can replace simple $\ell_1$ minimization by modified-CS \cite{modcsjournal} which requires fewer measurements for exact/accurate recovery as long as the previous support estimate, $\That_{t-1}$, is an accurate enough predictor of the current support, $T_t$. In our application, $\That_{t-1}$ is likely to contain a significant number of extras and in this case, a better idea is to solve the following weighted $\ell_1$ problem \cite{hassibi,friedlander}
\beq
{\min}_x \lambda \|x_T\|_1 +  \|x_{T^c}\|_1 \   \text{s.t.}   \ \|y_t - \Phi_{t} x\|_2 \le \xi, \ \ T:=  \That_{t-1}
\label{wCS}
\eeq 
with $\lambda < 1$ (modified-CS solves the above with $\lambda=0$). Denote its solution by $\Shat_{t,cs}$. One way to pick $\lambda$ is to let it be proportional to the estimate of the percentage of extras in $\That_{t-1}$. 
%
If slow support change does not hold, the previous support estimate is not a good predictor of the current support. In this case, doing the above is a bad idea and one should instead solve simple $\ell_1$, i.e. solve (\ref{wCS}) with $\lambda =1$. As explained in \cite{friedlander}, if the support estimate contains at least $50\%$ correct entries, then weighted $\ell_1$ is better than simple $\ell_1$.
We use the above criteria with true values replaced by estimates. Thus, if $\frac{|\hat{T}_{t-2}\cap \hat{T}_{t-1}|}{|\hat{T}_{t-2}|} > 0.5$, then we solve  (\ref{wCS}) with $\lambda = \frac{|\hat{T}_{t-2} \setminus \hat{T}_{t-1}|}{|\hat{T}_{t-1}|}$, else we solve it with $\lambda=1$.


\subsection{Improved support estimation} \label{supp_improve}
A simple way to estimate the support is by thresholding the solution of (\ref{wCS}). This can be improved by using the Add-LS-Del procedure for support and signal value estimation \cite{modcsjournal}. We proceed as follows. First we compute the set $\hat{T}_{t,\text{add}}$ by thresholding on $\Shat_{t,cs}$ in order to retain its $k$ largest magnitude entries.
We then compute a LS estimate of $S_t$ on $\hat{T}_{t,\text{add}}$ while setting it to zero everywhere else. As explained earlier, because of the LS step, $\hat{S}_{\text{t,add}}$ is a less biased estimate of $S_t$ than $\Shat_{t,cs}$.
We let $k = 1.4| \hat{T}_{t-1} |$ to allow for a small increase in the support size from $t-1$ to $t$. 
A larger value of $k$ also makes it more likely that elements of the set $(T_t \setminus \That_{t-1})$ are detected into the support estimate\footnote{Due to the larger weight on the $\|x_{(\That_{t-1}^c)}\|_1$ term as compared to that on the $\|x_{(\That_{t-1})}\|_1$ term, the solution of (\ref{wCS}) is biased towards zero on $\That_{t-1}^c$ and thus the solution values along $(T_t \setminus \That_{t-1})$ are smaller than the true ones.}.

The final estimate of the support, $\That_t$, is obtained by thresholding on $\hat{S}_{\text{t,add}}$ using a threshold $\omega$. If $\omega$ is appropriately chosen, this step helps to delete some of the extra elements from $\hat{T}_{\text{add}}$ and this ensures that the size of $\That_t$ does not keep increasing (unless the object's size is increasing).
An LS estimate computed on $\That_t$ gives us the final estimate of $S_t$, i.e. $\Shat_{t} = \text{LS}(y_t, A, \That_t)$.
We use $\omega = \sqrt{\| M_t \|^2/n}$ except in situations where $\|S_t\| \ll \|L_t\|$ - in this case we use $\omega = 0.25\sqrt{\| M_t\|^2/n}$.
An alternate approach is to let $\omega$ be proportional to the noise magnitude seen by the $\ell_1$ step, i.e. to let $\omega = q \|\betahat_t\|_\infty$, however this approach required different values of $q$ for different experiments (it is not possible to specify one $q$ that works for all experiments).

The complete algorithm with all the above steps is summarized in Algorithm \ref{PracReProCS}.

{ \begin{algorithm}
\caption{[$\Phat$, $\hat\Sigma$] = inc-SVD($\Phat$, $\hat\Sigma$, $D$)}  \label{incsvd}
\ben
\item set $D_{\parallel,proj} \leftarrow \Phat'D$ and $D_{\perp} \leftarrow (I - \Phat \Phat')D$
\item compute QR decomposition of $D_{\perp}$, i.e. $D_{\perp} \stackrel{QR}{=} J K$ (here $J$ is a {\em basis matrix} and $K$ is an upper triangular matrix)
\item compute the SVD: $\matr{\hat\Sigma \ & \ D_{\parallel,proj}}{0 \ & \ K} \stackrel{SVD}{=} \tilde{P} \tilde{\Sigma} \tilde{V}'$
\item update $\Phat \leftarrow [\Phat \ J] \tilde{P}$ and $\hat\Sigma \leftarrow \tilde{\Sigma}$
\een
Note: As explained in \cite{sequentialSVD}, due to numerical errors, step 4 done too often can eventually result in $\Phat$ no longer being a basis matrix. This typically occurs when one tries to use inc-SVD at every time $t$, i.e. when $D$ is a column vector. This can be addressed using the modified Gram Schmidt re-orthonormalization procedure whenever loss of orthogonality is detected \cite{sequentialSVD}.
\end{algorithm}
}

{ \subsection{Simplifying subspace update: simple recursive PCA} \label{recpca}
Even the practical version of p-PCA needs to set $K_{\min}$ and $K_{\max}$ besides also setting $b$ and $\alpha$. Thus, we also experiment with using PCA to replace p-PCA (it is difficult to prove a performance guarantee with PCA but that does not necessarily mean that the algorithm itself will not work). The simplest way to do this is to compute the top $\rhat$ left singular vectors of $[\Lhat_1, \Lhat_2, \dots \Lhat_t]$ either at each time $t$ or every $\alpha$ frames. While this is simple, its complexity will keep increasing with time $t$ which is not desirable. Even if we use the last $d$ frames instead of all past frames, $d$ will still need to be large compared to $\rhat$ to get an accurate estimate. To address this issue, we can use the recursive PCA (incremental SVD) algorithm given in Algorithm \ref{incsvd}.  We give the complete algorithm that uses this and a rank $\rhat$ truncation step every $d$ frames (motivated by \cite{sequentialSVD}) in Algorithm \ref{pracrep_rpca}.
}
%
%

\begin{algorithm}[h]
\caption{Practical ReProCS-Recursive-PCA                                                                                                                                                                                         }\label{pracrep_rpca}
\textbf{Input:} $M_t$;  \textbf{Output:} $\That_t$, $\hat{S}_t$, $\hat{L}_t$; \textbf{Parameters:} $q, b, \alpha$, set $\alpha=20$ in all experiments, set $q,b$ as explained in Algorithm \ref{PracReProCS}.

\textbf{Initialization: } $[\Phat_{0}, \Sigma_0] \leftarrow \abasis([M_1, \dots M_{t_\train}],b\%)$,  $\hat{r} \leftarrow \rank(\Phat_0)$, $d \leftarrow 3 \hat{r}$,
initialize  $\Phat_{tmp} \leftarrow \Phat_0$, $\hat\Sigma_{tmp} \leftarrow \Sigma_0$, $\hat{P}_{(t_\train)} \leftarrow \hat{P}_0$ and  $\hat{T}_t \leftarrow [.]$.
\textbf{For $t > t_{\train}$ do}
\begin{enumerate}
\item Perpendicular Projection: do as in Algorithm \ref{PracReProCS}.
\item Sparse Recovery: do as in Algorithm \ref{PracReProCS}.
\item Estimate $L_t$: do as in Algorithm \ref{PracReProCS}.
\item Update $\hat{P}_{t}$: recursive PCA
\ben
\item If $\text{mod}(t-t_\train, \alpha)=0$,
\ben
\item $[\Phat_{tmp}$, $\hat\Sigma_{tmp}]$ $\leftarrow$ inc-SVD($\Phat_{tmp}$, $\hat\Sigma_{tmp}$, $[\Lhat_{t-\alpha+1}, \dots \Lhat_t]$) where inc-SVD is given in Algorithm \ref{incsvd}.
\item $\Phat_{t} \leftarrow (\Phat_{tmp})_{1:\rhat}$
\een
Else $\Phat_{t} \leftarrow \Phat_{t-1}$.

\item If $\text{mod}(t-t_\train, d)=0$,
\ben
\item $\Phat_{tmp} \leftarrow (\Phat_{tmp})_{1:\rhat}$ and  $\hat\Sigma_{tmp} \leftarrow (\hat\Sigma_{tmp})_{1:\rhat,1:\rhat}$
\een

\een
\end{enumerate}
\end{algorithm}

\section{Compressive Measurements: Recovering $S_t$} \label{compressive}
Consider the problem of recovering $S_t$ from $$M_t := A S_t + B L_t$$ when $A$ and $B$ are $m \times n$ and $m \times n_2$ matrices, $S_t$ is an $n$ length vector and $L_t$ is an $n_2$ length vector. In general $m$ can be larger, equal or smaller than $n$ or $n_2$. In the compressive measurements' case, $m<n$.
{ To specify the assumptions needed in this case, we need to define
the basis matrix for $\Span(B P_{(j)})$ and we need to define a generalization of the denseness coefficient.
\begin{definition}
Let $Q_j:=\text{basis}(B P_{(j)})$ and let $Q_{j,\new}:=\text{basis}((I - Q_{j-1} Q_{j-1}') B P_{(j)})$.
\end{definition}
\begin{definition}
For a matrix or a vector $M$, define
\beq
\kappa_{s,A}(M)=\kappa_{s,A}(\Span(M)) : = \max_{|T| \le s} \|{A_T}' \text{basis}(M)\|_2
\eeq
where $\|.\|_2$ is the vector or matrix $2$-norm. This quantifies the incoherence between the subspace spanned by any set of $s$ columns of $A$ and the range of $M$.
\end{definition}

We assume the following. 
\ben
\item $L_t$ and $S_t$ satisfy the assumptions of Sec \ref{lowdim}, \ref{suppchange}.
\item The matrix $A$ satisfies the restricted isometry property \cite{candes_rip}, i.e. $\delta_s(A) \le \delta_* \ll 1$.
\item The denseness assumption is replaced by:
$\kappa_{2s,A}(Q_j) \le \kappa_* $, $\kappa_{2s,A}(Q_{j,\new}) \le \kappa_{\new} < \kappa_*$ for a $\kappa_*$ that is small compared to one. Notice that this depends on  the $L_t$'s and on the matrices $A$ and $B$.
\een
}

Assume that we are given an initial training sequence that satisfies $M_t = B L_t$ for $t=1,2, \dots t_\train$. The goal is to recover $S_t$ at each time $t$. It is not possible to recover $L_t$ unless $B$ is time-varying (this case is studied in \cite{zhan_icassp}). In many imaging applications, e.g. undersampled fMRI or single-pixel video imaging, $B=A$ ($B=A$ is a partial Fourier matrix for MRI and is a random Gaussian or Rademacher matrix for single-pixel imaging). On the other hand, if $L_t$ is large but low-dimensional sensor noise, then $B=I$ (identity matrix), while $A$ is the measurement matrix.

Let $\tilde{L}_t:= B L_t$. It is easy to see that if $L_t$ lies in a slowly changing low-dimensional subspace, the same is also true for the sequence $\tilde{L}_t$. Consider the problem of recovering $S_t$ from $M_t := A S_t +\tilde{L}_t$ when an initial training sequence $M_t := \tilde{L}_t$ for $t=1,2, \dots t_\train$ is available. Using this sequence, it is possible to estimate its approximate basis $\Qhat_0$ as explained earlier. If we then project $M_t$ into the subspace orthogonal to $\Span(\Qhat_0)$, the resulting vector $y_t:=\Phi_{t} M_t$ satisfies $$ y_t = (\Phi_{t} A) S_t + \beta_t$$ where $\beta_t = \Phi_{t} B L_t$ is small noise for the same reasons explained earlier. Thus, one can still recover $S_t$ from $y_t$ by $\ell_1$ or weighted $\ell_1$ minimization followed by support recovery and LS. Then, $\tilde{L}_t$ gets recovered as $\hat{\tilde{L}}_t \leftarrow M_t - A \hat{S}_t$ and this is used for updating its subspace estimate. We summarize the resulting algorithm in Algorithm \ref{uPracReProCS}. This is being analyzed in ongoing work \cite{rrpcp_globalsip}.

{ The following lemma explains why some of the extra assumptions are needed for this case.
\begin{lem}  \cite{rrpcp_globalsip} \label{delta_kappa_comp}
For a {\em basis matrix}, $Q$,
$$\delta_s( (I-QQ')A) \le \kappa_{s,A}(Q)^2 + \delta_s(A)$$
\end{lem}
Using the above lemma with $Q \equiv Q_j$, it is clear that incoherence of $Q_j$ w.r.t. any set of $2s$ columns of $A$ along with RIP of $A$ ensures that any $s$ sparse vector $x$ can be recovered from $y := (I-Q_j Q_j')A x$ by $\ell_1$ minimization. In compressive ReProCS, the measurement matrix uses $\hat{Q}_j$ instead of $Q_j$ and also involves small noise. With more work, these arguments can be extended to this case as well [see \cite{rrpcp_globalsip}].
}

\begin{algorithm}[t]
\caption{Compressive ReProCS                                                                                                                                                                                         }\label{uPracReProCS}
Use Algorithm \ref{PracReProCS} or \ref{pracrep_rpca} with the following changes.
\bi
\item Replace $\Phi_{t}$ in step 2  by $\Phi_{t} A$.
\item Replace step 3 by $\hat{\tilde{L}}_t \leftarrow M_t - A \hat{S}_t$.
\item Use $\hat{\tilde{L}}_t$ in place of $\Lhat_t$ and $\Qhat$ in place of $\Phat$ everywhere.
\ei
\end{algorithm}

\section{Model Verification} \label{modelverify}

\subsection{Low-dimensional and slow subspace change assumption}
We used two background image sequence datasets. The first was a video of lake water motion. 
For this sequence, $n = 6480$ and the number of images were 1500. 
The second was an indoor video of window curtains moving due to the wind. There was also some lighting variation. 
The latter part of this sequence also contains a foreground (various persons coming in, writing on the board and leaving).
For this sequence, the image size was $n = 5120$ and the number of background-only images were 1755. 
Both sequences are posted at \url{http://www.ece.iastate.edu/~hanguo/PracReProCS.html}. 

First note that any given background image sequence will never be exactly low-dimensional, but only be approximately so. Secondly, in practical data, the subspace does not just change as simply as in the model of Sec \ref{lowdim}. Typically there are some changes to the subspace at every time $t$. Moreover, with just one training sequence of a given type, it is not possible to estimate the covariance matrix of $L_t$ at each $t$ and thus one cannot detect the subspace change times. The only thing one can do is to assume that there may be a change every $\tau$ frames, and that during these $\tau$ frames the $L_t$'s are stationary and ergodic; estimate the covariance matrix of $L_t$ for this period using a time average; compute its eigenvectors corresponding to $b\%$ energy (or equivalently compute the $b\%$ approximate basis of $[L_{t-\tau+1}, \dots L_t]$) and use these as $P_{(j)}$. These can be used to test our assumptions.  

\begin{figure*}
				\centering				
				\begin{subfigure}{0.33\textwidth}
              \includegraphics[scale=0.44]{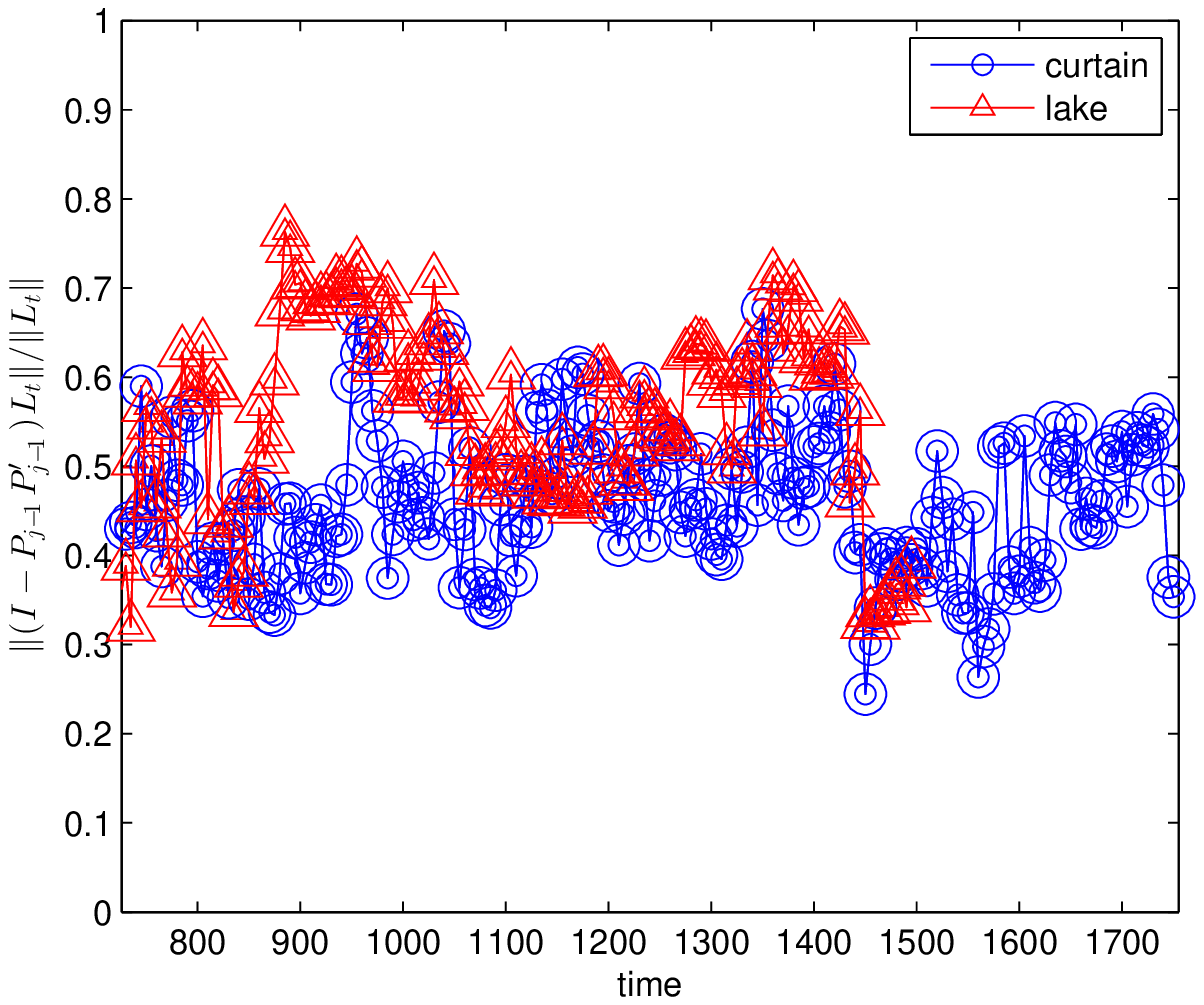}
					\caption{}
				\label{SubspaceVeri}
               \end{subfigure}%
				\begin{subfigure}{0.33\textwidth}
					\includegraphics[scale=0.44]{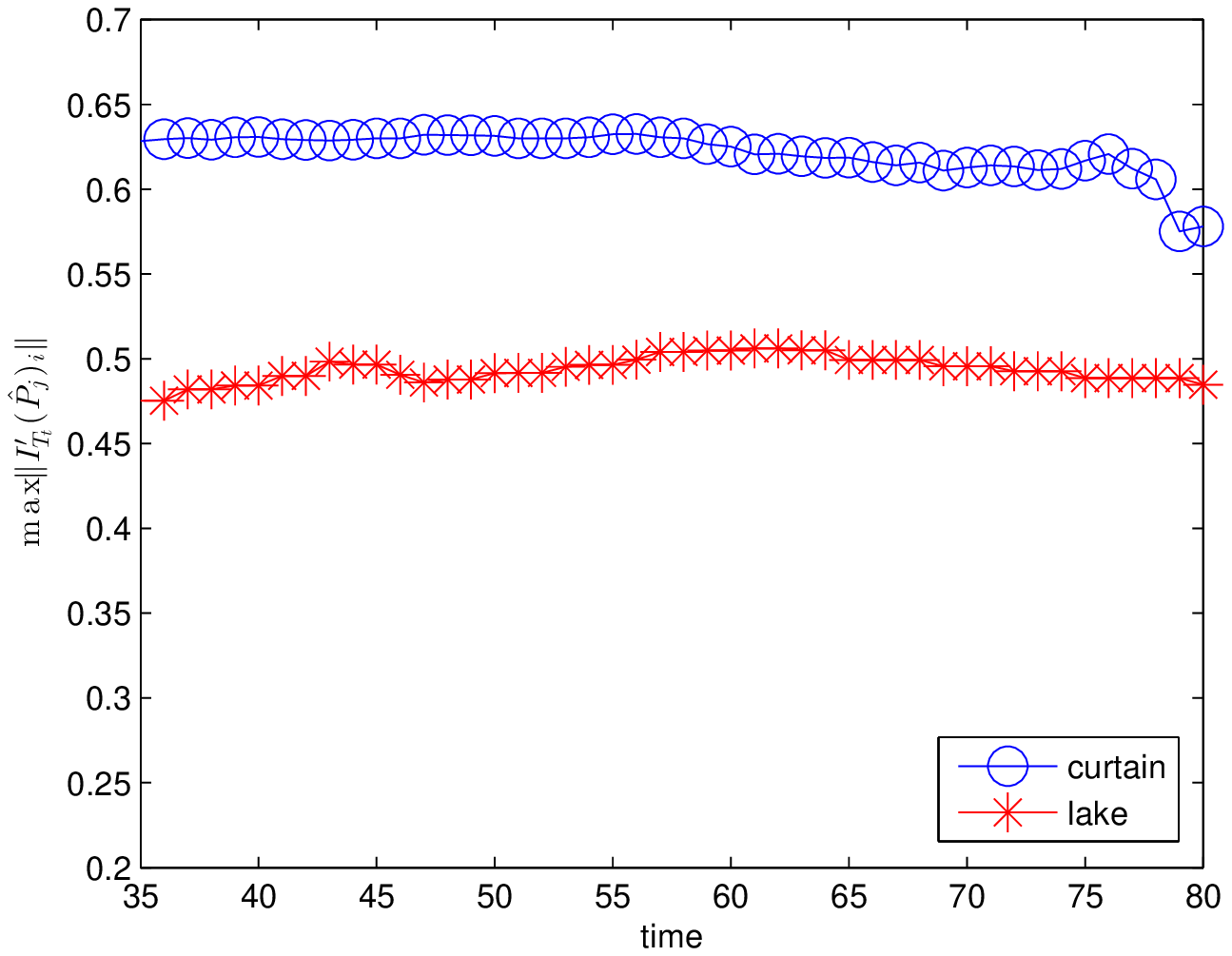}
					\caption{}
                 \label{DenseVeri}
				\end{subfigure}%
				\begin{subfigure}{0.33\textwidth}
					\includegraphics[scale=0.44]{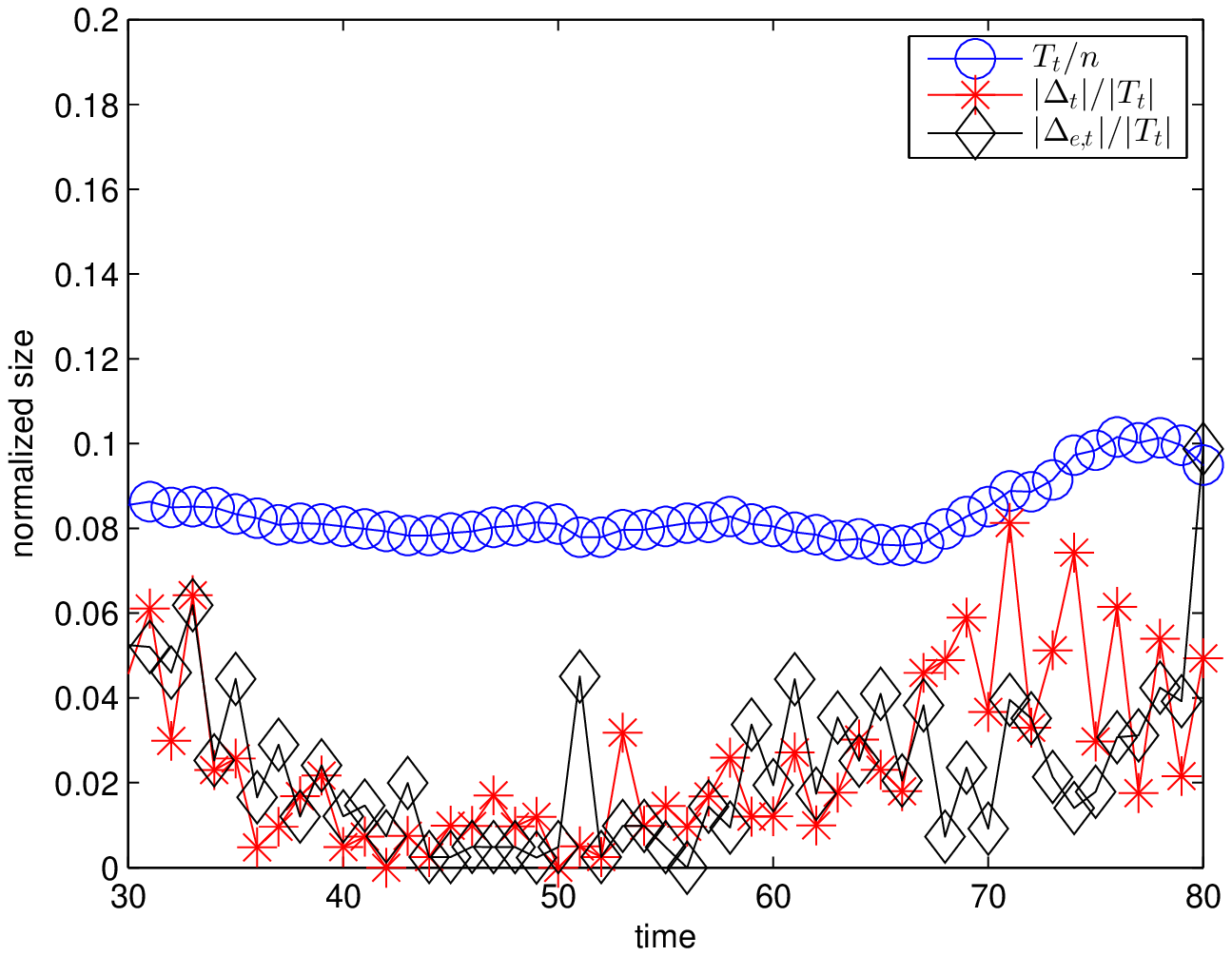}
					\caption{}
                  \label{SuppVeri}
				\end{subfigure}
				\caption{\small{(a) Verification of slow subspace change assumption. (b) Verification of denseness assumption. (c) Verification of small support size, small support change}}
			\end{figure*}

Testing for slow subspace change can be done in various ways. In \cite[Fig 6]{rrpcp_perf}, we do this after low-rankifying the video data first. This helps to very clearly demonstrate slow subspace change, but then it is not checking what we really do on real video data. In this work, we proceed without low-rankifying the data. We let  $t_{0} = 0$ and $t_j=t_0 + j \tau$ with $\tau = 725$.
Let $L_t$ denote the mean subtracted background image sequence, i.e. $L_t=B_t-\mu$ where $\mu = (1/t_1) \sum_{t=0}^{t_1} B_t$. We computed $P_{(j)}$ as $P_{(j)}=\abasis([L_{t_{j}}, ... L_{t_{j+1}-1}],95\%)$. We observed that $\rank(P_{(j)})\le 38$ for curtain sequence, while $\rank(P_{(j)})\le 33$ for lake sequence. In other words,  95\% of the energy is contained in only 38 or lesser directions in either case, i.e. both sequences are approximately low-dimensional. Notice that the dimension of the matrix $[L_{t_{j}}, ... L_{t_{j+1}-1}]$ is $n \times \tau$ and $\min(n,\tau)=\tau=725$ is much larger than 38.
To test for slow subspace change, in Fig. \ref{SubspaceVeri},  we plot $\|(I - P_{(j-1)} P_{(j-1)}') L_t\|_2 / \|L_t\|_2$ when $t \in [t_j, t_{j+1})$. Notice that, after every change time ($t_j = 725,1450$), this quantity is initially small for the first 100-150 frames and then increases gradually. It later decreases also but that is allowed (all we need is that it be small initially and increase slowly).

\subsection{Denseness assumption} \label{dense_assu}
Exactly verifying the denseness assumption is impossible since computing $\kappa_s(.)$ has exponential complexity  (one needs to check all sets $T$ of size $s$).
Instead, to get some idea if it holds even just for $T$ replaced by $T_t$, in Fig. \ref{DenseVeri}, we plot $\max_i \|{I_{T_t}}' (P_{(j)})_i\|_2$ where $T_t$ is the true or estimated support of $S_t$ at time $t$. 
For the lake sequence, $T_t$ is simulated and hence known. For the curtain sequence, we select a part of the sequence in which the person is wearing a black shirt (against a white curtains' background). This part corresponds to $t=35$ to $t=80$. For this part, ReProCS returns a very accurate estimate of $T_t$, and we use this estimated support as a proxy for the true support $T_t$.

\subsection{Support size, support change and slow support change}
For real video sequences, it is not possible to get the true foreground support. Thus we used $\That_t$ for the part of the curtain sequence described above in Sec \ref{dense_assu} as a proxy for $T_t$. 
We plot the support size normalized by the image size $|T_t|/n$, and we plot the number of additions and removals normalized by the support size, i.e. $|\Delta_t|/|T_t|$ and $|\Delta_{e,t}|/|T_t|$ in Fig. \ref{SuppVeri}.
Notice from the figure that the support size is at most 10.2\% of the image size. Notice also that at least at every 3 frames, there is at least a 1\% support change. Thus there is some support change every few frames, thus exposing the part of the background behind the foreground. Finally notice that the maximum number of support changes is only 9.9\% of the support size, i.e. slow support change holds for this piece.

\section{Simulation and Experimental Results} \label{expts}

In this section, we show comparisons of ReProCS with other batch and recursive algorithms for robust PCA. For implementing the $\ell_1$ or weighted $\ell_1$ minimizations, we used  the YALL1 $\ell_1$ minimization toolbox \cite{yall1}, its code is available at \url{http://yall1.blogs.rice.edu/}.

{\bf Code and data for our algorithms and for all experiments given below is available at} \url{http://www.ece.iastate.edu/~hanguo/ReProCS_demo.rar}.

{ 
{\em Simulated Data. }
In this experiment, the measurement at time $t$, $M_t := L_t +S_t$, is an $n \times 1$ vector with $n=100$. We generated $L_t$ using the autoregressive model described in \cite{rrpcp_allerton} with auto-regression parameter 0.1, and the decay parameter $f_d=0.1$. The covariance of a direction decayed to zero before being removed. 
There was one change time $t_1$. For $t < t_1$, $P_{t}=P_0$ was a rank $r_0=20$ matrix and $Cov(a_t)$ was a diagonal matrix with entries $10^4, 0.7079\times10^4, 0.7079^2 \times 10^4, \cdots, 14.13$. At $t=t_1$, $c=c_{1,\new}=2$ new directions, $P_{1,\new}$, got added with $Cov(a_{t,\new})$ being diagonal with entries 60 and 50. Also, the variance along two existing directions started to decay to zero exponentially. We used $t_{\train}=2000$ and $t_1 = t_{\train}+5$. The matrix $[P_0 \ P_{1,\new}]$ was generated as the first 22 columns of  an $n \times n$ random orthonormal matrix  (generated by first generating an $n \times n$ matrix random Gaussian matrix and then orthonormalizing it).
For $1 \le t \le t_\train$, $S_t=0$ and hence $M_t=L_t$. For $t>t_\train$, the support set, $T_t$, was generated in a correlated fashion: $S_t$ contained one block of size 9 or 27 (small and large support size cases). The block stayed static with probability $0.8$ and move up or down by one pixel with probability $0.1$ each independently at each time. Thus the support sets were highly correlated. The magnitude of the nonzero elements of $S_t$ is fixed at either 100 (large) or 10 (small).


For the small magnitude $S_t$ case, $\|L_t\|_2$ ranged from 150 to 250 while $\|S_t\|_2$ was equal to 30 and 52, i.e. in this case $\|S_t\|_2 \ll \|L_t\|_2$. For the large magnitude case, $\|S_t\|_2$ was 300 and 520. We implemented ReProCS (Algorithm \ref{PracReProCS})  with $b=99.99$ since this data is exactly low-rank. We used $q=0.25$ for the small magnitude $S_t$ case and $q=1$ for the other case. We compared with three recursive robust PCA methods -- incremental robust subspace learning (iRSL) \cite{Li03anintegrated} and adapted (outlier-detection enabled) incremental SVD (adapted-iSVD) \cite{sequentialSVD} and GRASTA \cite{grass_undersampled} -- and with two batch methods -- Principal Components' Pursuit (PCP) \cite{rpca} \footnote{We use the Accelerated Proximal Gradient algorithm\cite{apg} and Inexact ALM algorithm \cite{alm} (designed for large scale problems) to solve PCP (\ref{PCP}). The code is available at \url{http://perception.csl.uiuc.edu/matrix-rank/sample_code.html}.} and robust subspace learning (RSL)\footnote{The code of RSL is available at http://www.salleurl.edu/$\sim$ftorre/papers/rpca/rpca.zip.} \cite{Torre03aframework}. Results are shown in Table \ref{tablesim}.
}

From these experiments, we can conclude that ReProCS is able to successfully recover both small magnitude and fairly large support-sized $S_t$'s; iRSL has very bad performance in both cases; RSL, PCP and GRASTA work to some extent in certain cases, though not as well as ReProCS.
ReProCS operates by first approximately nullifying $L_t$, i.e. computing $y_t$ as in (\ref{defyt}), and then recovering $S_t$ by exploiting its sparsity.
iRSL and RSL  also compute $y_t$ the same way, but they directly use $y_t$ to detect or soft-detect (and down-weight) the support of $S_t$ by thresholding. Recall that $y_t$ can be rewritten as $y_t = S_t + (-\hat{P}_{t-1}\hat{P}_{t-1}'S_t) + \beta_t$. As the support size of $S_t$ increases, the interference due to $(-\hat{P}_{t-1}\hat{P}_{t-1}'S_t)$ becomes larger, resulting in wrong estimates of $S_t$. For the same reason, direct thresholding is also difficult when some entries of $S_t$ are small while others are not.
Adapted-iSVD is our adaptation of iSVD \cite{sequentialSVD} in which we use the approach of iRSL described above to provide the outlier locations to iSVD (iSVD is an algorithm for recursive PCA with missing entries or what can be called recursive low-rank matrix completion).  It fills in the corrupted locations of $L_t$ by imposing  that $L_t$ lies in $\Span(\Phat_{t-1})$. We used a threshold of $0.5 \min_{i \in T_t} |(S_t)_i|$ for both iRSL and adapted-iSVD (we also tried various other options for thresholds but with similar results).
Since adapted-iSVD and iRSL are recursive methods, a wrong $\Shat_t$, in turn, results in wrong subspace updates, thus also causing $\beta_t$ to become large and finally causing the error to blow up.

RSL works to some extent for larger support size of $S_t$'s but fails when the magnitude of the nonzero $S_t$'s is small. PCP fails since the support sets are generated in a highly correlated fashion and the support sizes are large (resulting in the matrix ${\cal S}_t$ being quite rank deficient also).
%
GRASTA \cite{grass_undersampled} is a recent recursive method from 2012. It was implemented using code posted at \url{https://sites.google.com/site/hejunzz/grasta}. We tried two versions of GRASTA: the demo code as it is and the demo code modified so that it used as much information as ReProCS used (i.e. we used all available frames for training instead of just 100; we used all measurements instead of just 20\% randomly selected pixels; and we used $\hat{r}$ returned by ReProCS as the rank input instead of using rank=5 always). In this paper, we show the latter case. Both experiments are shown on our supplementary material page \url{http://www.ece.iastate.edu/~hanguo/PracReProCS.html}.

{\em Partly Simulated Video: Lake video with simulated foreground. }
In the comparisons shown next, we only compare with PCP, RSL and GRASTA. To address a reviewer comment, we also compare with the batch algorithm of \cite{mateos_anomaly2,mateos_anomaly} (referred to as MG in the figures) implemented using code provided by the authors. There was not enough information in the papers or in the code to successfully implement the recursive algorithm.

We implemented ReProCS (Algorithm \ref{PracReProCS} and Algorithm \ref{pracrep_rpca})  with $b=95$ since the videos are only approximately low-rank and we used $q=1$ since the magnitude of $S_t$ is not small compared to that of $L_t$. The performance of both ReProCS-pPCA (Algorithm \ref{PracReProCS}) and ReProCS-recursive-PCA (Algorithm \ref{pracrep_rpca}) was very similar. Results with using the latter are shown in Fig. \ref{reprocs_rec_pca}.

\begin{figure*}
\begin{center}
\begin{subfigure}{0.5\textwidth}
	\centering
	\includegraphics[scale=0.5]{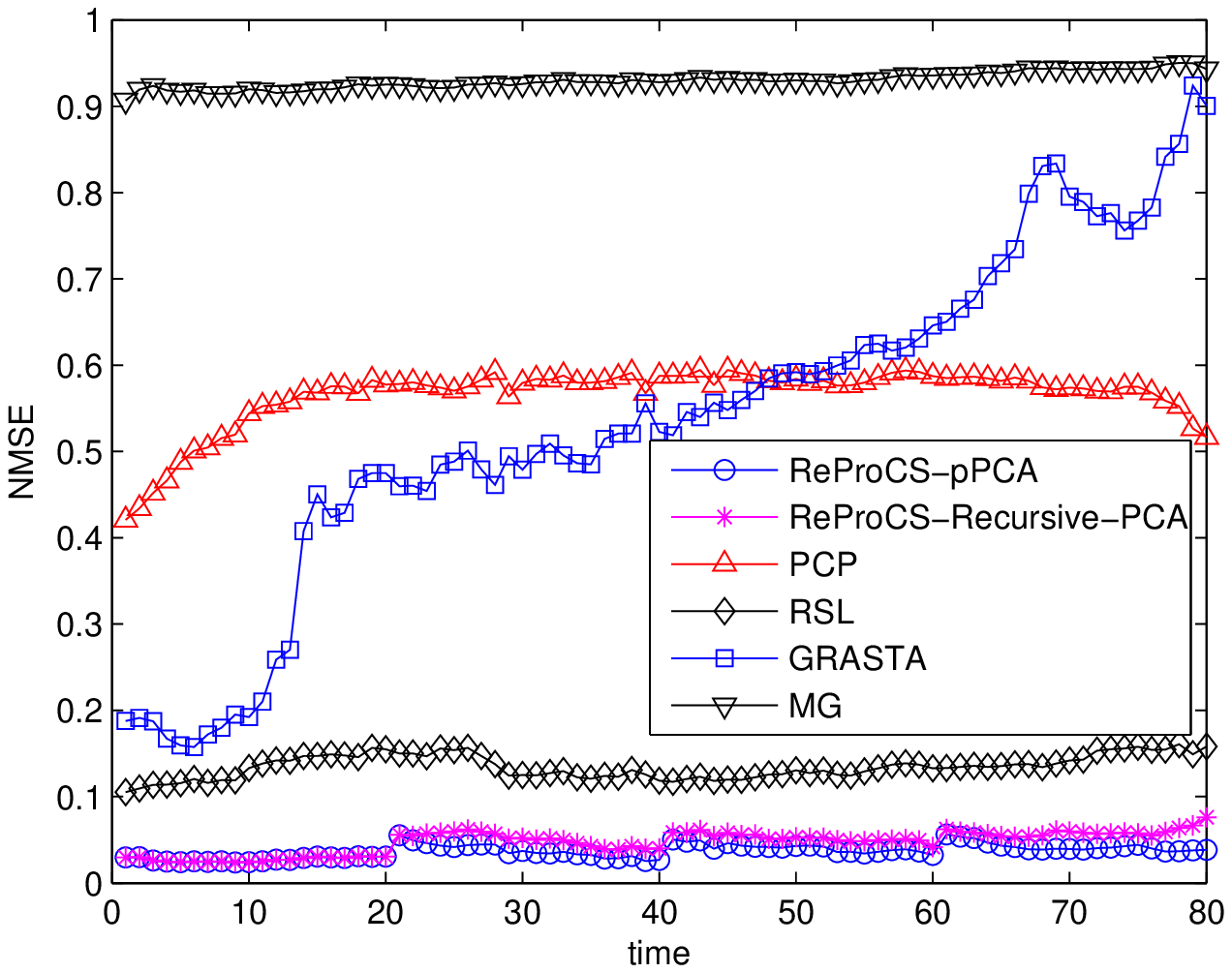}
	\caption{}\label{NMSE}
\end{subfigure}%
\hfill
\begin{subfigure}{0.5\textwidth}
	\centering
	\includegraphics[scale=0.5]{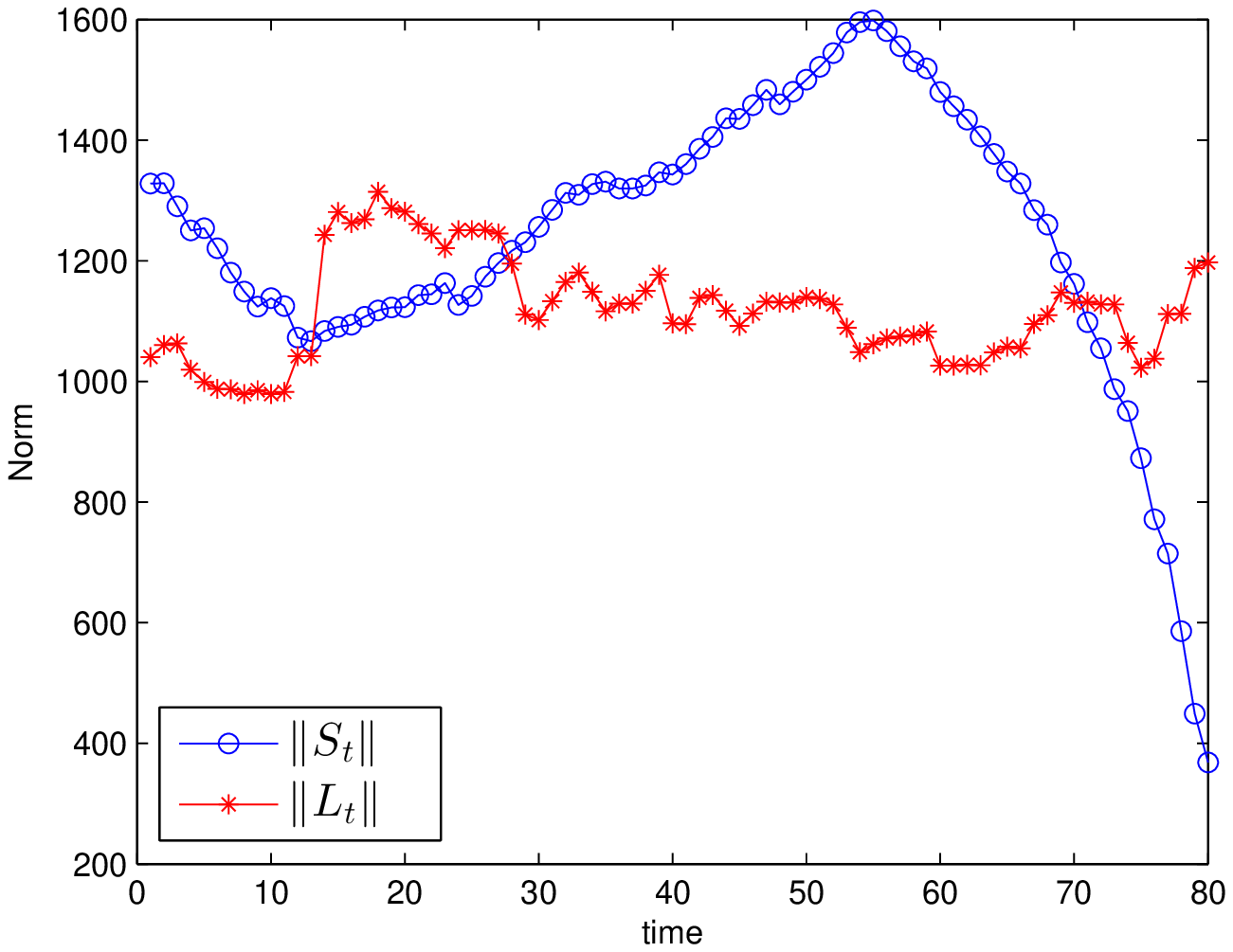}
	\caption{}\label{NormCompare}
\end{subfigure}
\end{center}
\caption{\small{Experiments on partly simulated video. (a) Normalized mean squared error in recovering $S_t$ for realizations. (b) Comparison of $\| S_t \|_2$ and $\| L_t \|$ for one realization.
MG refers to the batch algorithm of \cite{mateos_anomaly2,mateos_anomaly} implemented using code provided by the authors. There was not enough information in the papers or in the code to successfully implement the recursive algorithm.
}}
\end{figure*}

We used the lake sequence described earlier to serve as a real background sequence.  Foreground consisting of a rectangular moving object was overlaid on top of it using (\ref{overlay}). The use of a real background sequence allows us to evaluate performance for data that only approximately satisfies the low-dimensional and slow subspace change assumptions. The use of the simulated foreground allows us to control its intensity so that the resulting $S_t$ is small or of the same order as $L_t$ (making it a difficult sequence), see Fig. \ref{NormCompare}.

The foreground $F_t$ was generated as follows.  For $1 \le t \le t_\train$,  $F_t = 0$. For $t>t_\train$, $F_t$ consists of a $45 \times 25$ moving block whose centroid moves horizontally according to a constant velocity model with small random acceleration \cite[Example V.B.2]{poor_book}.  To be precise, let  $p_t$ be the horizontal location of the block's centroid at time $t$, let $v_t$ denote its horizontal velocity. 
Then $g_t := \left[ \begin{array}{c}p_t\\ v_t \\ \end{array} \right]$ satisfies
$
g_t = G g_{t-1} + \left[
                    \begin{array}{c}
                      0 \\
                      n_t \\
                    \end{array}
                  \right]  \label{motion_model} \nn
$
where $G := \left[ \begin{array}{cc} 1 & \ 1 \\ 0 & \ 1 \\ \end{array}\right]$ and $n_t$ is a zero mean truncated Gaussian with variance $Q$ and with $-2\sqrt{Q}<|n_t|<2\sqrt{Q}$. The  nonzero pixels' intensity is i.i.d. over time and space and distributed as $\text{uniform}(b_1,b_2)$, i.e. $(F_t)_i \sim \text{uniform}(b_1,b_2)$ for $i \in T_t$.
%
%
%
In our experiments, we generated the data with $t_\train=1420$, $b_1=170$, $b_2=230$, $p_{t_0+1}=27$, $v_{t_0+1}=0.5$ and $Q=0.02$. With these values of $b_1,b_2$, as can be seen from Fig. \ref{NormCompare}, $\|S_t\|_2$ is roughly equal or smaller than $\|L_t\|_2$  making it a difficult sequence. Since it is not much smaller, ReProCS used $q=1$; since background data is approximately low-rank it used $b=95$.

\begin{figure*}
	\centering
	\begin{tabular}{cc}
		\includegraphics[width=15mm]{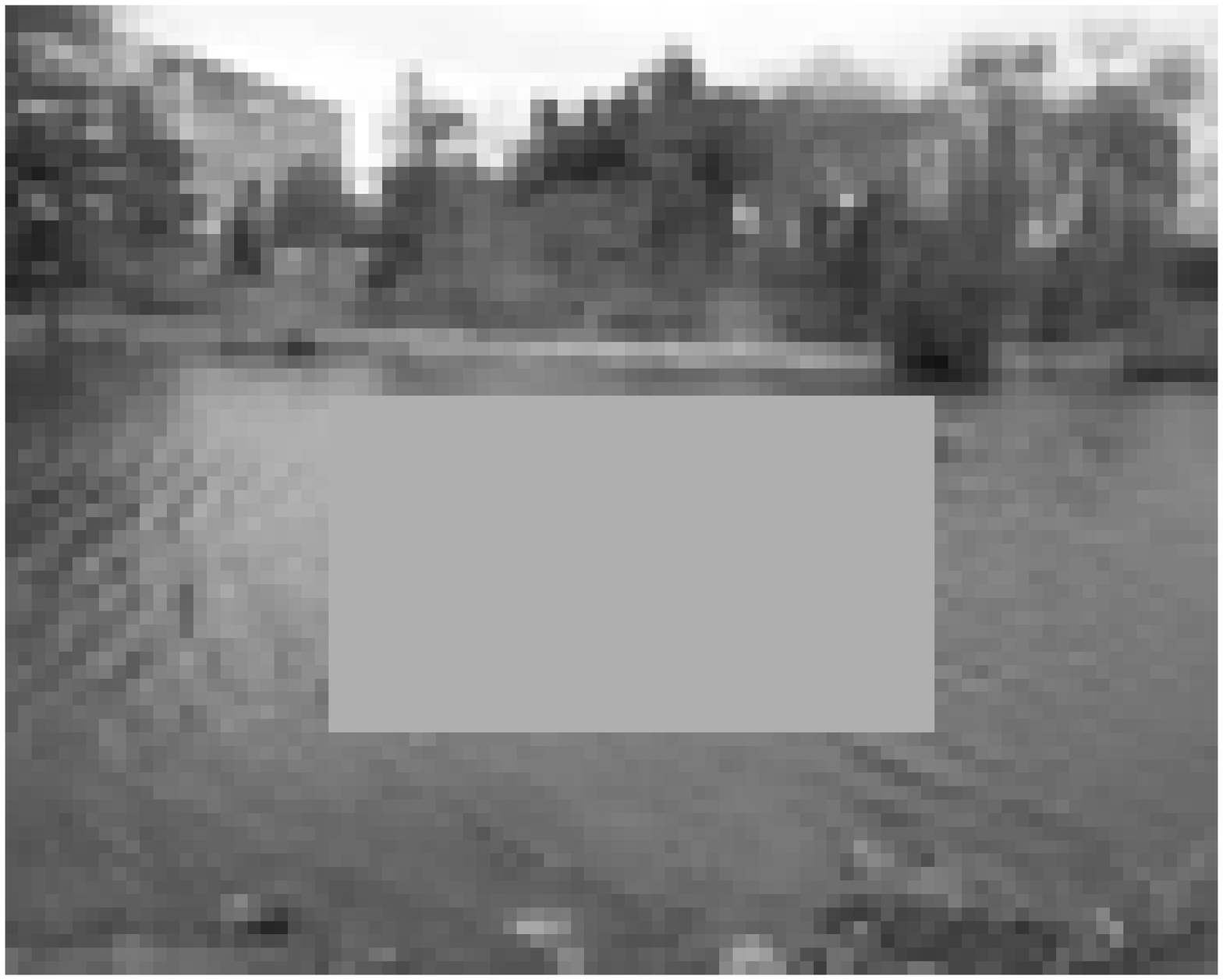}
		\includegraphics[width=15mm]{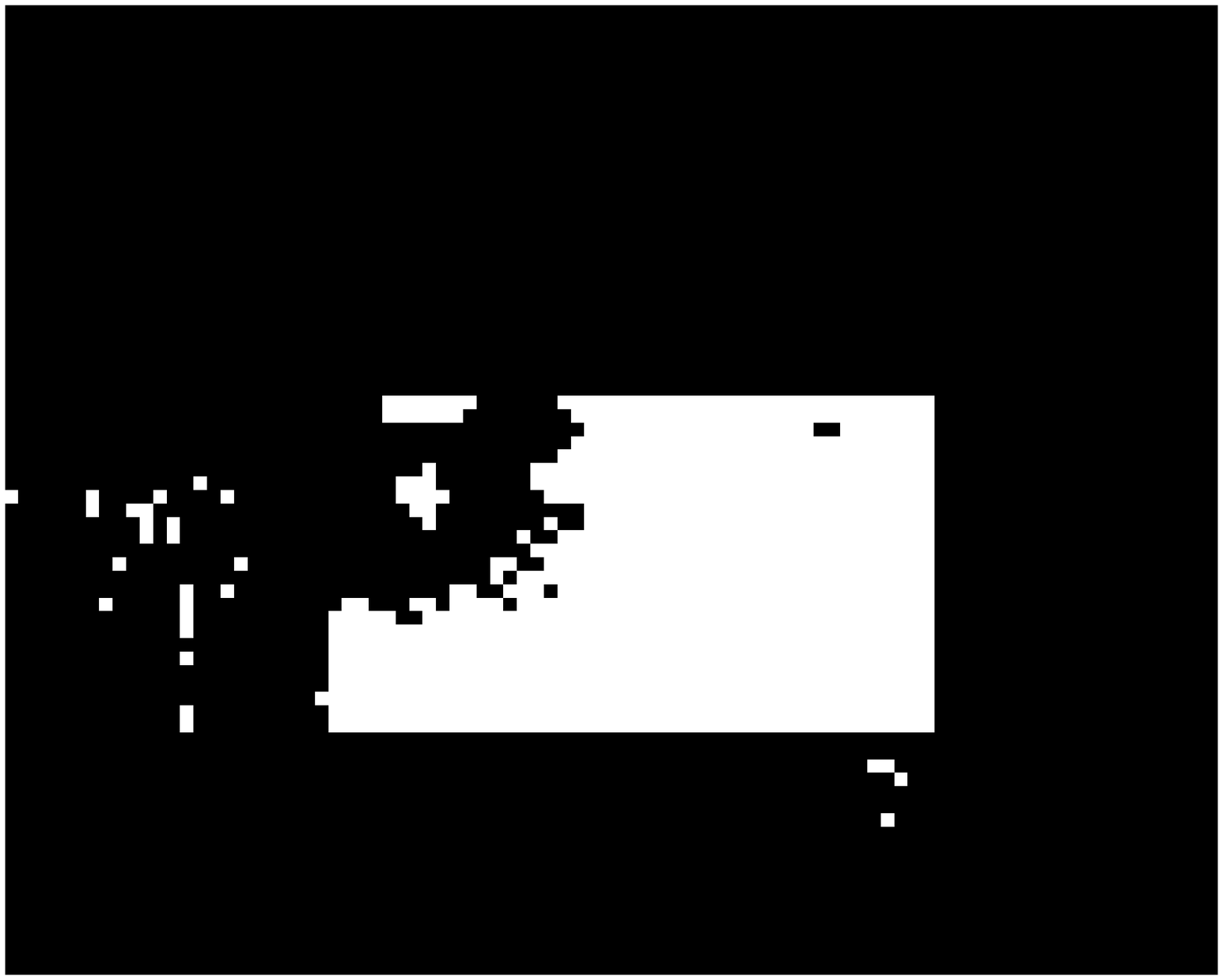}
		\includegraphics[width=15mm]{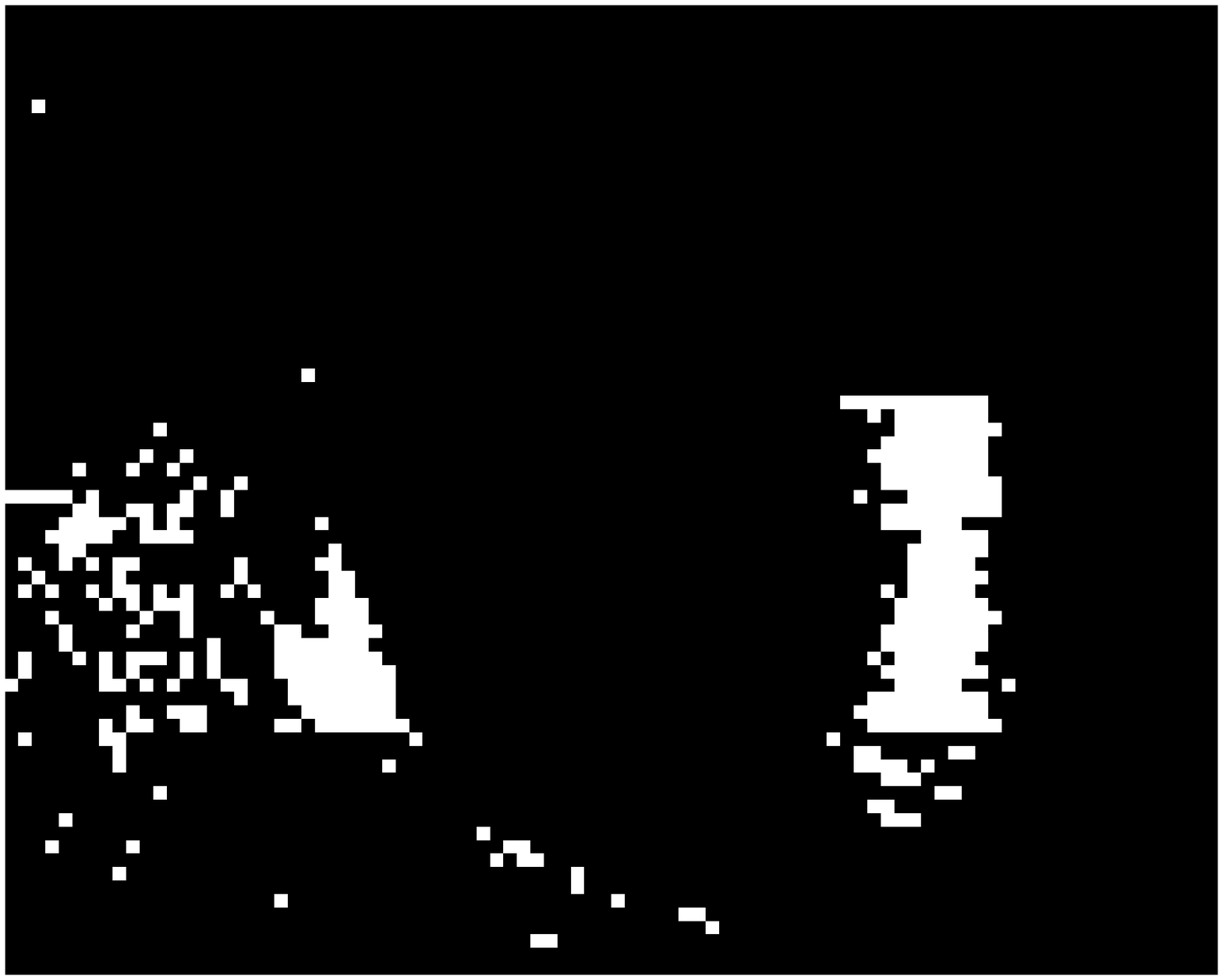}
		\includegraphics[width=15mm]{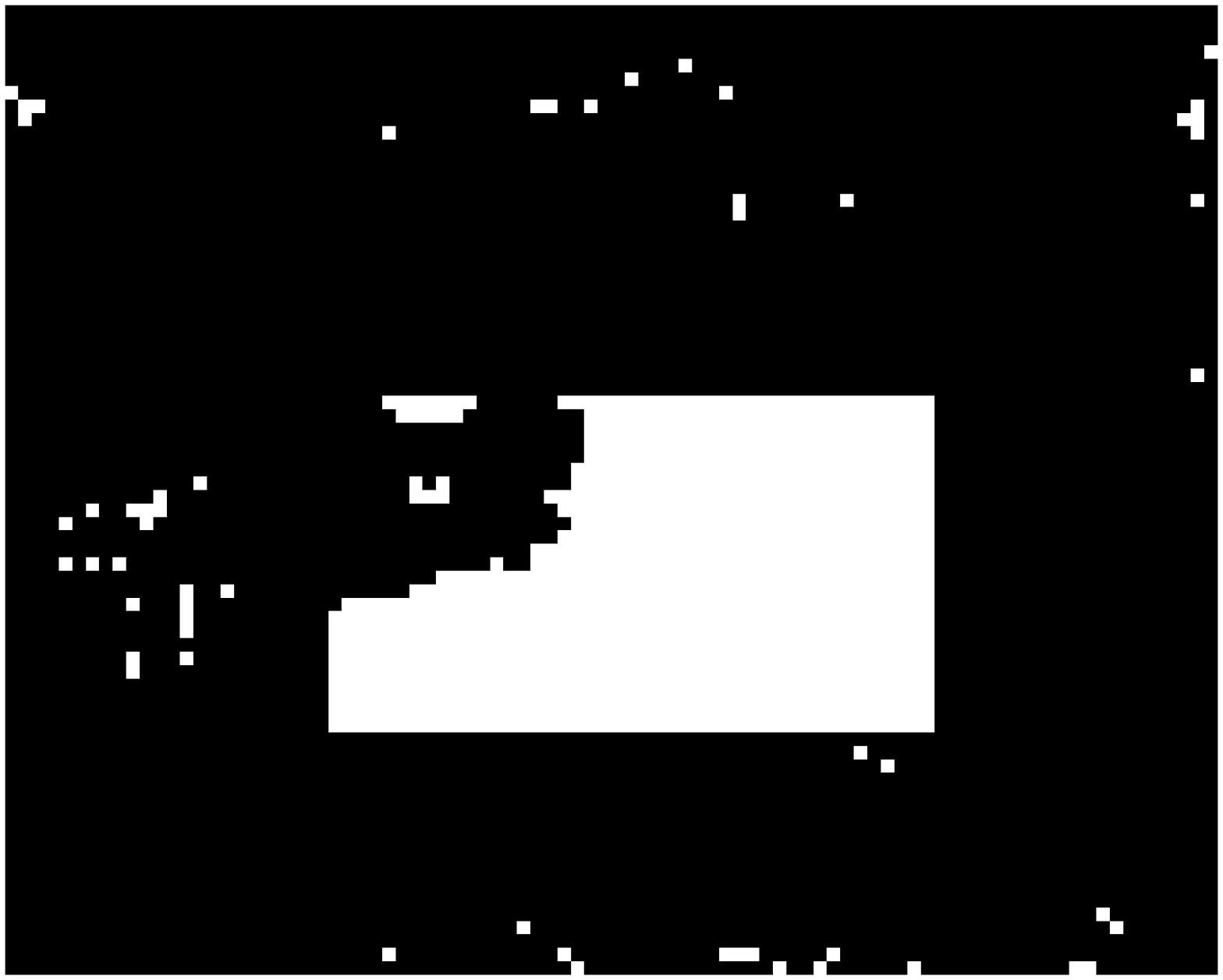}
		\includegraphics[width=15mm]{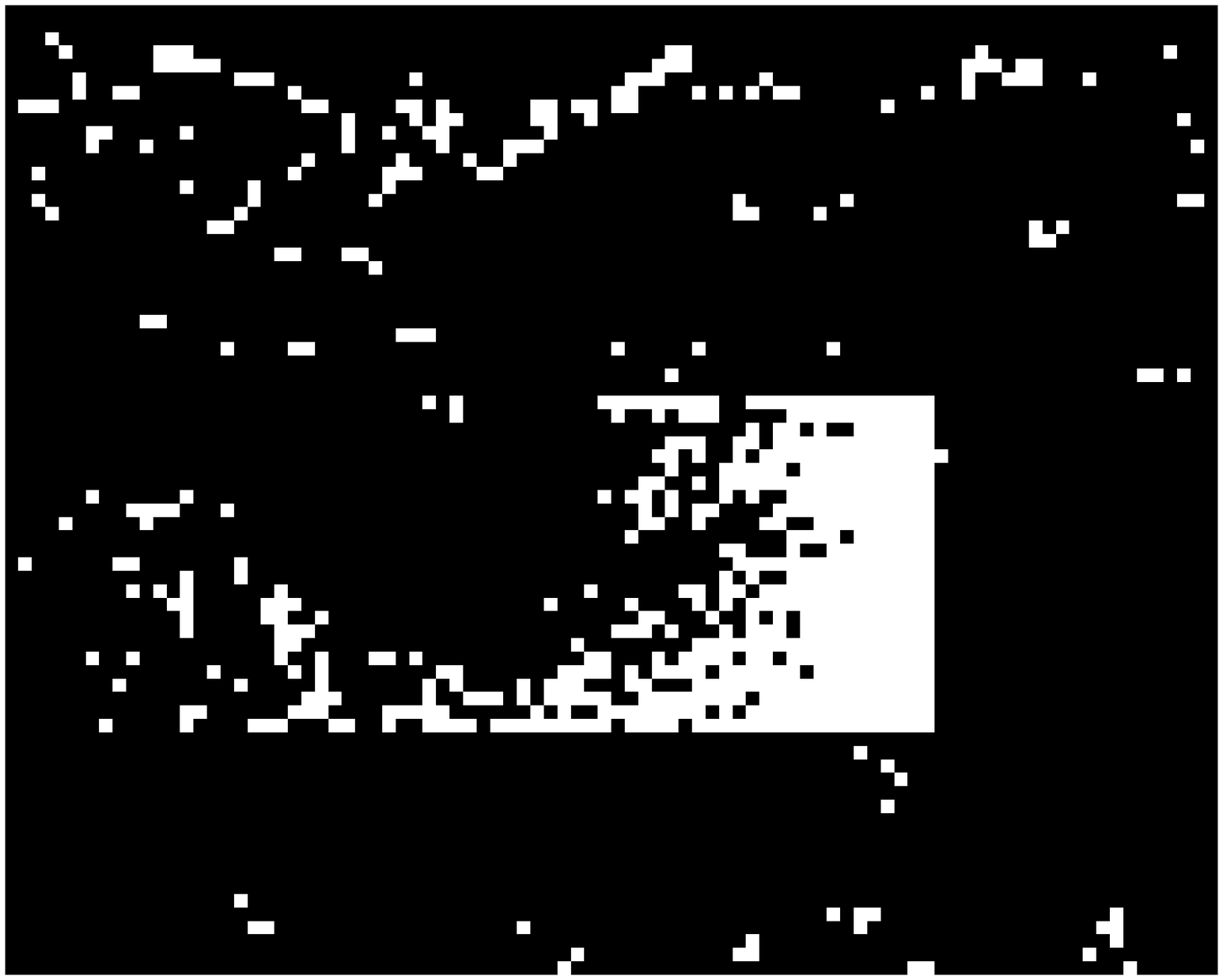}
		\includegraphics[width=15mm]{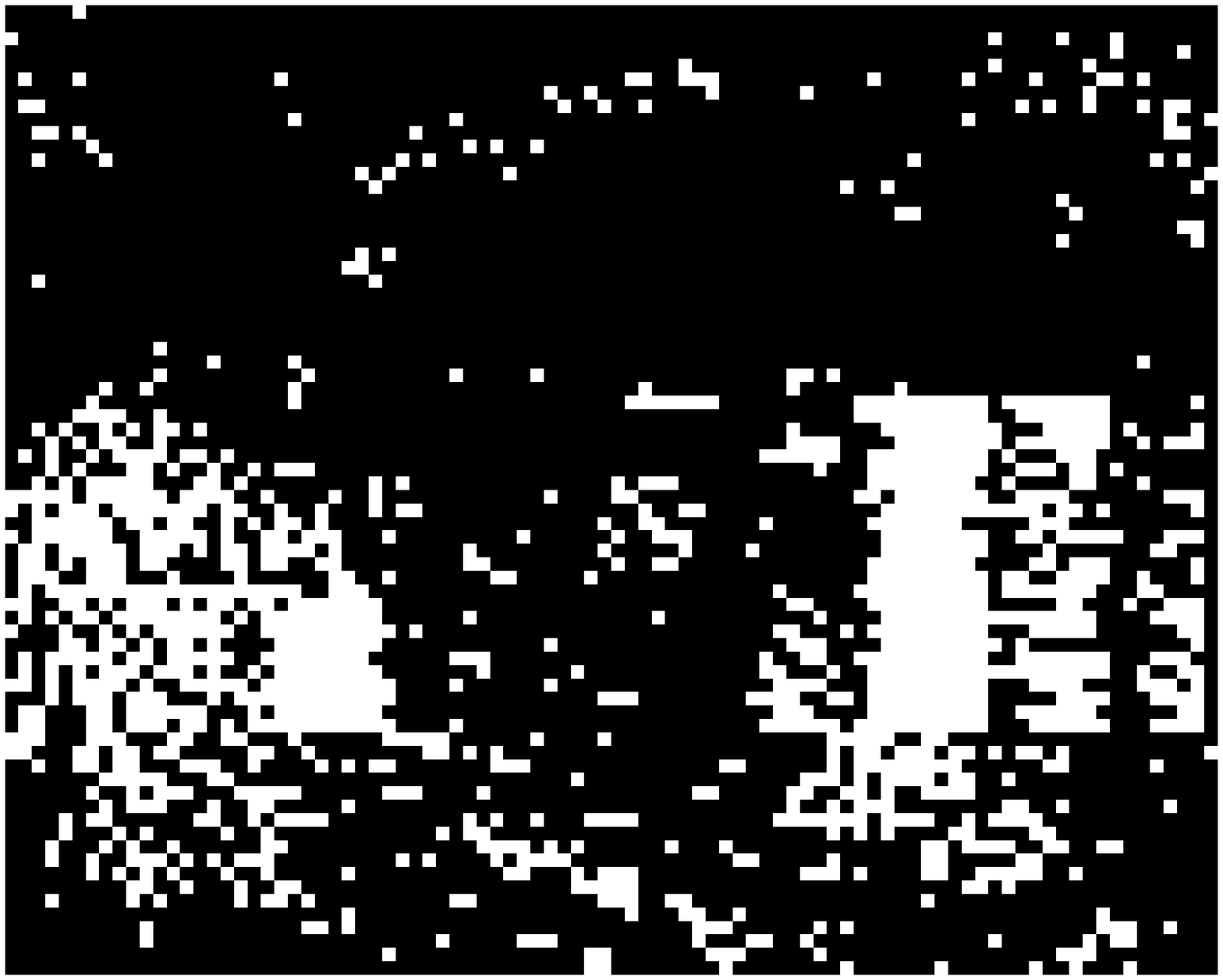}
		\includegraphics[width=15mm]{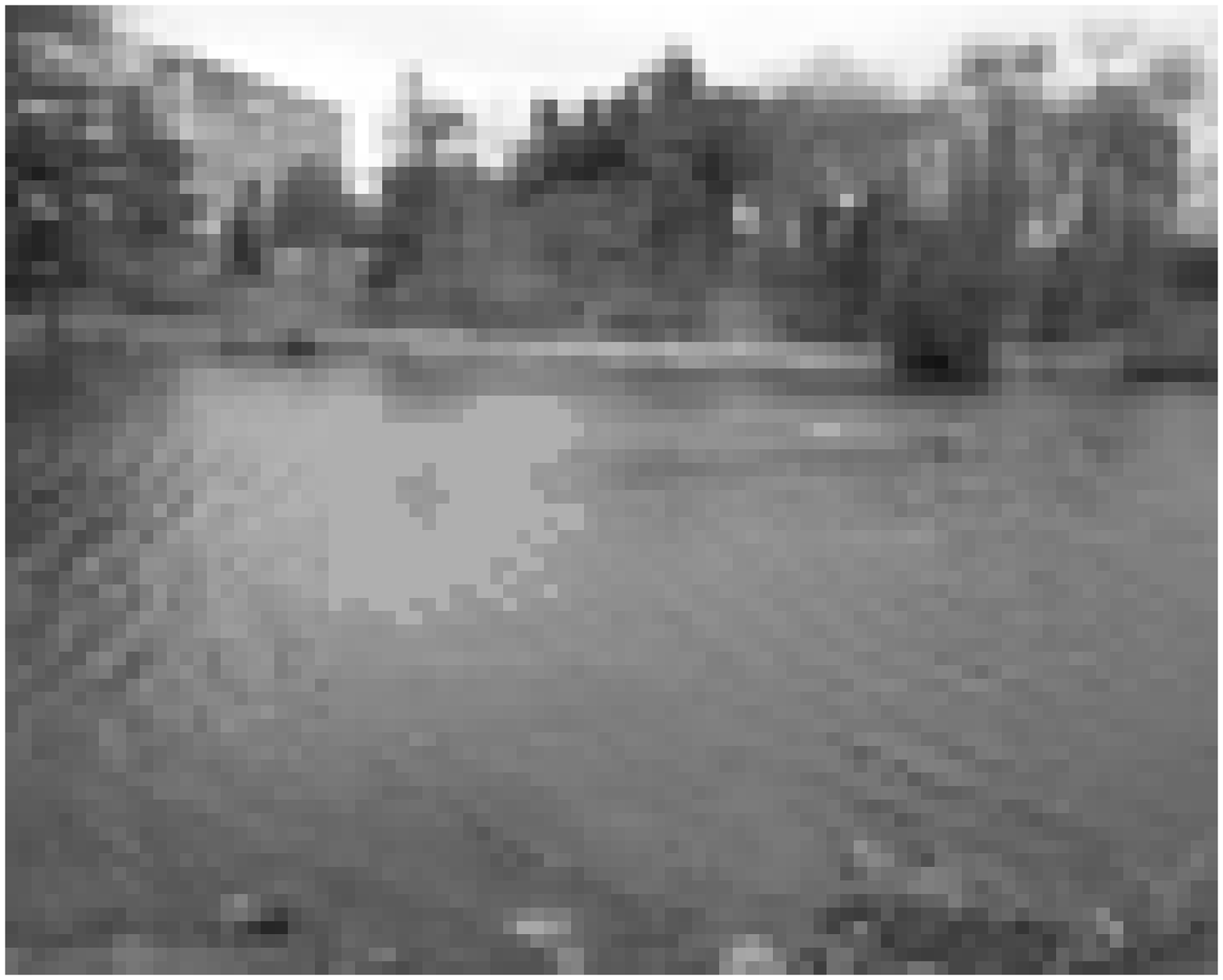}
		\includegraphics[width=15mm]{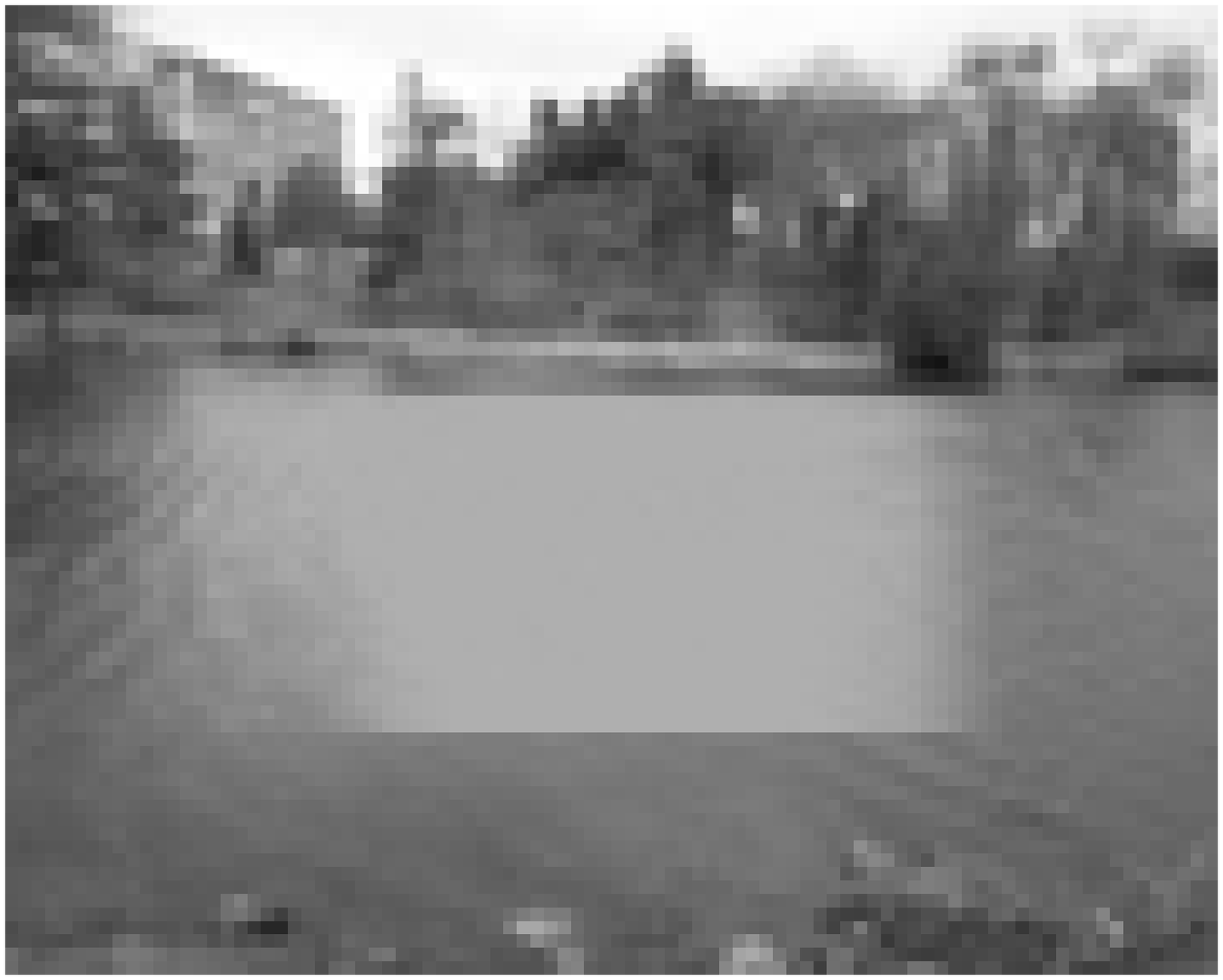}
		\includegraphics[width=15mm]{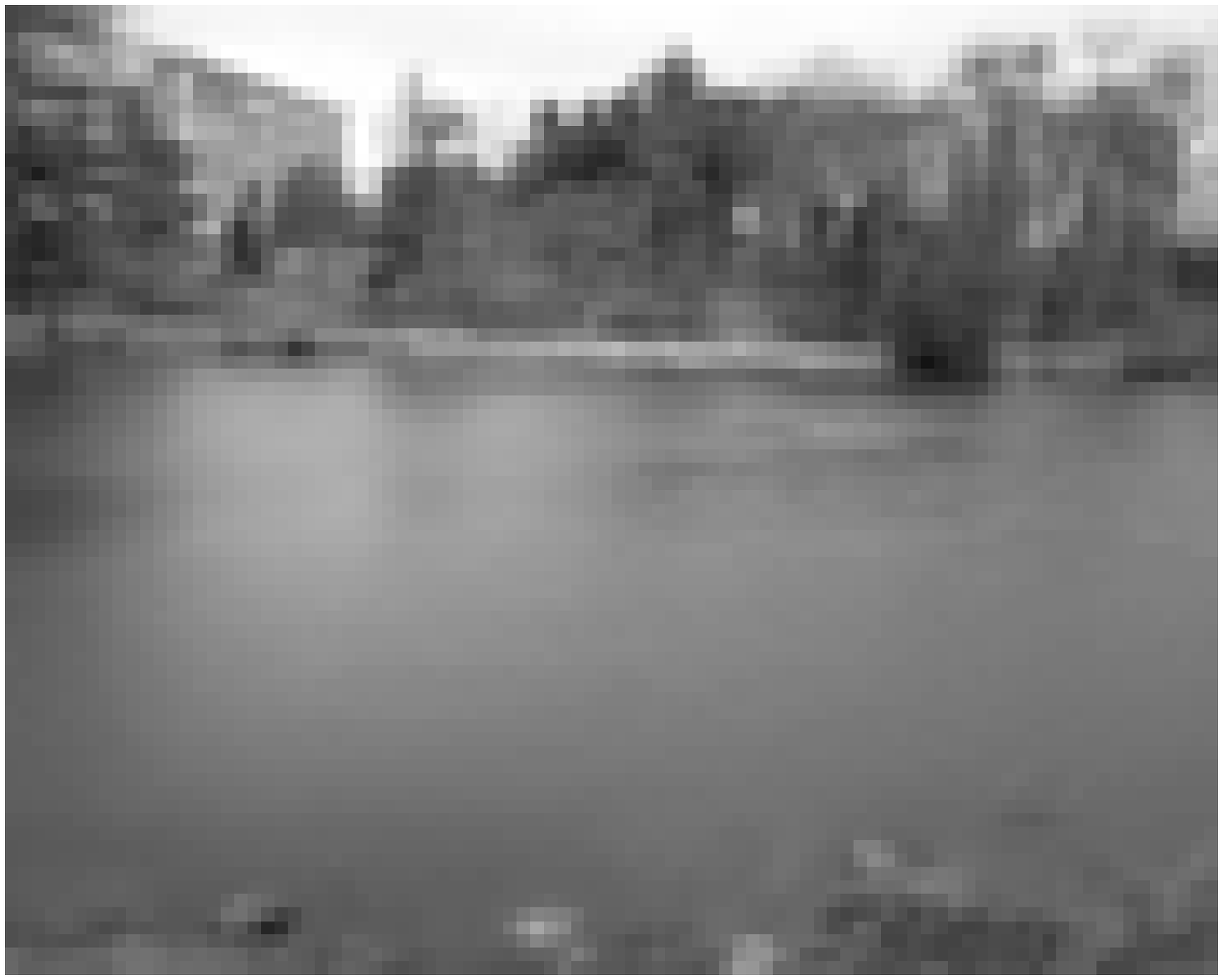}
		\includegraphics[width=15mm]{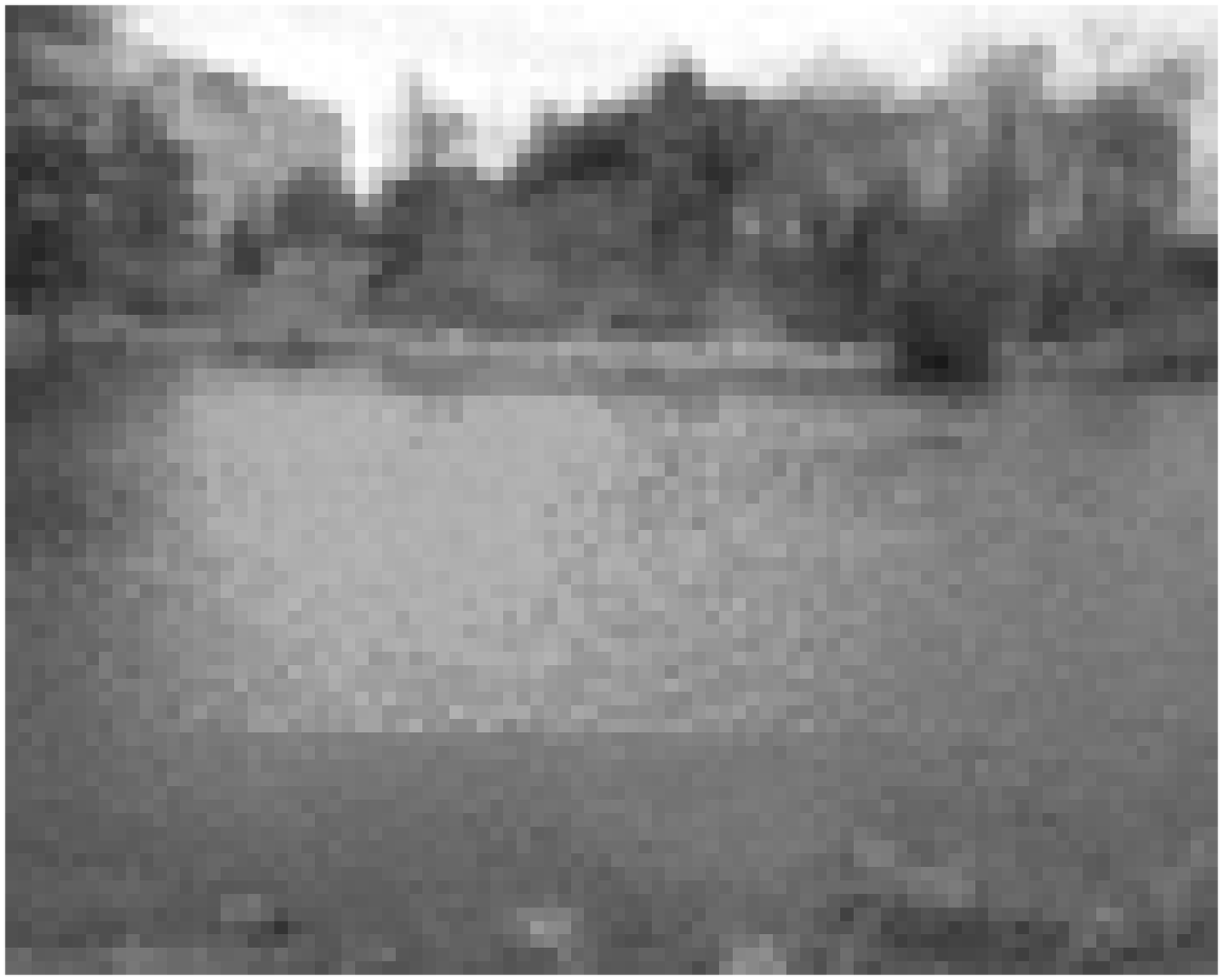}
		\includegraphics[width=15mm]{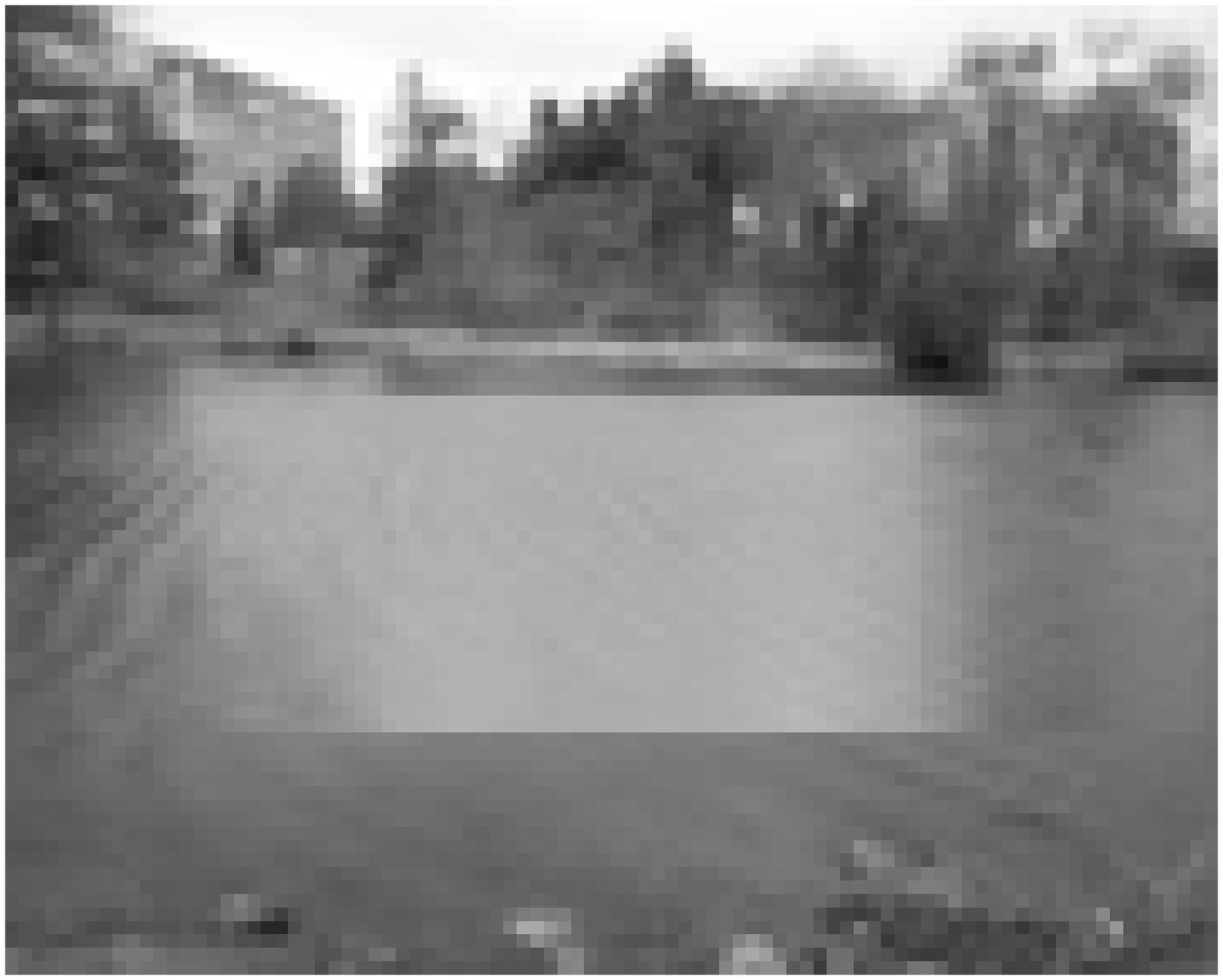}\\

		\includegraphics[width=15mm]{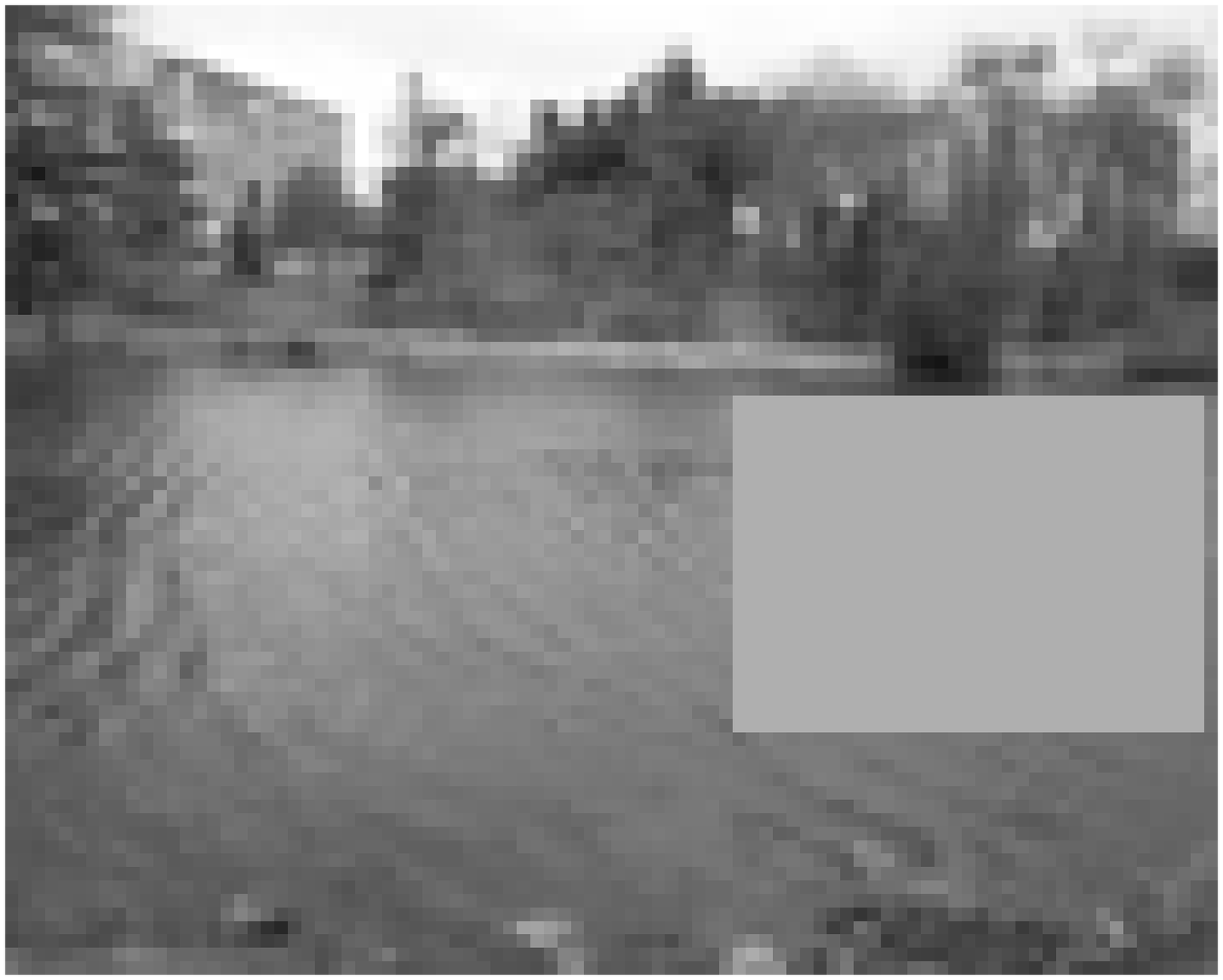}
		\includegraphics[width=15mm]{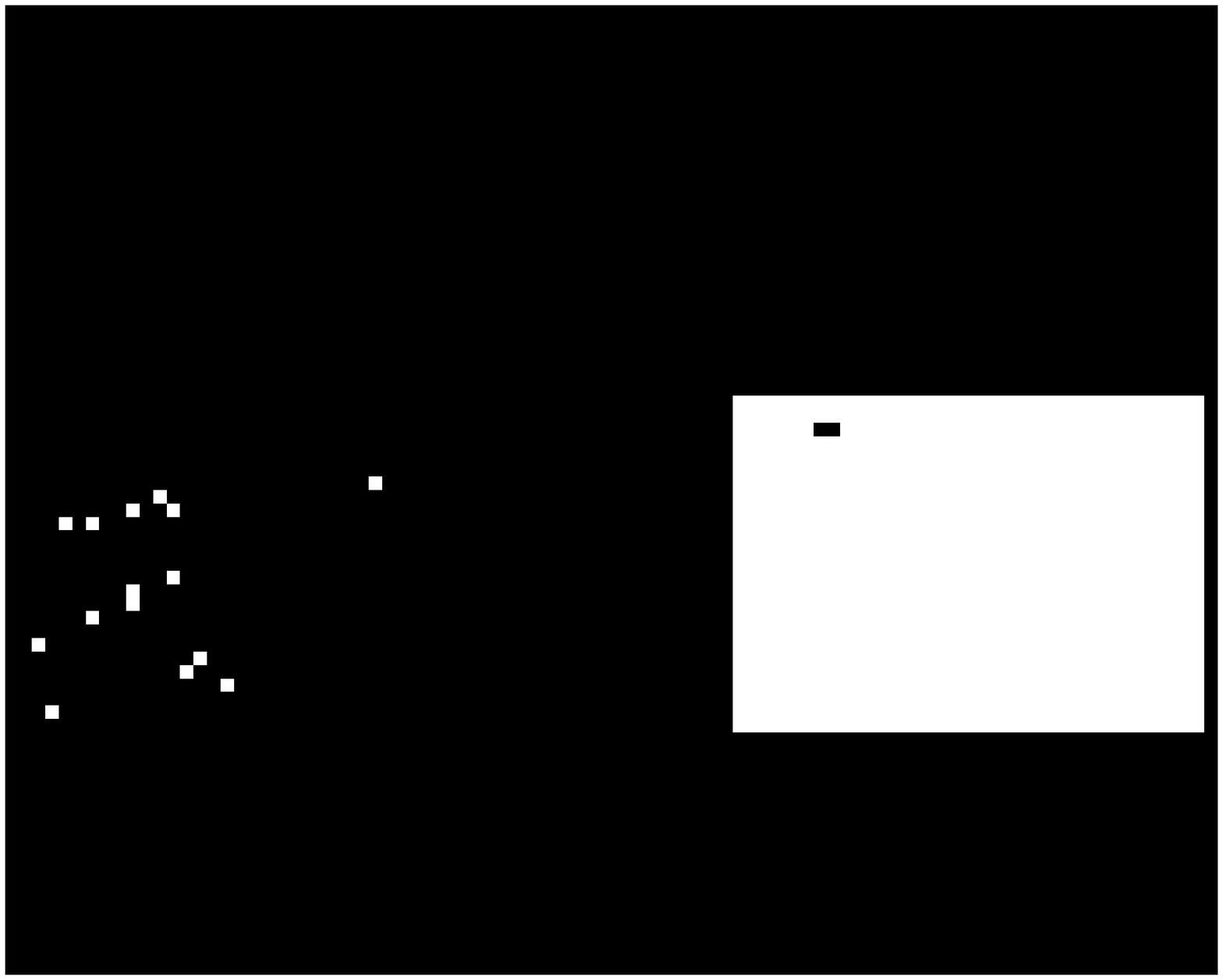}
		\includegraphics[width=15mm]{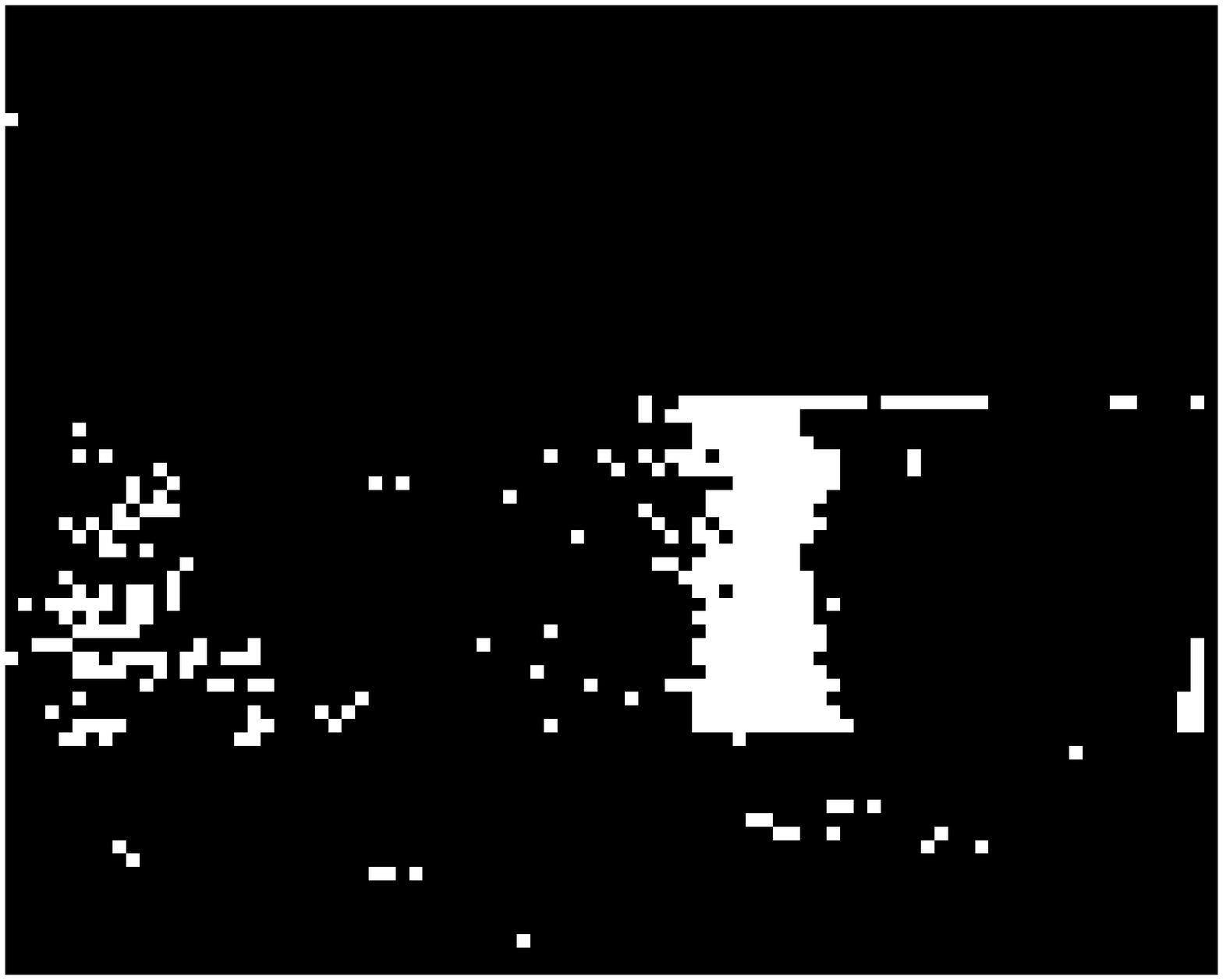}
		\includegraphics[width=15mm]{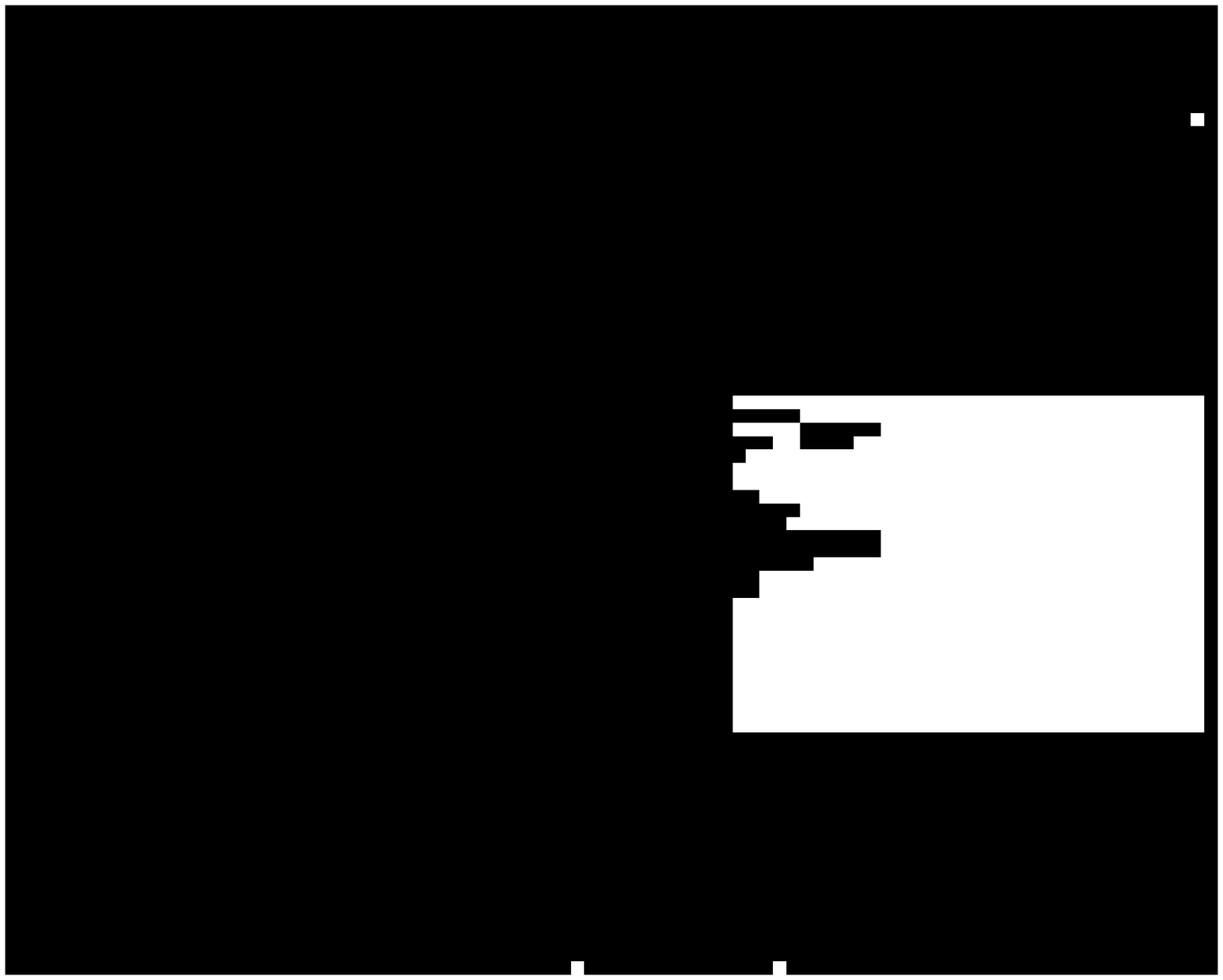}
		\includegraphics[width=15mm]{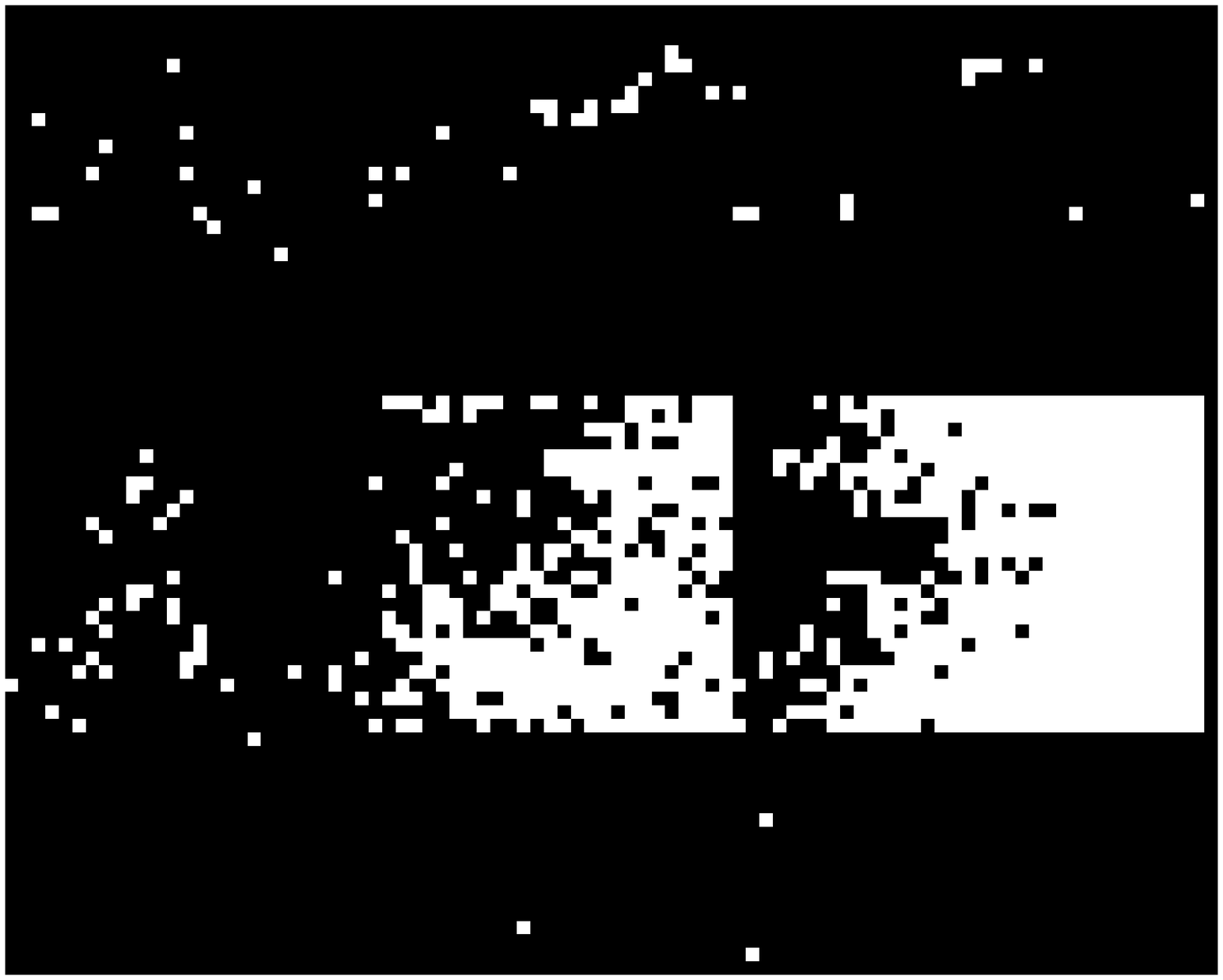}
		\includegraphics[width=15mm]{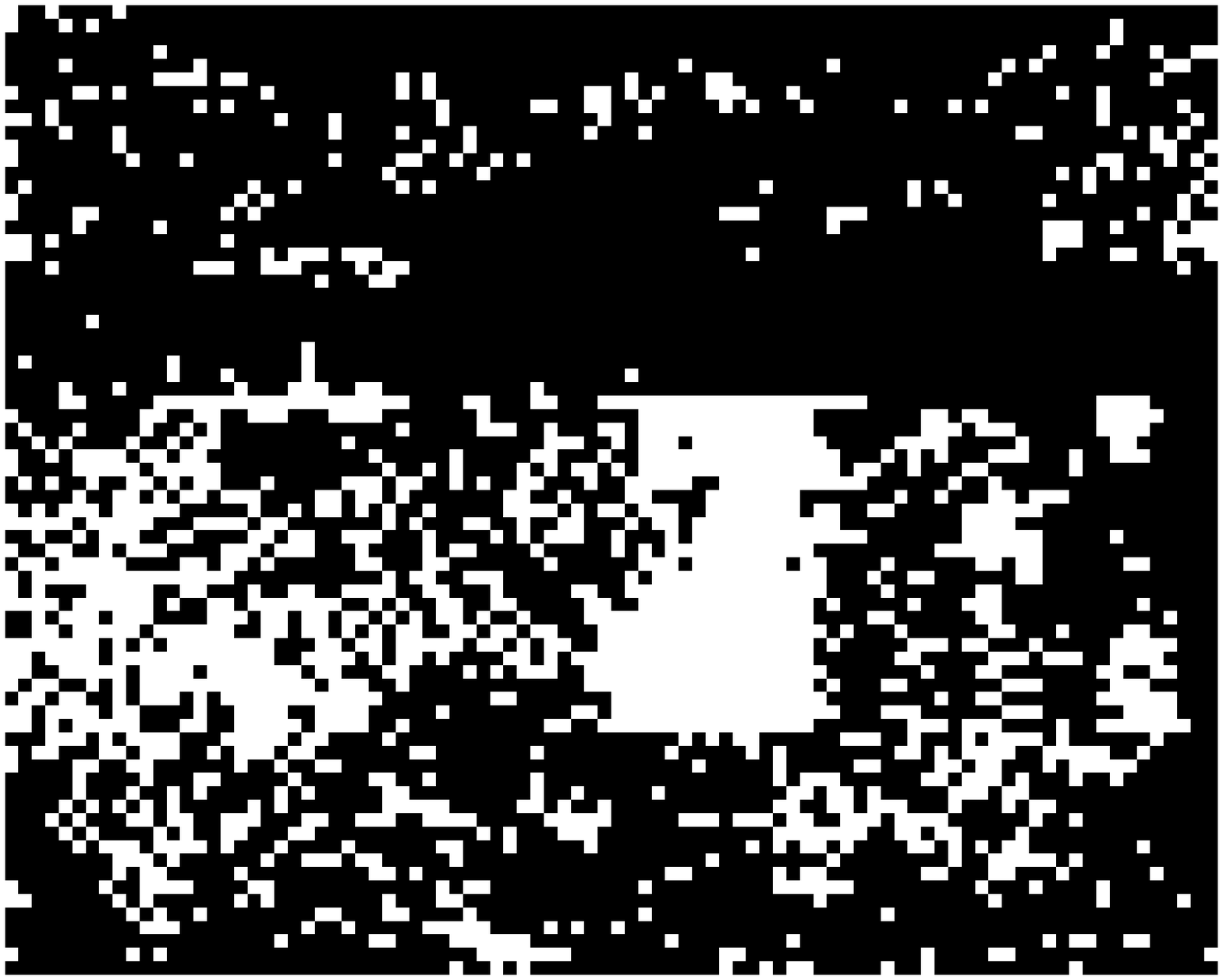}
		\includegraphics[width=15mm]{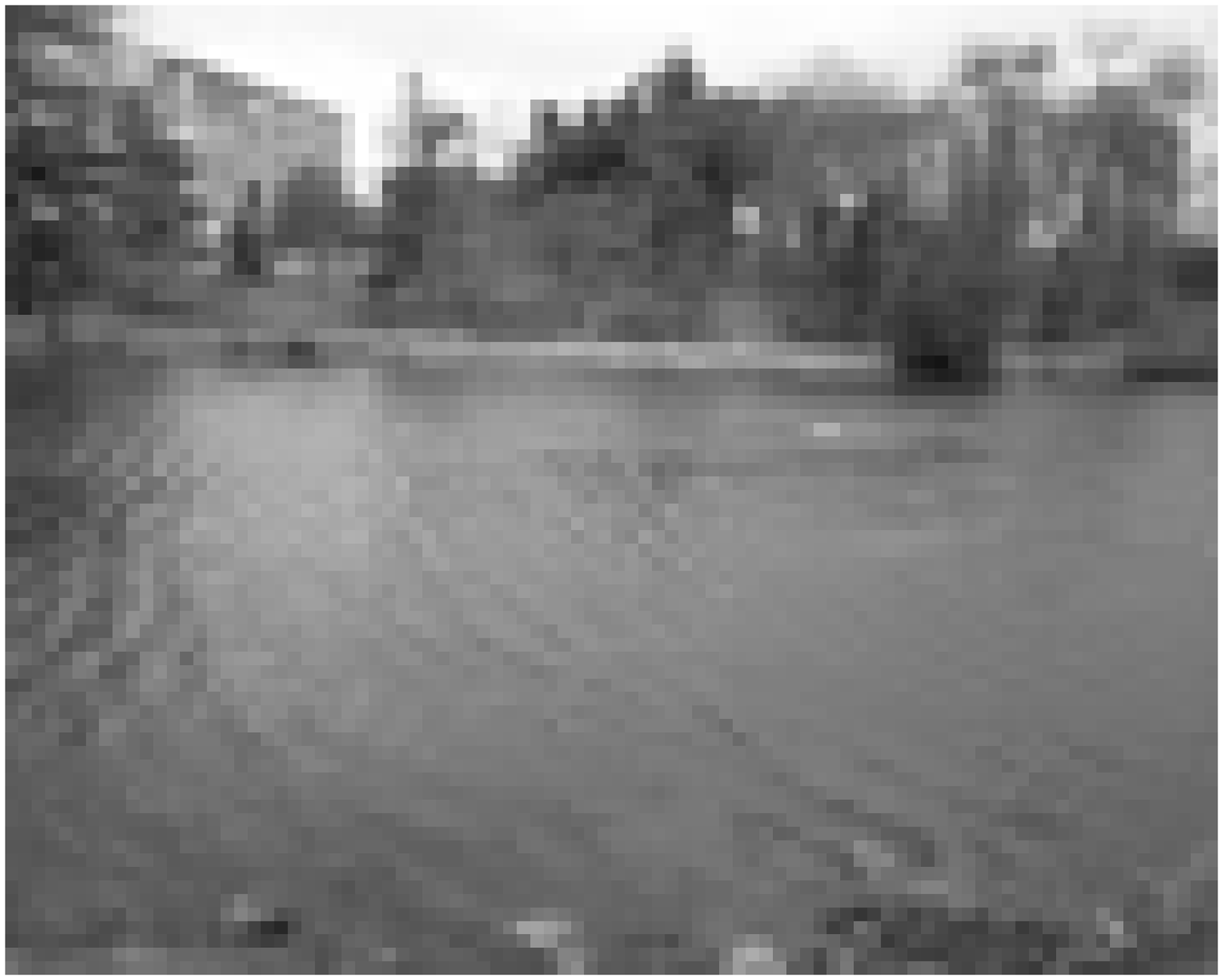}
		\includegraphics[width=15mm]{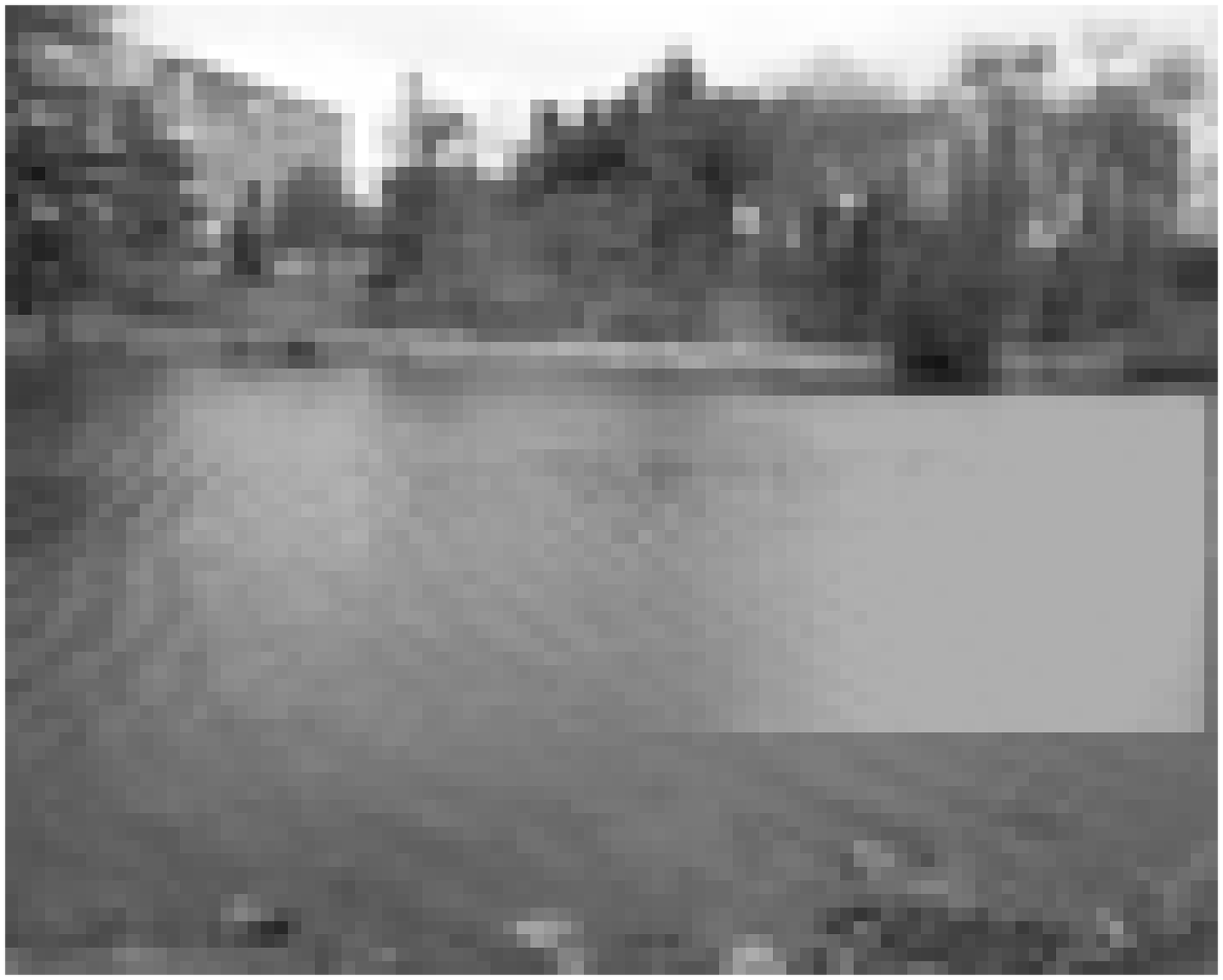}
		\includegraphics[width=15mm]{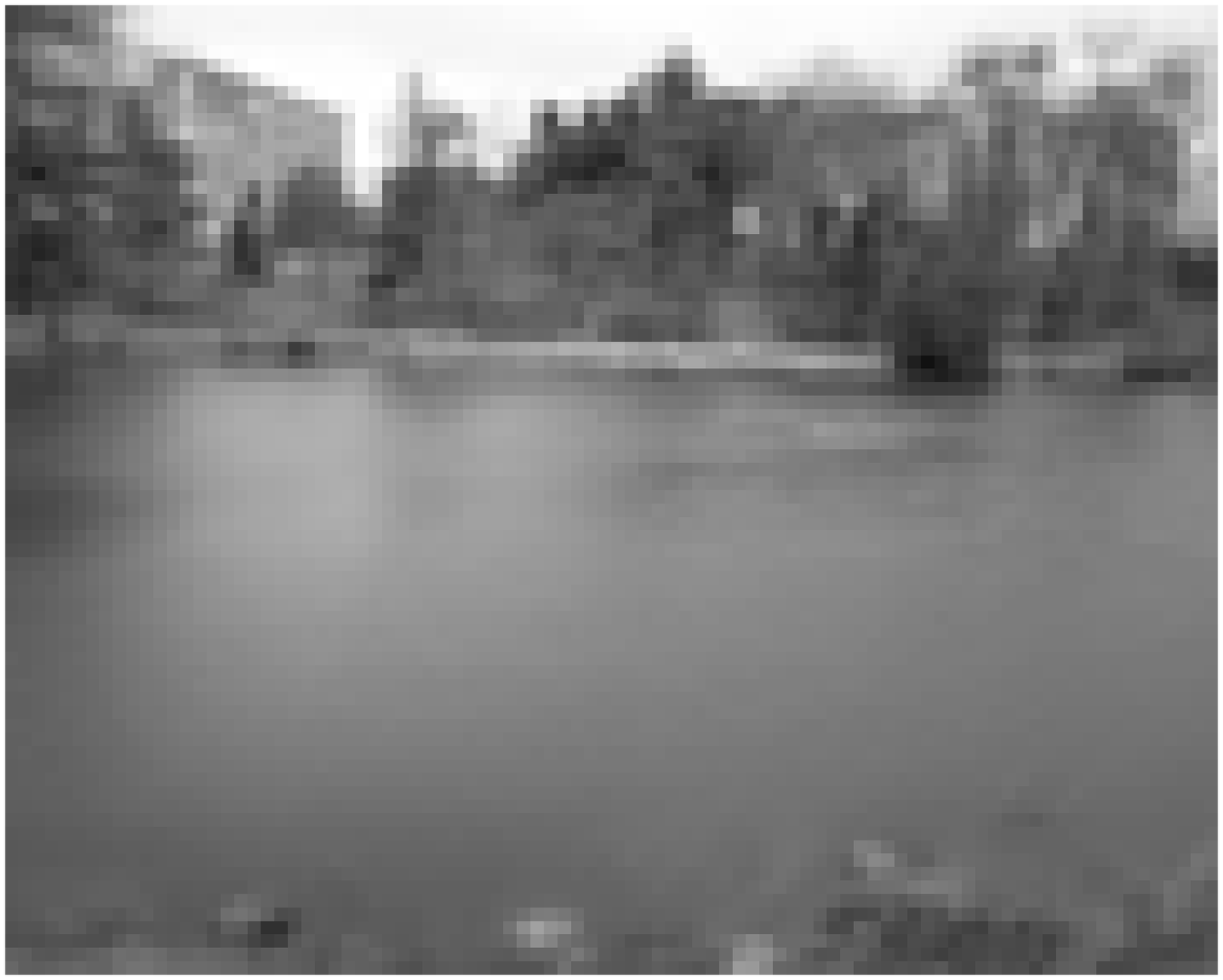}
		\includegraphics[width=15mm]{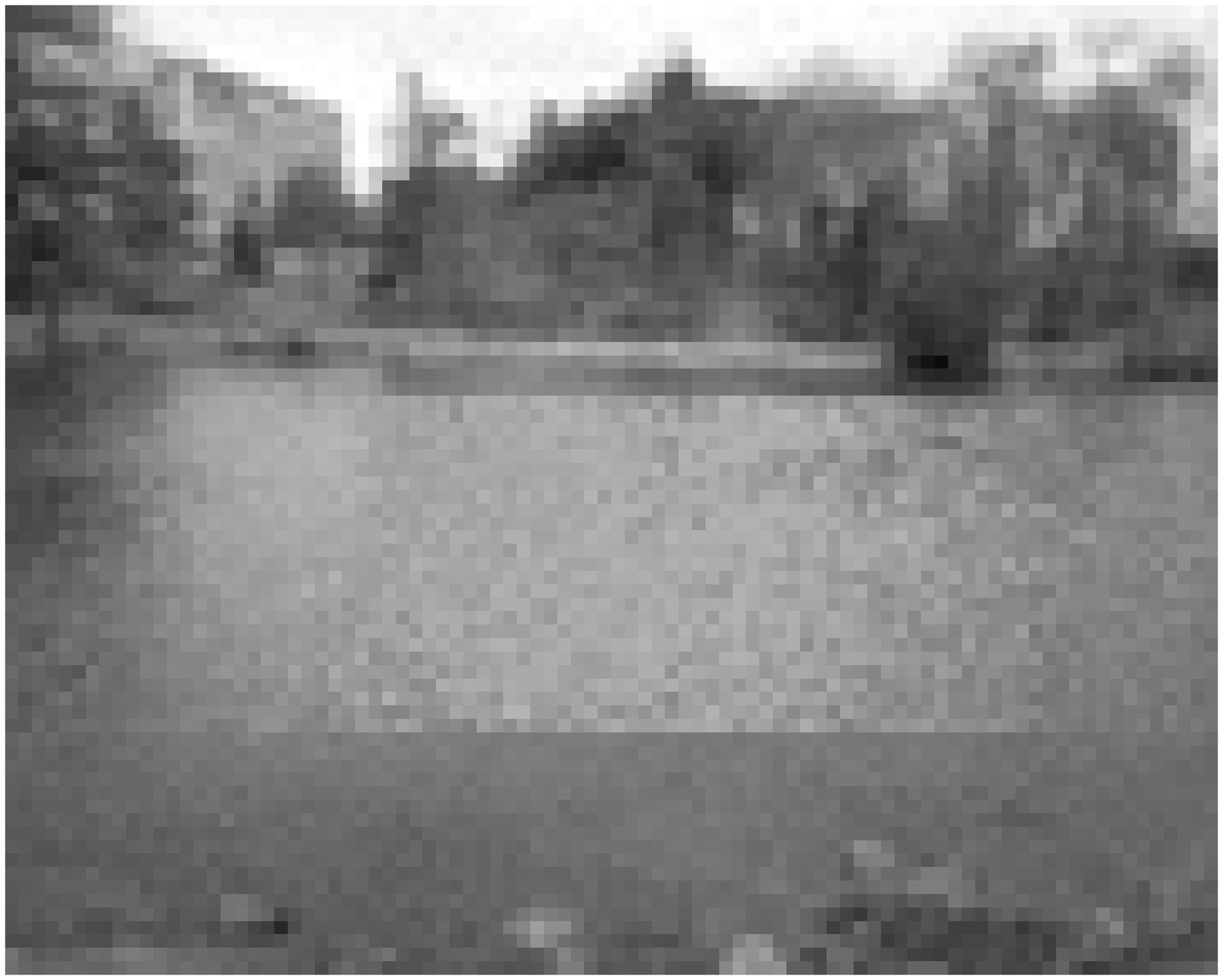}
		\includegraphics[width=15mm]{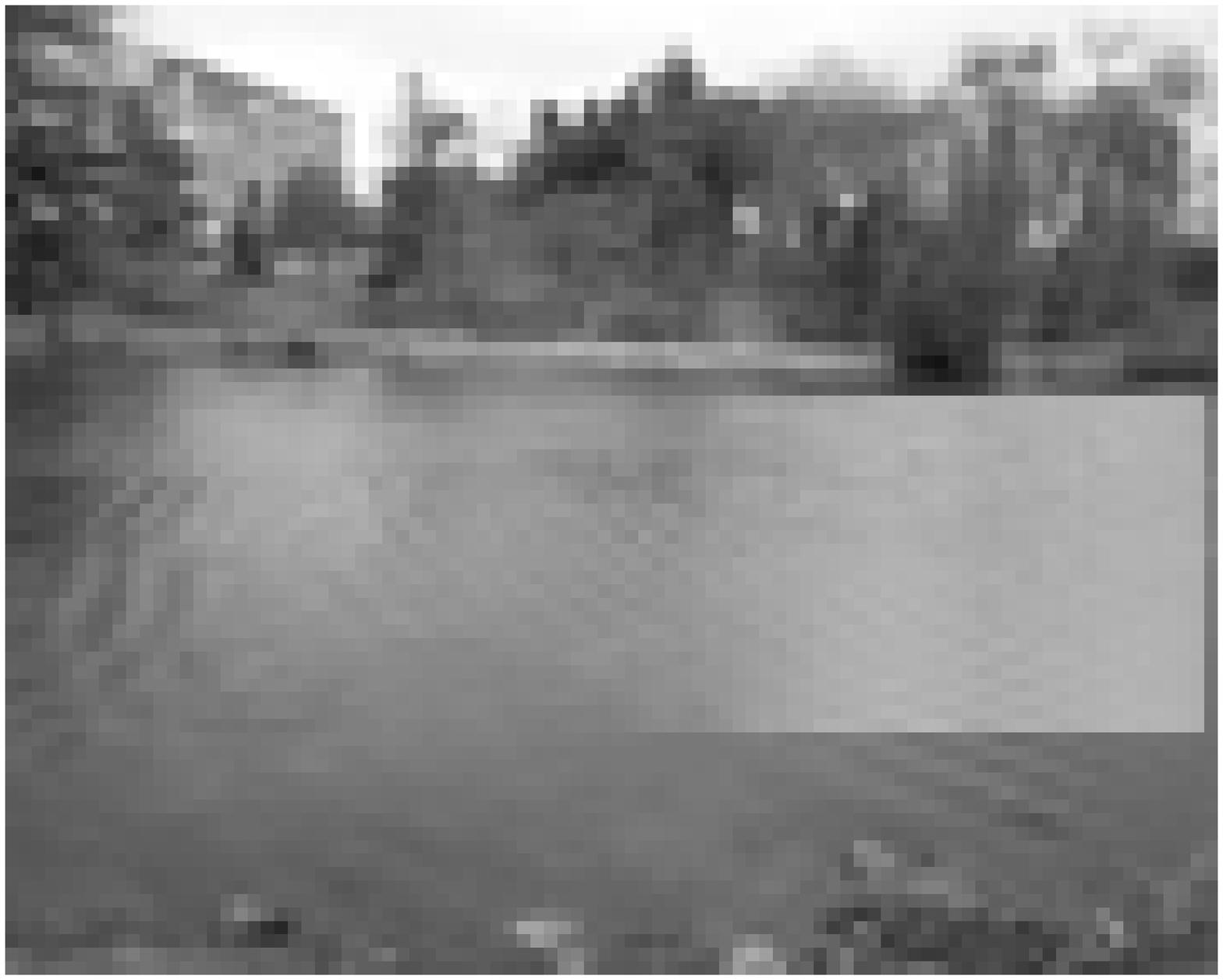}\\
		\includegraphics[width=15mm]{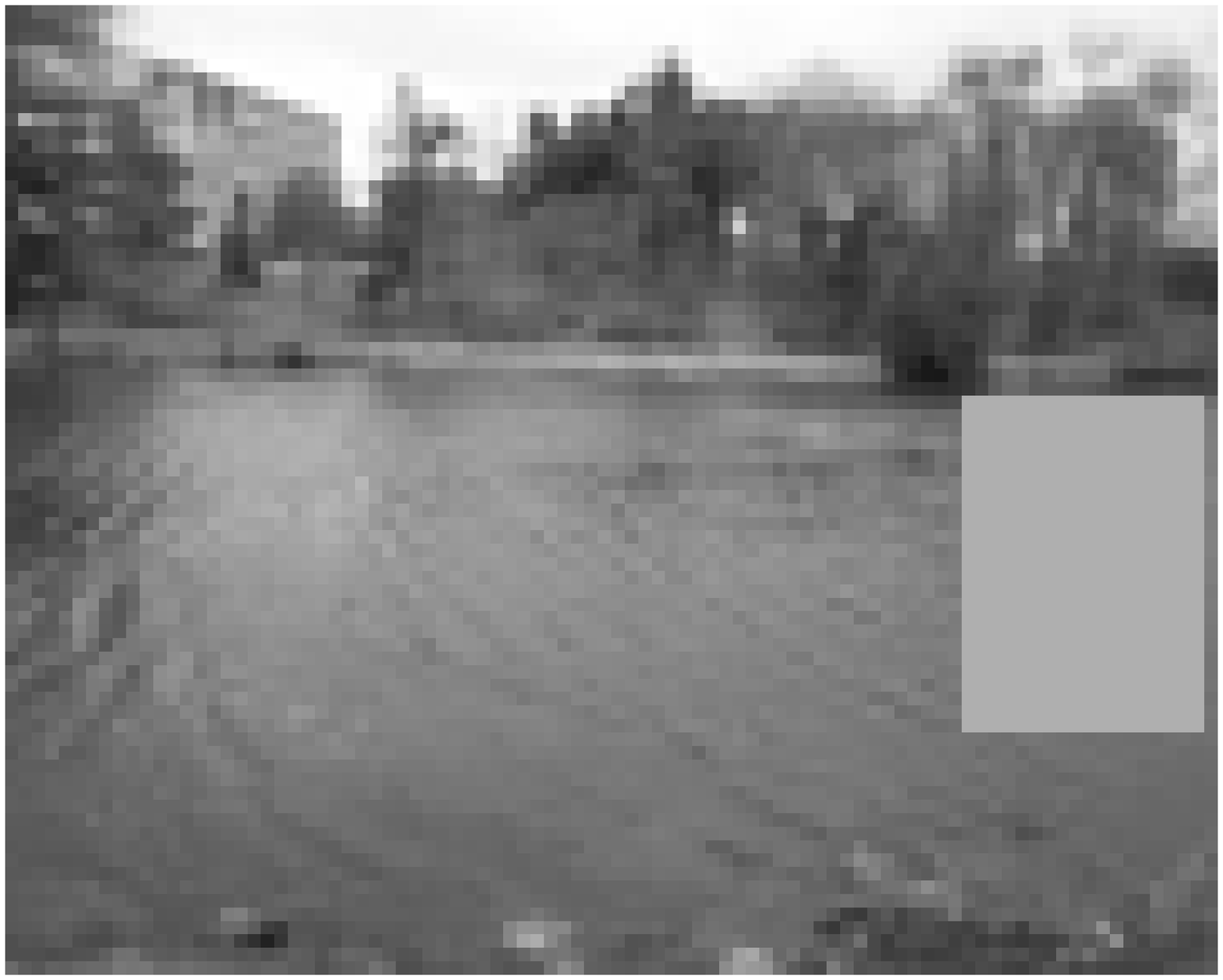}
		\includegraphics[width=15mm]{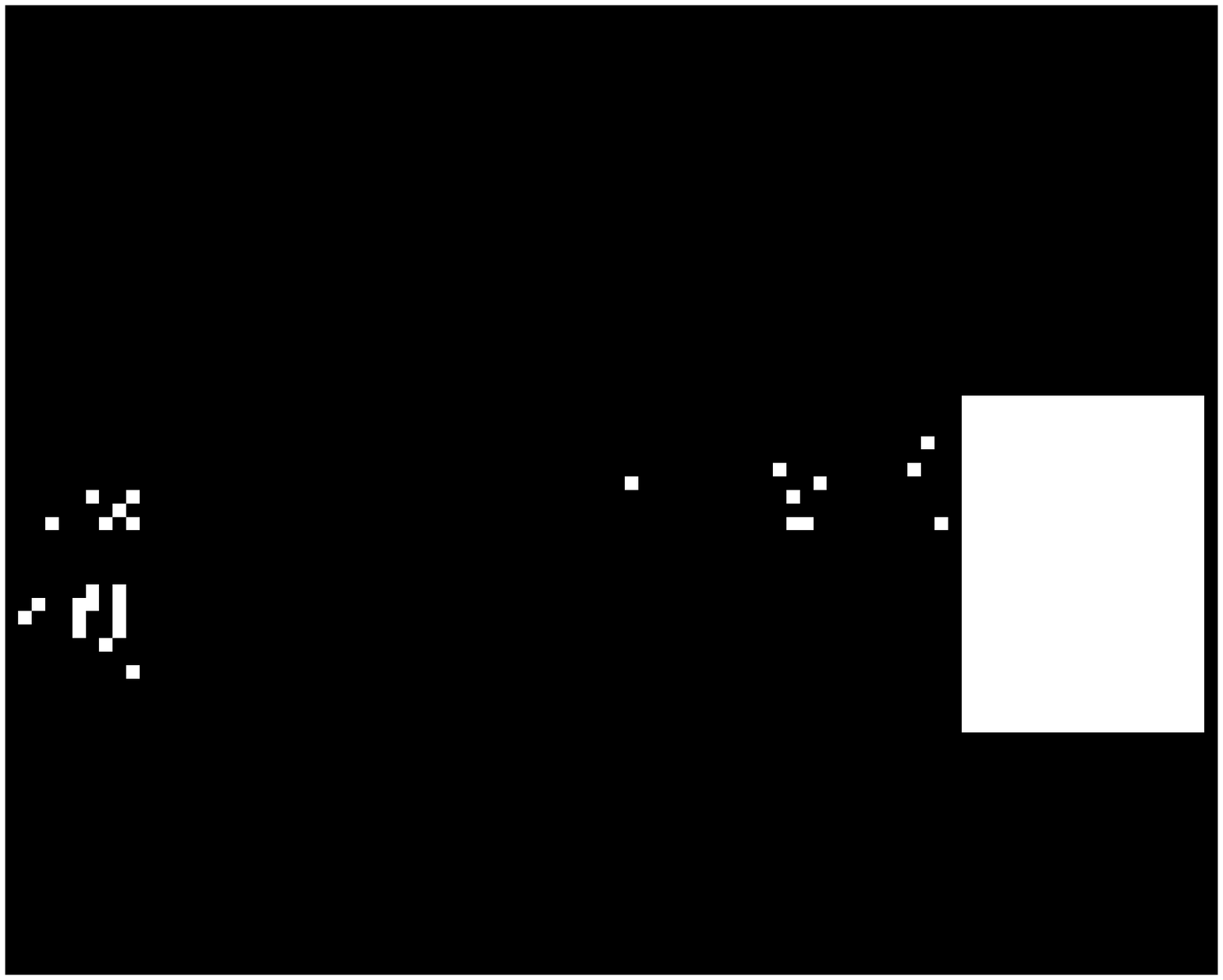}
		\includegraphics[width=15mm]{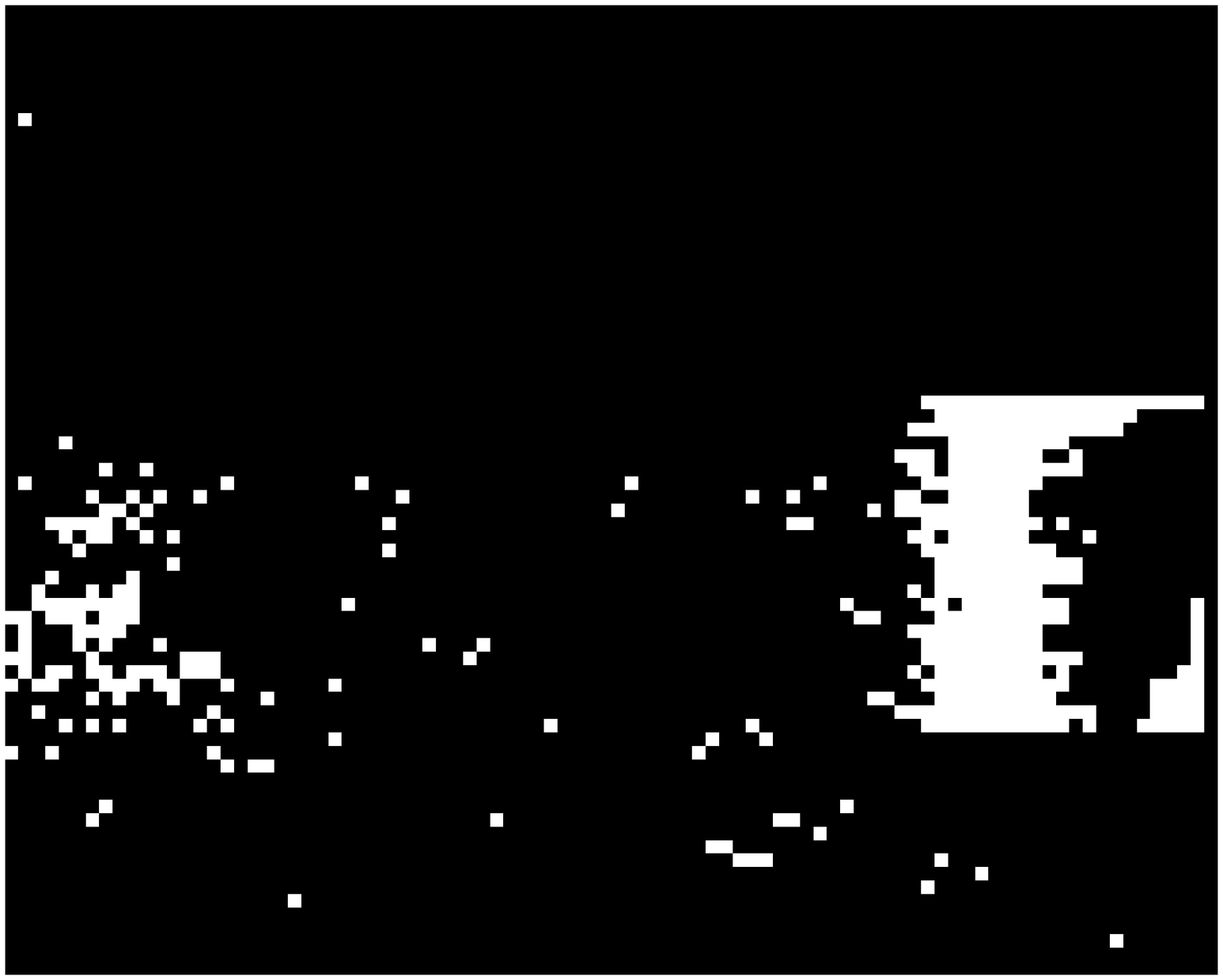}
		\includegraphics[width=15mm]{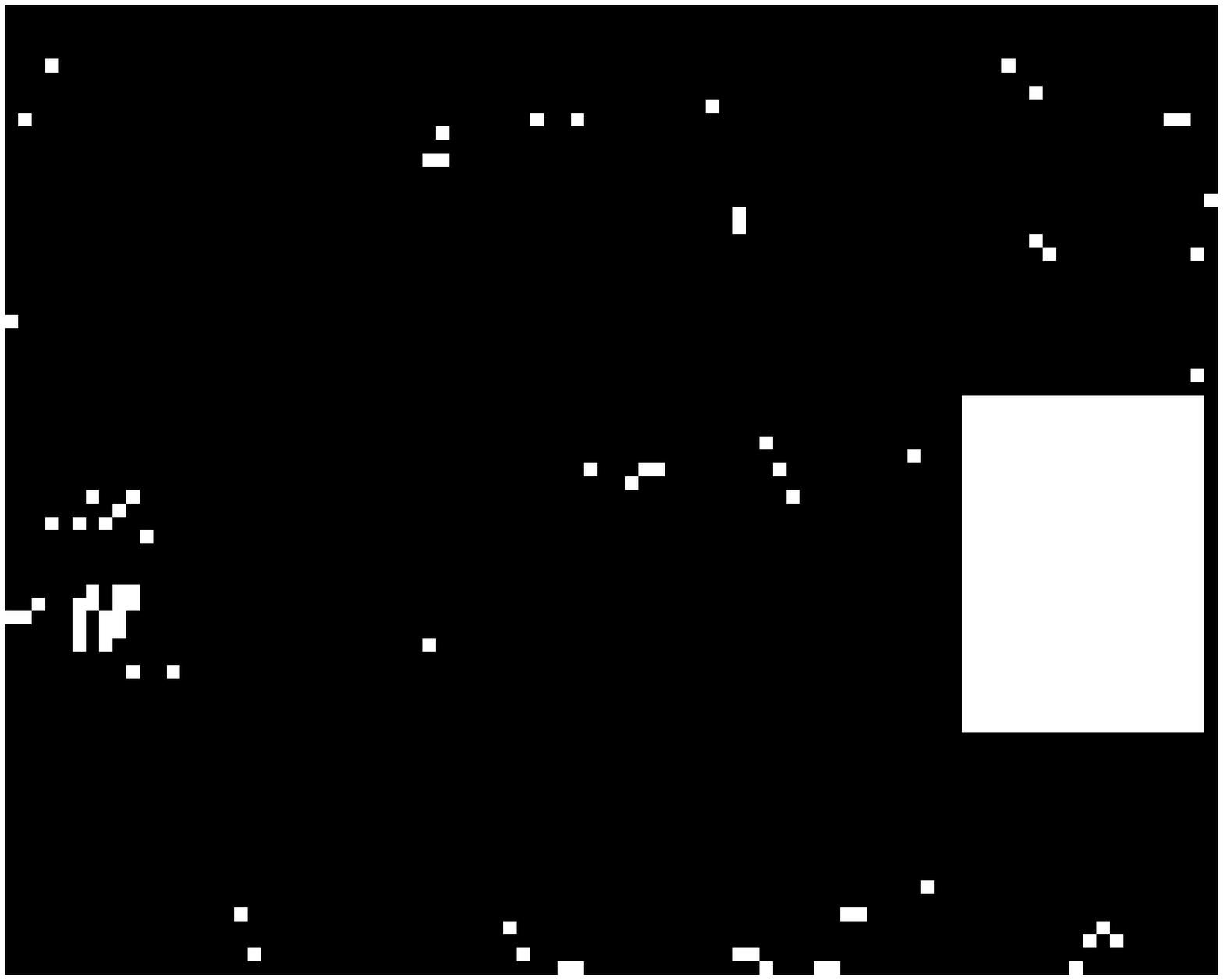}
		\includegraphics[width=15mm]{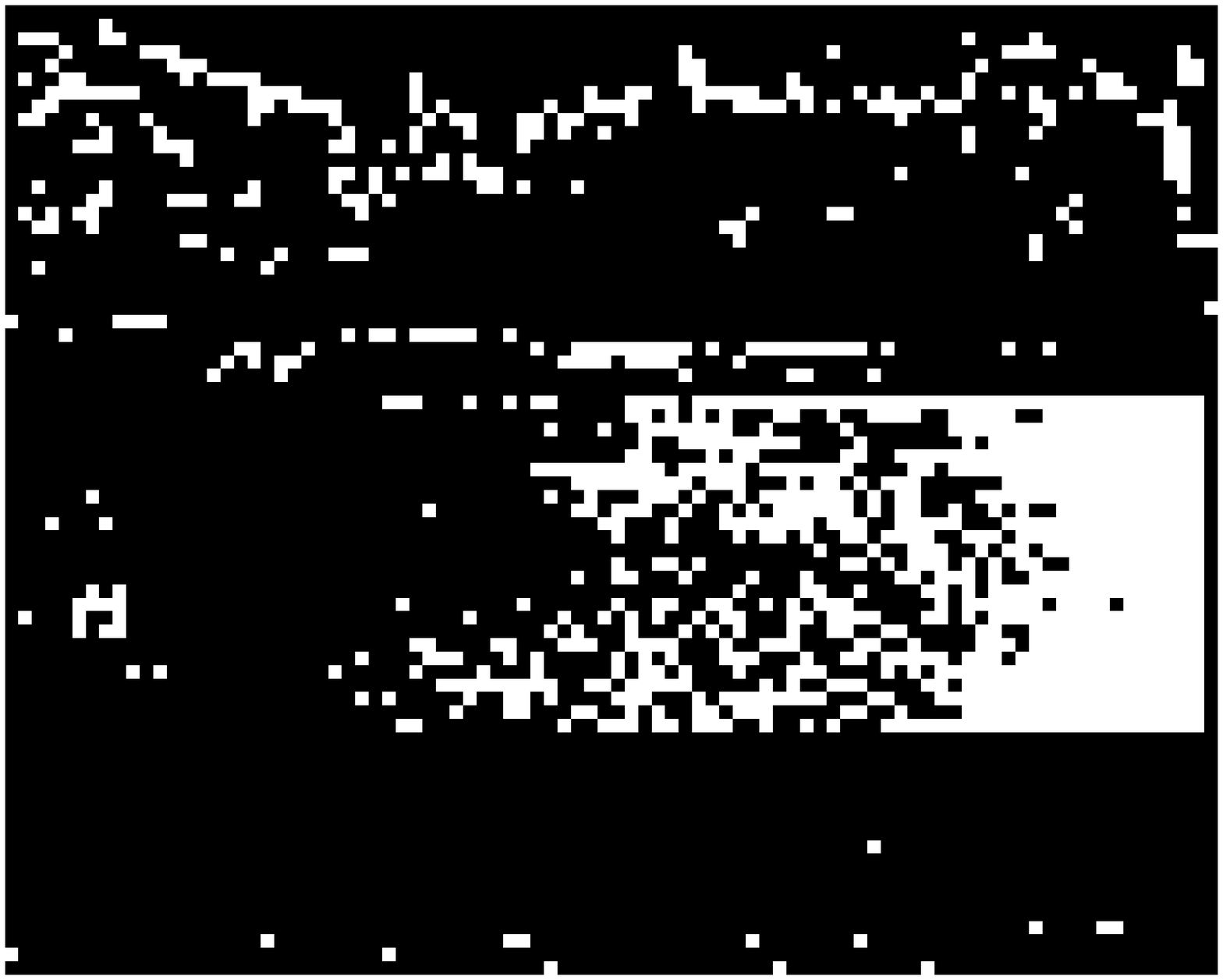}
		\includegraphics[width=15mm]{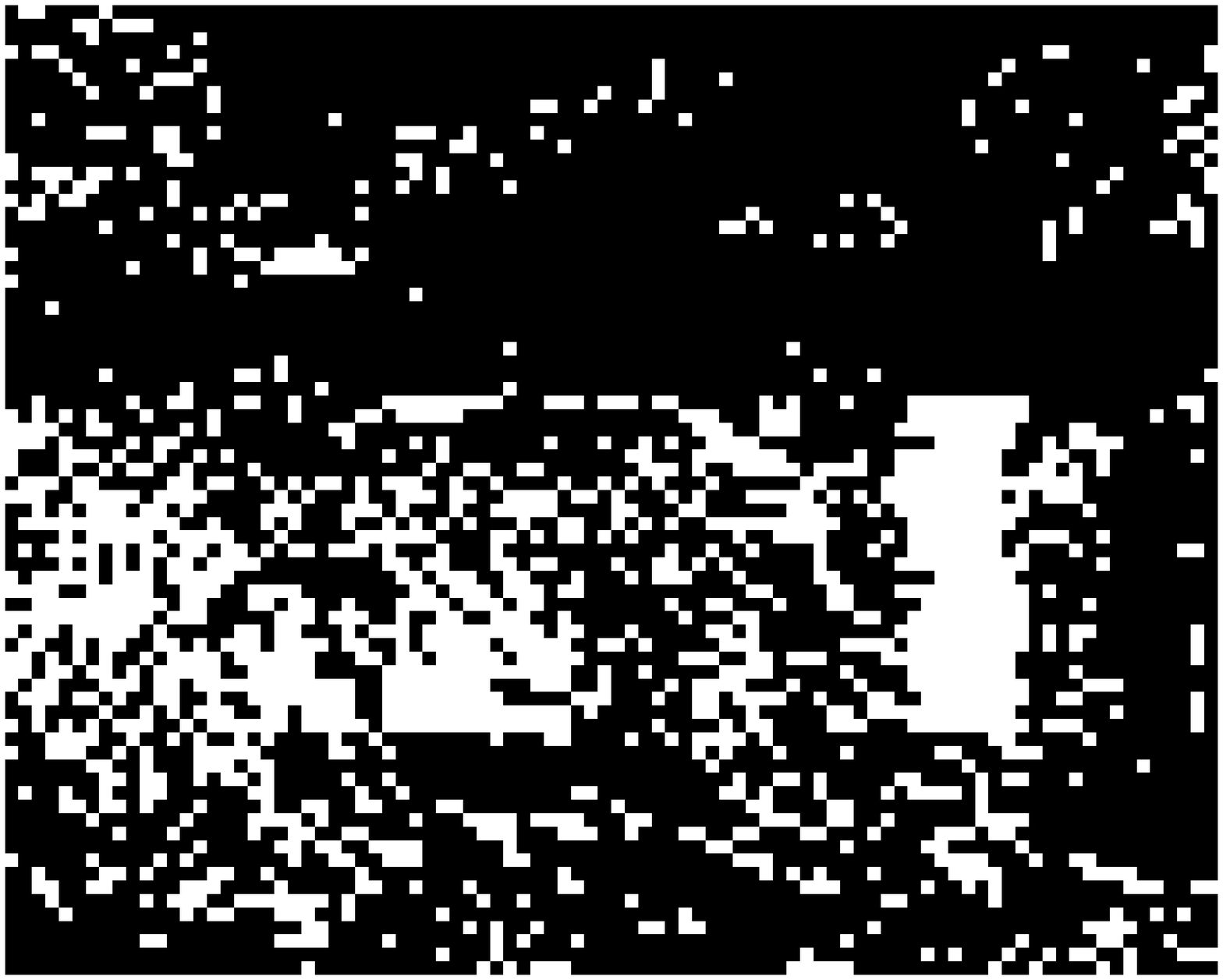}
		\includegraphics[width=15mm]{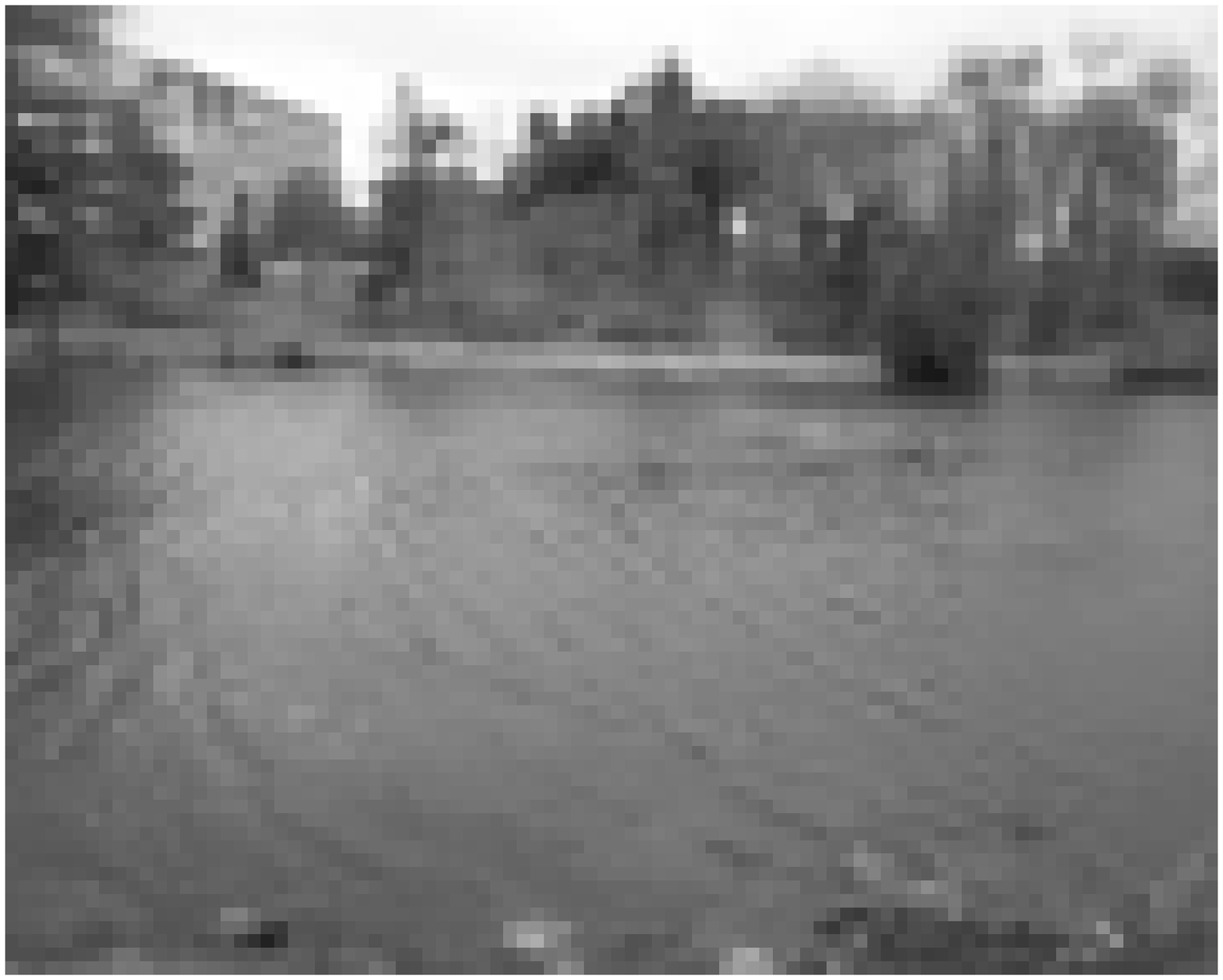}
		\includegraphics[width=15mm]{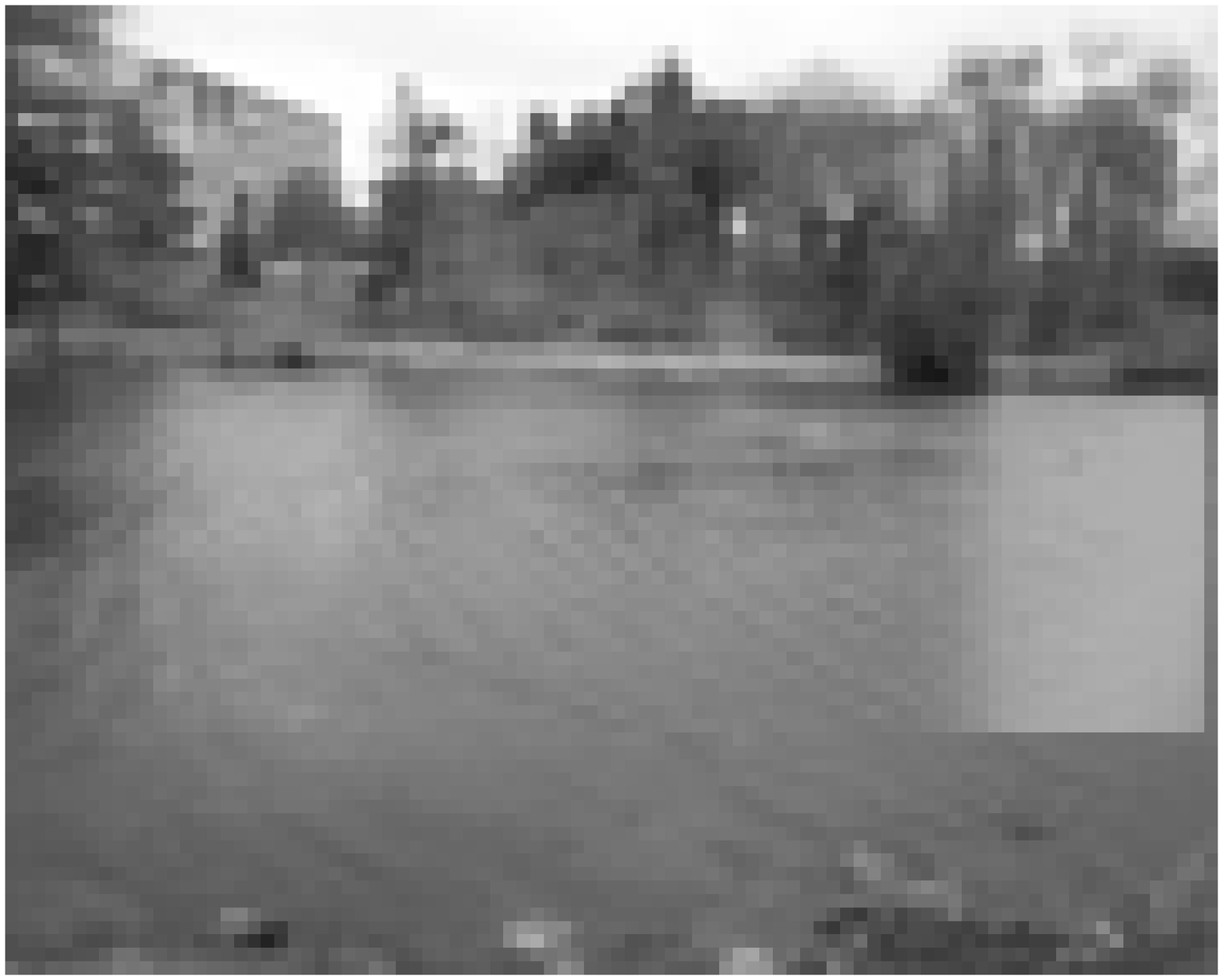}
		\includegraphics[width=15mm]{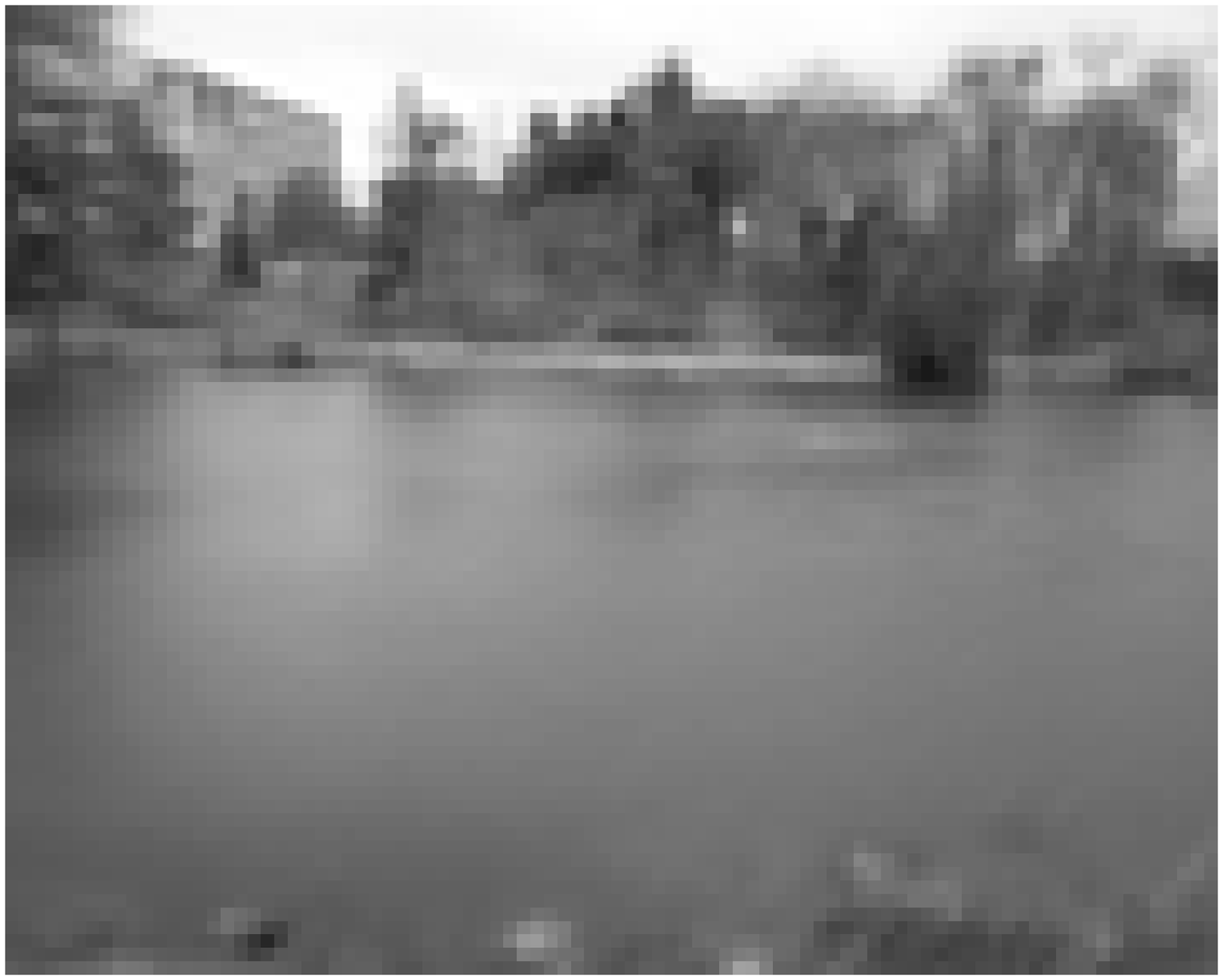}
		\includegraphics[width=15mm]{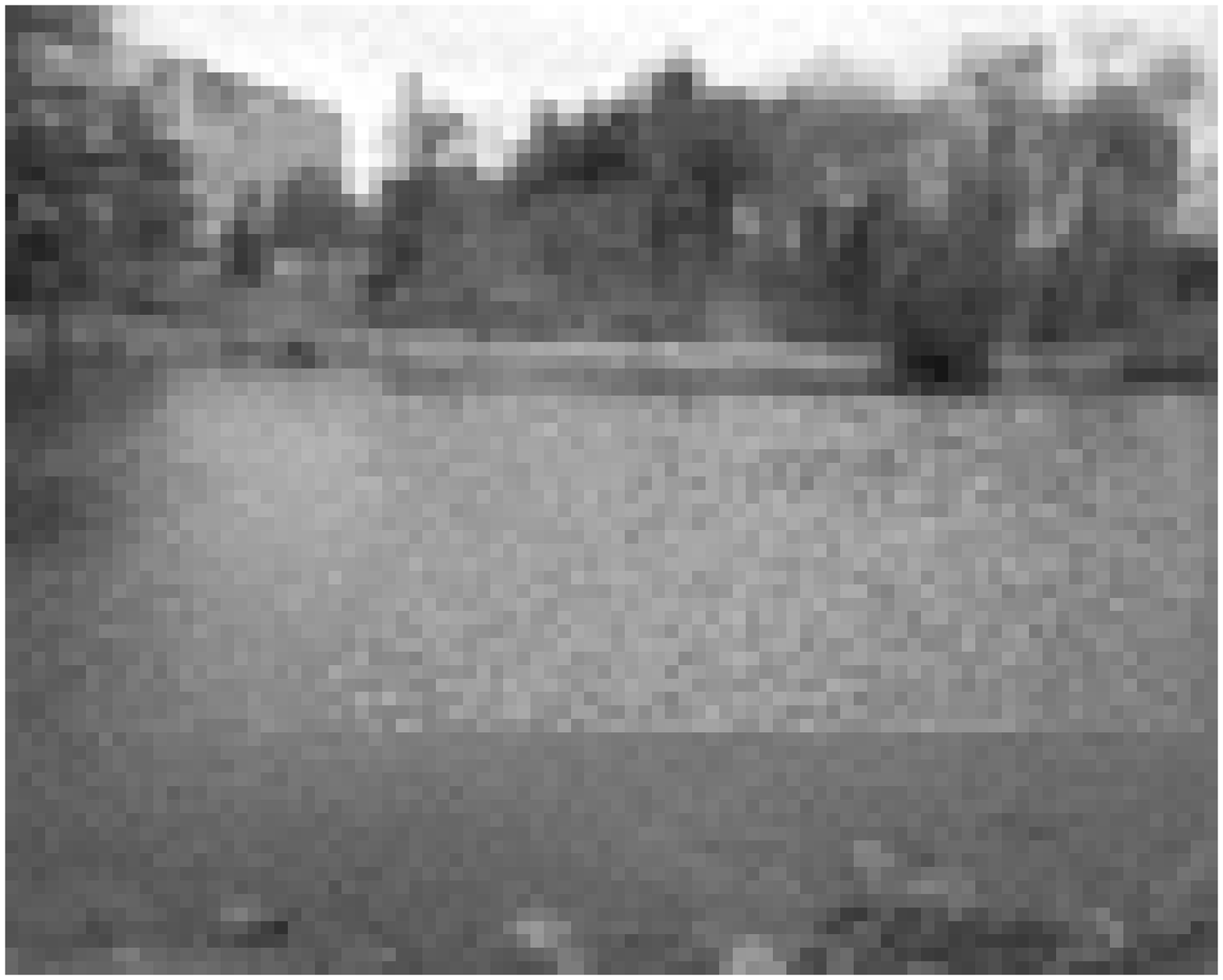}
		\includegraphics[width=15mm]{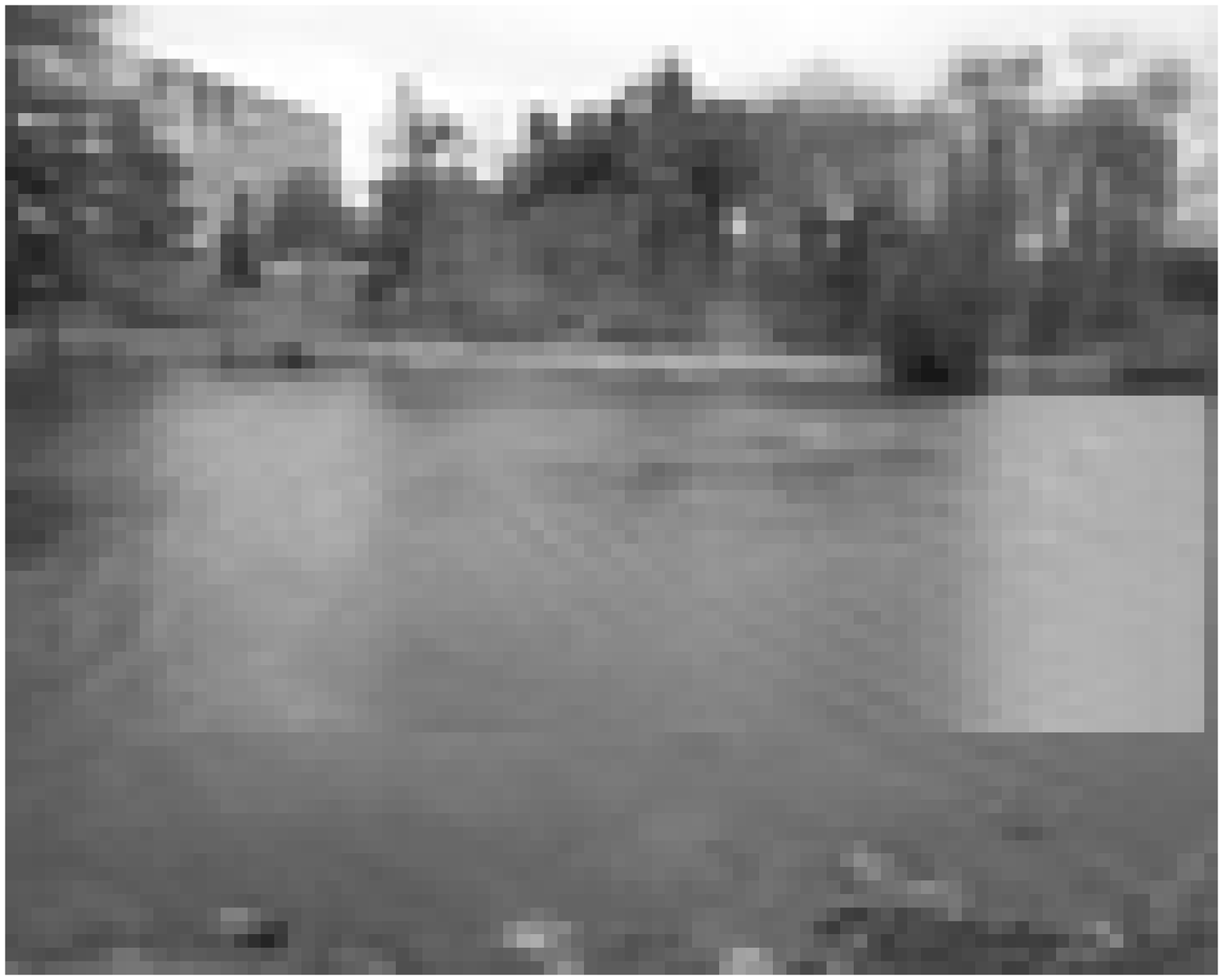}\\
		\hspace*{\fill}\makebox[0pt]{original }\hspace*{\fill}
		\hspace*{\fill}\makebox[0pt]{ReProCS}\hspace*{\fill}
		\hspace*{\fill}\makebox[0pt]{PCP}\hspace*{\fill}
		\hspace*{\fill}\makebox[0pt]{RSL}\hspace*{\fill}
		\hspace*{\fill}\makebox[0pt]{GRASTA}\hspace*{\fill}
		\hspace*{\fill}\makebox[0pt]{MG}\hspace*{\fill}
		\hspace*{\fill}\makebox[0pt]{ReProCS}\hspace*{\fill}
		\hspace*{\fill}\makebox[0pt]{PCP}\hspace*{\fill}
		\hspace*{\fill}\makebox[0pt]{RSL}\hspace*{\fill}
		\hspace*{\fill}\makebox[0pt]{GRASTA}\hspace*{\fill}
		\hspace*{\fill}\makebox[0pt]{MG}\hspace*{\fill}\\
		\hspace*{\fill}\makebox[0pt]{}\hspace*{\fill}
		\hspace*{\fill}\makebox[0pt]{(fg)}\hspace*{\fill}
		\hspace*{\fill}\makebox[0pt]{(fg)}\hspace*{\fill}
		\hspace*{\fill}\makebox[0pt]{(fg)}\hspace*{\fill}
		\hspace*{\fill}\makebox[0pt]{(fg)}\hspace*{\fill}
		\hspace*{\fill}\makebox[0pt]{(fg)}\hspace*{\fill}
		\hspace*{\fill}\makebox[0pt]{(bg)}\hspace*{\fill}
		\hspace*{\fill}\makebox[0pt]{(bg)}\hspace*{\fill}
		\hspace*{\fill}\makebox[0pt]{(bg)}\hspace*{\fill}
		\hspace*{\fill}\makebox[0pt]{(bg)}\hspace*{\fill}
		\hspace*{\fill}\makebox[0pt]{(bg)}\hspace*{\fill}
	\end{tabular}
\caption{\small{Original video at $t=t_\train+30,60,70$ and its foreground (fg) and background (bg) layer recovery results using ReProCS (ReProCS-pCA) and other algorithms. 
MG refers to the batch algorithm of \cite{mateos_anomaly2,mateos_anomaly} implemented using code provided by the authors. There was not enough information in the papers or in the code to successfully implement the recursive algorithm. For fg, we only show the fg support in white for ease of display.
}}
\label{LakeCompare}
\end{figure*}

We generated 50 realizations of the video sequence using these parameters and compared all the algorithms to estimate $S_t$, $L_t$ and then the foreground and the background sequences. We show comparisons of the normalized mean squared error (NMSE) in recovering $S_t$ in Fig. \ref{NMSE}. Visual comparisons of both foreground and background recovery for one realization are shown in Fig. \ref{LakeCompare}. The recovered foreground image is shown as a white-black image showing the foreground support: pixels in the support estimate are white.
PCP gives large error for this sequence since the object moves in a highly correlated fashion and occupies a large part of the image. GRASTA also does not work. 
RSL is able to recover a large part of the object correctly, however it also recovers many more extras than ReProCS. The reason is that the magnitude of the nonzero entries of $S_t$ is quite small (recall that $(S_t)_i = (F_t-B_t)_i$ for $i \in T_t$) and is such that $\|L_t\|_2$ is about as large as $\|S_t\|_2$ or sometimes larger (see Fig. \ref{NormCompare}).

\begin{figure*}
	\centering
	\begin{tabular}{cc}
		\includegraphics[width=15mm]{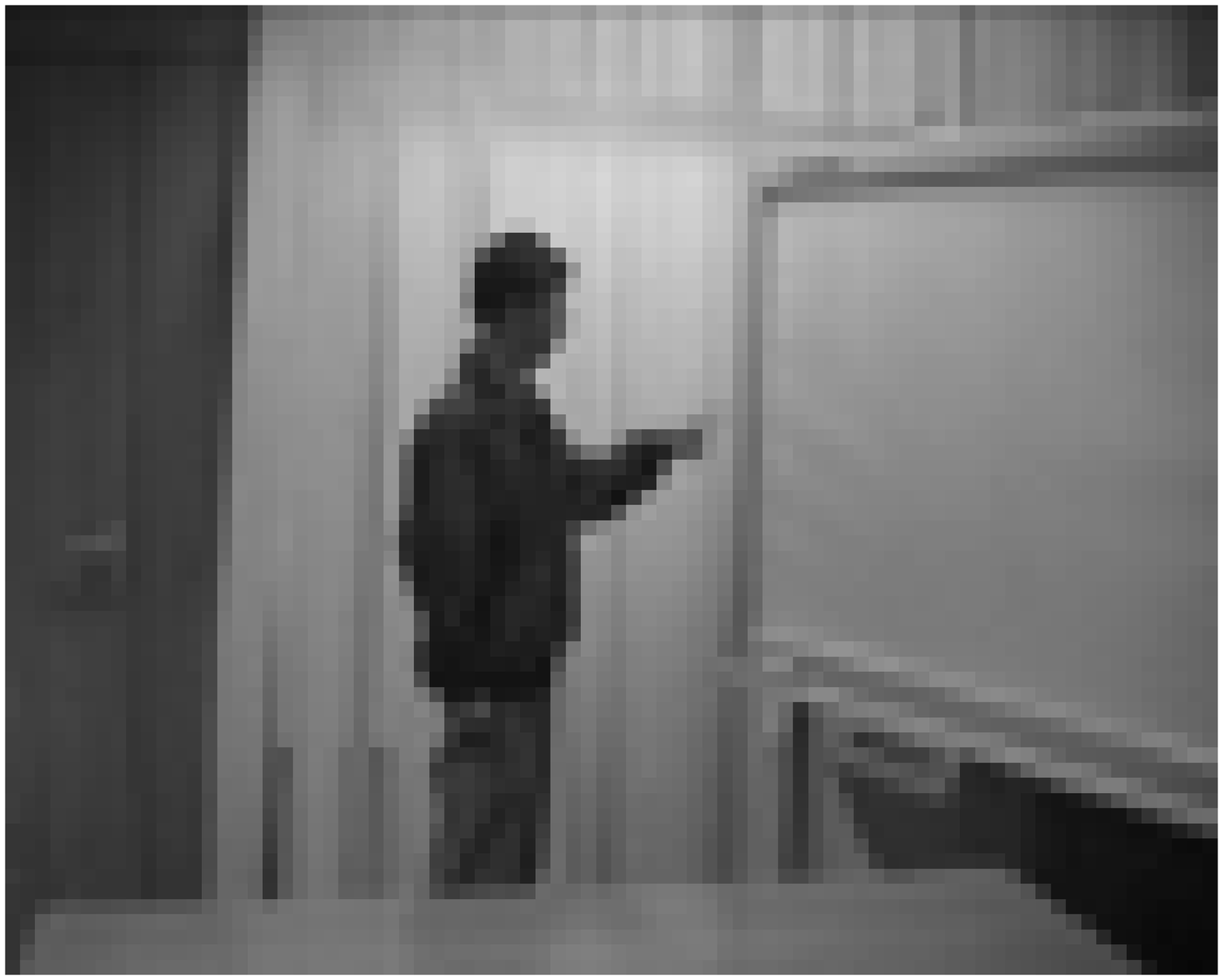}
		\includegraphics[width=15mm]{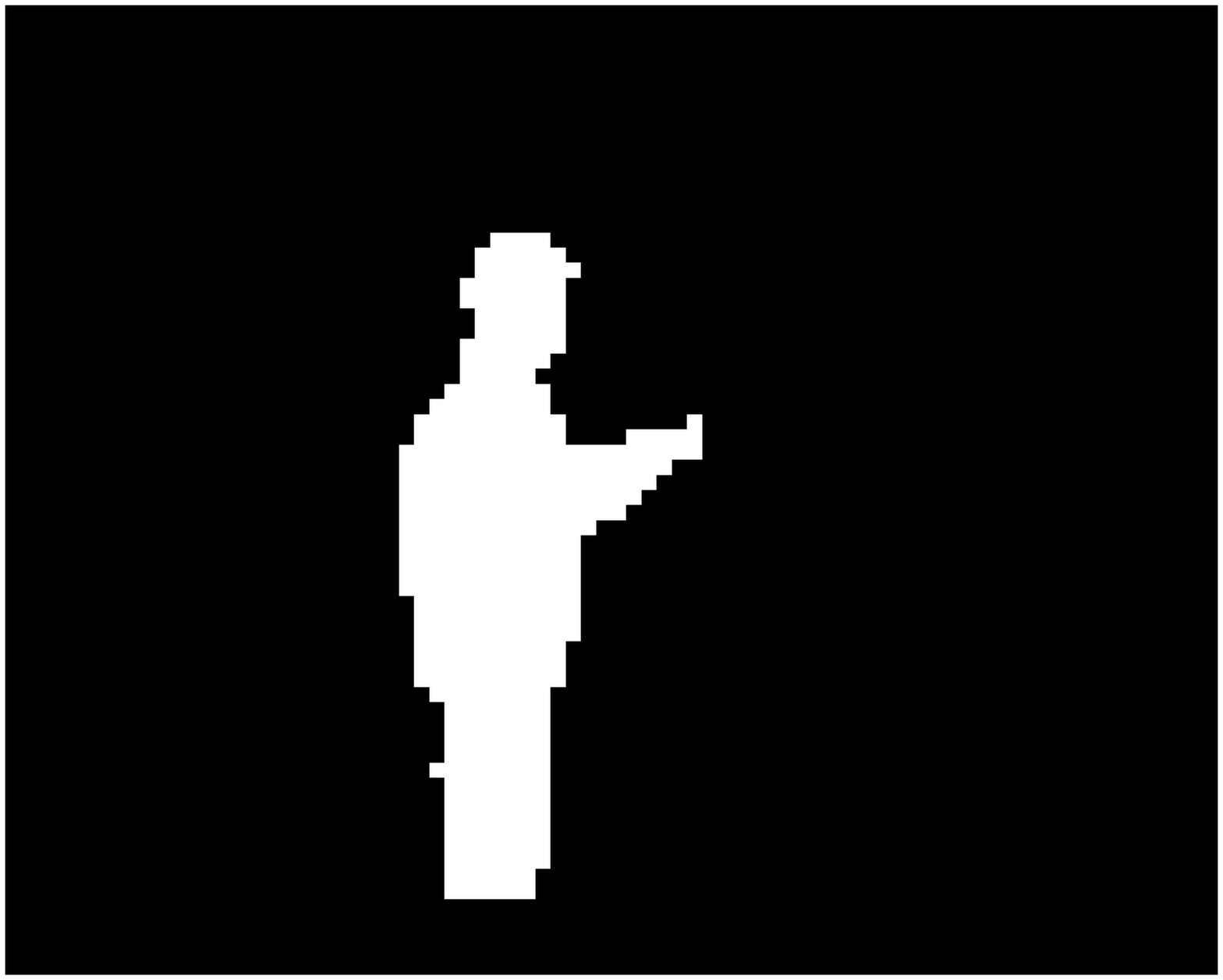}
		\includegraphics[width=15mm]{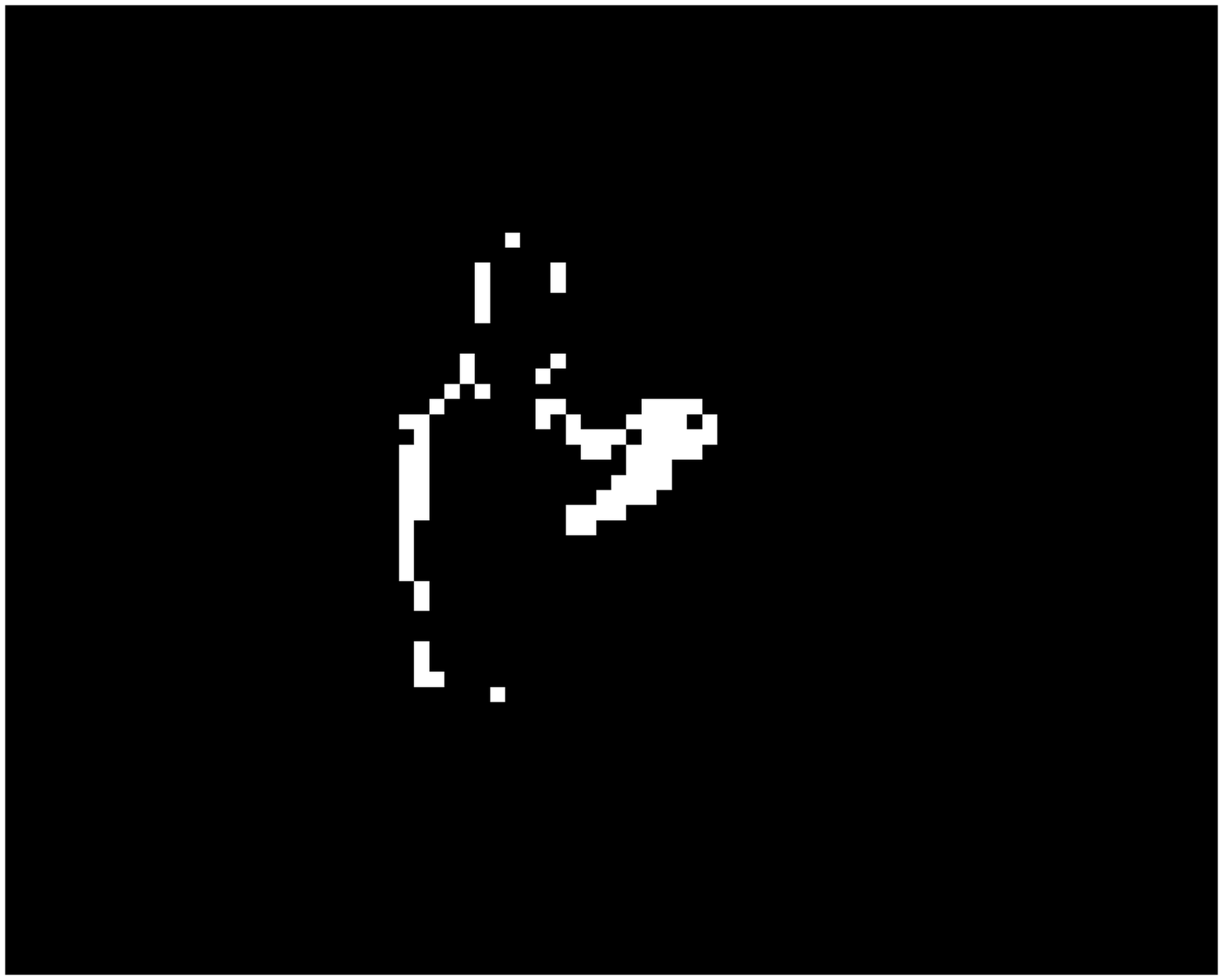}
		\includegraphics[width=15mm]{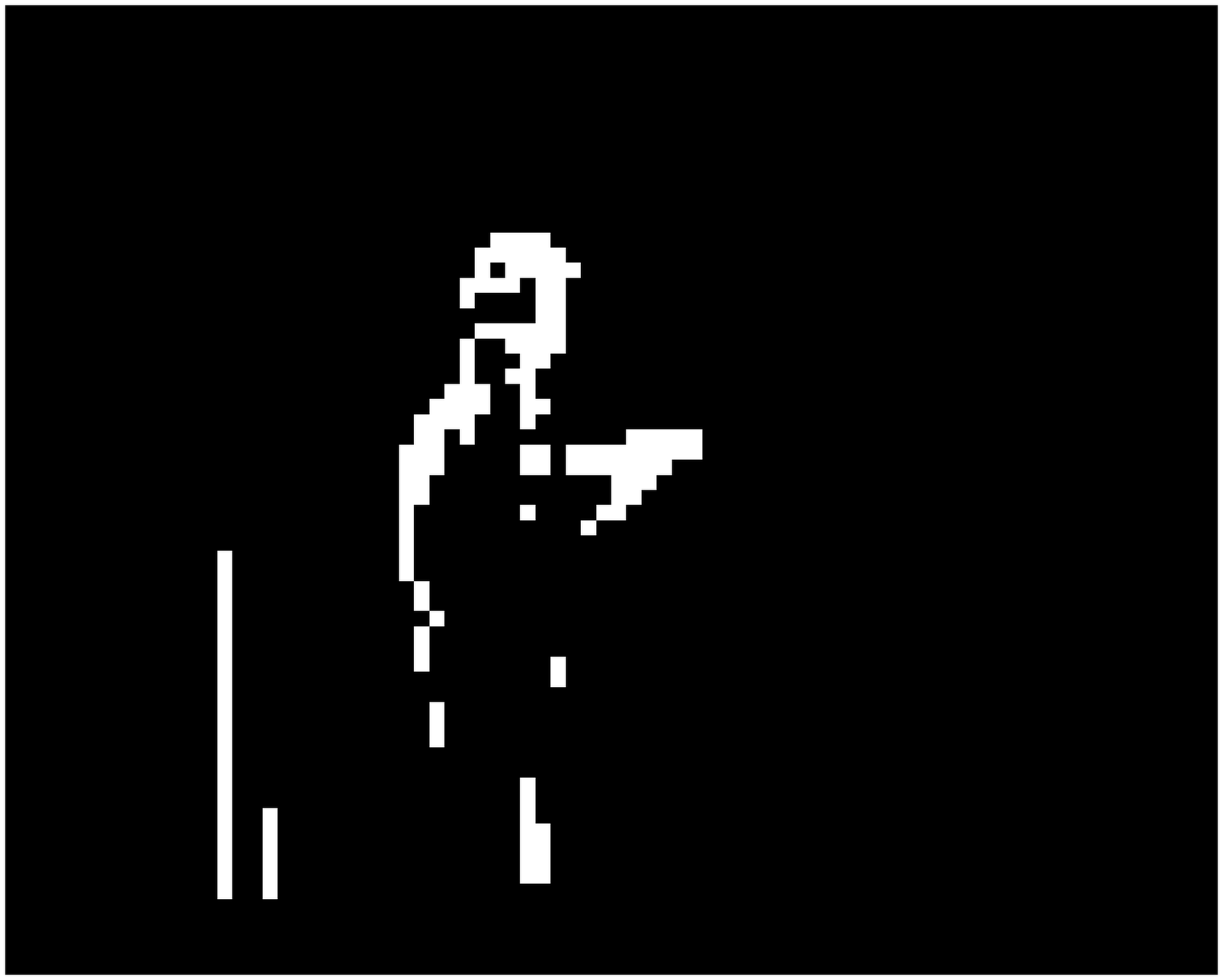}
		\includegraphics[width=15mm]{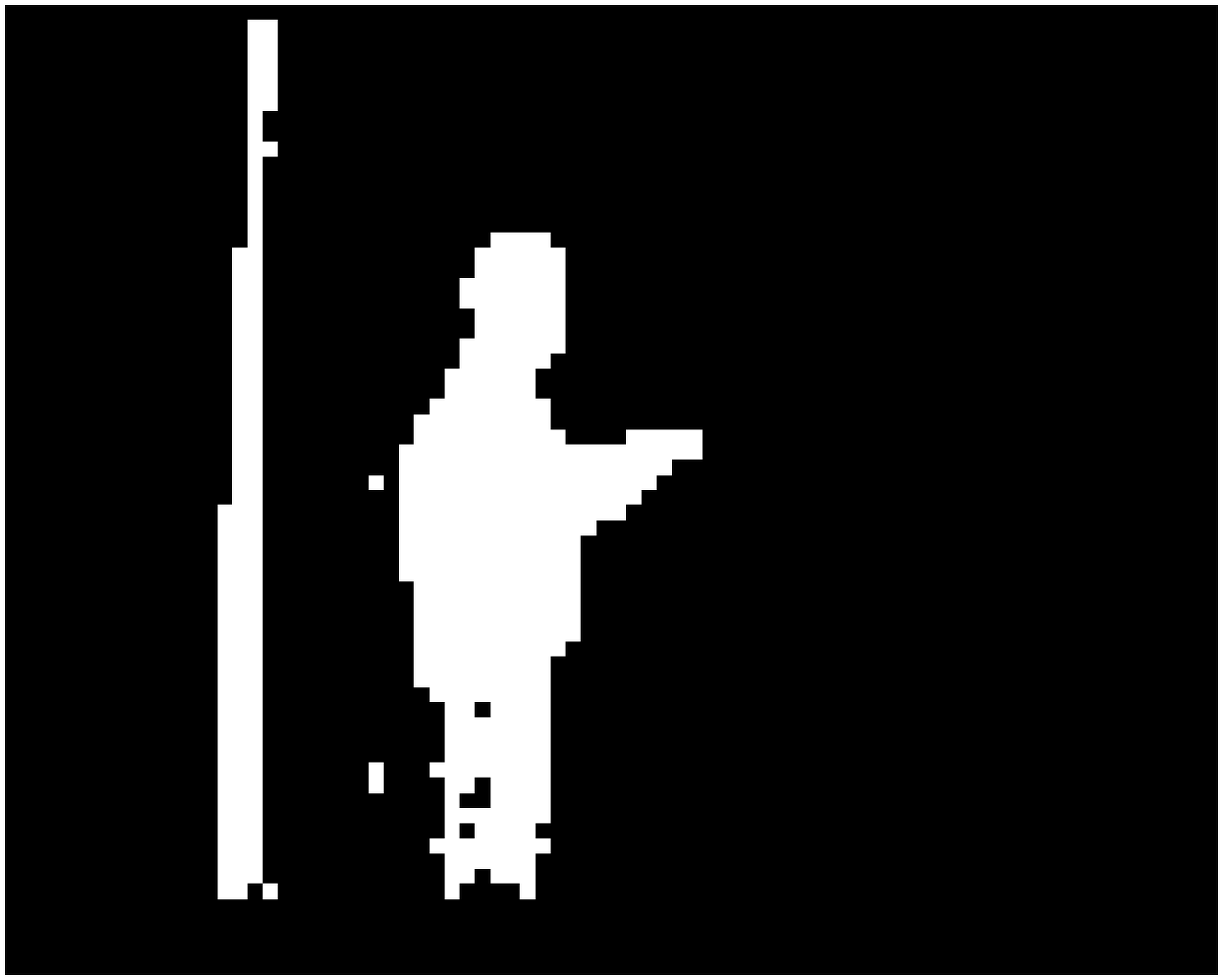}
		\includegraphics[width=15mm]{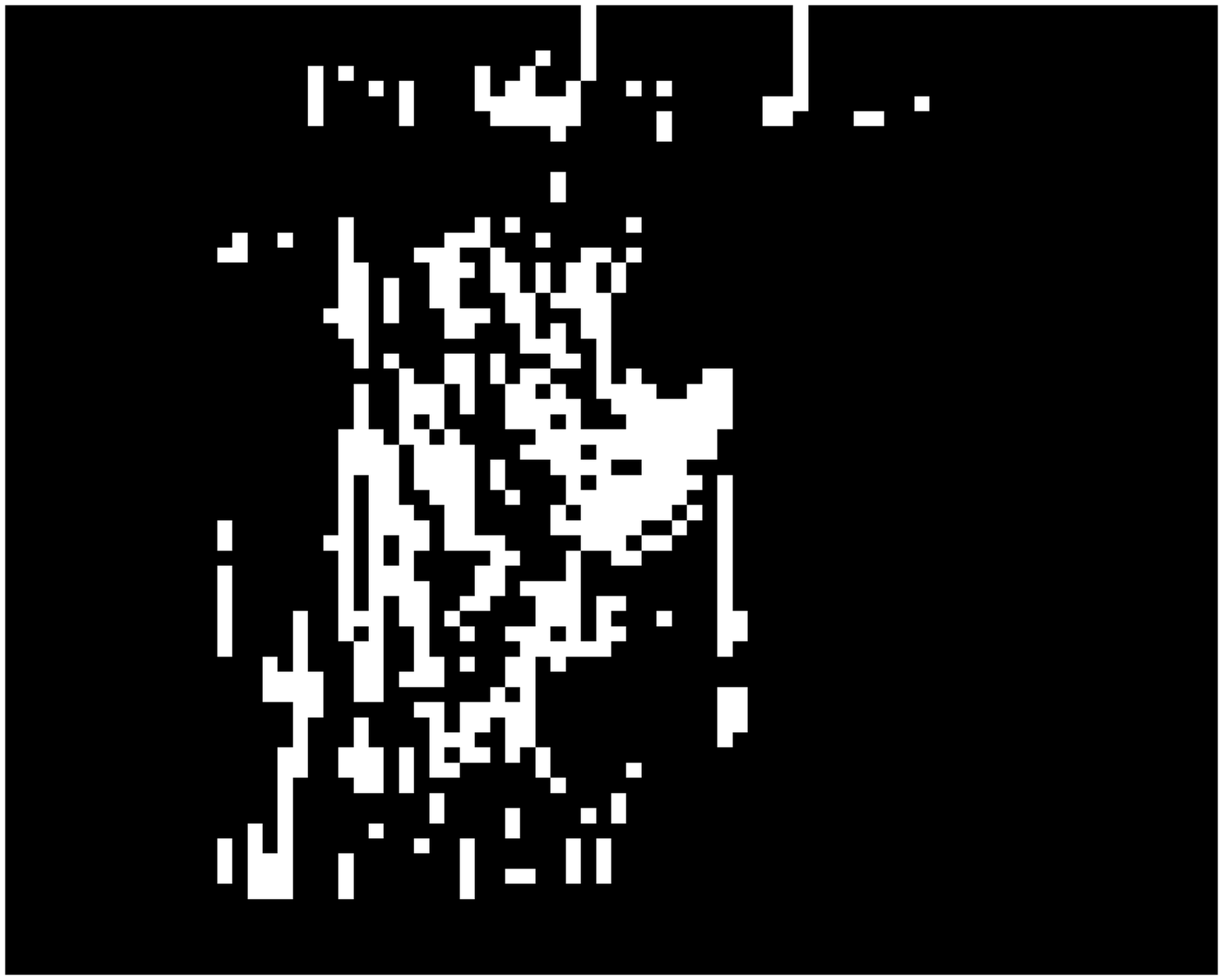}
		\includegraphics[width=15mm]{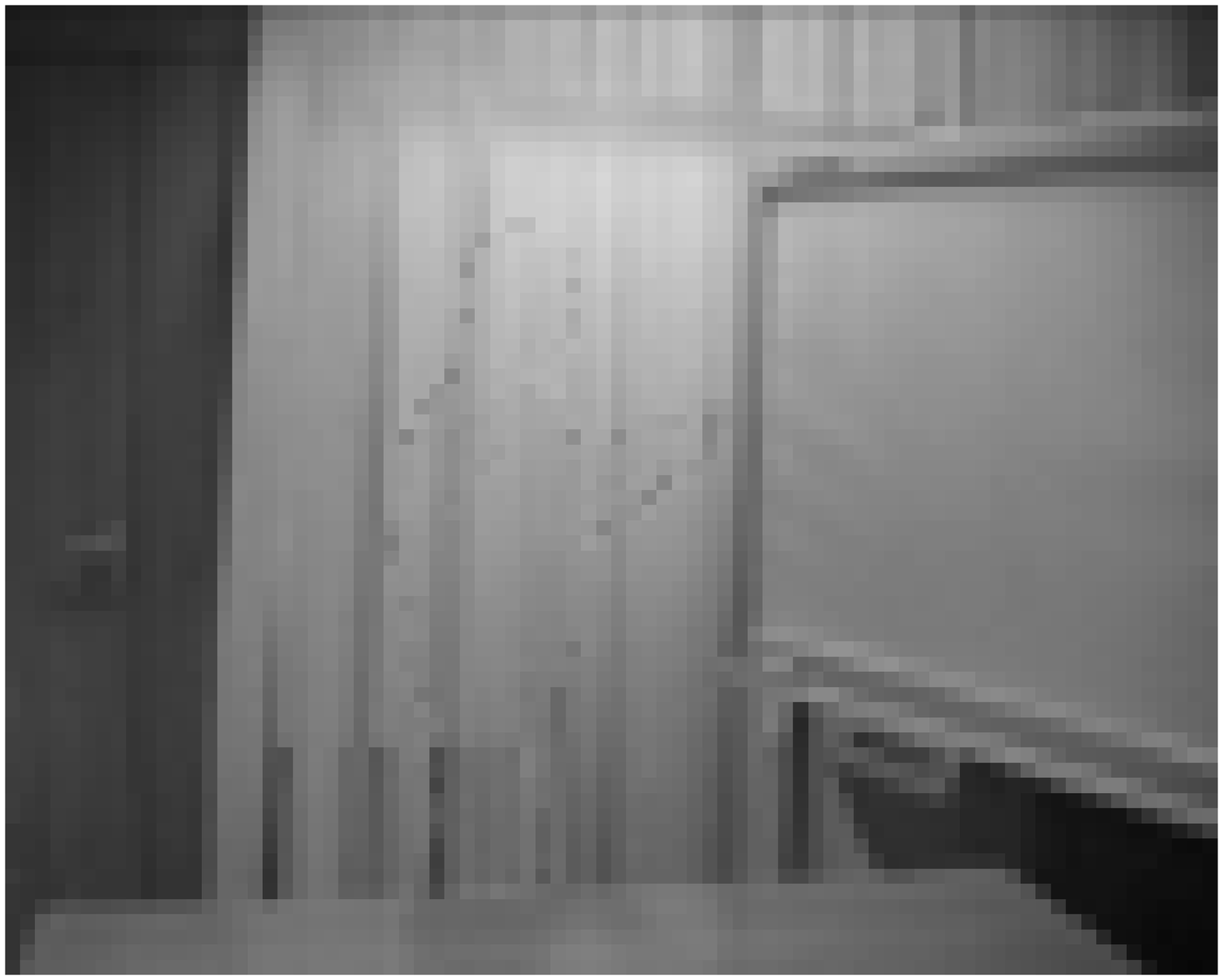}
		\includegraphics[width=15mm]{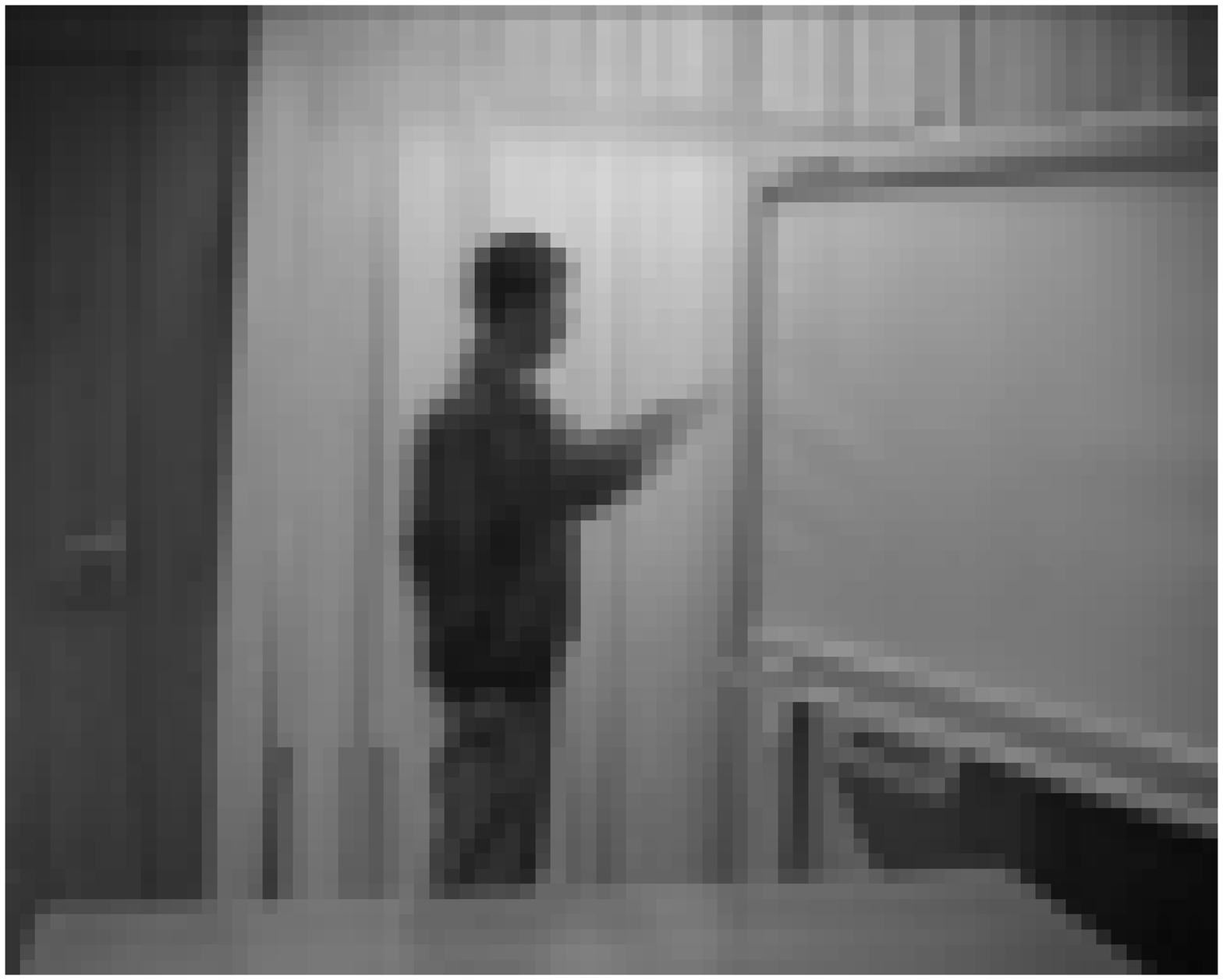}
		\includegraphics[width=15mm]{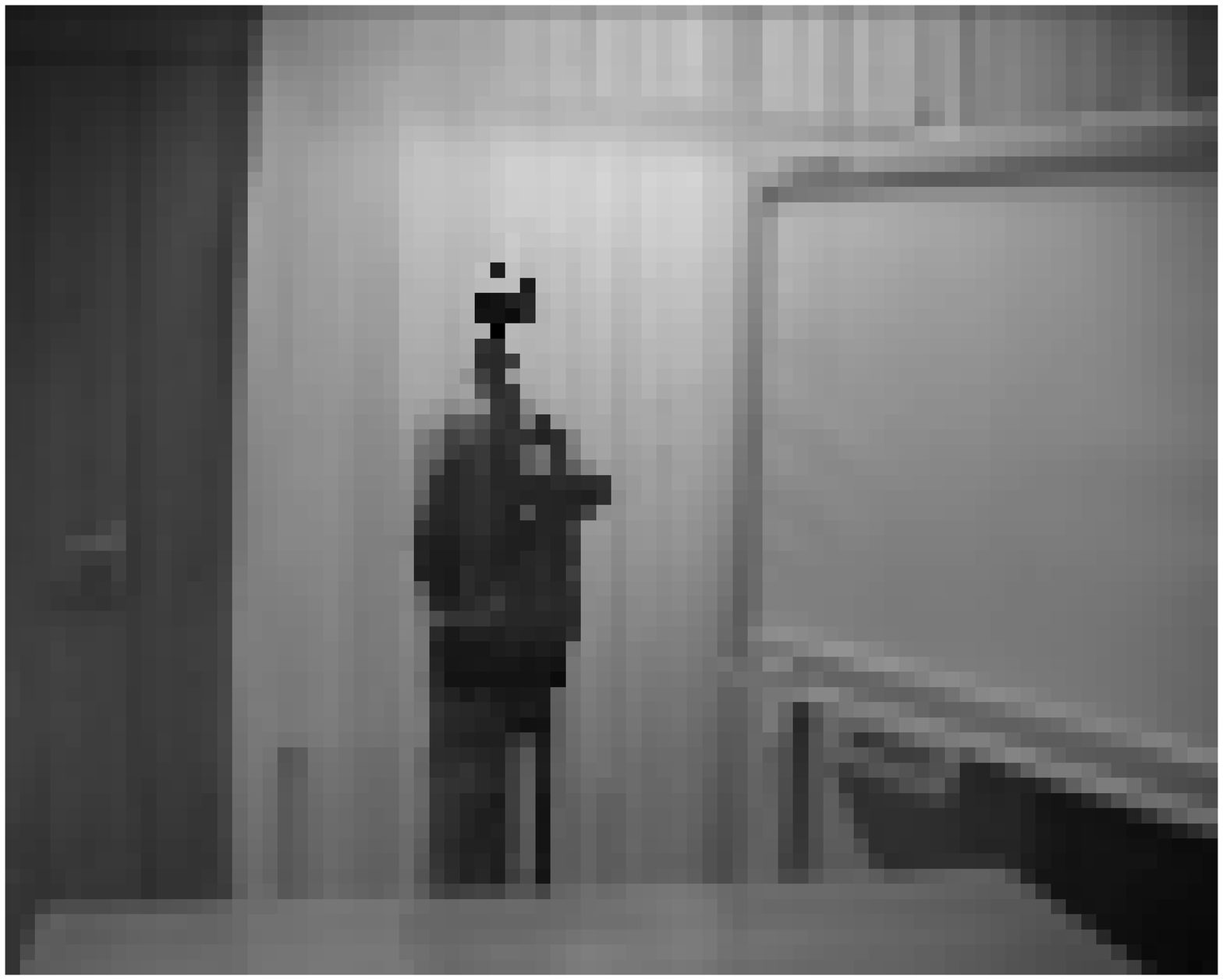}
		\includegraphics[width=15mm]{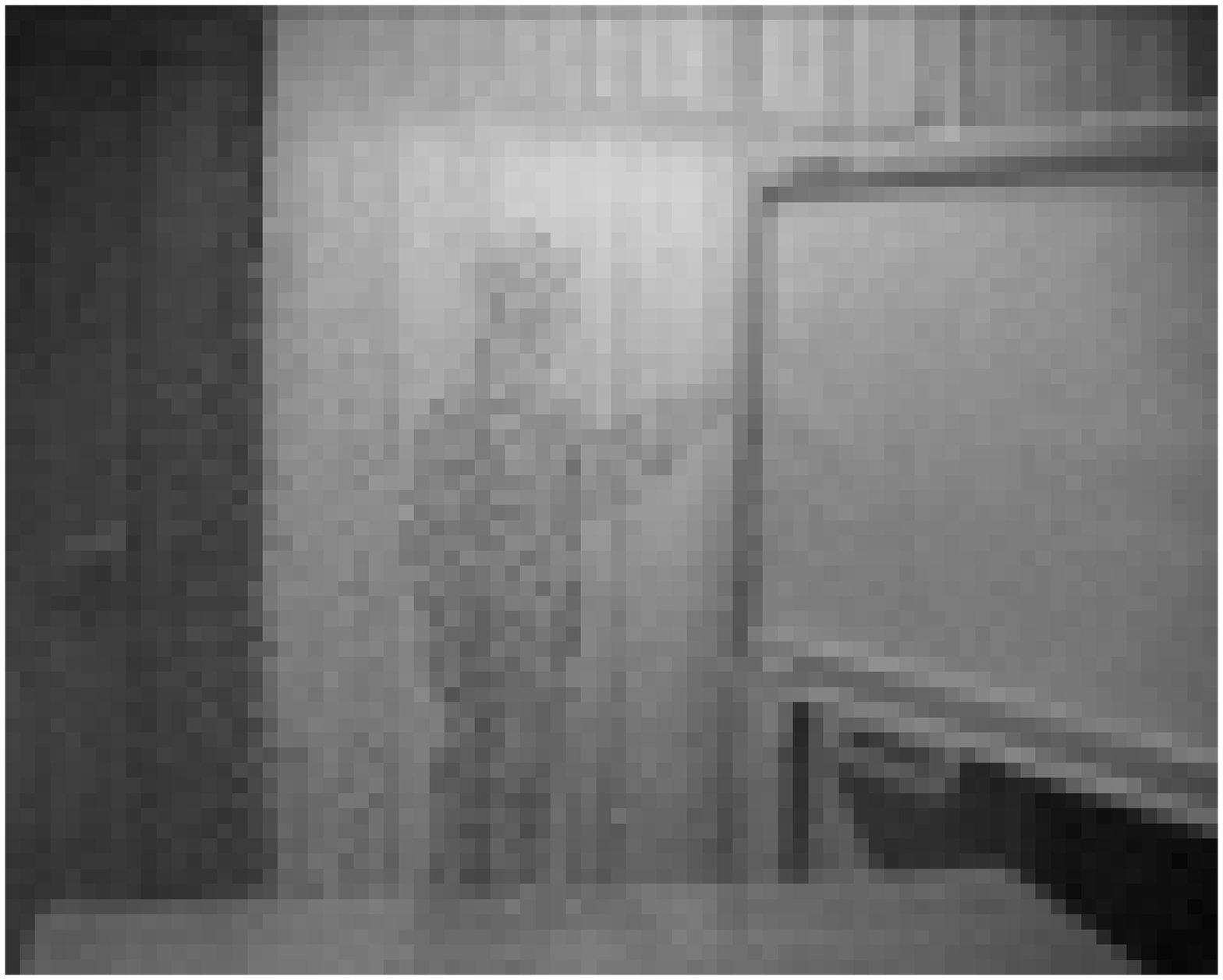}
		\includegraphics[width=15mm]{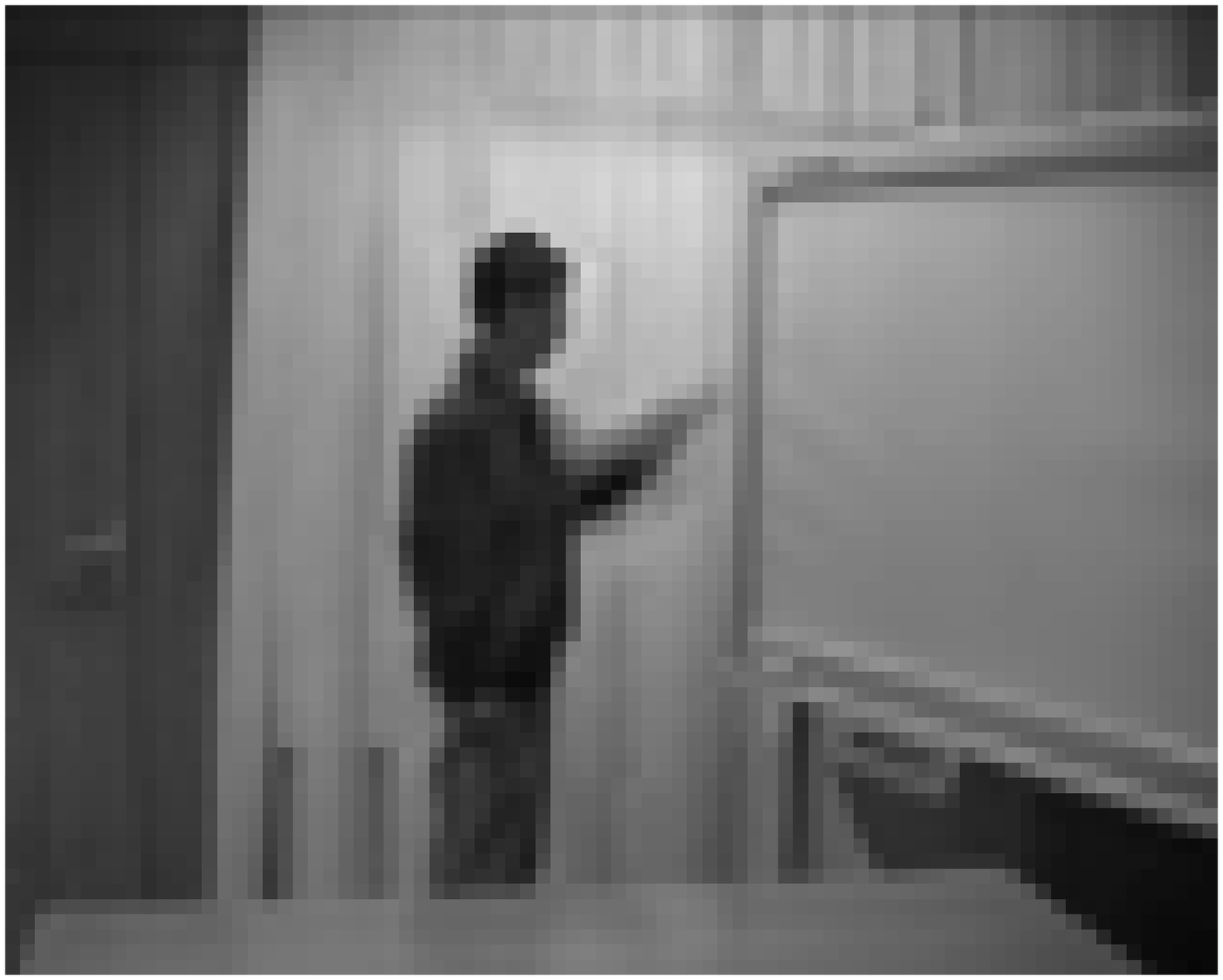}\\
		\includegraphics[width=15mm]{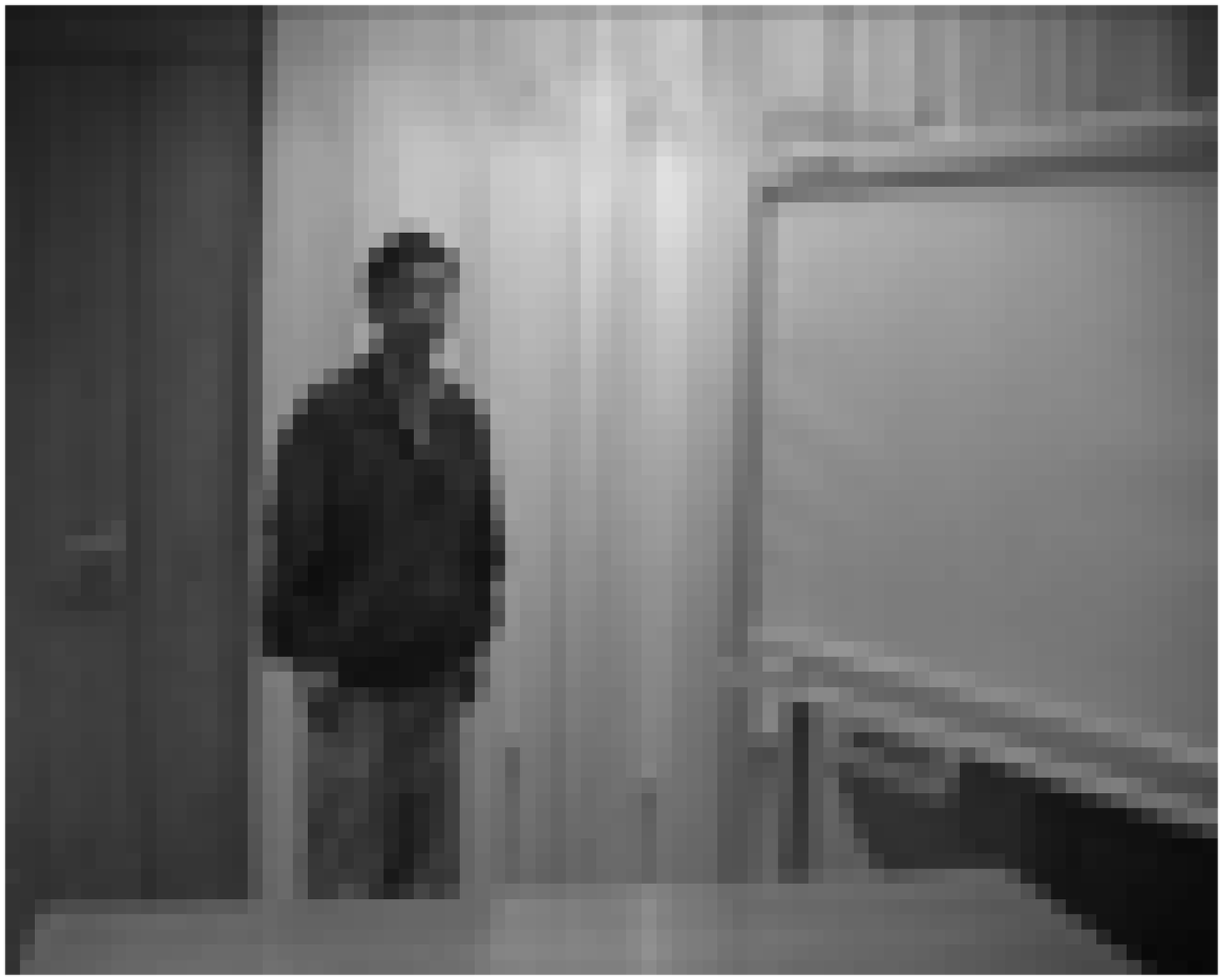}
		\includegraphics[width=15mm]{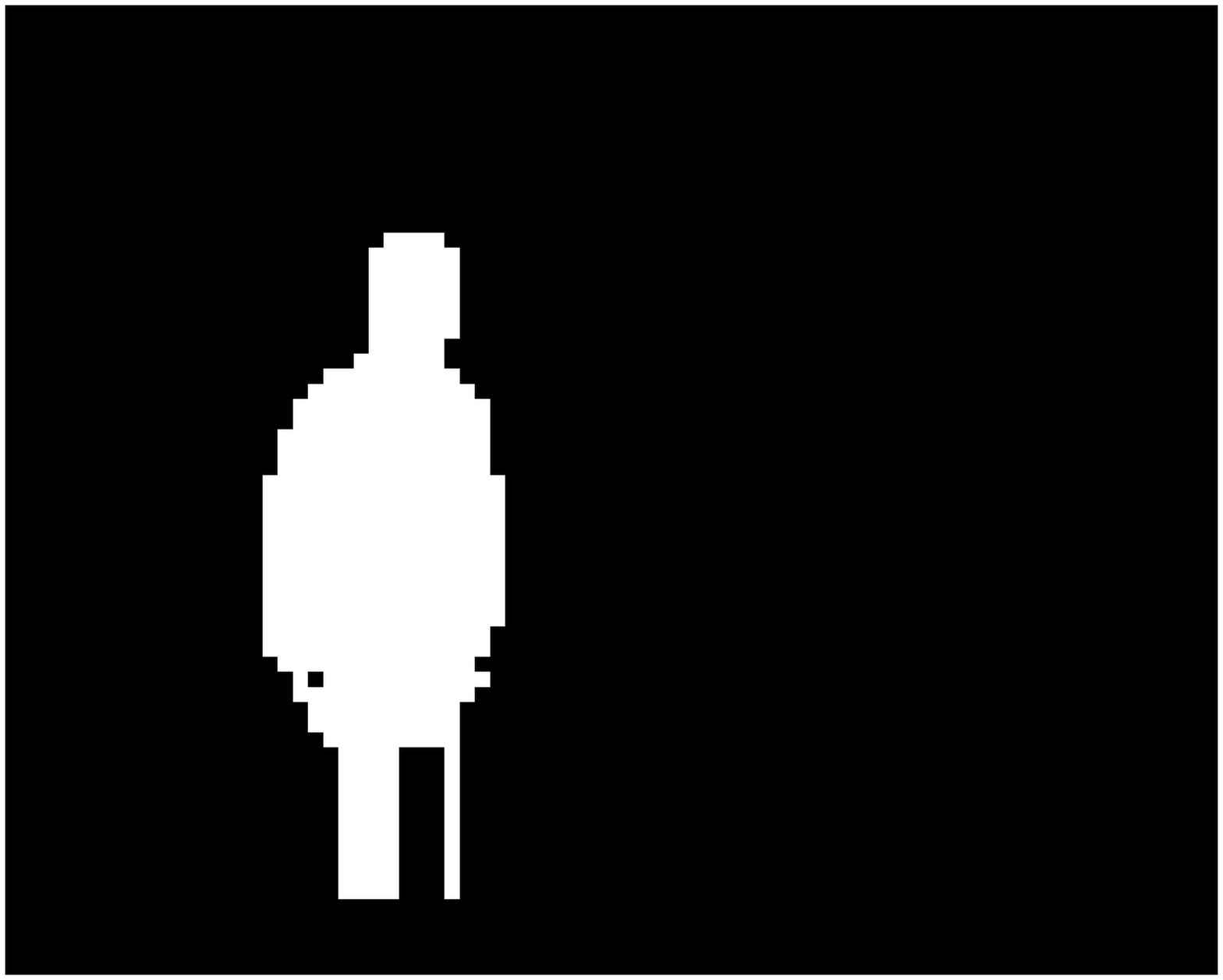}
		\includegraphics[width=15mm]{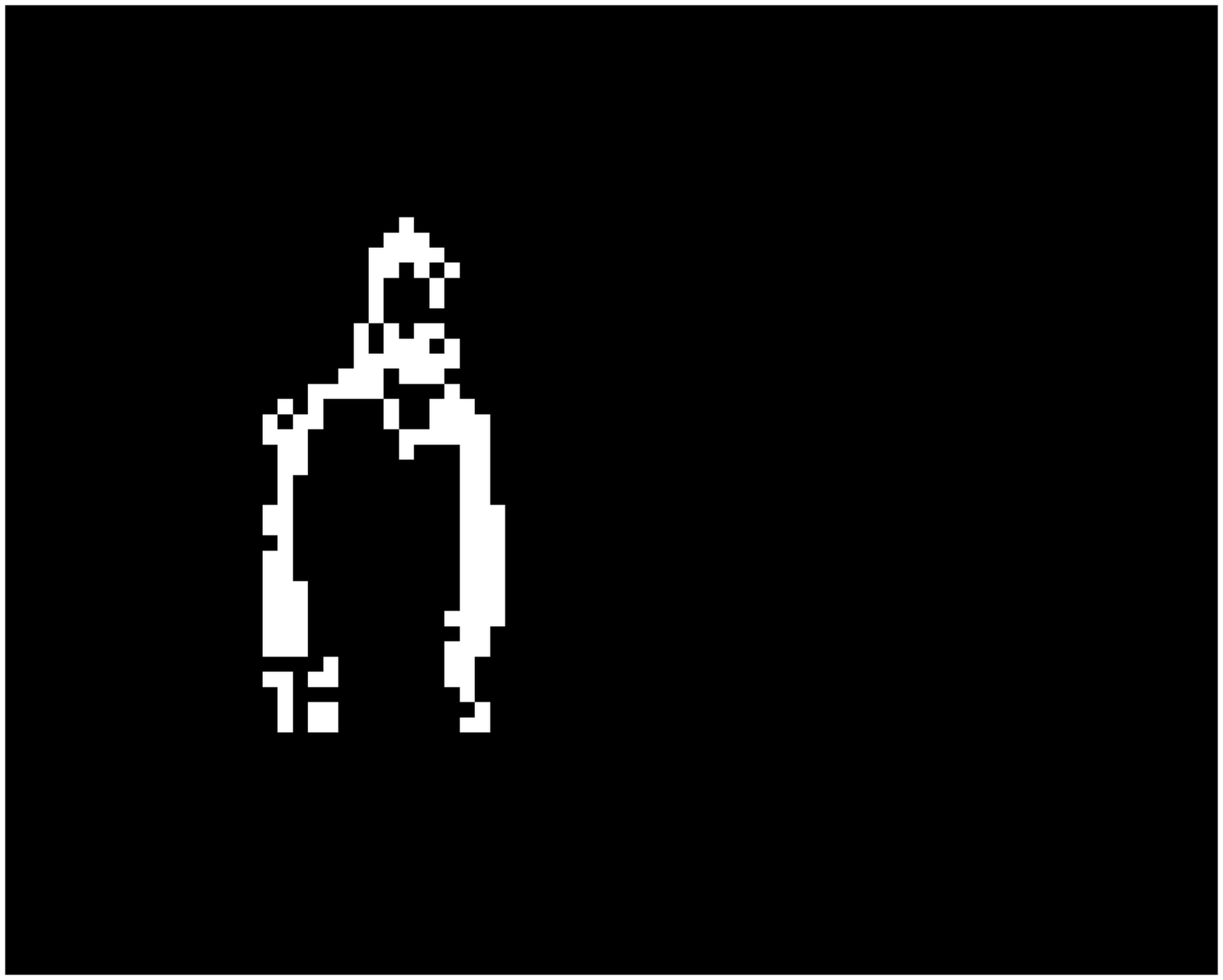}
		\includegraphics[width=15mm]{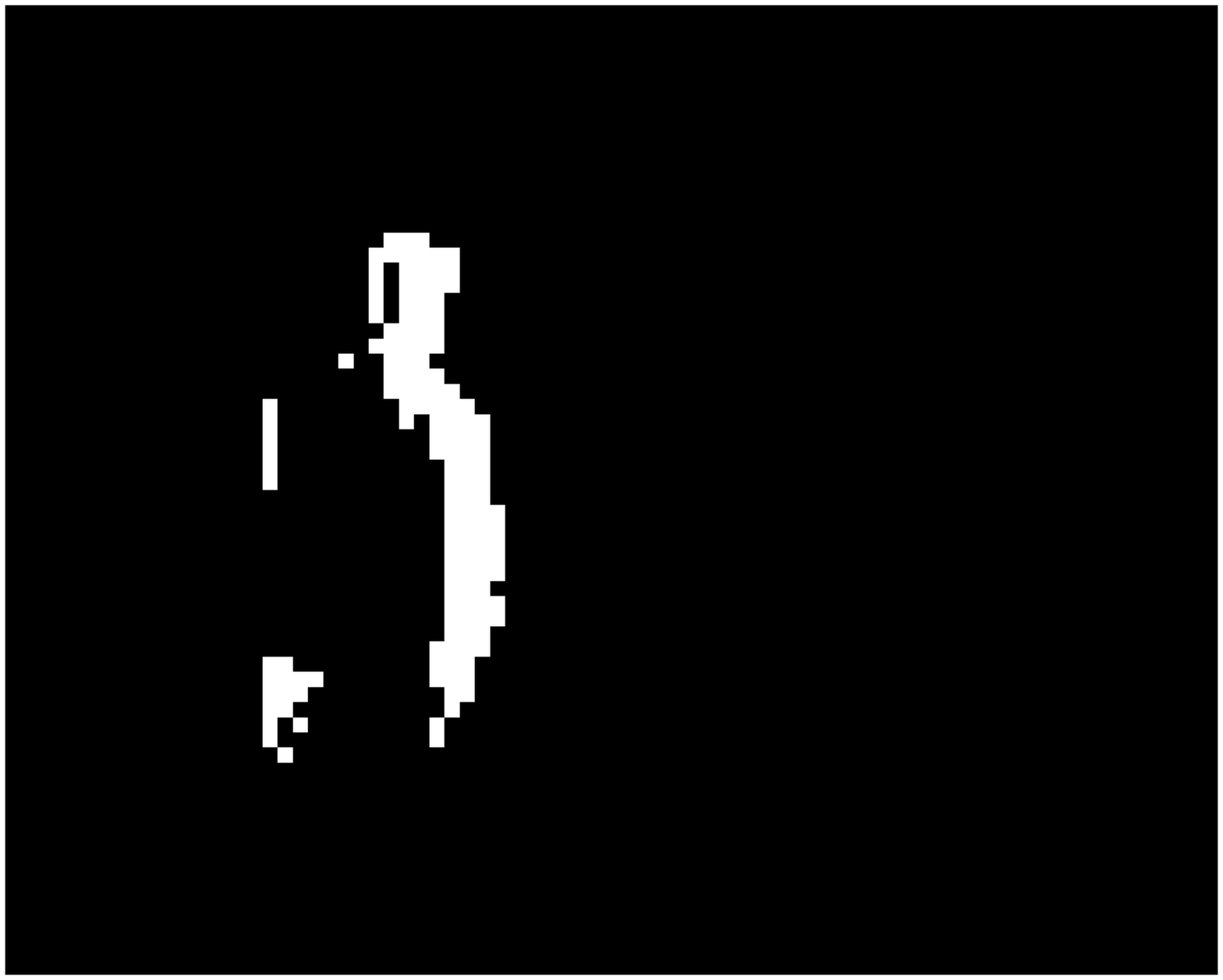}
		\includegraphics[width=15mm]{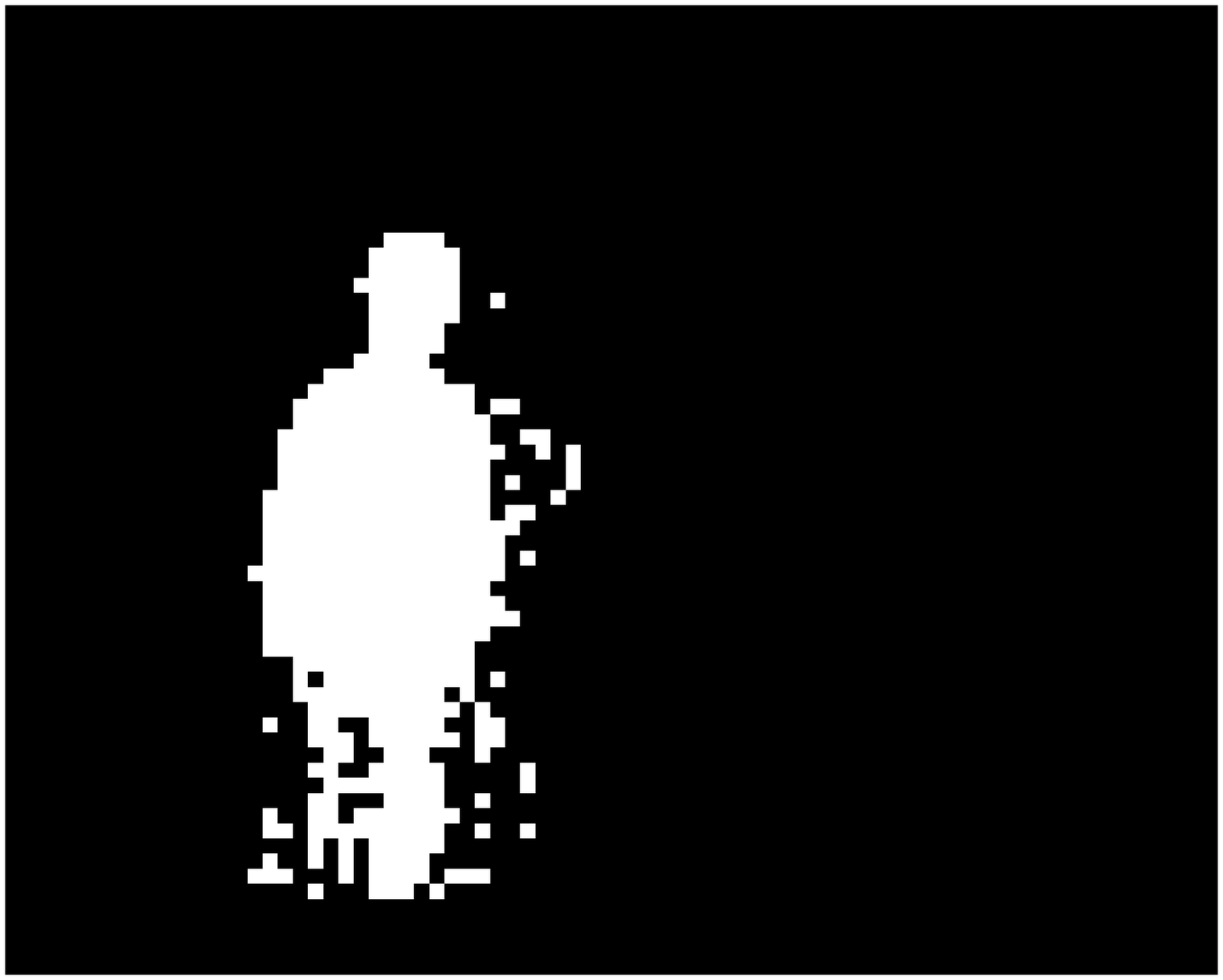}
		\includegraphics[width=15mm]{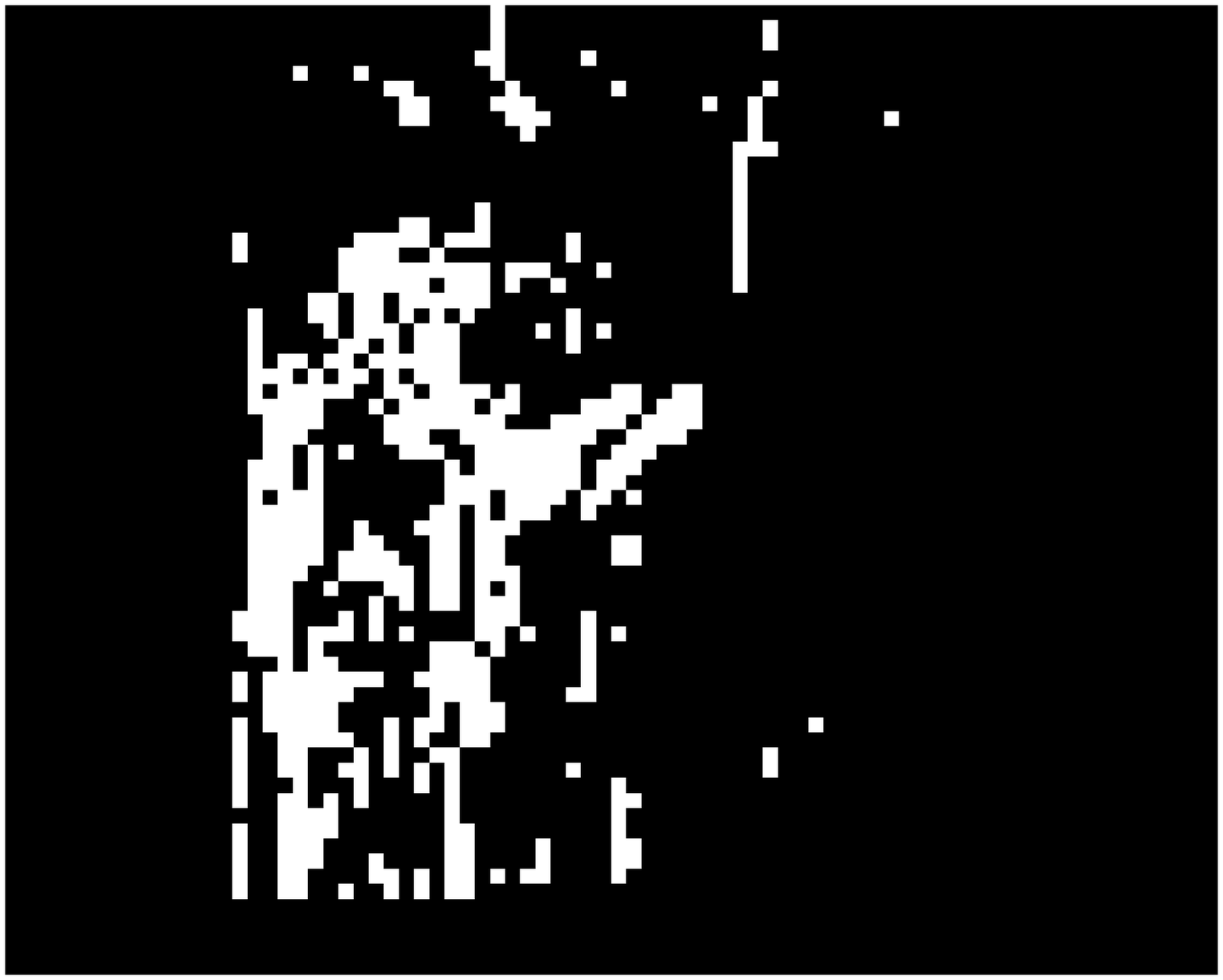}
		\includegraphics[width=15mm]{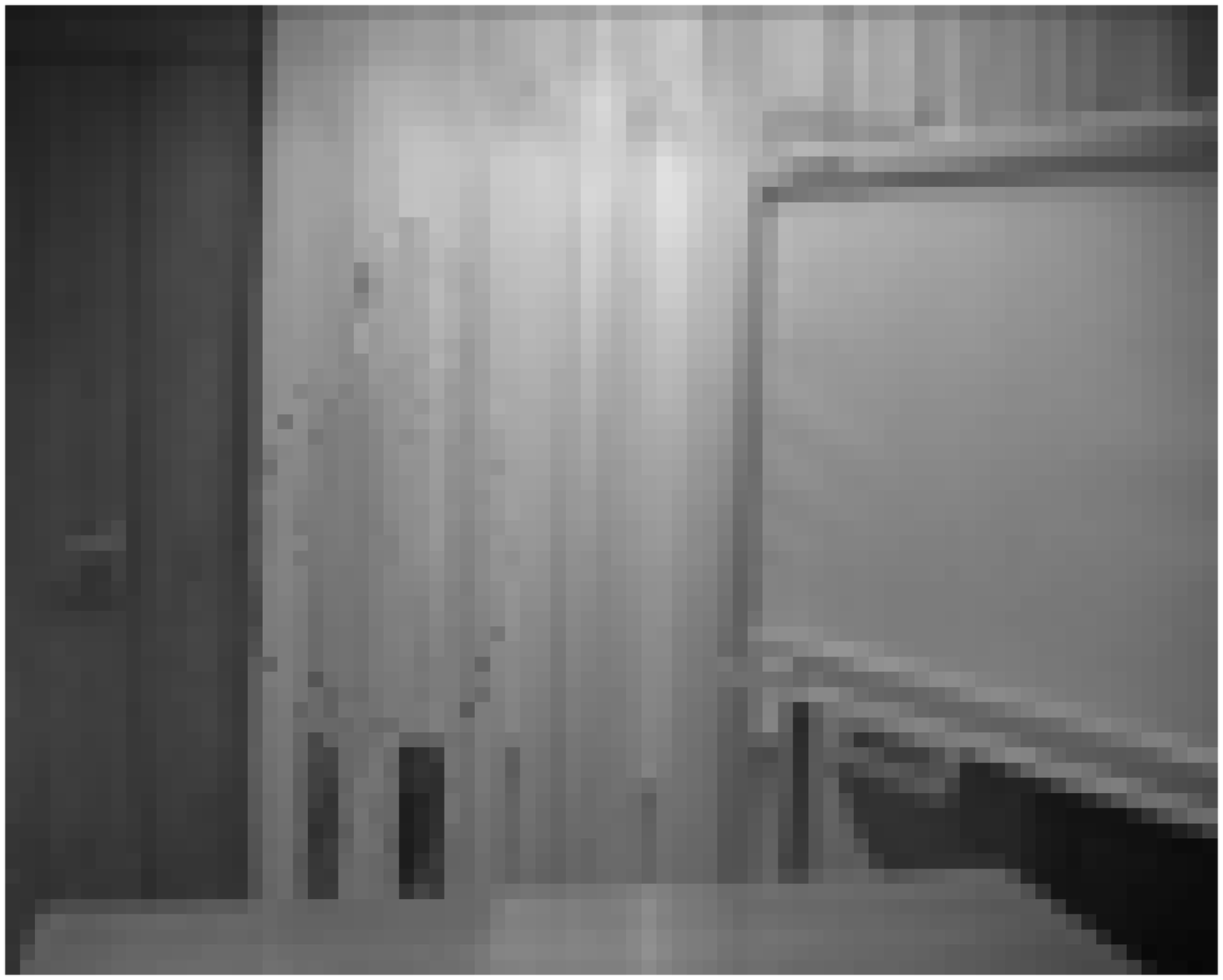}
		\includegraphics[width=15mm]{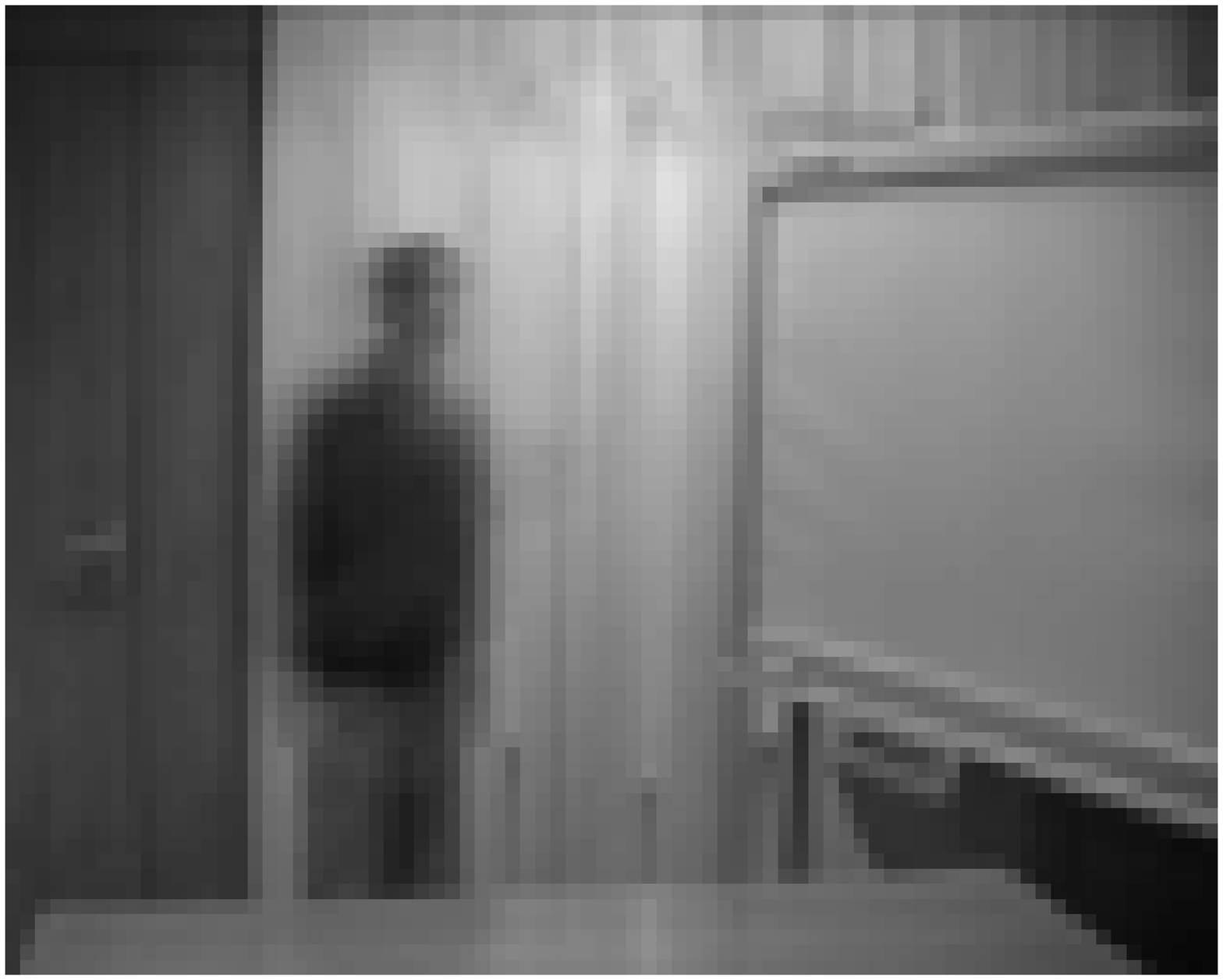}
		\includegraphics[width=15mm]{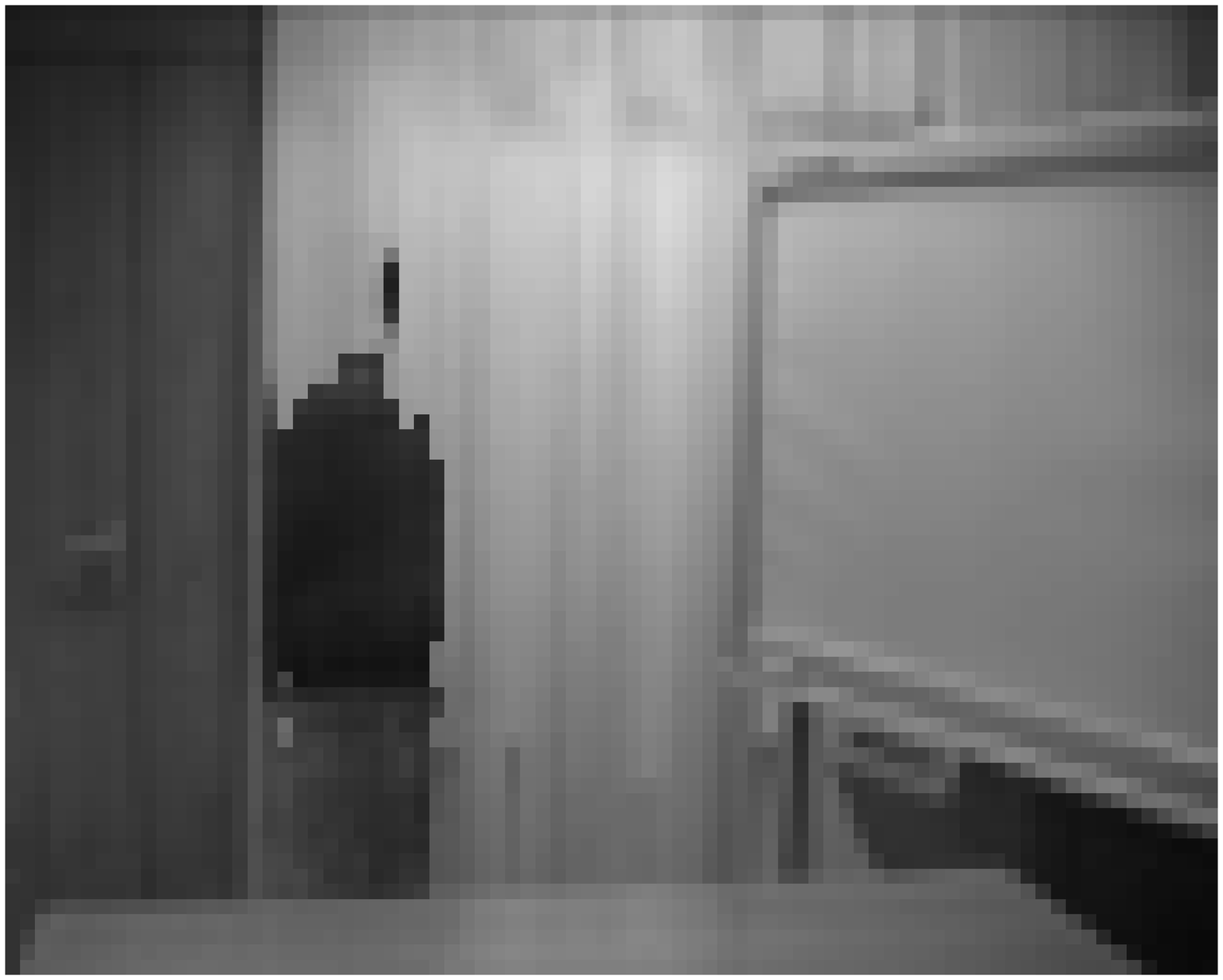}
		\includegraphics[width=15mm]{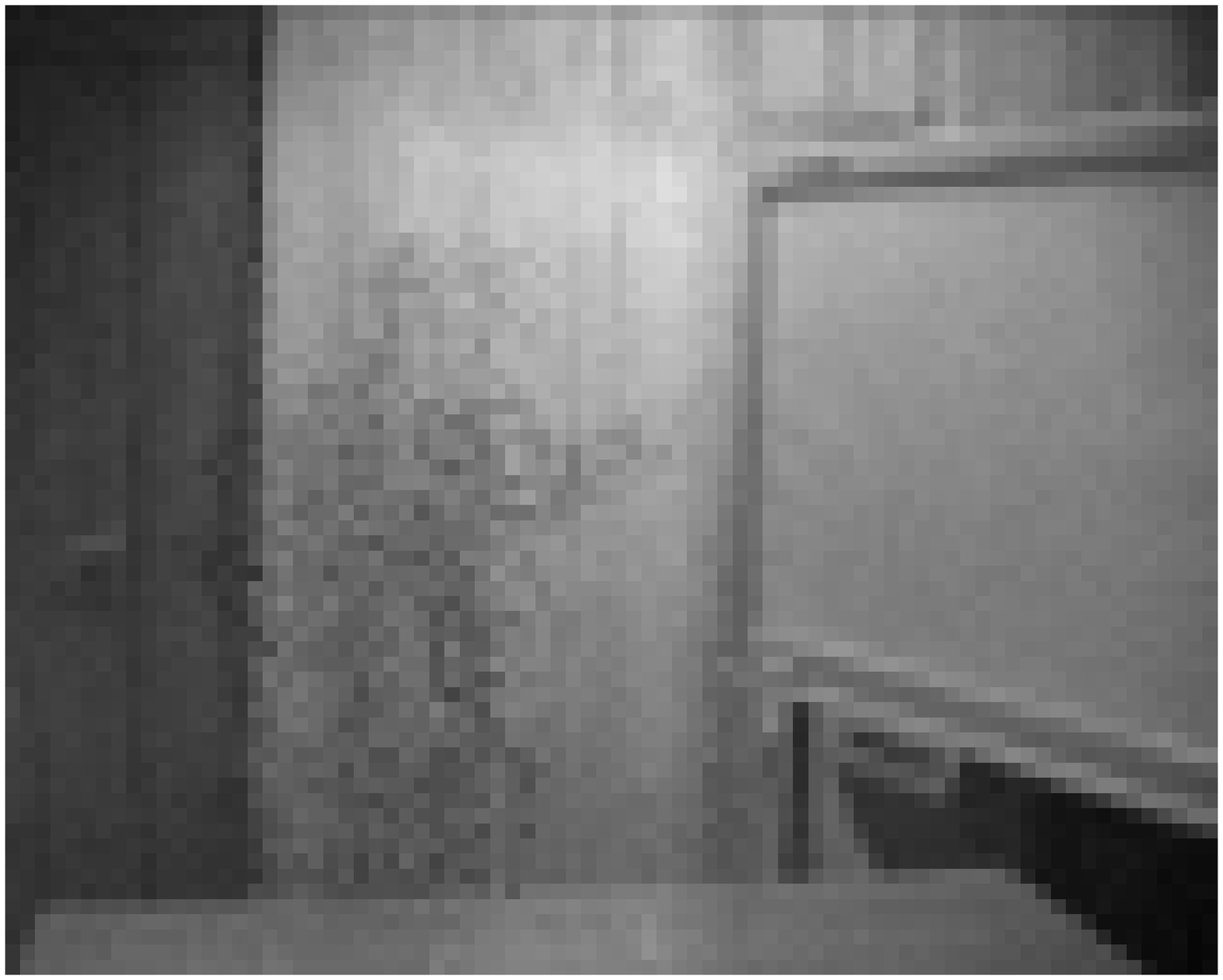}
		\includegraphics[width=15mm]{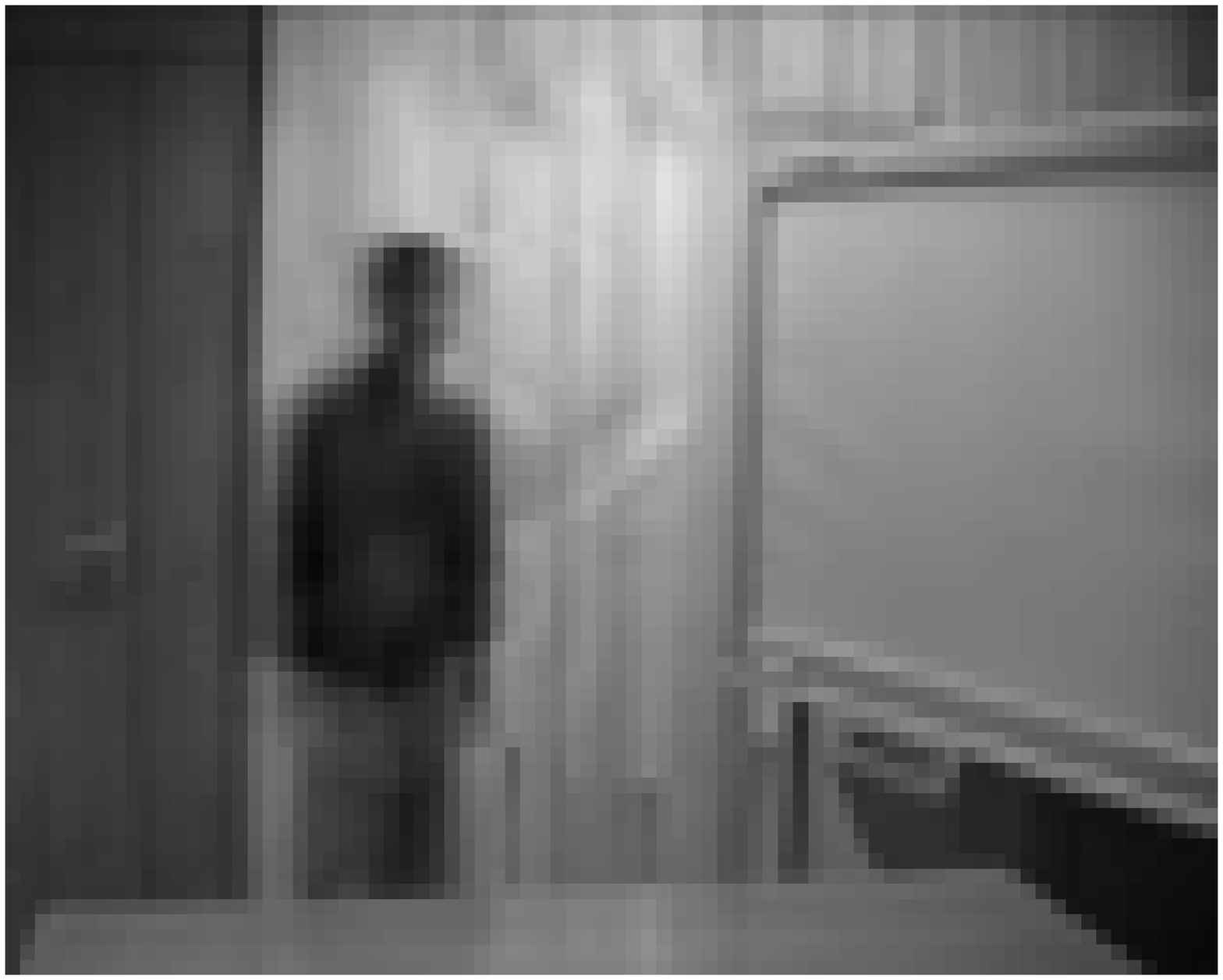}\\
		\includegraphics[width=15mm]{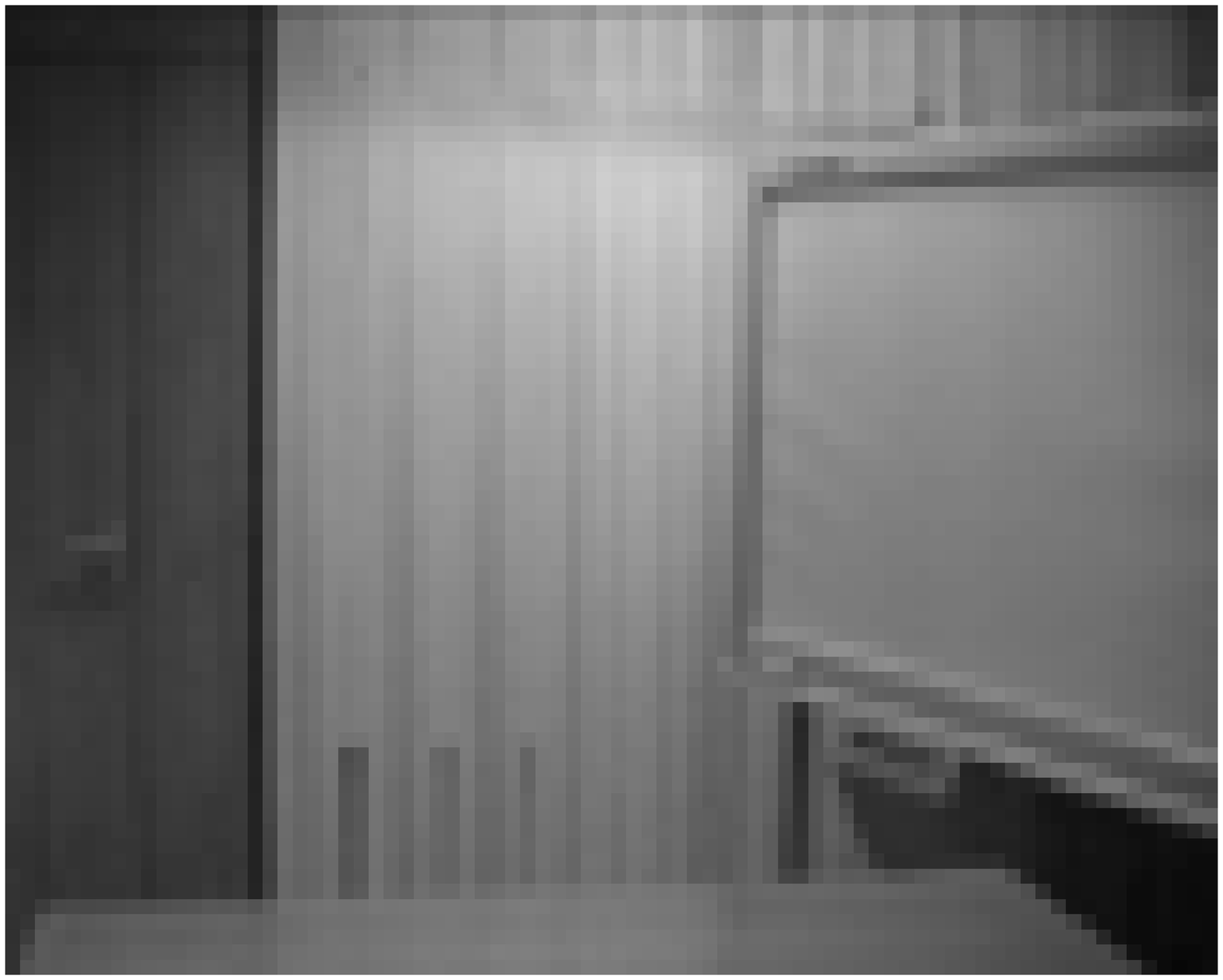}
		\includegraphics[width=15mm]{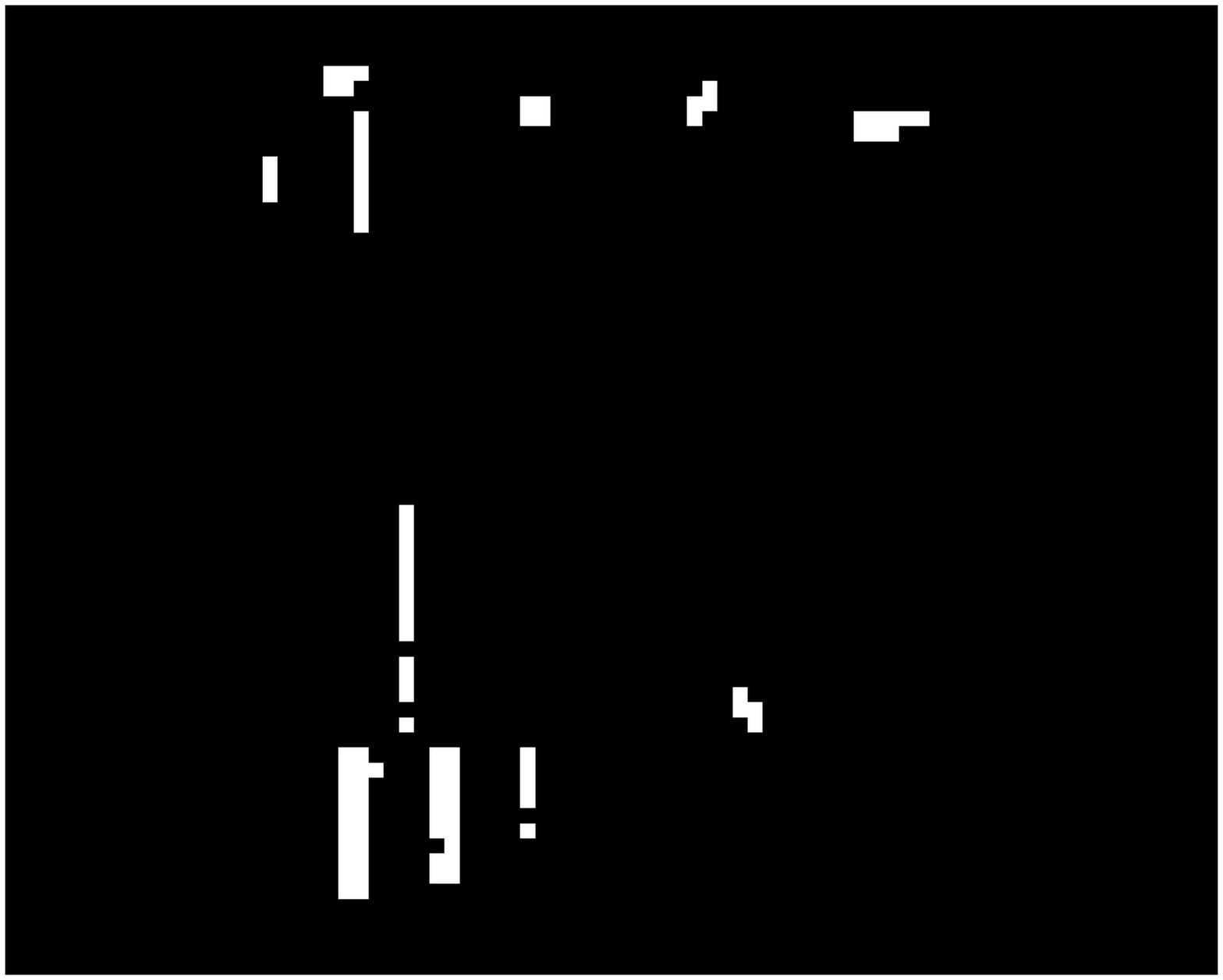}
		\includegraphics[width=15mm]{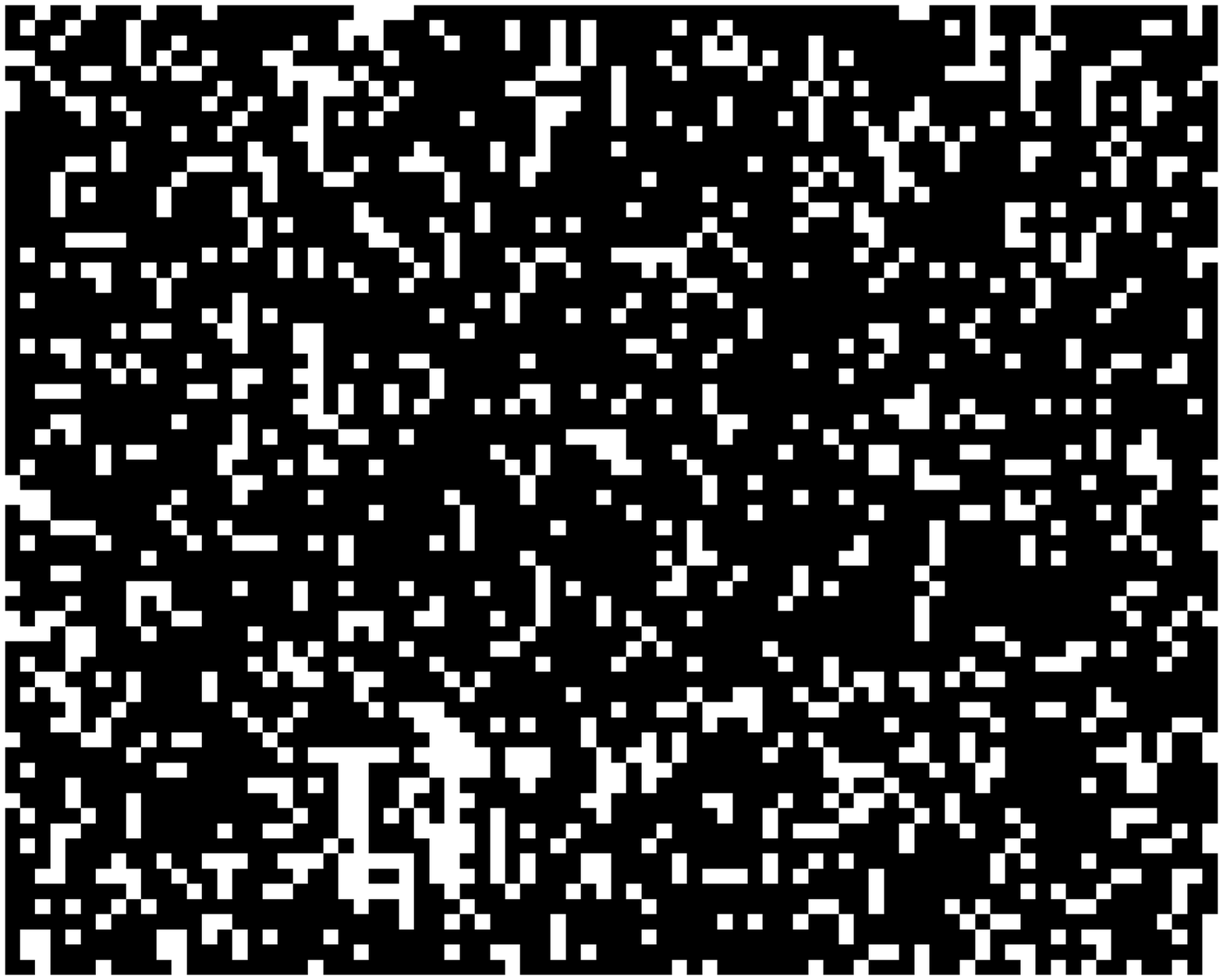}
		\includegraphics[width=15mm]{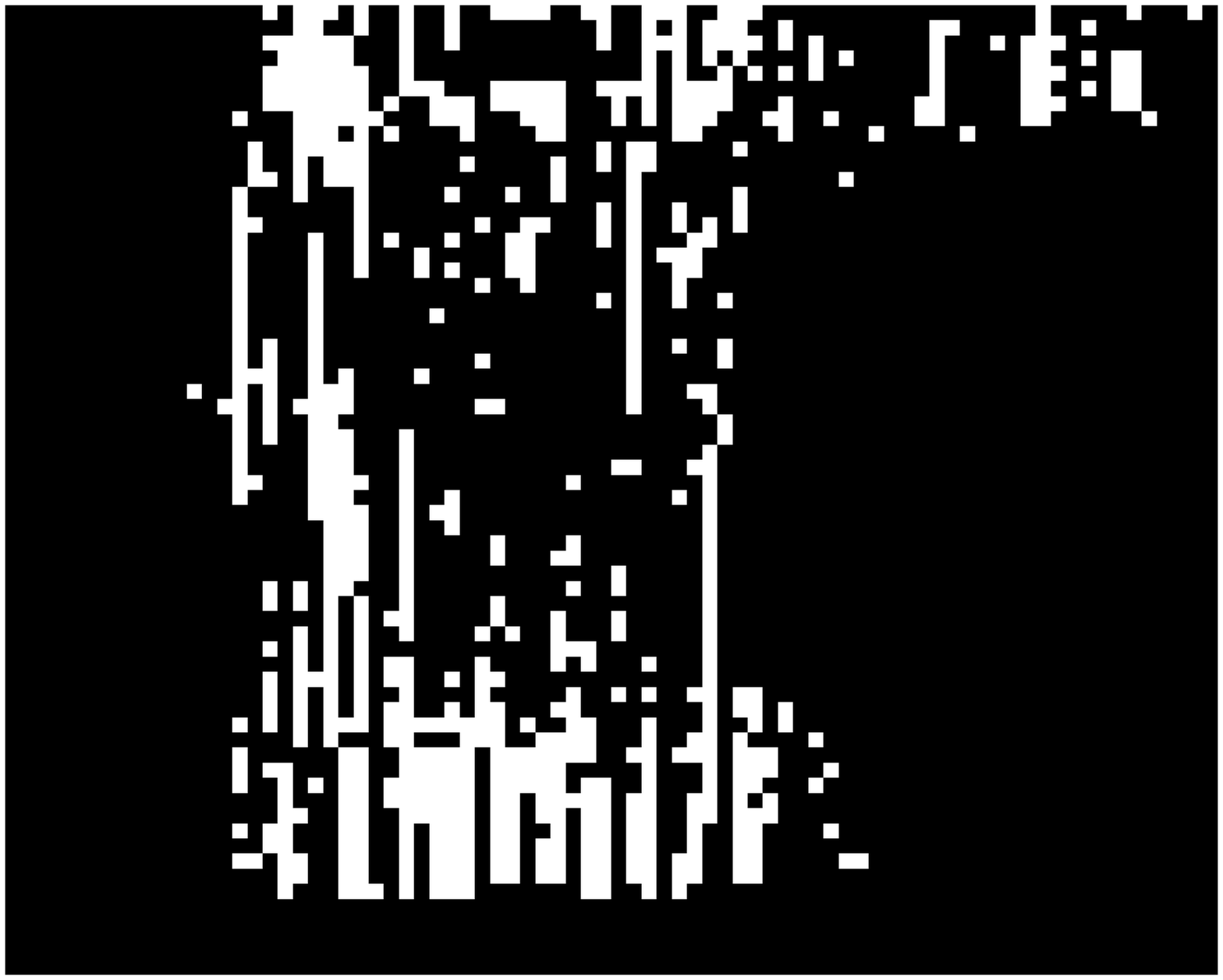}
		\includegraphics[width=15mm]{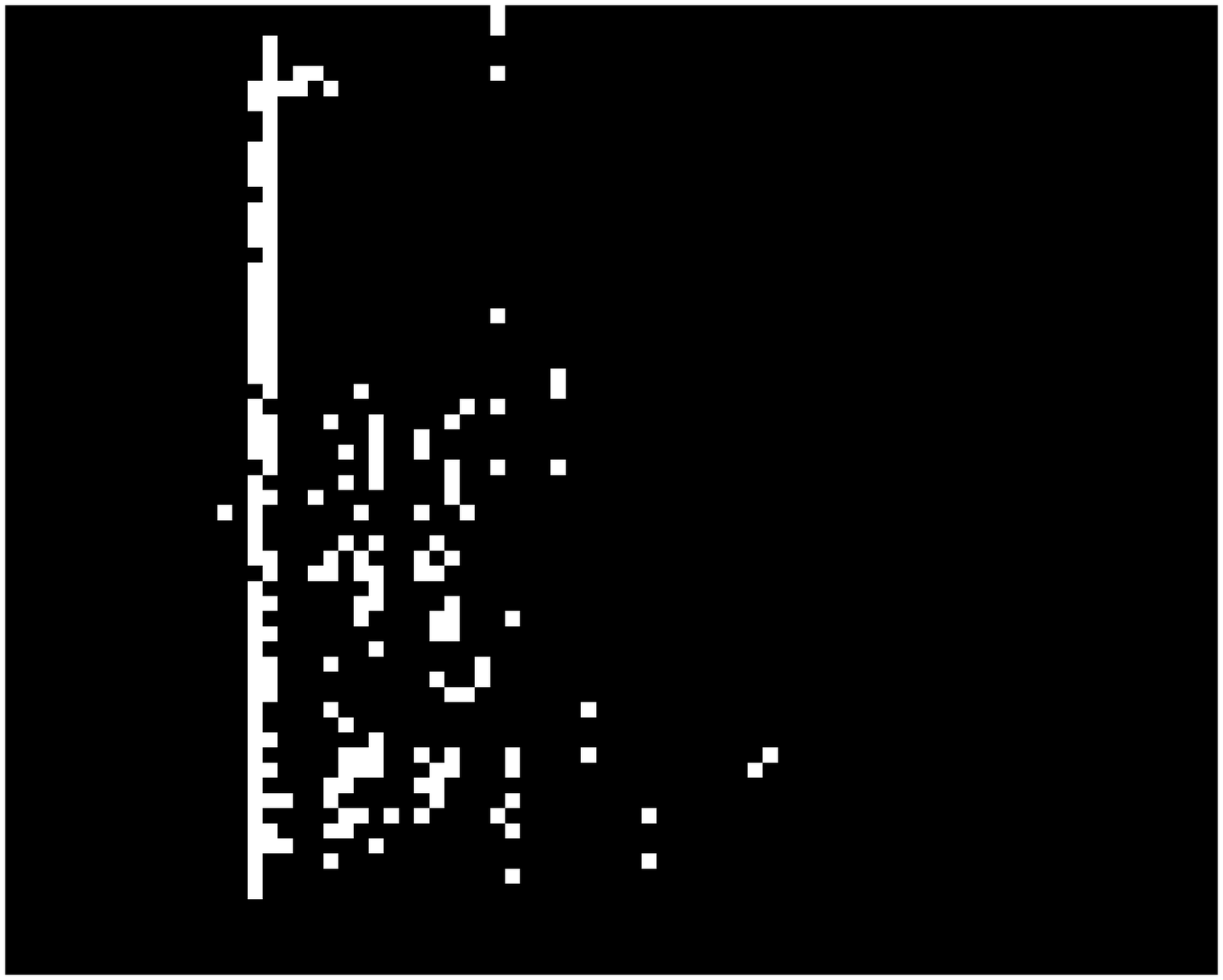}
		\includegraphics[width=15mm]{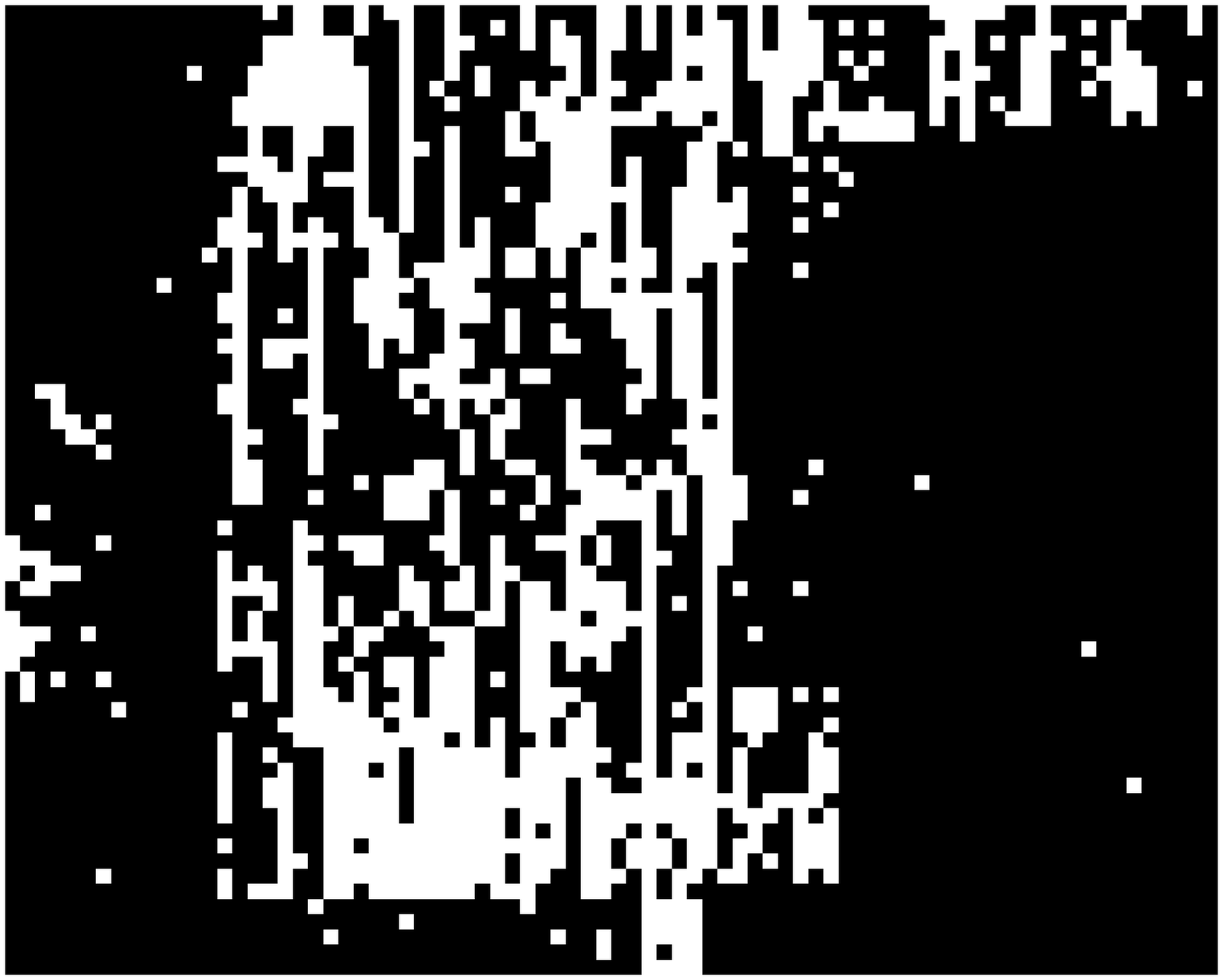}
		\includegraphics[width=15mm]{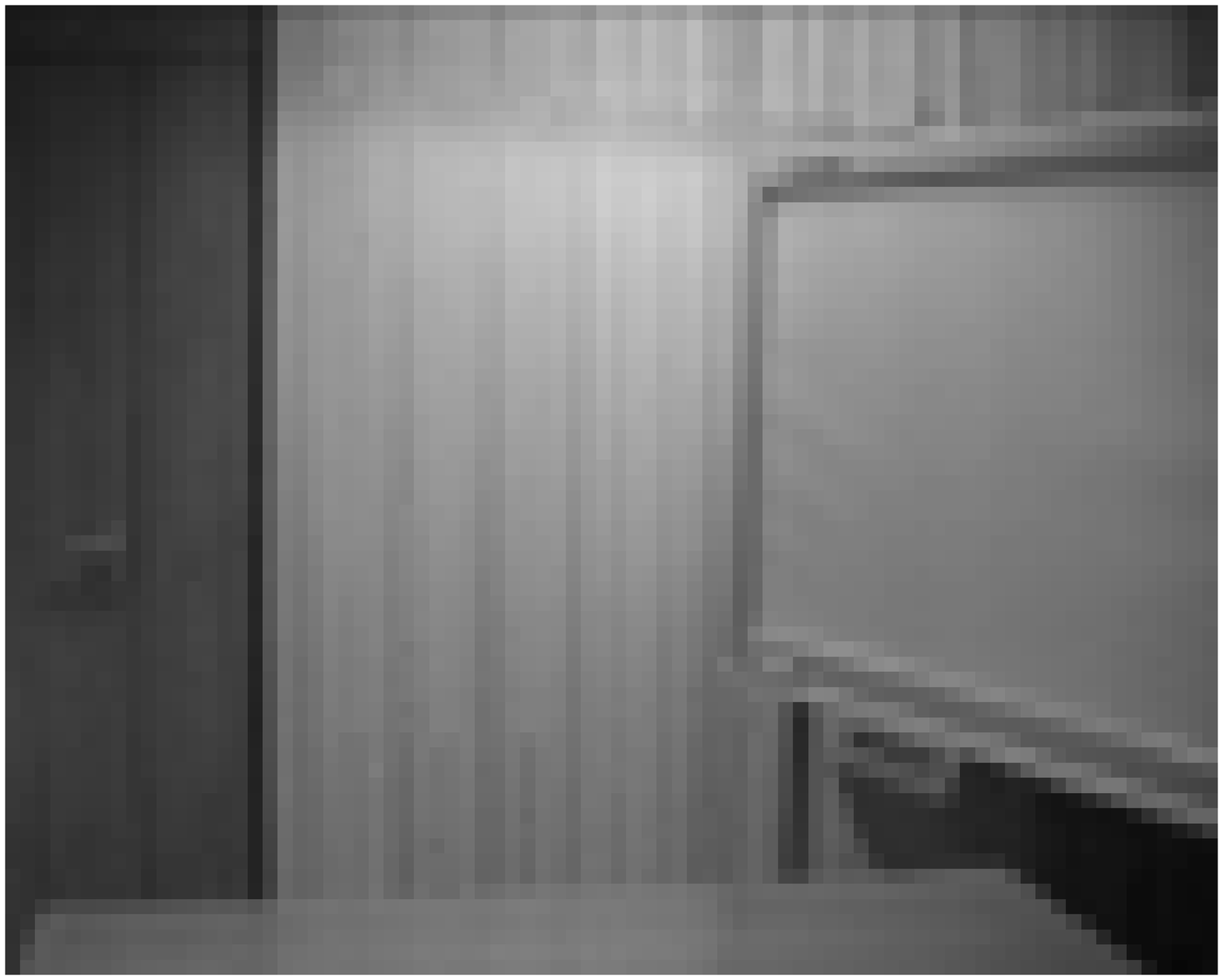}
		\includegraphics[width=15mm]{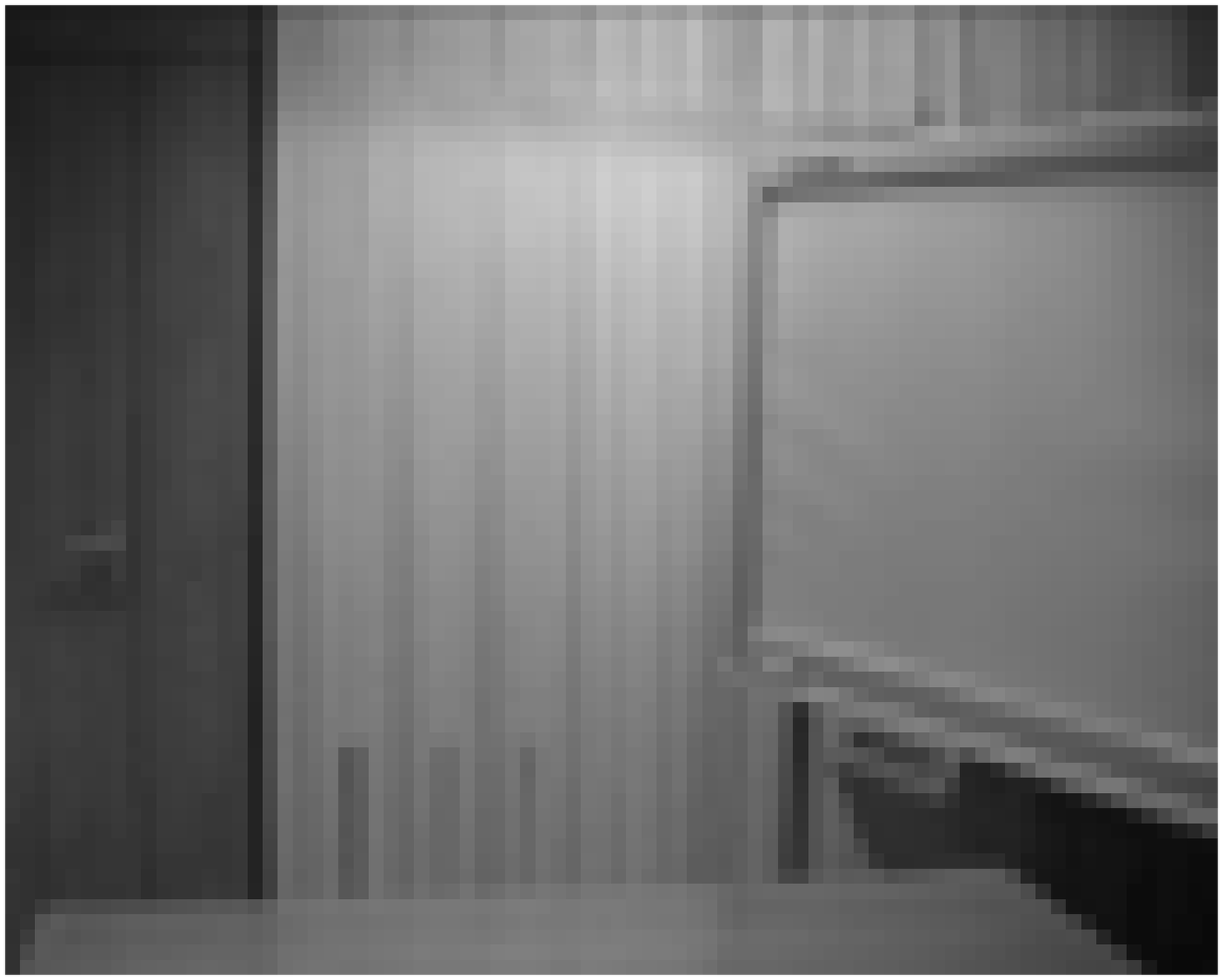}
		\includegraphics[width=15mm]{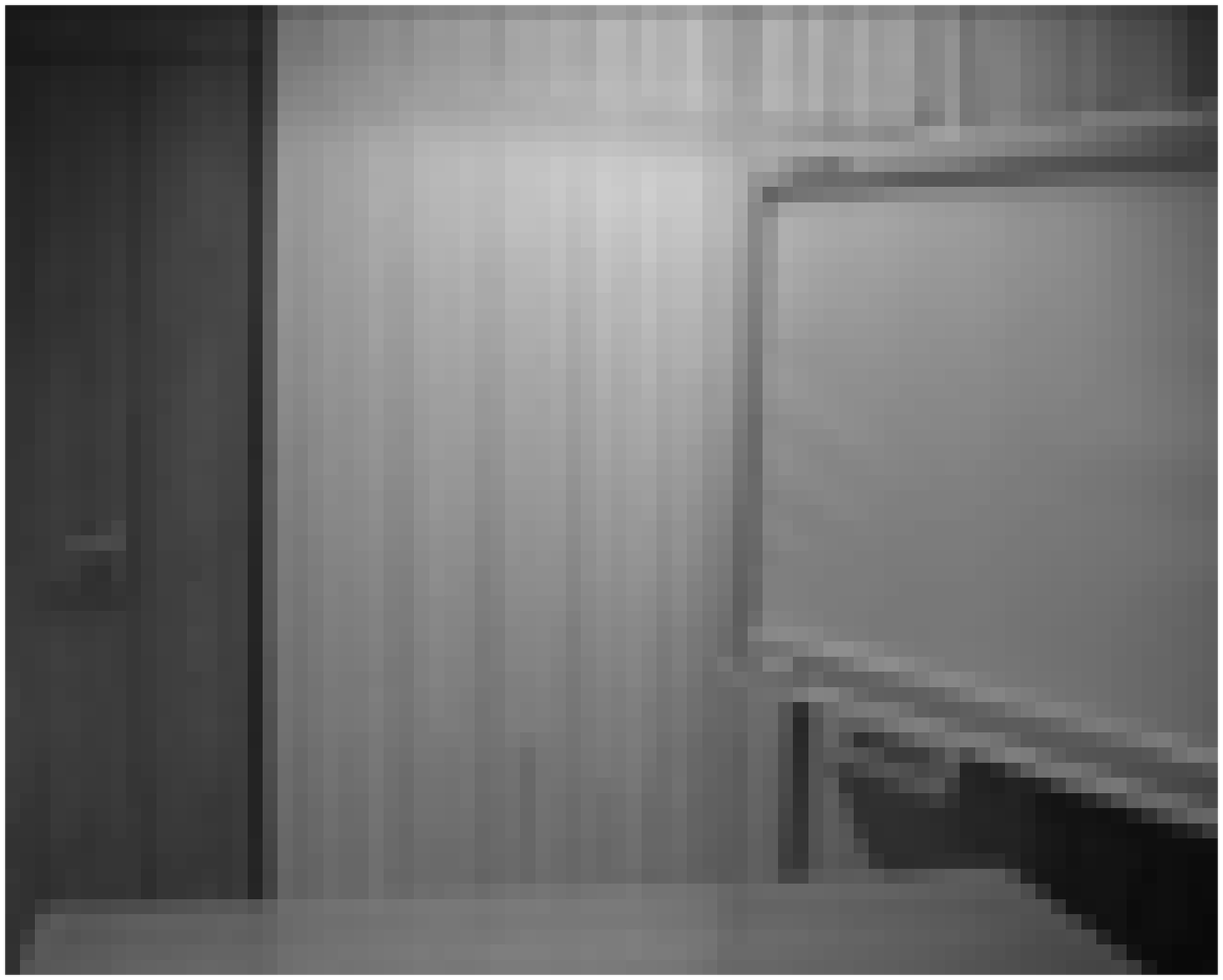}
		\includegraphics[width=15mm]{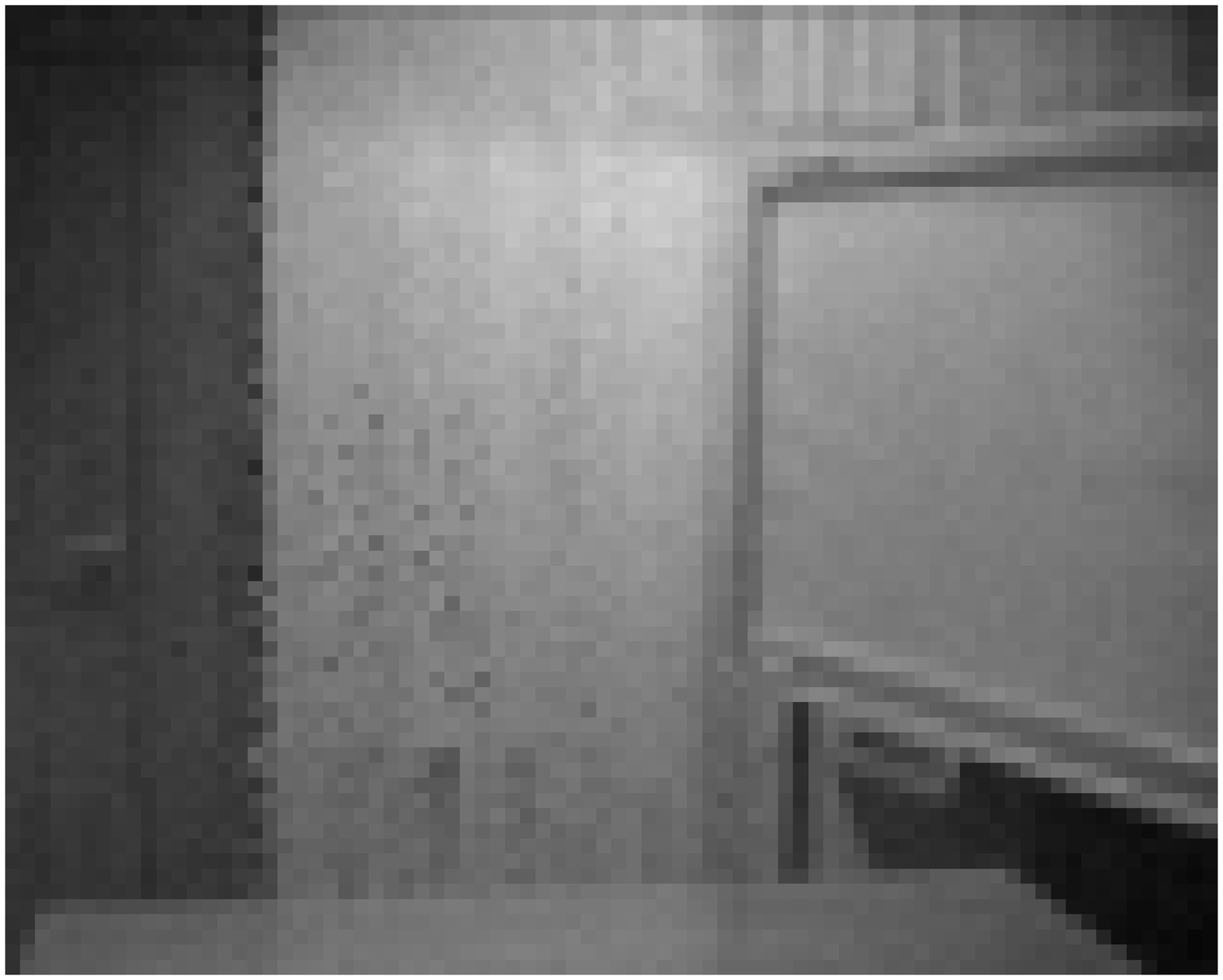}
		\includegraphics[width=15mm]{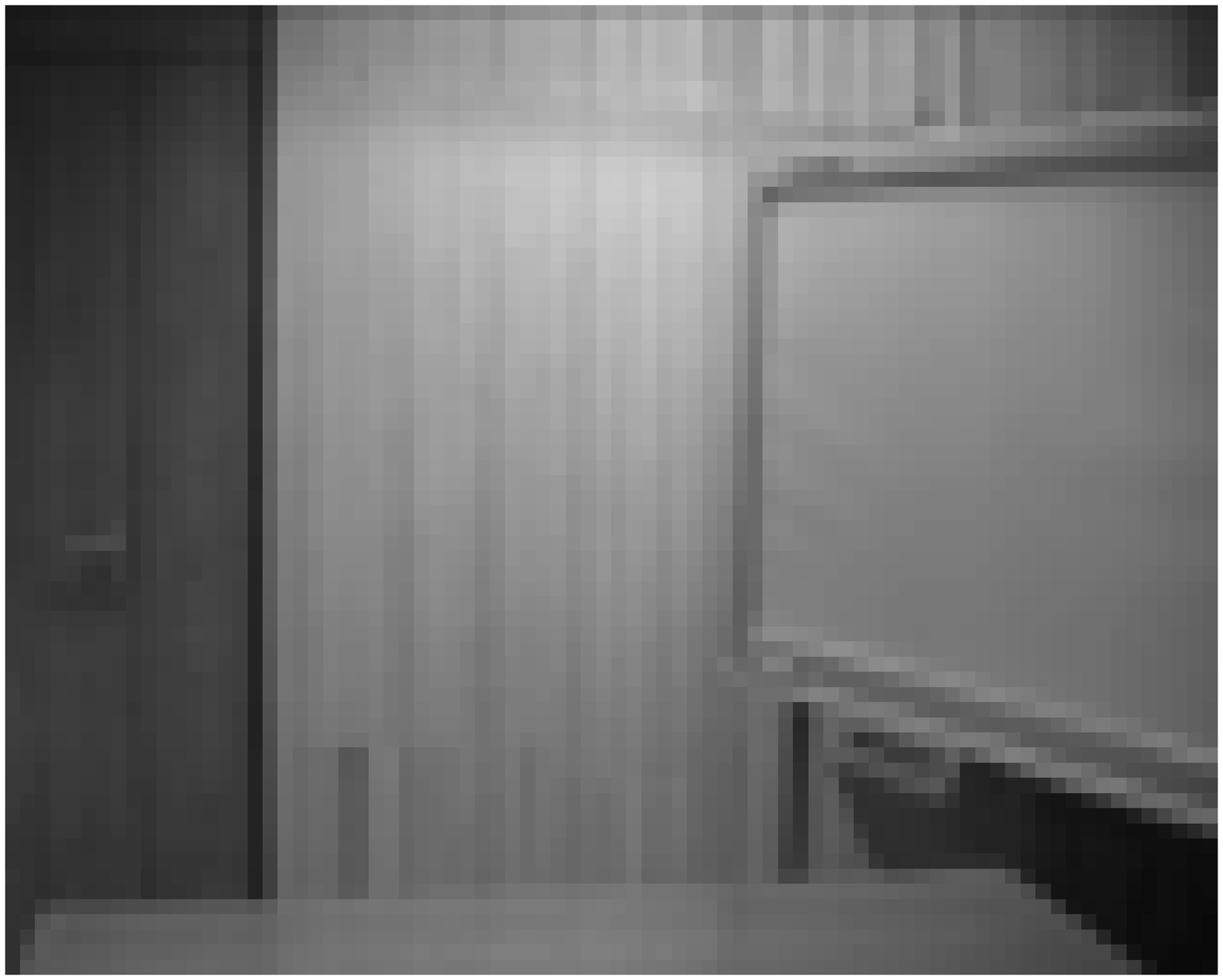}\\
		\includegraphics[width=15mm]{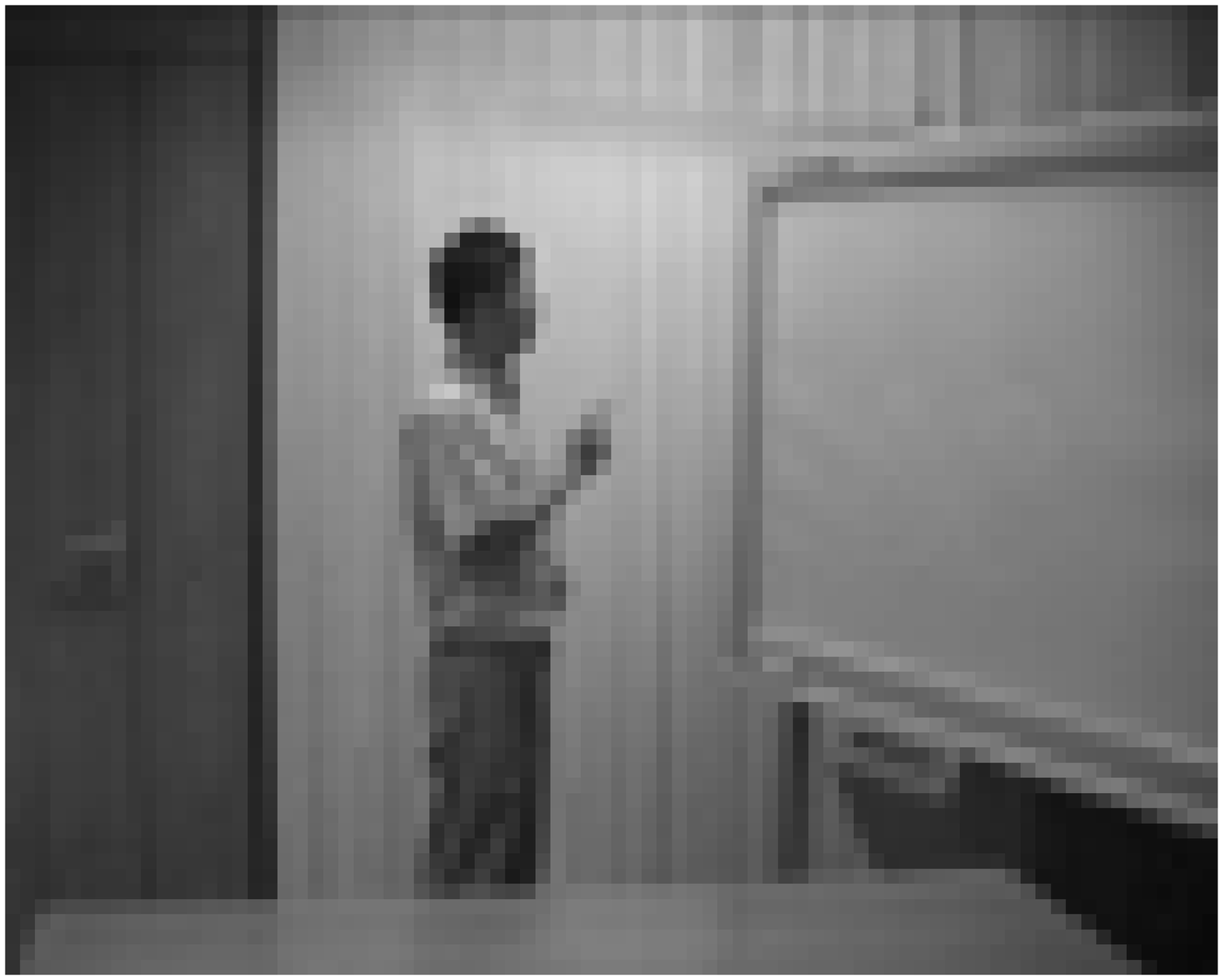}
		\includegraphics[width=15mm]{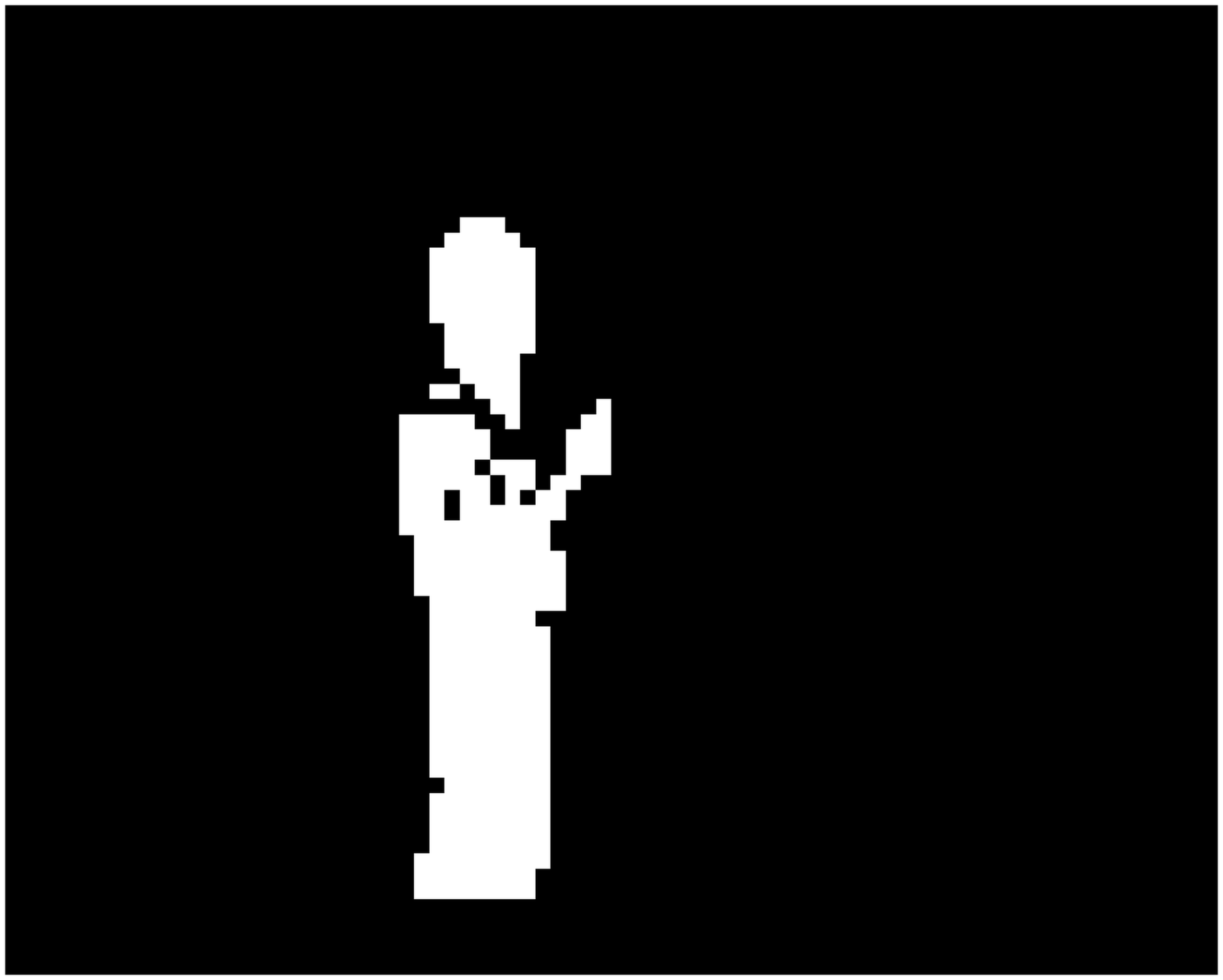}
		\includegraphics[width=15mm]{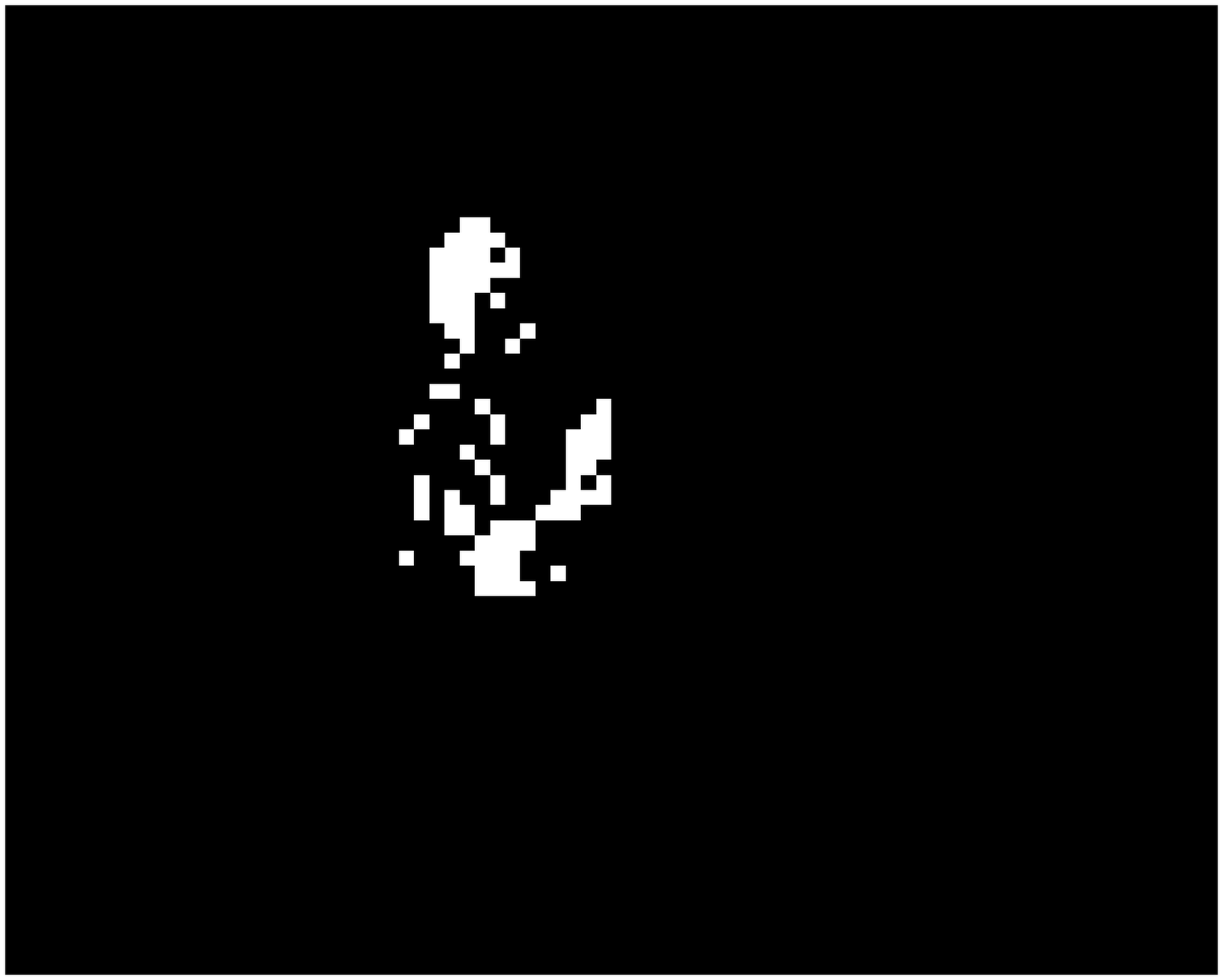}
		\includegraphics[width=15mm]{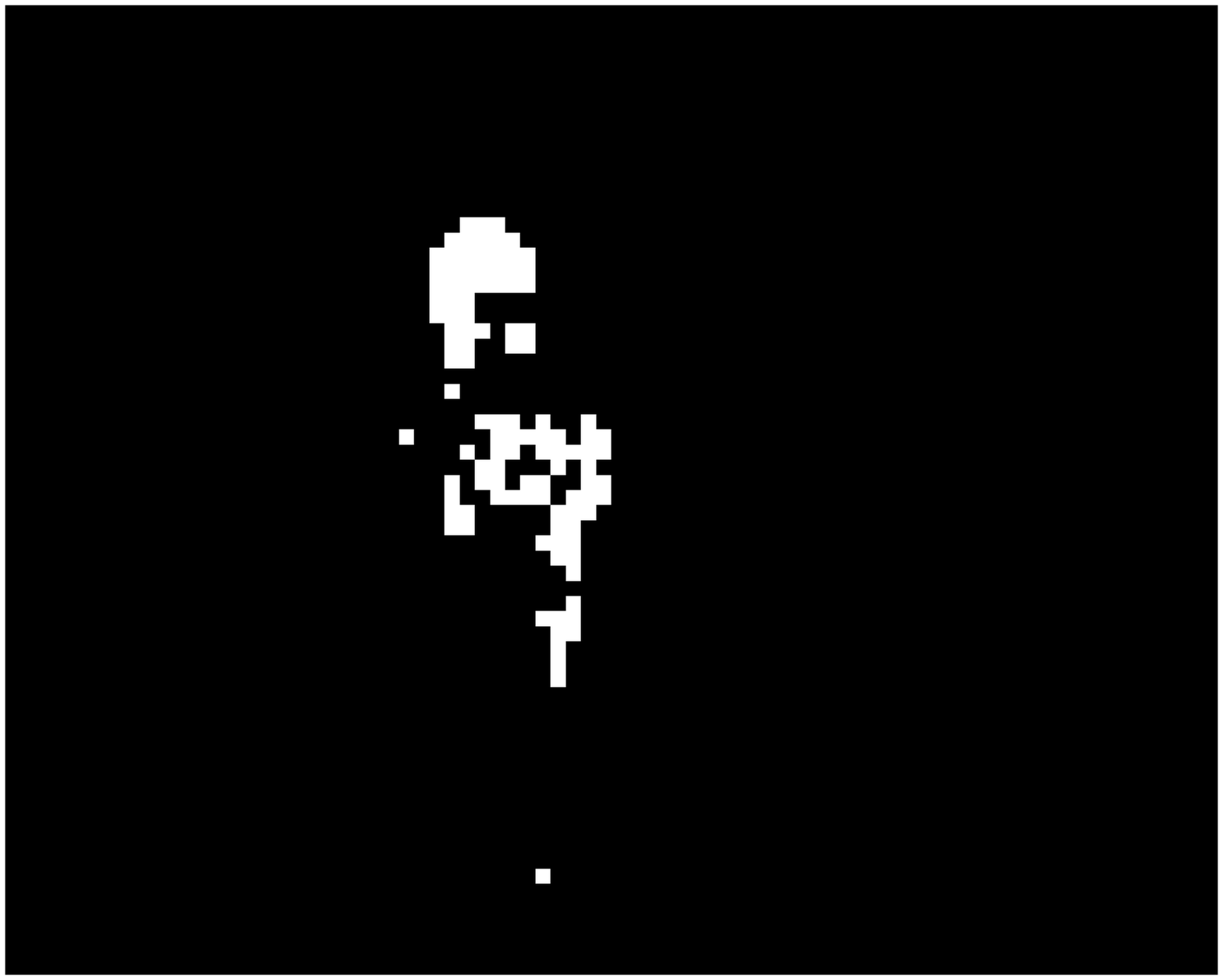}
		\includegraphics[width=15mm]{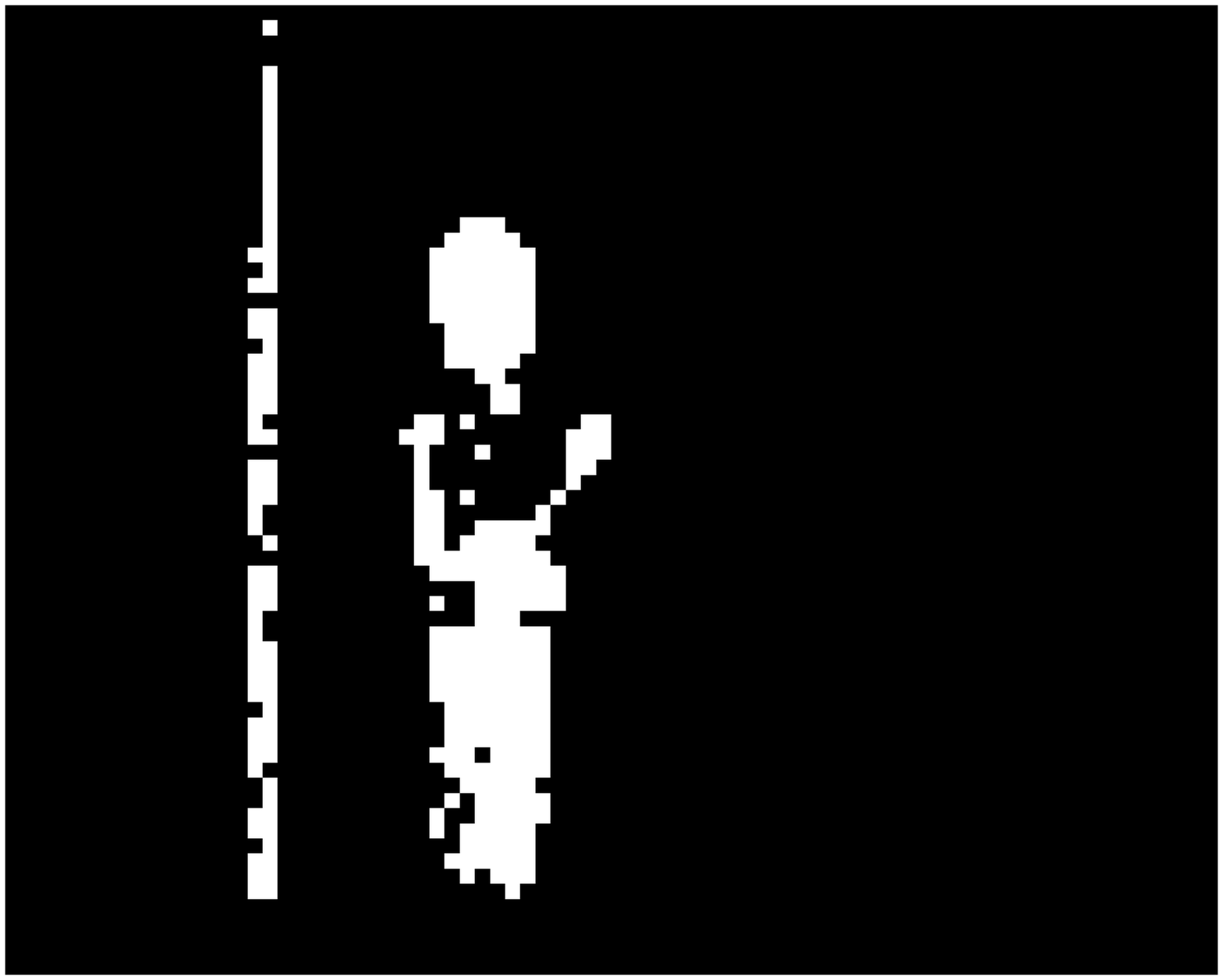}
		\includegraphics[width=15mm]{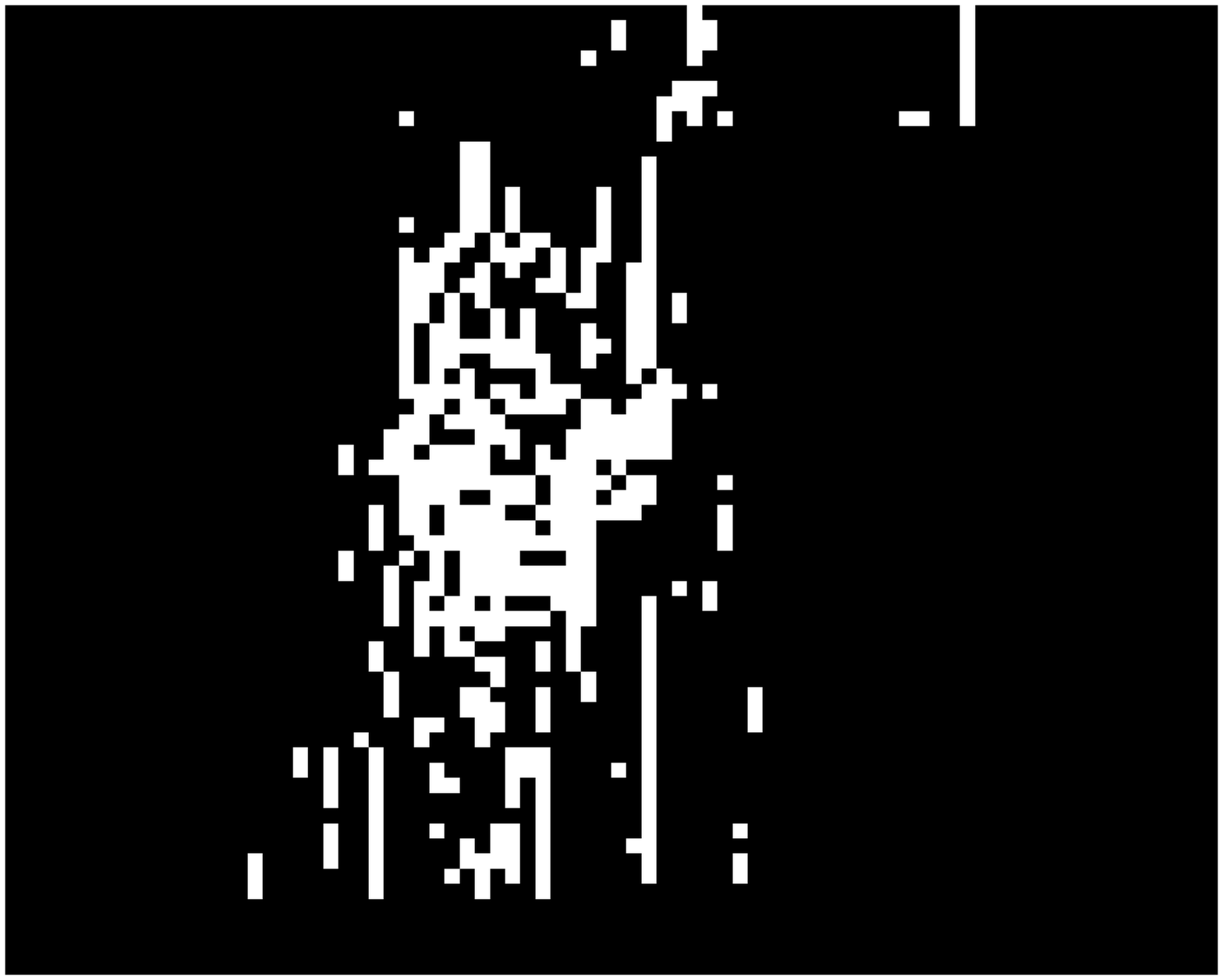}
		\includegraphics[width=15mm]{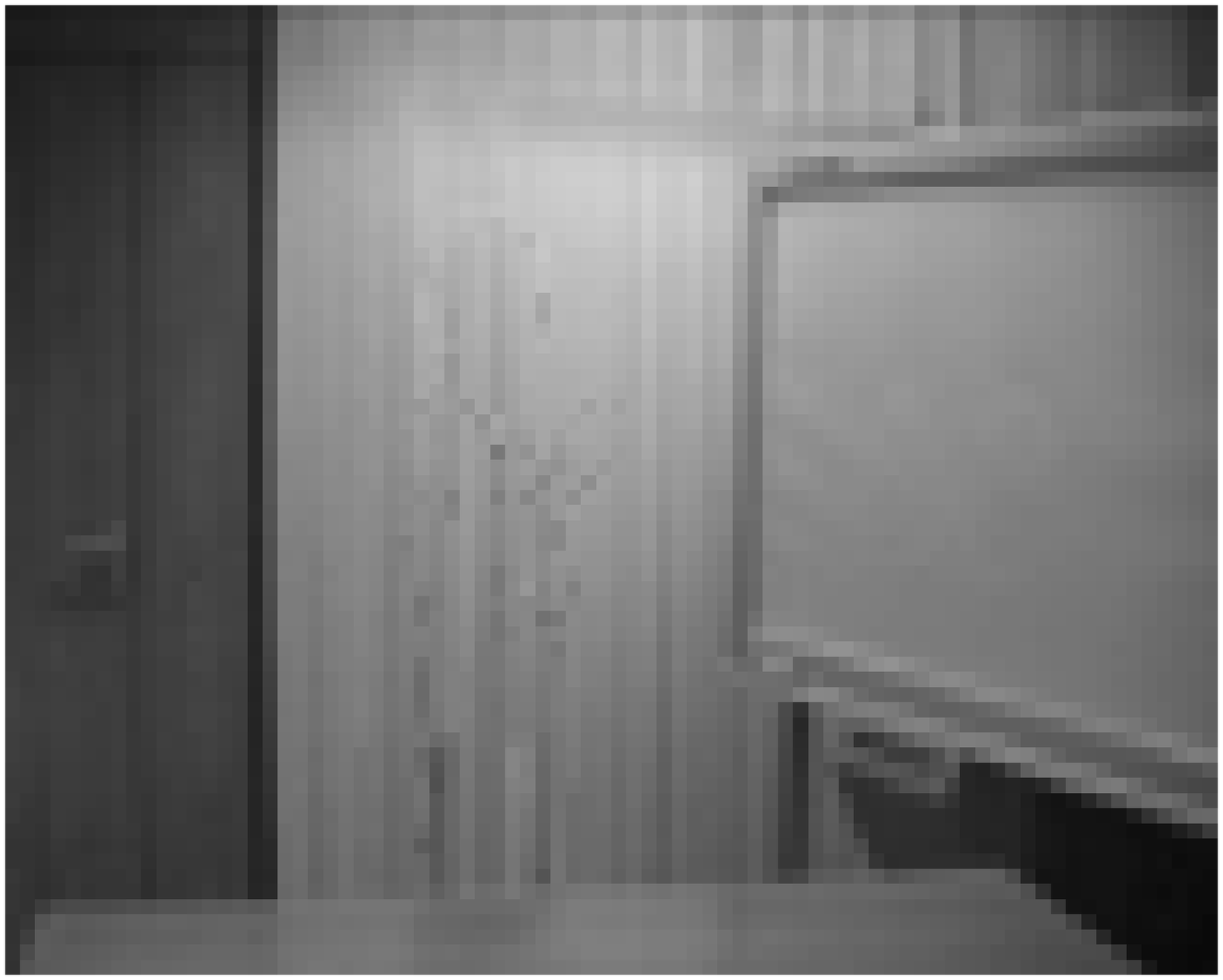}
		\includegraphics[width=15mm]{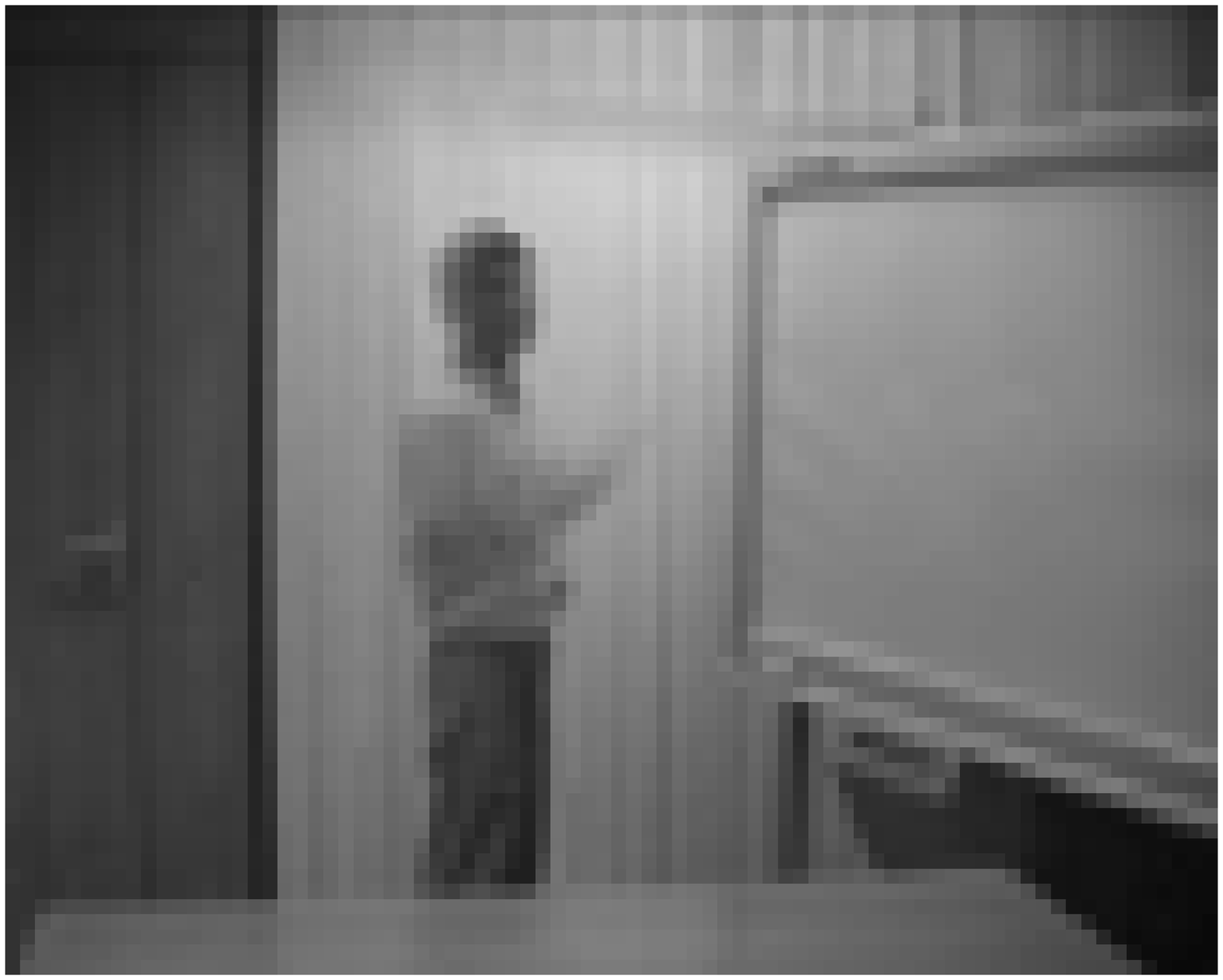}
		\includegraphics[width=15mm]{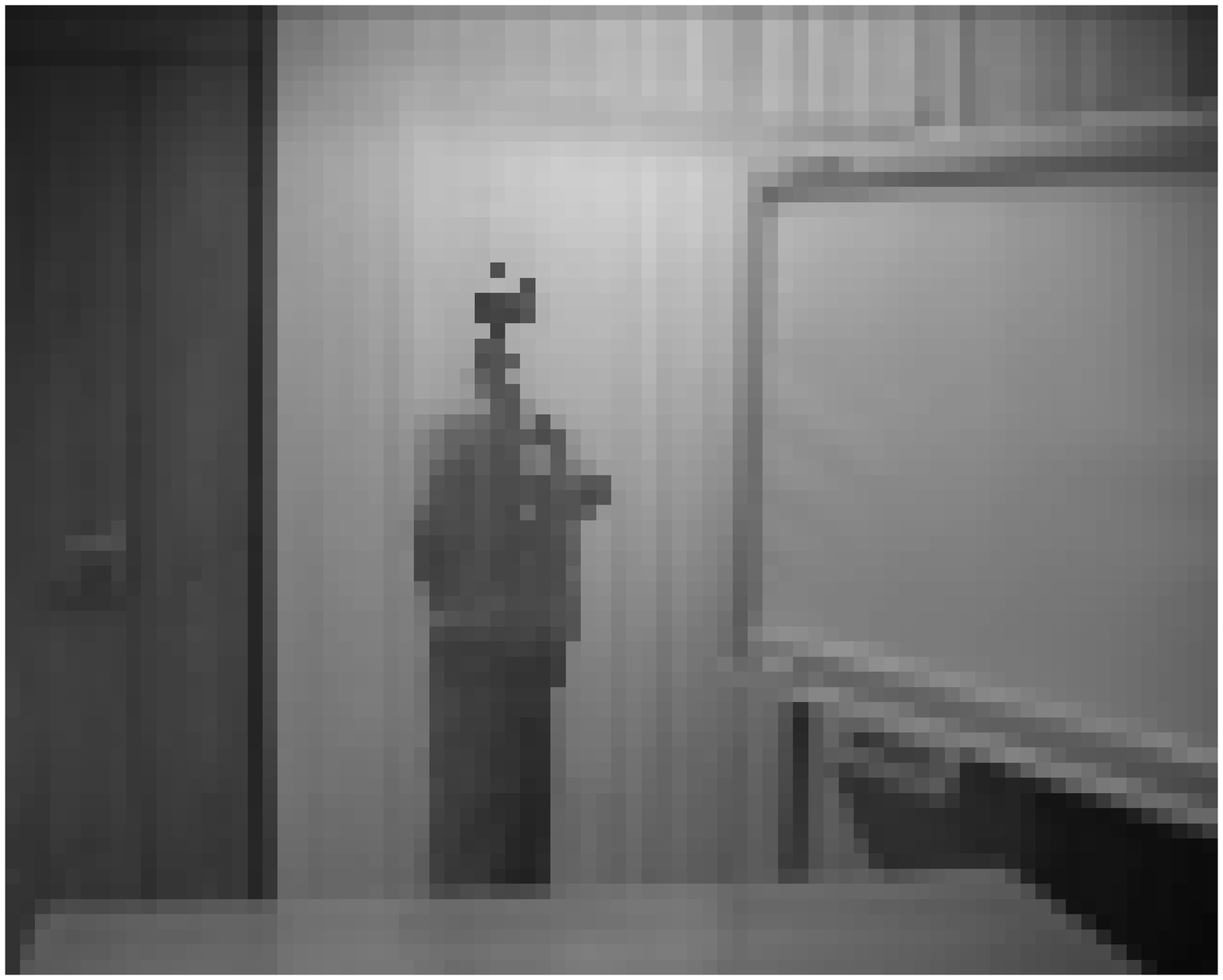}
		\includegraphics[width=15mm]{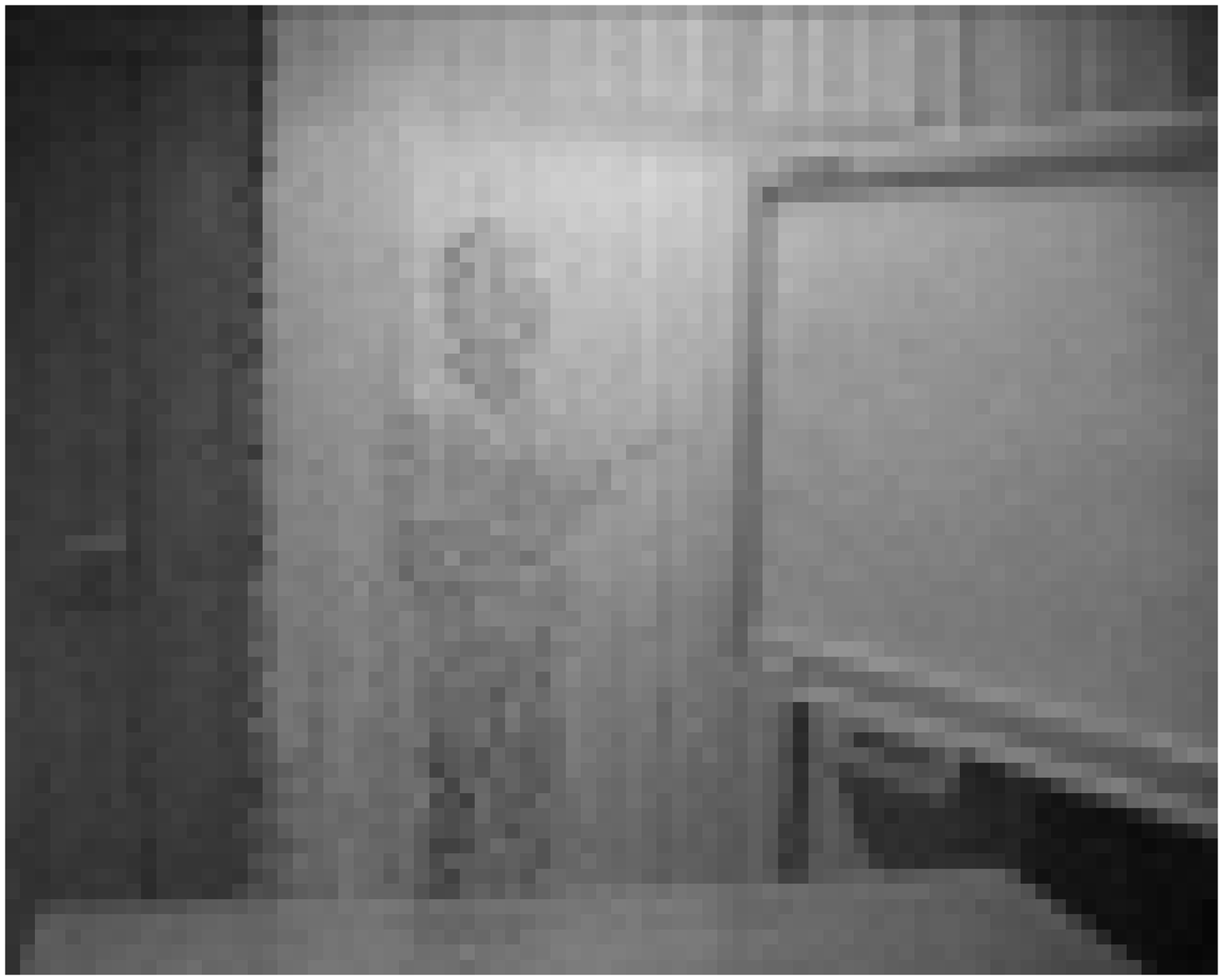}
		\includegraphics[width=15mm]{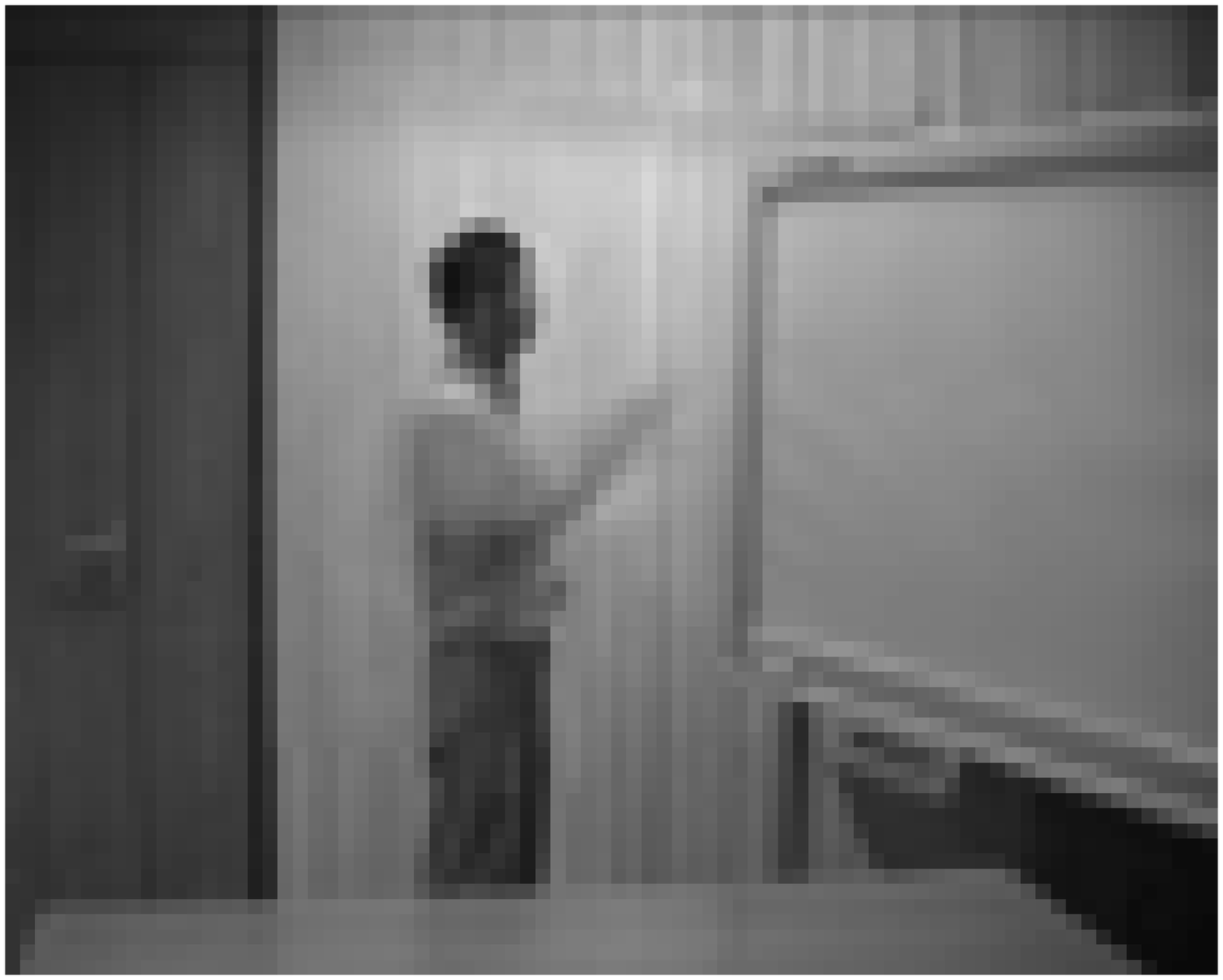}\\
		\includegraphics[width=15mm]{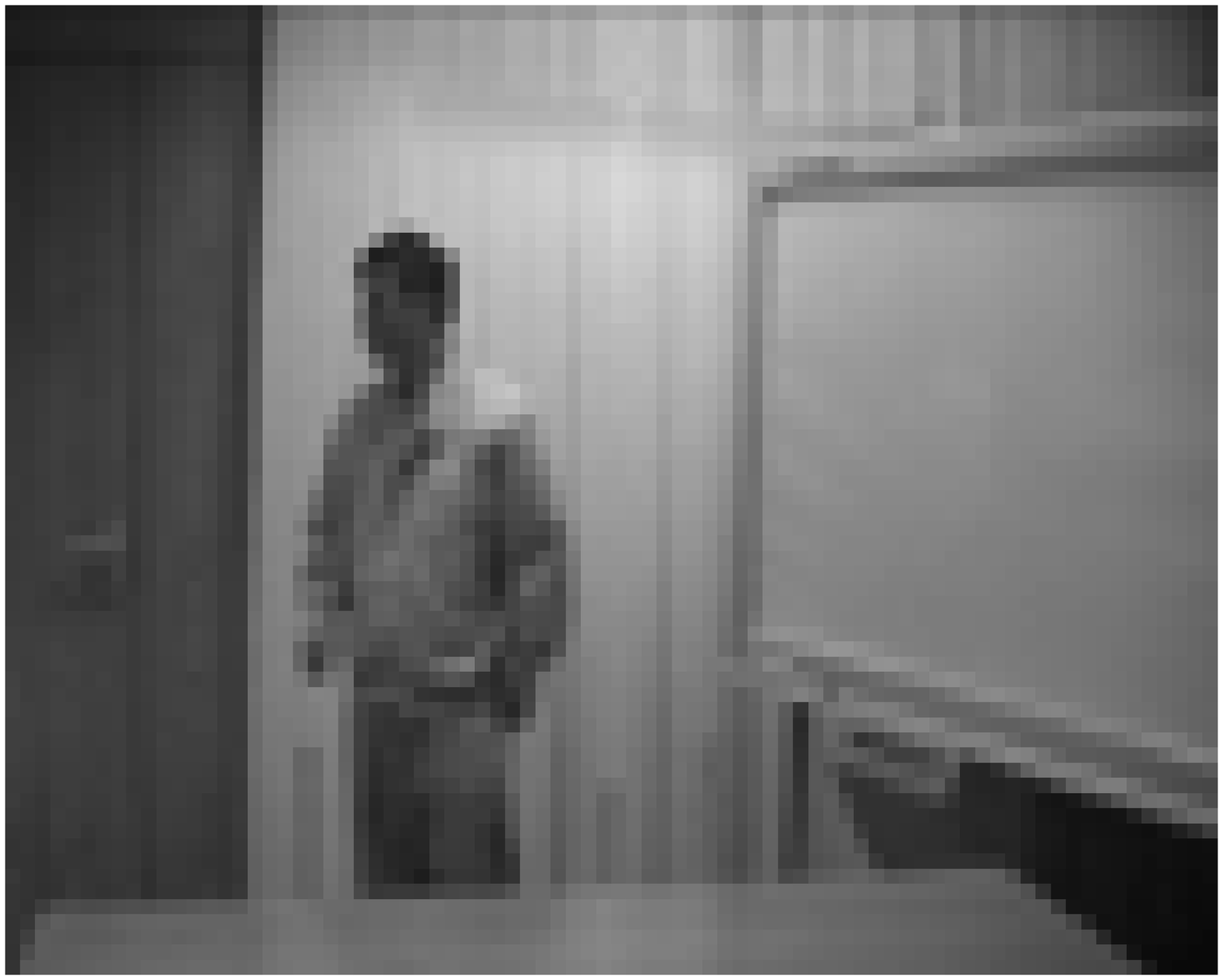}
		\includegraphics[width=15mm]{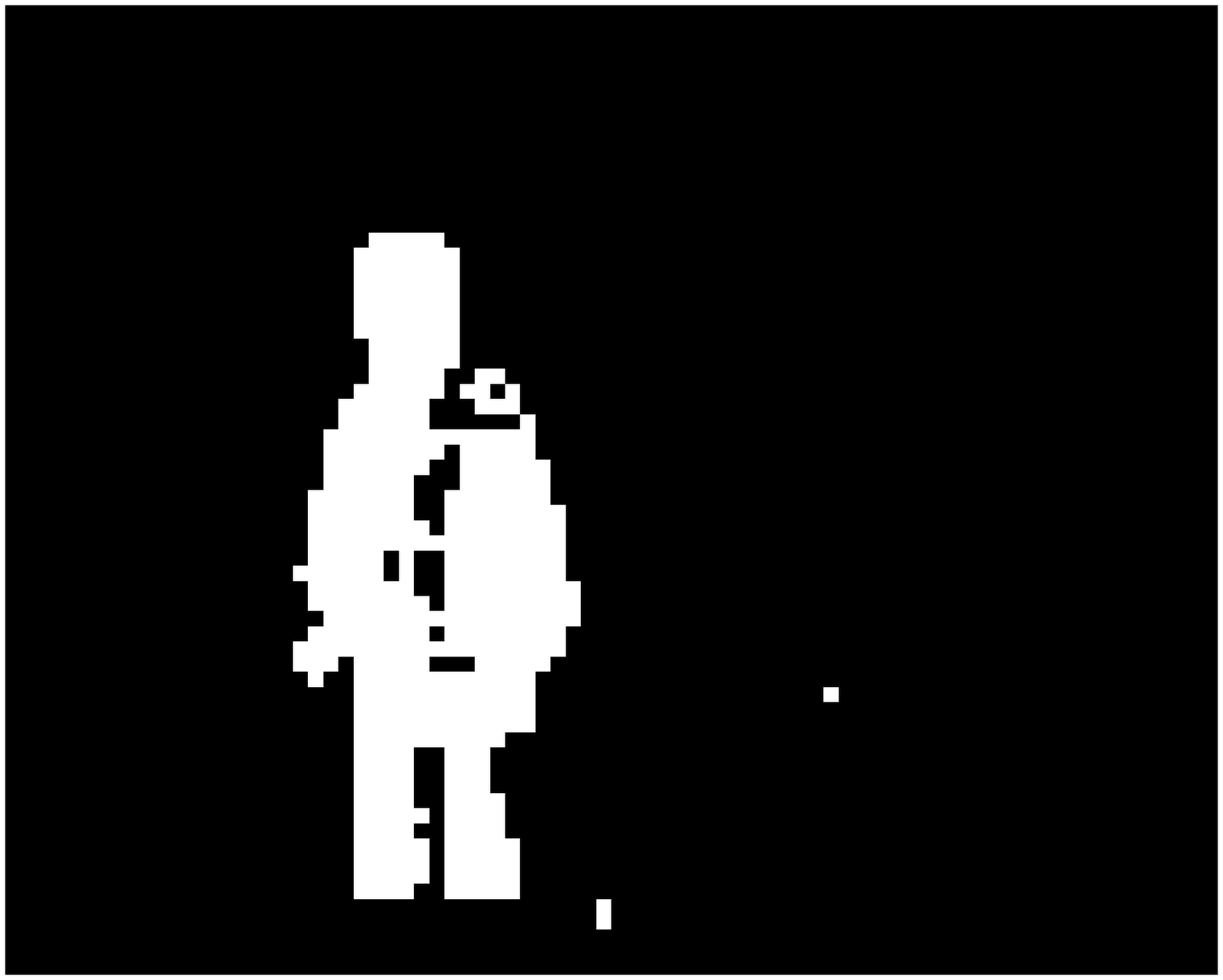}
		\includegraphics[width=15mm]{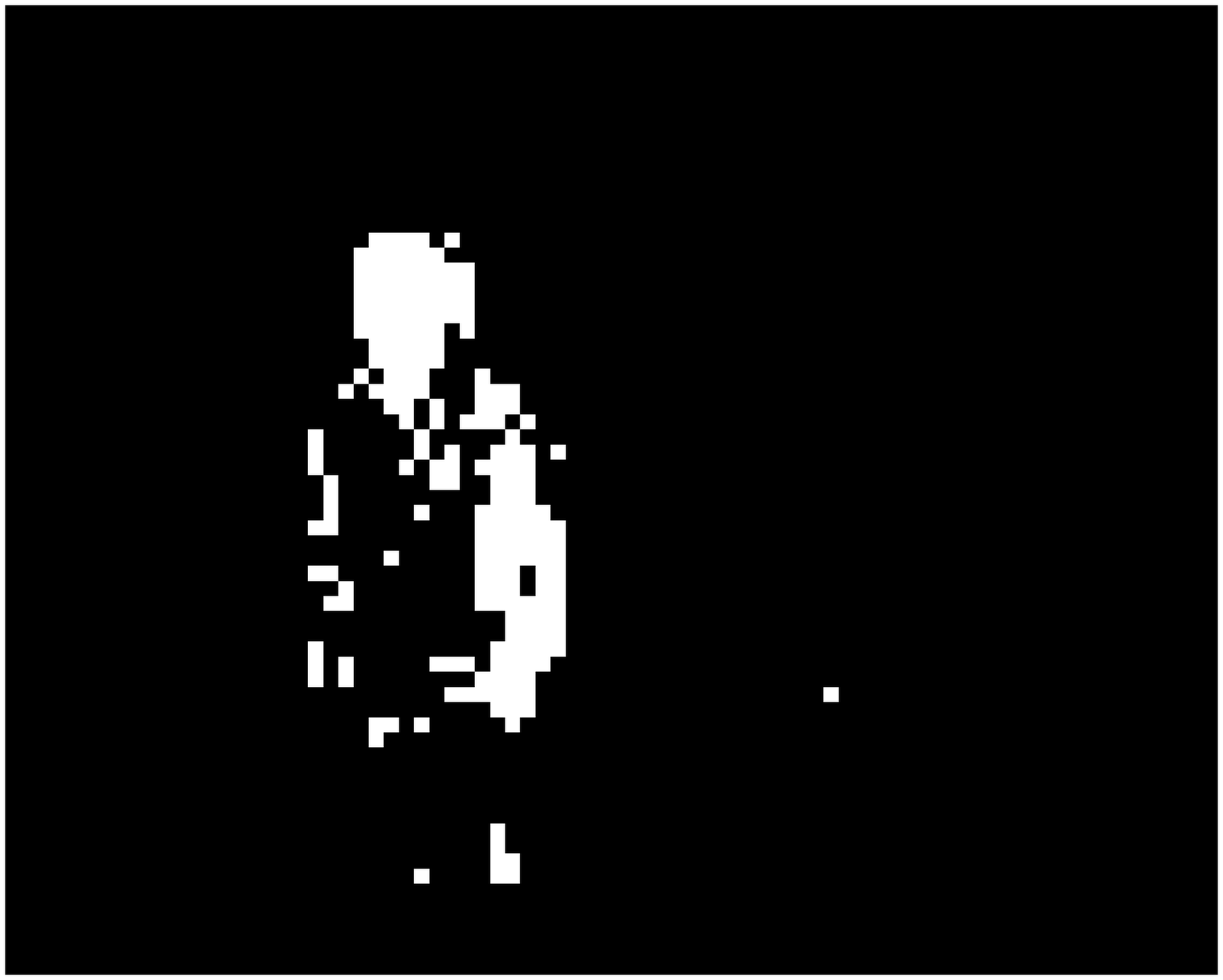}
		\includegraphics[width=15mm]{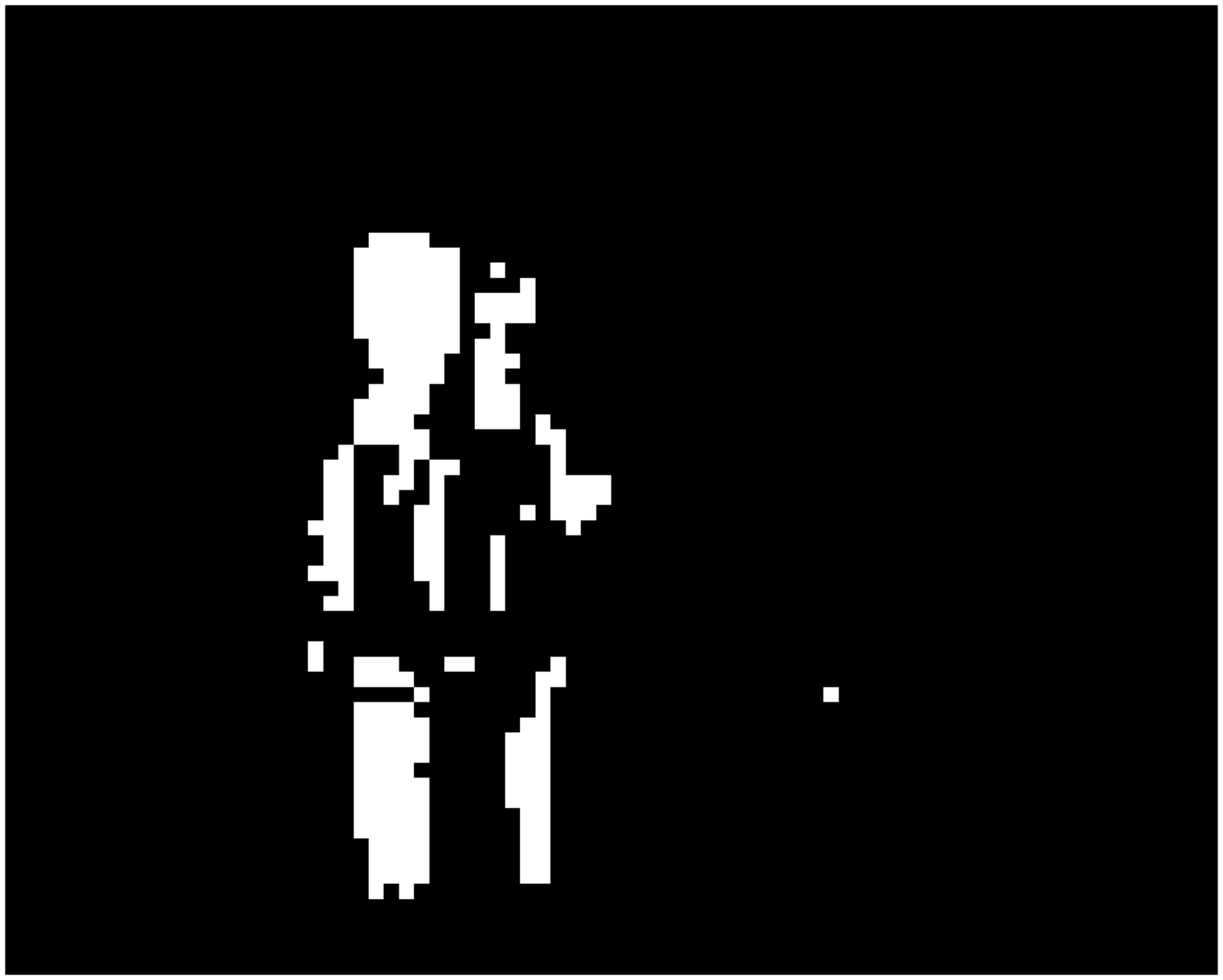}
		\includegraphics[width=15mm]{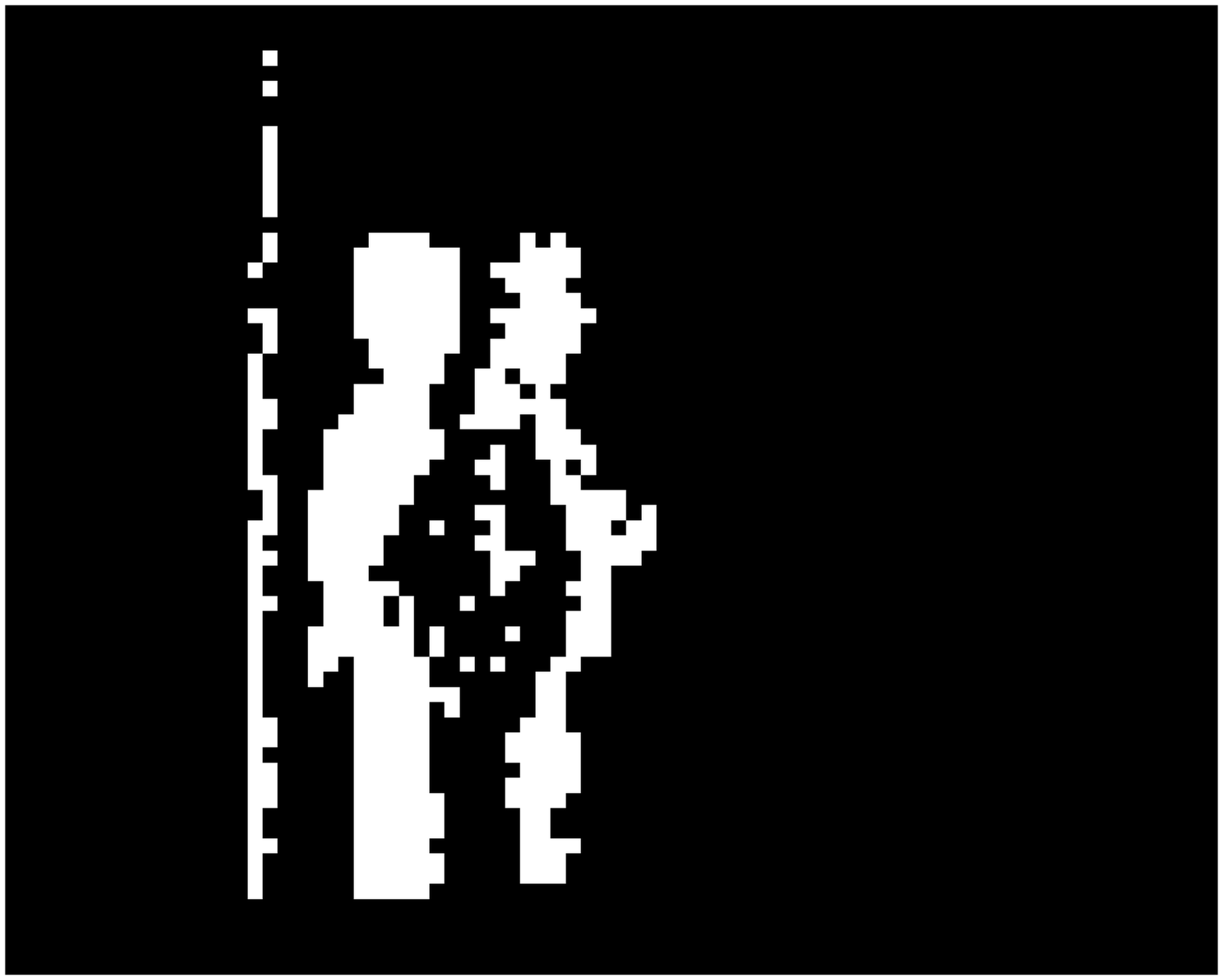}
		\includegraphics[width=15mm]{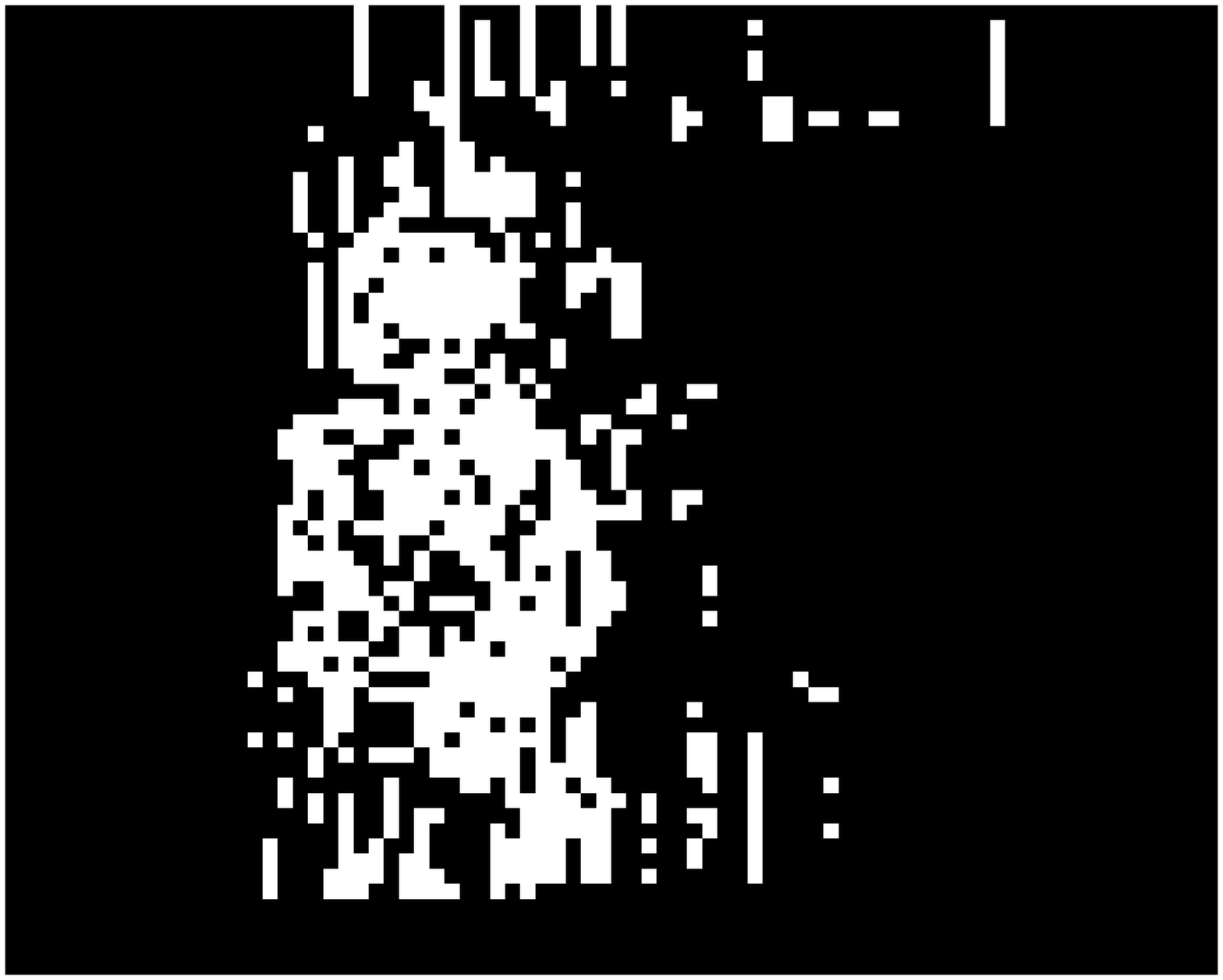}
		\includegraphics[width=15mm]{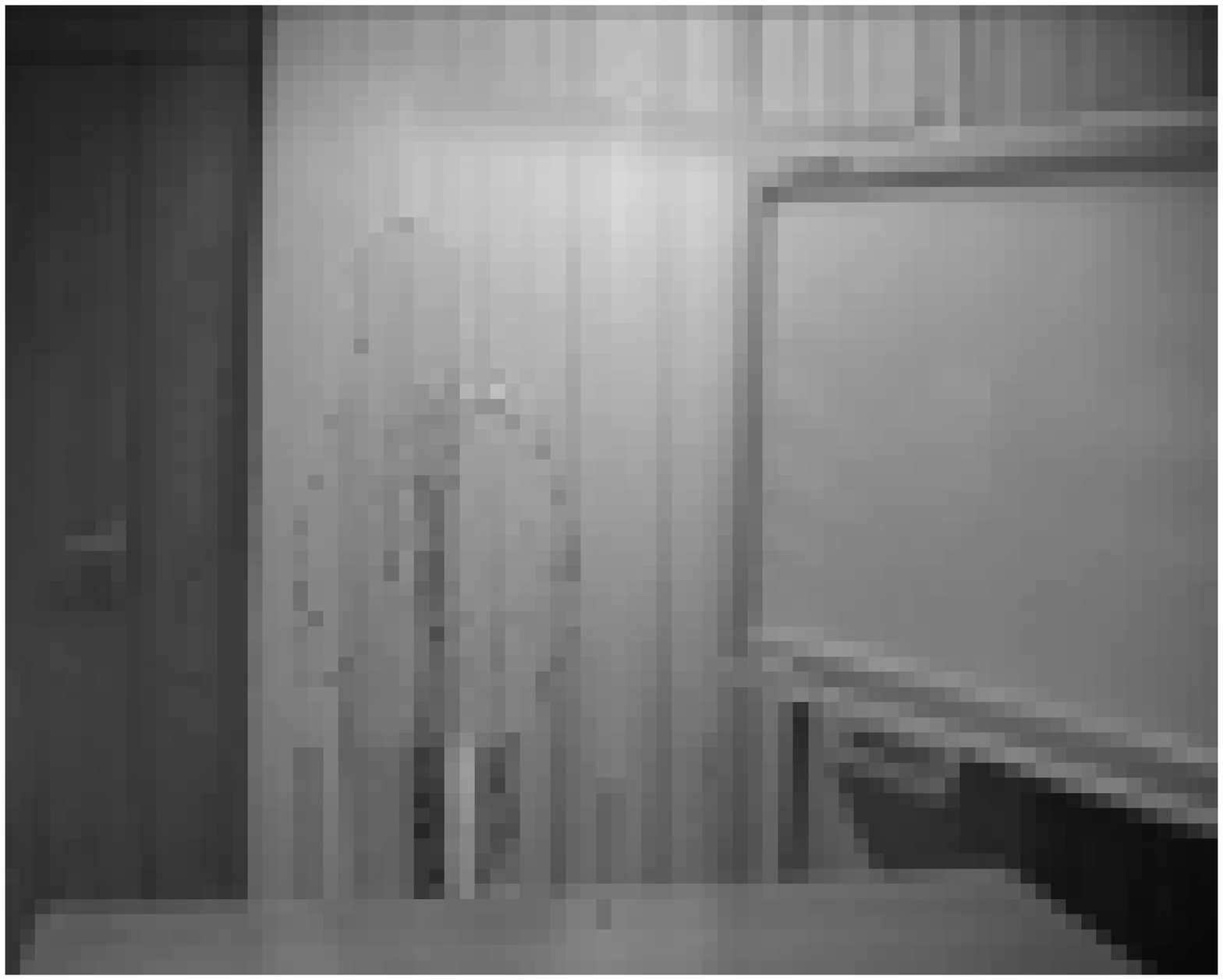}
		\includegraphics[width=15mm]{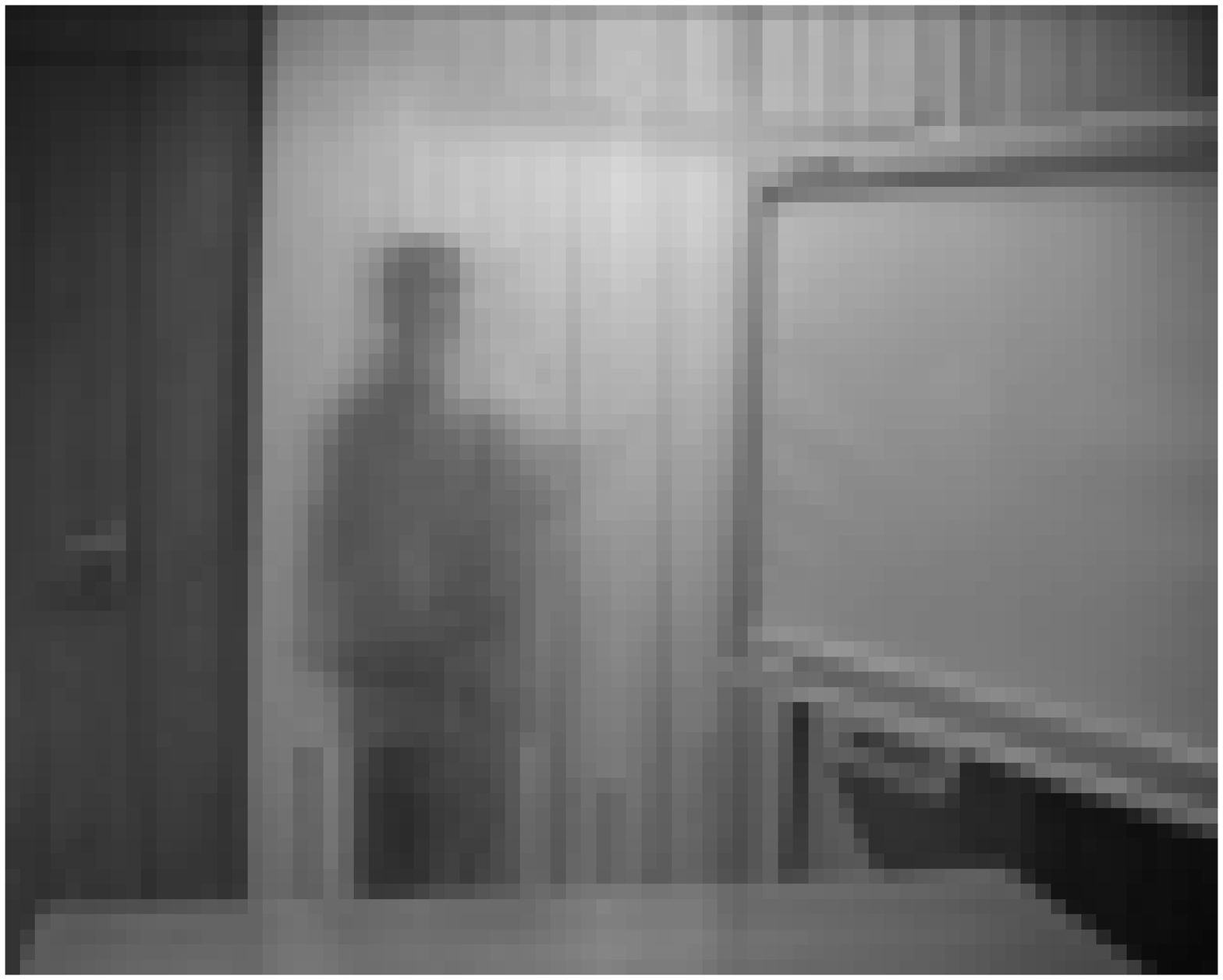}
		\includegraphics[width=15mm]{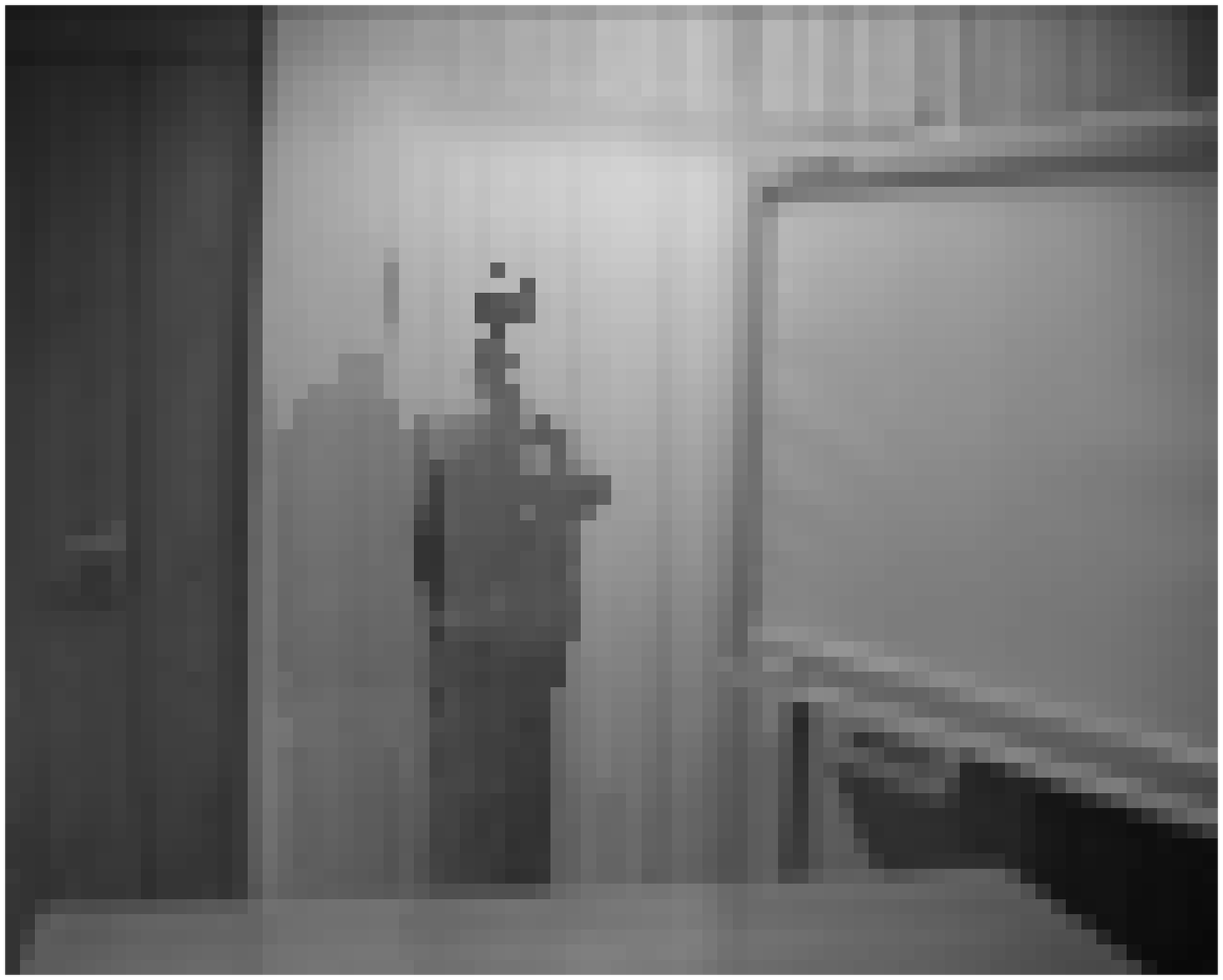}
		\includegraphics[width=15mm]{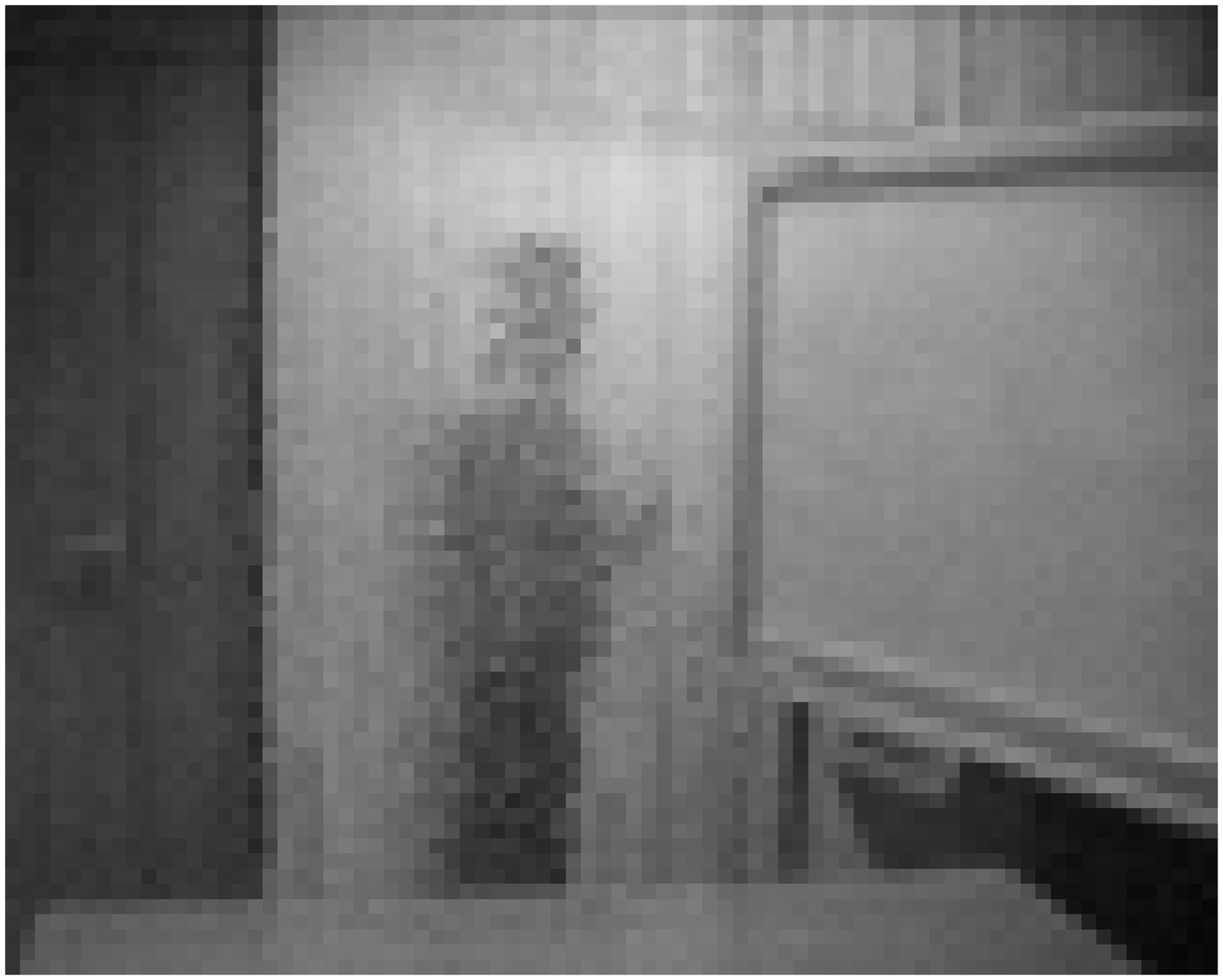}
		\includegraphics[width=15mm]{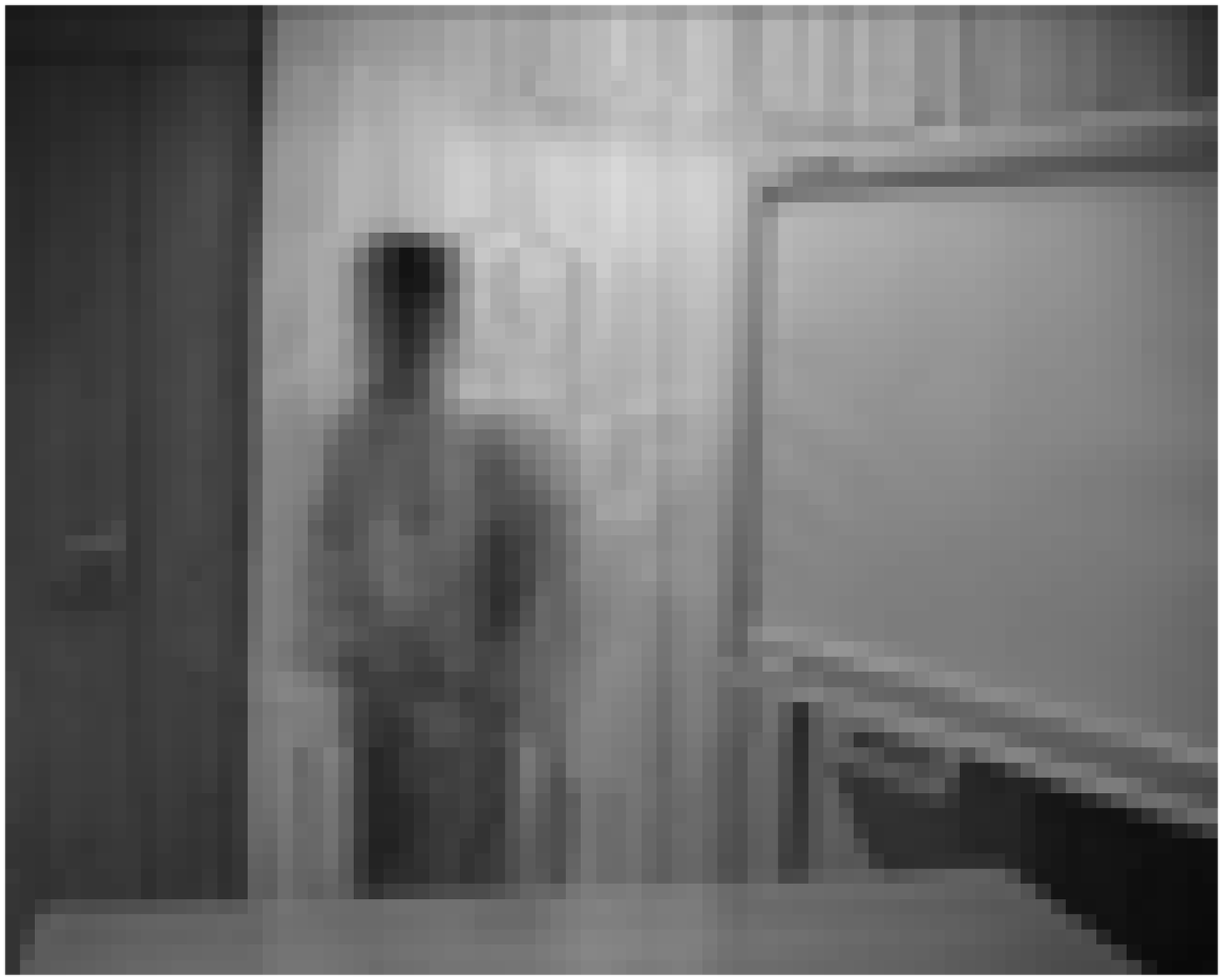}\\
		\hspace*{\fill}\makebox[0pt]{original }\hspace*{\fill}
		\hspace*{\fill}\makebox[0pt]{ReProCS}\hspace*{\fill}
		\hspace*{\fill}\makebox[0pt]{PCP}\hspace*{\fill}
		\hspace*{\fill}\makebox[0pt]{RSL}\hspace*{\fill}
		\hspace*{\fill}\makebox[0pt]{GRASTA}\hspace*{\fill}
		\hspace*{\fill}\makebox[0pt]{MG}\hspace*{\fill}
		\hspace*{\fill}\makebox[0pt]{ReProCS}\hspace*{\fill}
		\hspace*{\fill}\makebox[0pt]{PCP}\hspace*{\fill}
		\hspace*{\fill}\makebox[0pt]{RSL}\hspace*{\fill}
		\hspace*{\fill}\makebox[0pt]{GRASTA}\hspace*{\fill}
		\hspace*{\fill}\makebox[0pt]{MG}\hspace*{\fill}\\
		\hspace*{\fill}\makebox[0pt]{}\hspace*{\fill}
		\hspace*{\fill}\makebox[0pt]{(fg)}\hspace*{\fill}
		\hspace*{\fill}\makebox[0pt]{(fg)}\hspace*{\fill}
		\hspace*{\fill}\makebox[0pt]{(fg)}\hspace*{\fill}
		\hspace*{\fill}\makebox[0pt]{(fg)}\hspace*{\fill}
		\hspace*{\fill}\makebox[0pt]{(fg)}\hspace*{\fill}
		\hspace*{\fill}\makebox[0pt]{(bg)}\hspace*{\fill}
		\hspace*{\fill}\makebox[0pt]{(bg)}\hspace*{\fill}
		\hspace*{\fill}\makebox[0pt]{(bg)}\hspace*{\fill}
		\hspace*{\fill}\makebox[0pt]{(bg)}\hspace*{\fill}
		\hspace*{\fill}\makebox[0pt]{(bg)}\hspace*{\fill}
	
	\end{tabular}
\caption{\small{Original video sequence at $t=t_\train+60, 120, 199, 475, 1148$ and its foreground (fg) and background (bg) layer recovery results using ReProCS (ReProCS-pCA) and other algorithms.
For fg, we only show the fg support in white for ease of display.
}}
\label{CurtainCompare}
\end{figure*}

{\em Real Video Sequences. }
Next we show comparisons on two real video sequences. These are originally taken from \url{http://perception.i2r.a-star.edu.sg/bk_model/bk_index.html} and \url{http://research.microsoft.com/en-us/um/people/jckrumm/wallflower/testimages.htm}, respectively. 
The first is the curtain sequence described earlier. For $t>1755$, in the foreground, a person with a black shirt walks in, writes on the board and then walk out, then a second person with a white shirt does the same and then a third person with a white shirt does the same. This video is challenging because (i) the white shirt color and the curtains' color is quite similar, making the corresponding $S_t$ small in magnitude; and (ii) because the background variations are quite large while the foreground person moves slowly. As can be seen from Fig. \ref{CurtainCompare}, ReProCS's performance is significantly better than that of the other algorithms for both foreground and background recovery. This is most easily seen from the recovered background images. One or more frames of the background recovered by PCP, RSL and GRASTA contains the person, while none of the ReProCS ones does.

\begin{figure*}
	\centering
	\begin{tabular}{cc}
		\includegraphics[width=15mm]{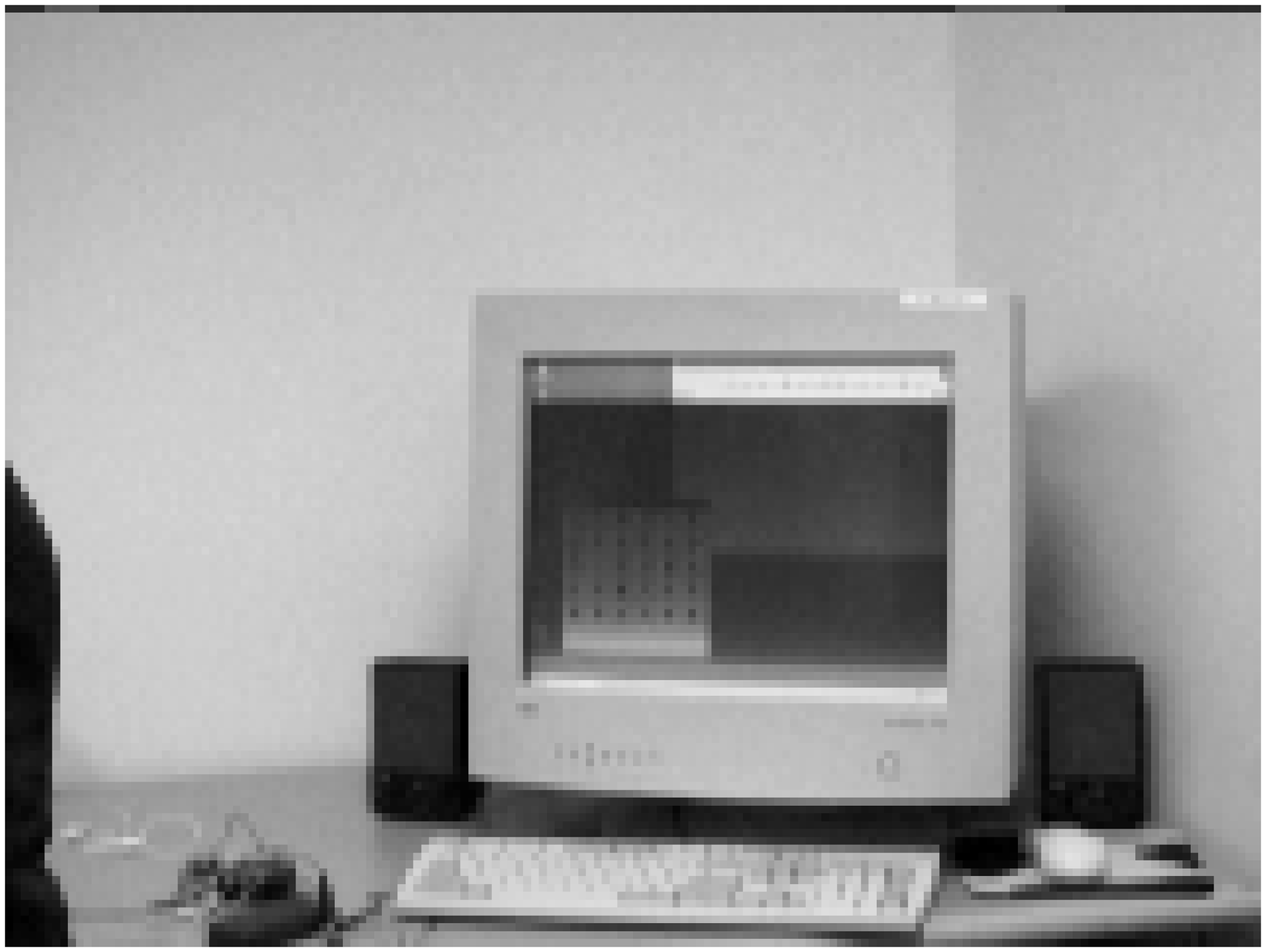}
		\includegraphics[width=15mm]{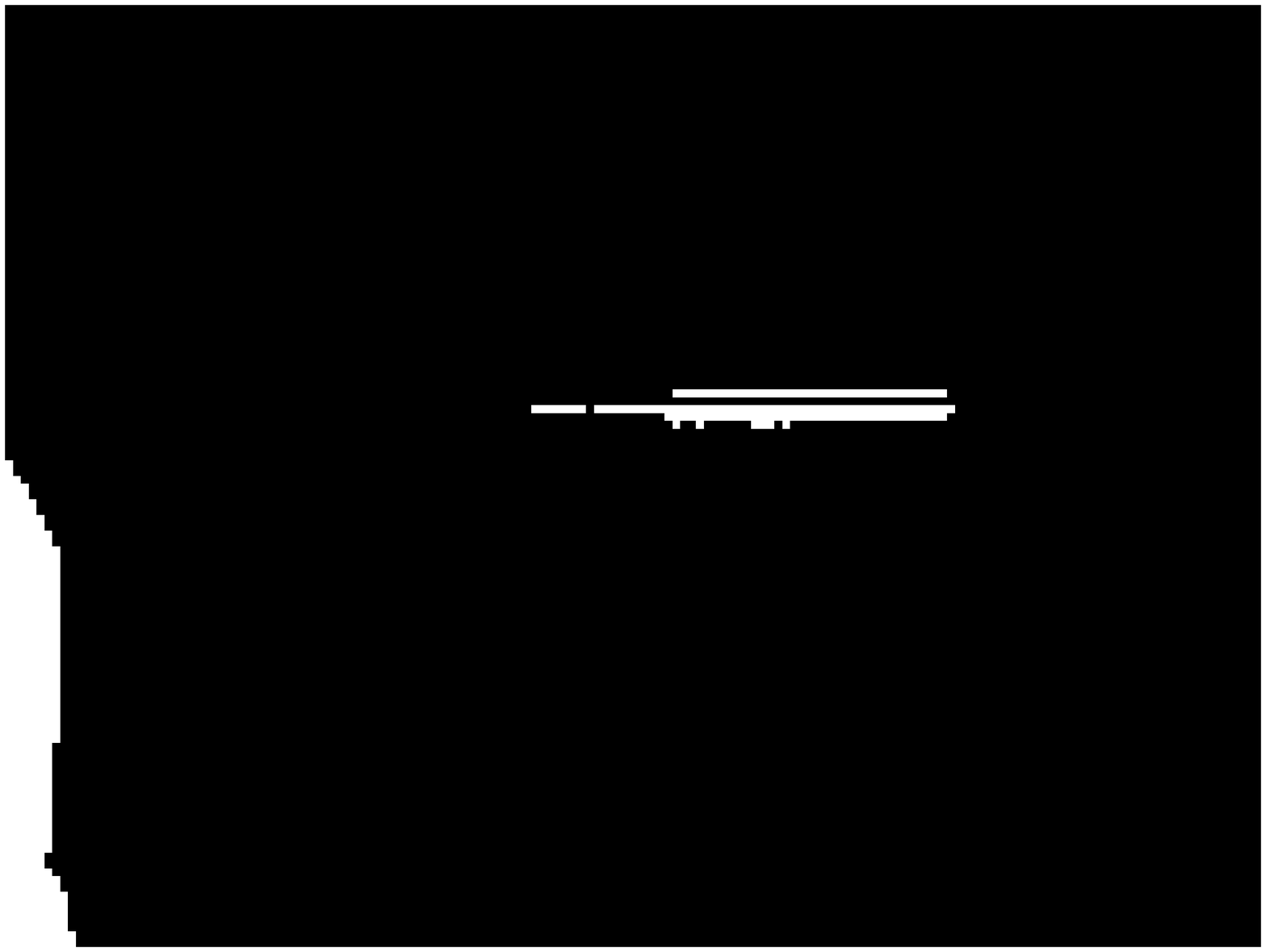}
		\includegraphics[width=15mm]{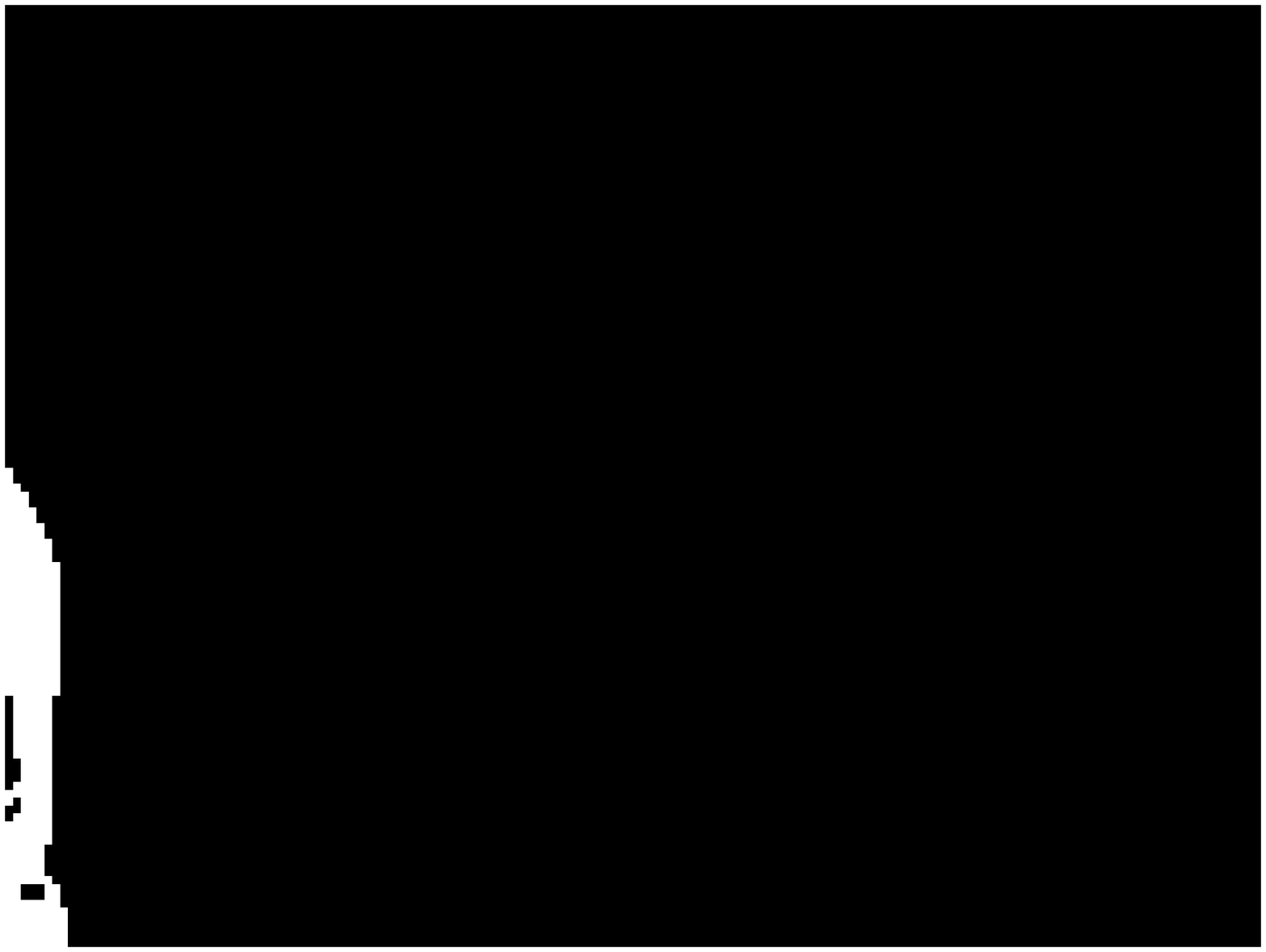}
		\includegraphics[width=15mm]{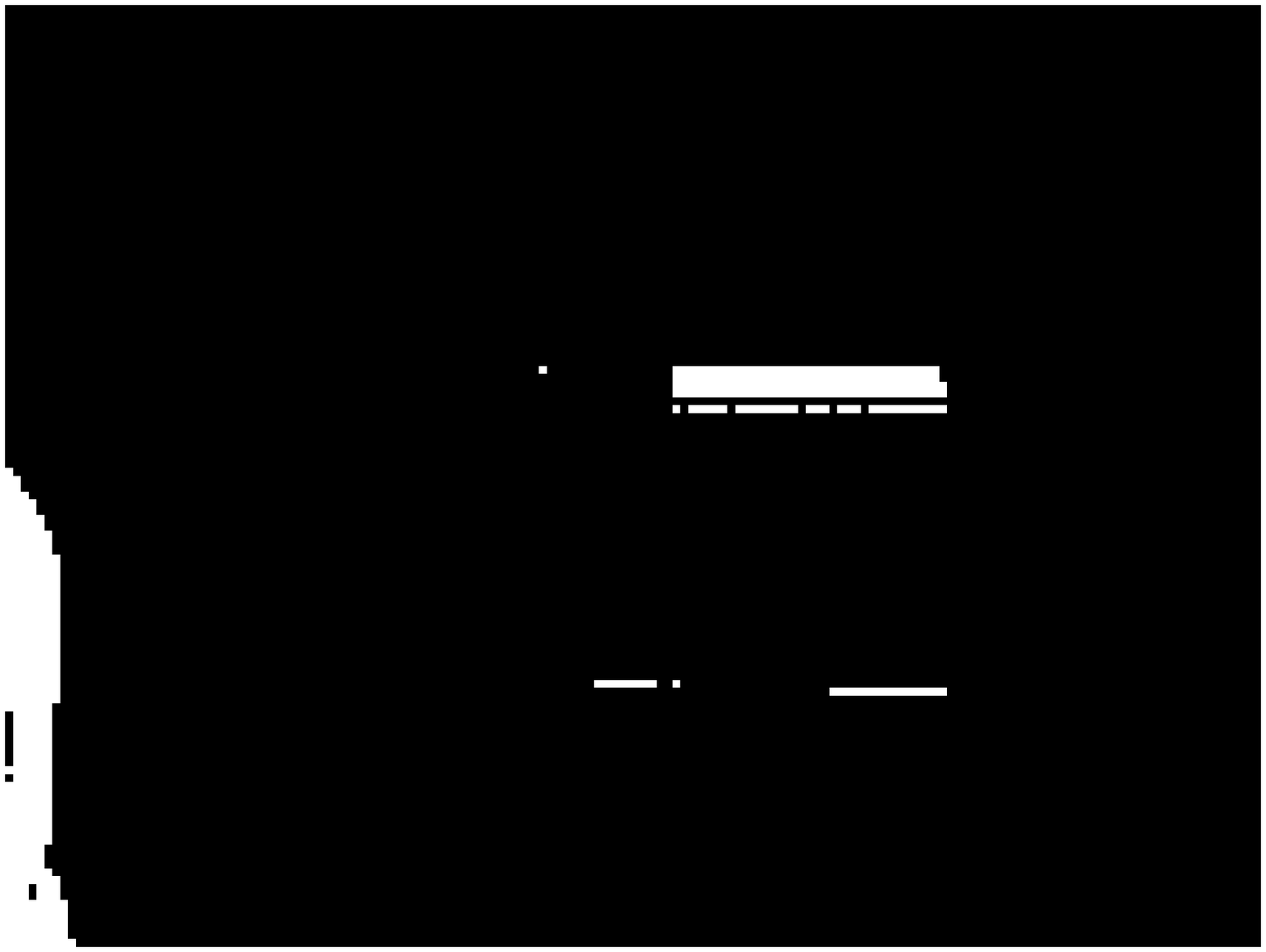}
		\includegraphics[width=15mm]{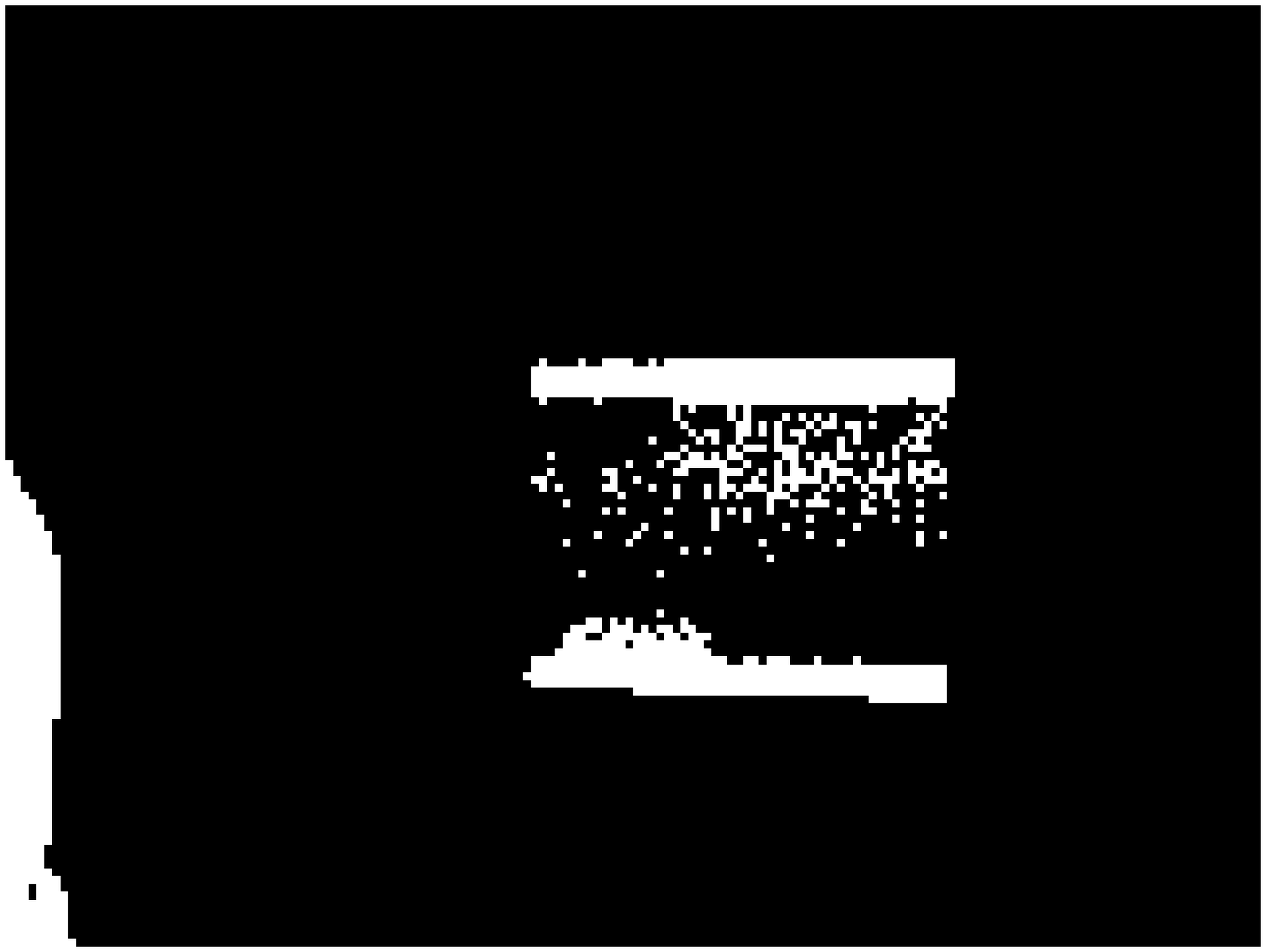}
		\includegraphics[width=15mm]{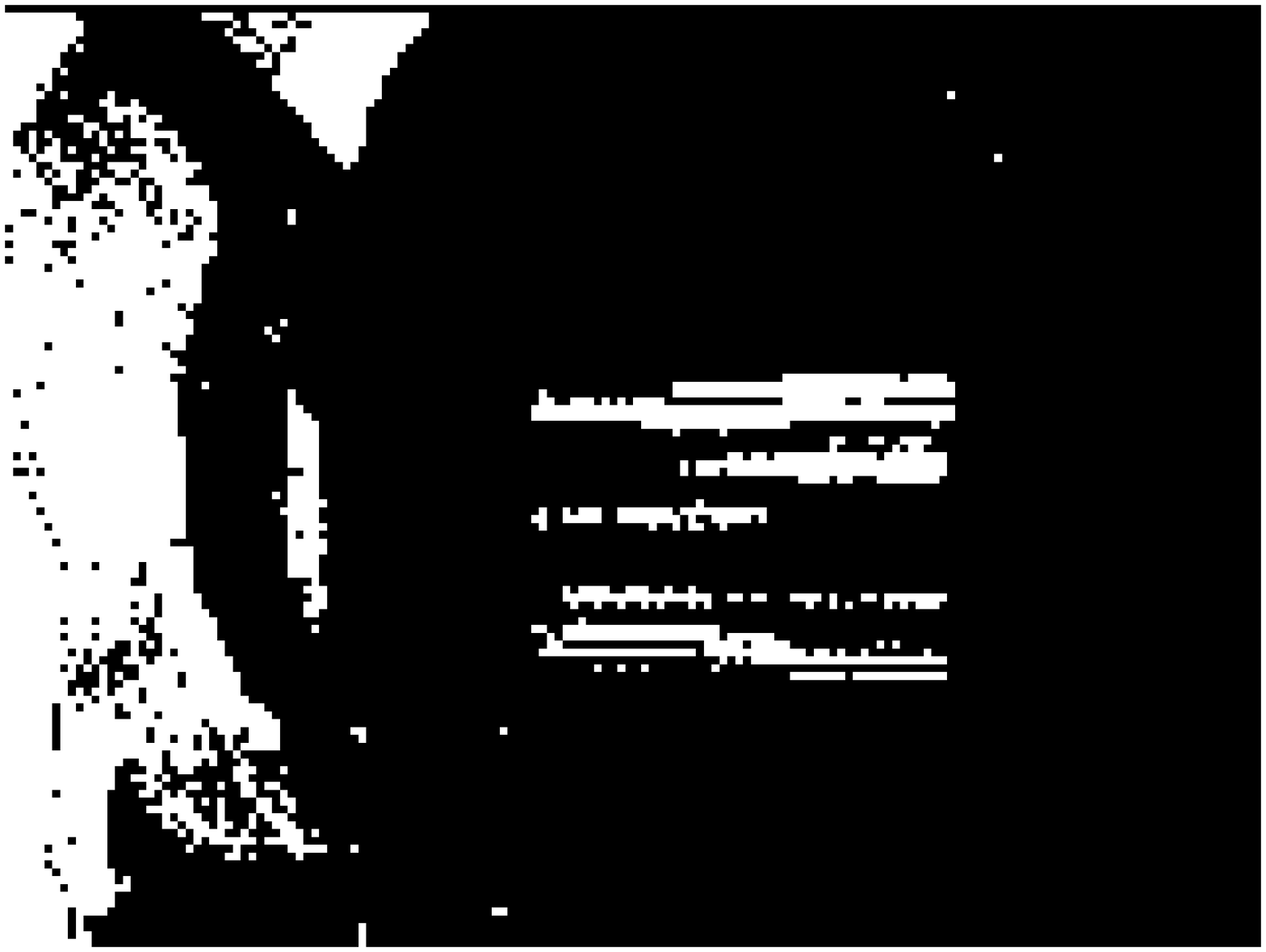}
		\includegraphics[width=15mm]{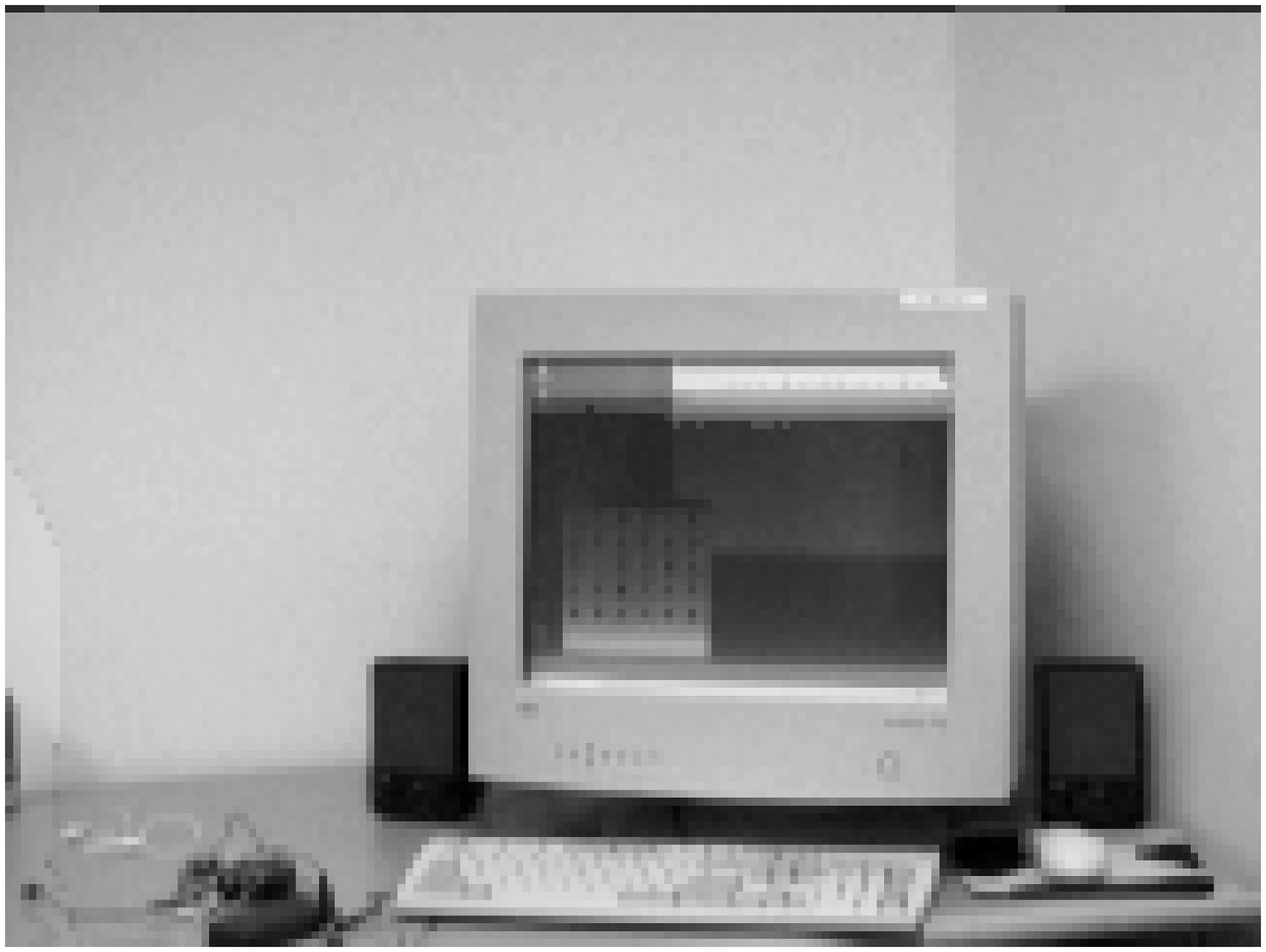}
		\includegraphics[width=15mm]{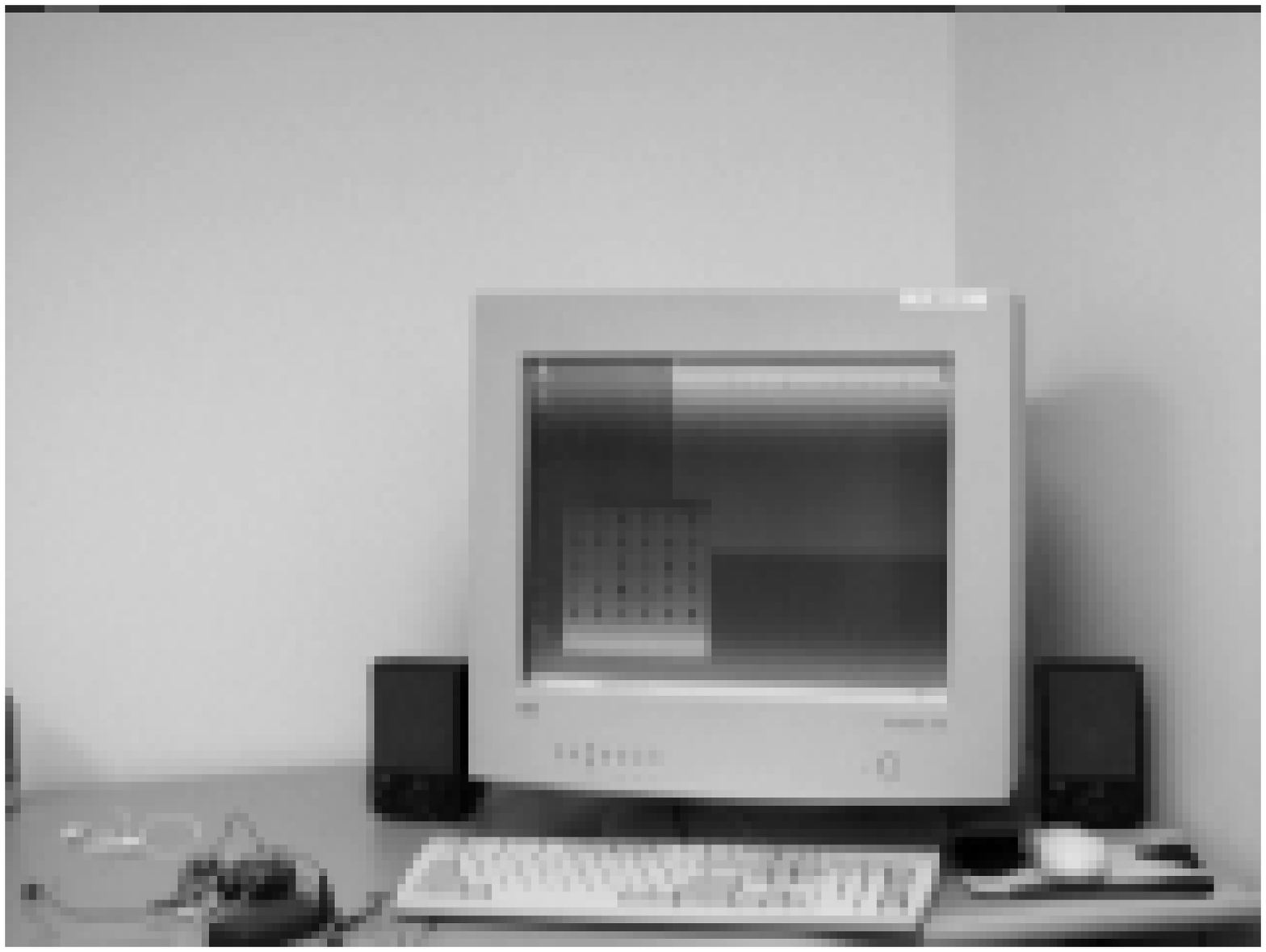}
		\includegraphics[width=15mm]{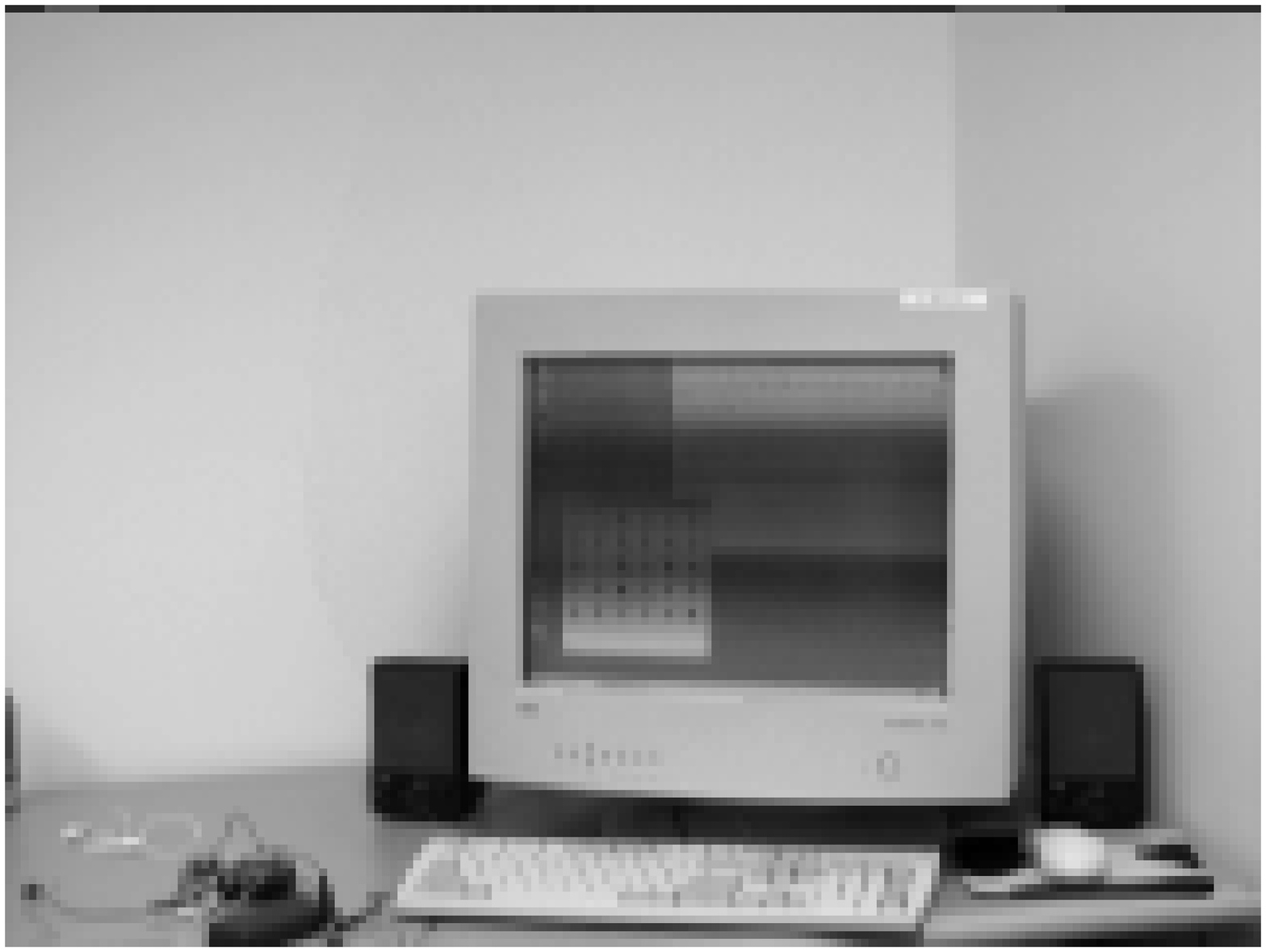}
		\includegraphics[width=15mm]{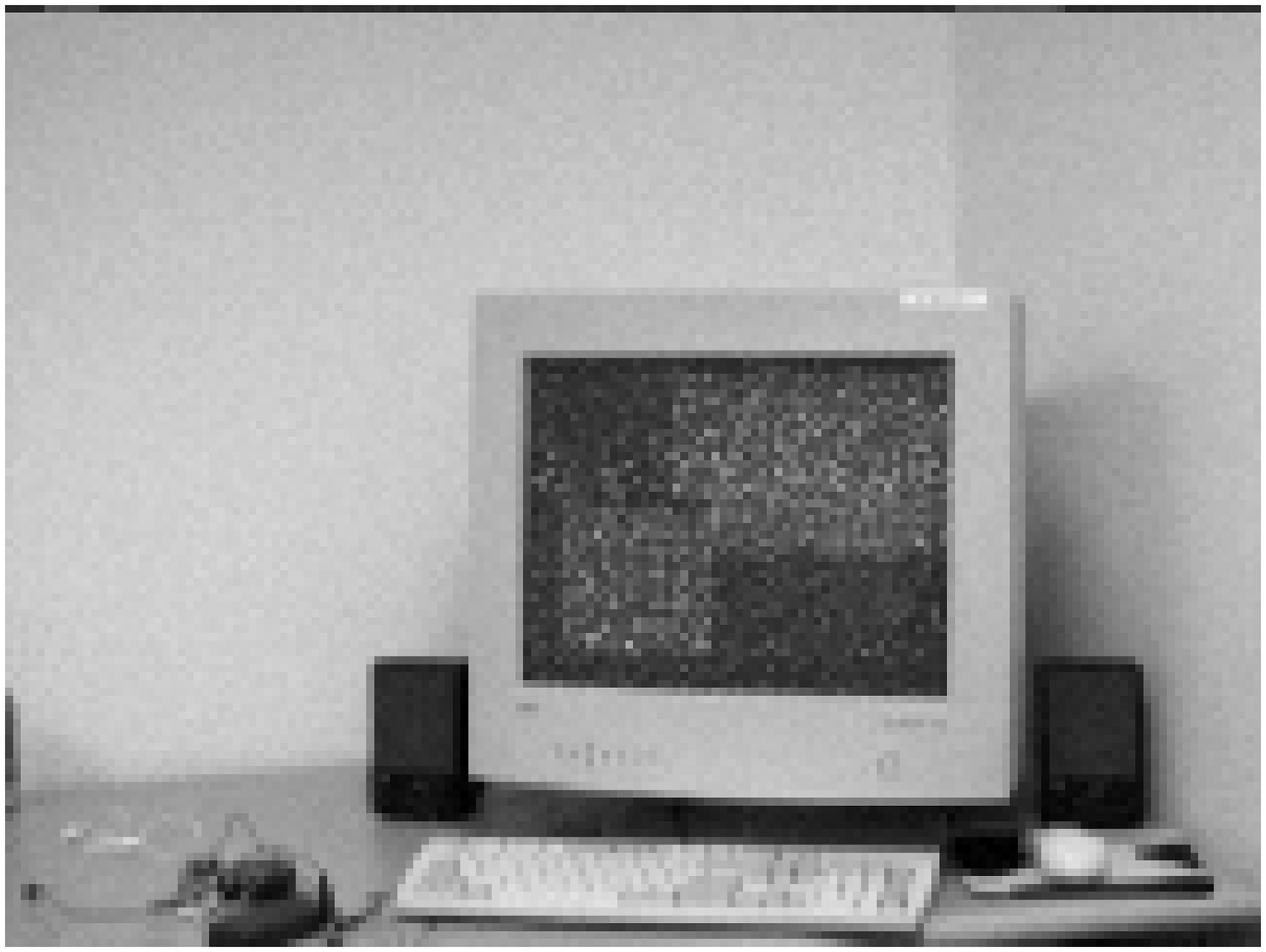}
		\includegraphics[width=15mm]{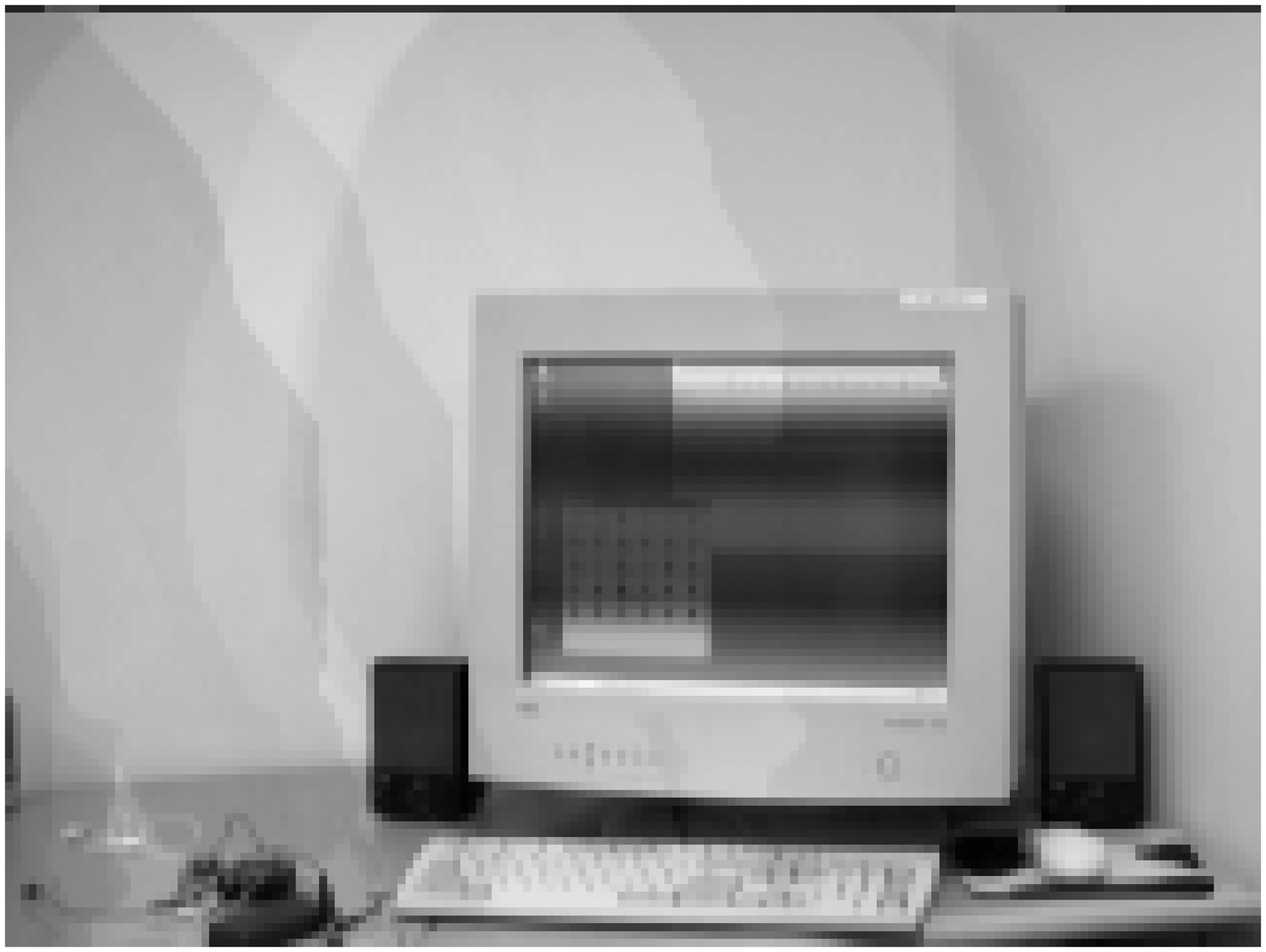}\\
		\includegraphics[width=15mm]{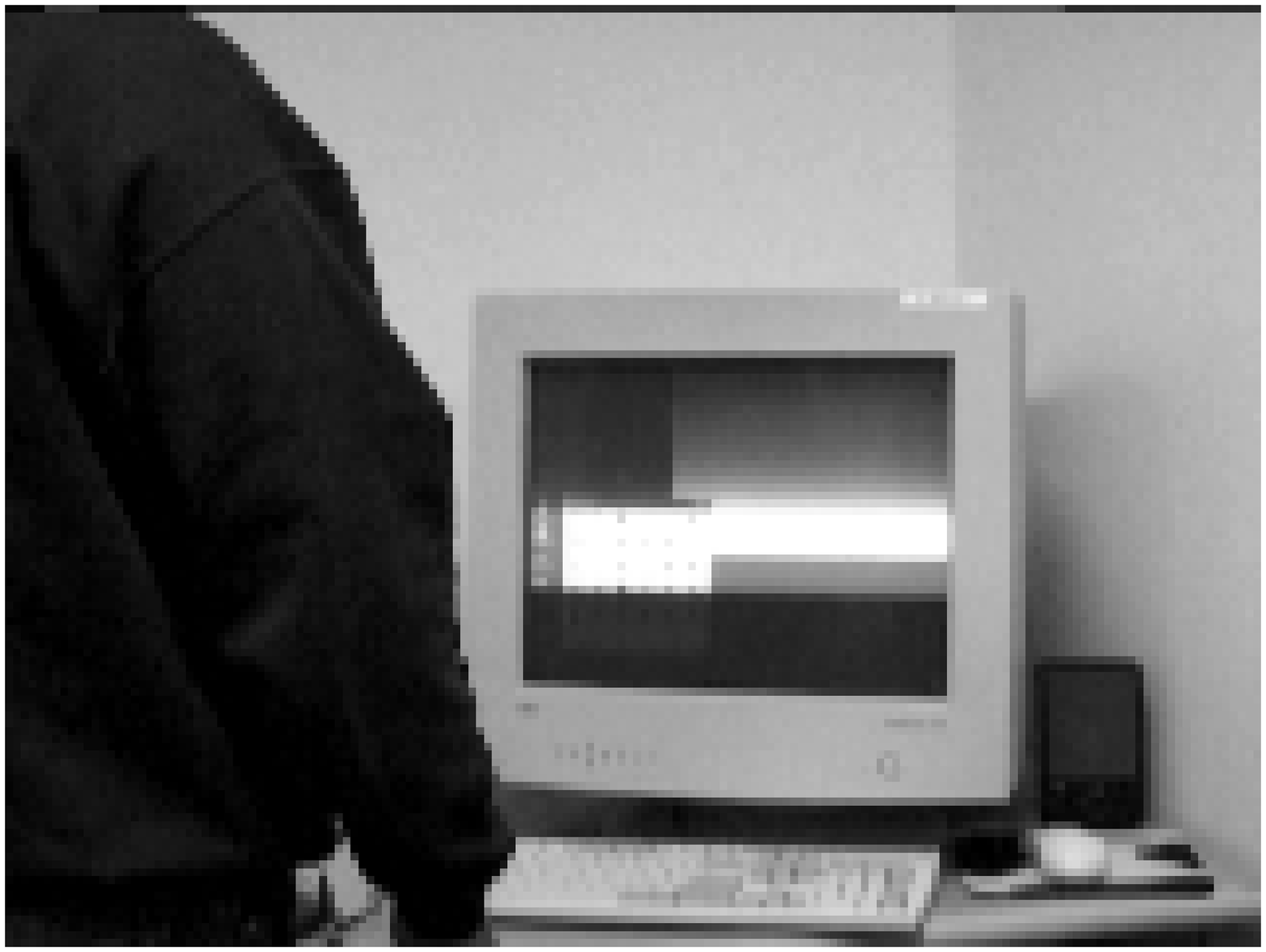}
		\includegraphics[width=15mm]{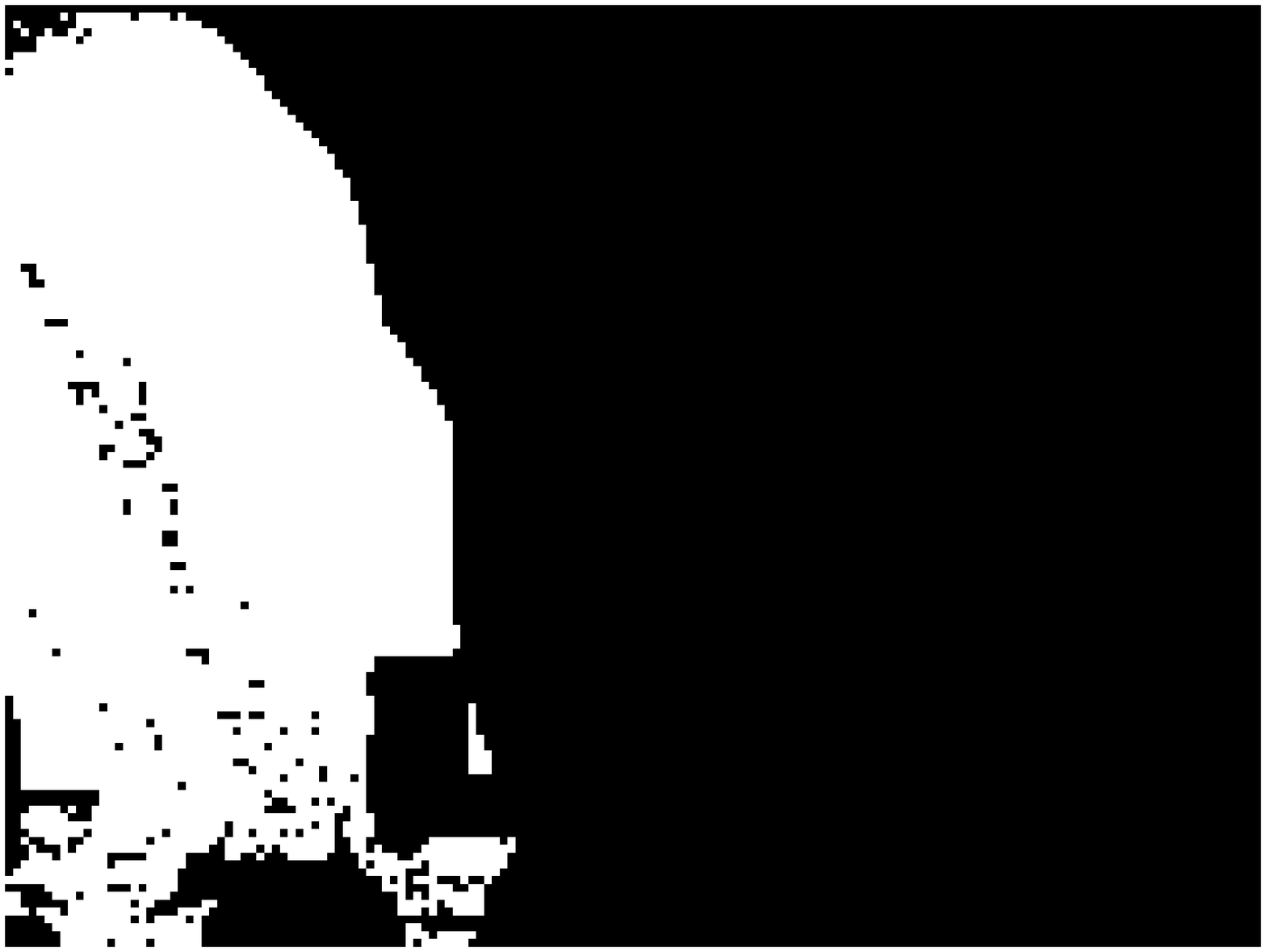}
		\includegraphics[width=15mm]{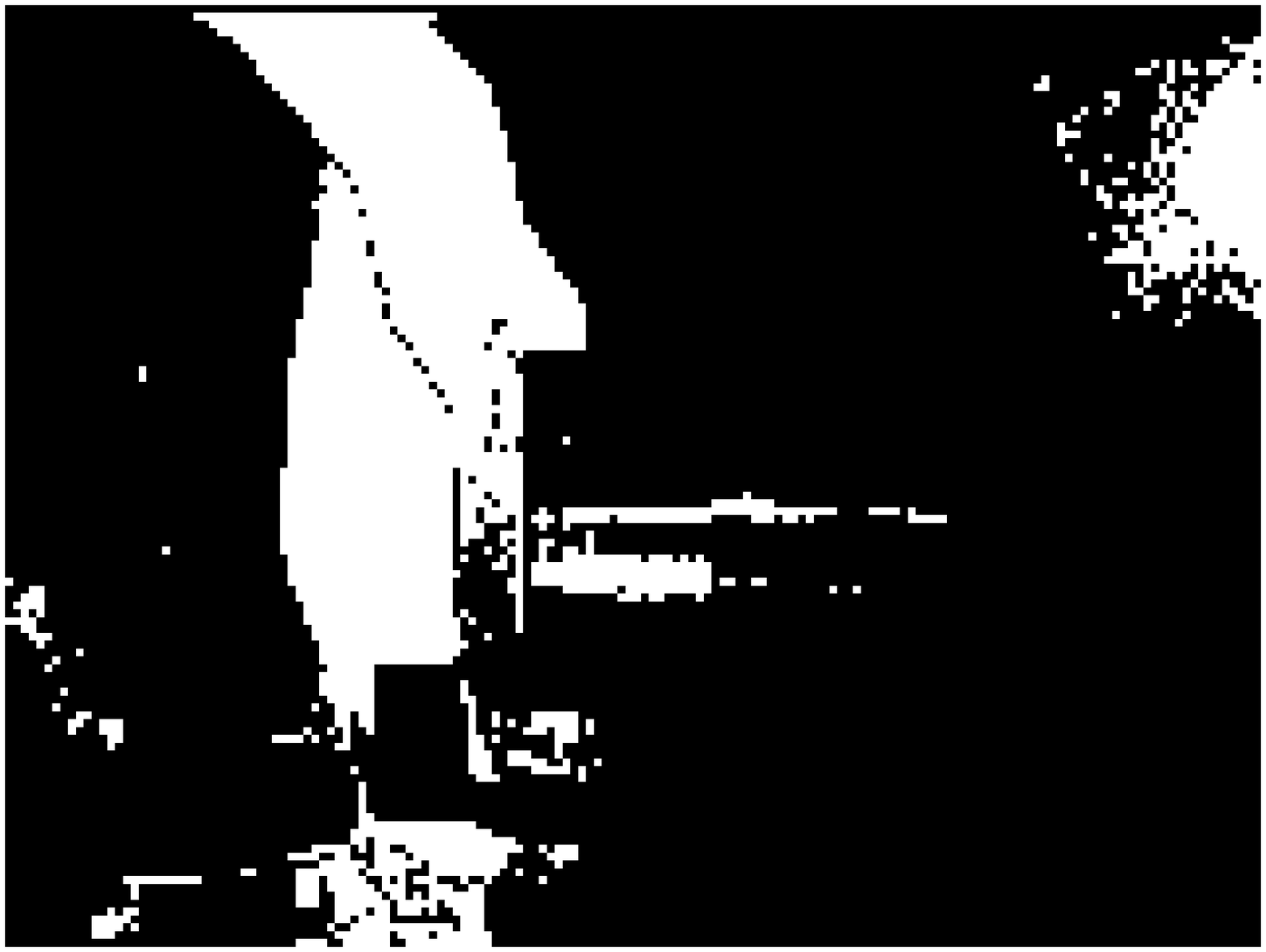}
		\includegraphics[width=15mm]{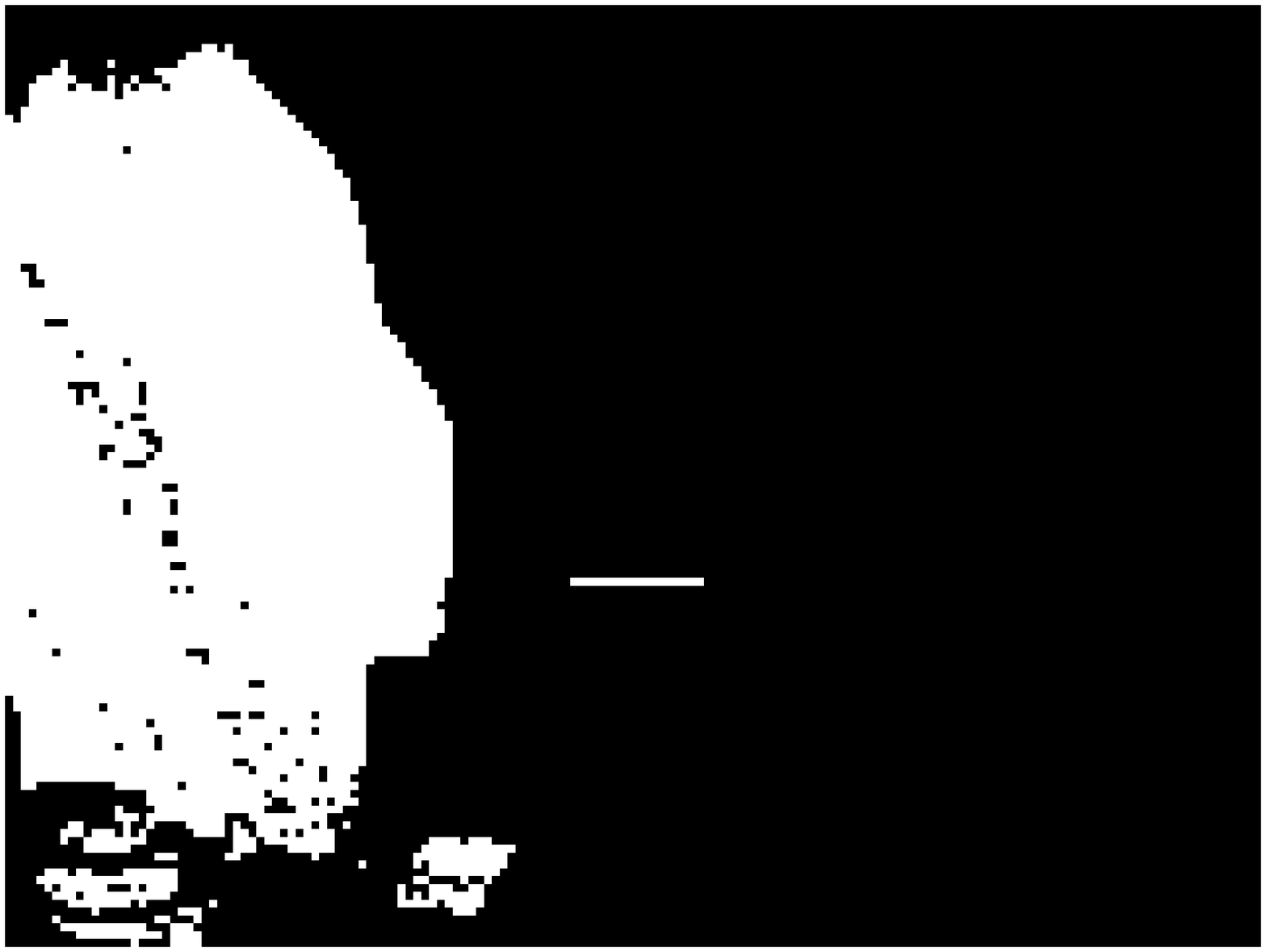}
		\includegraphics[width=15mm]{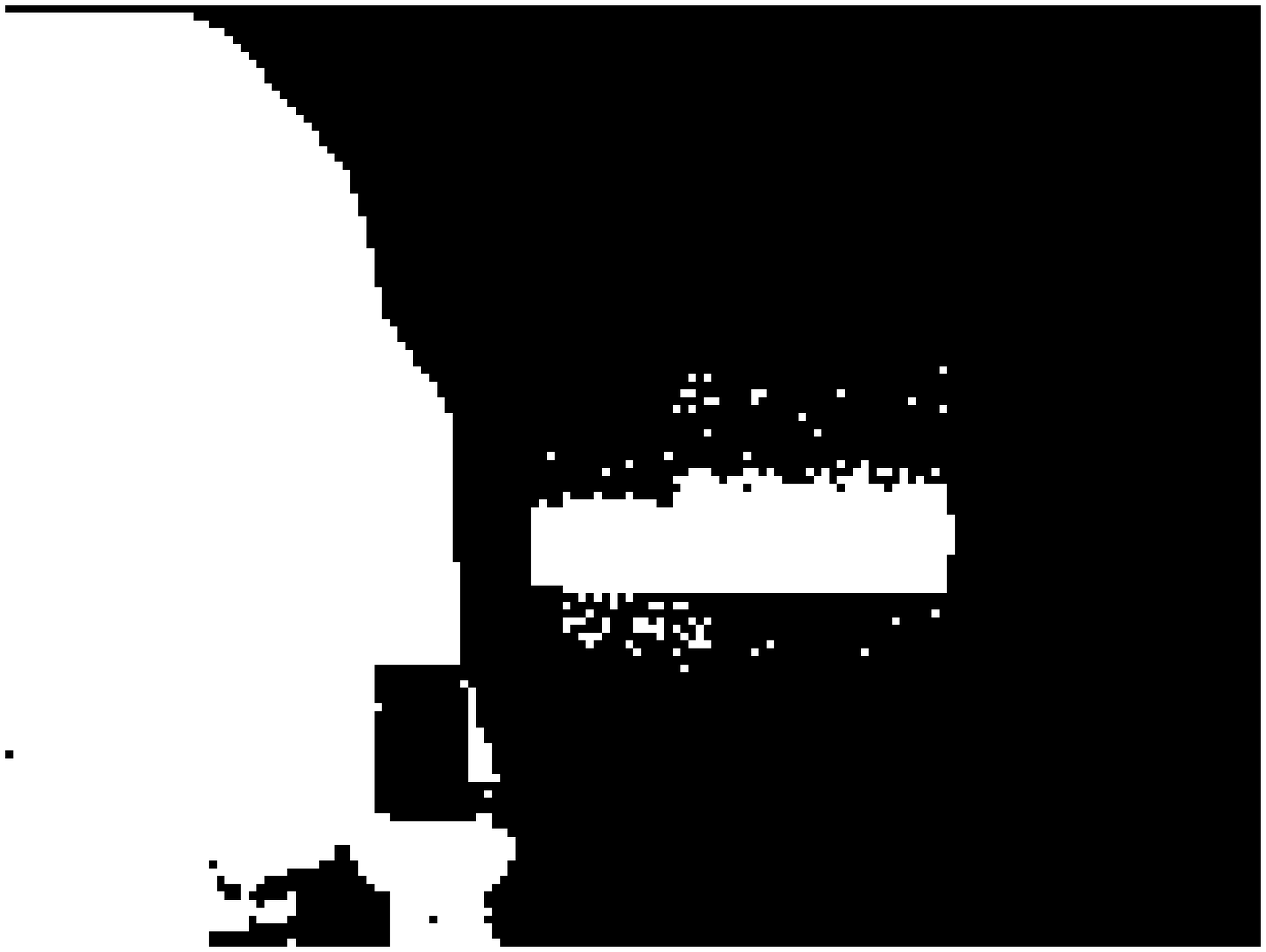}
		\includegraphics[width=15mm]{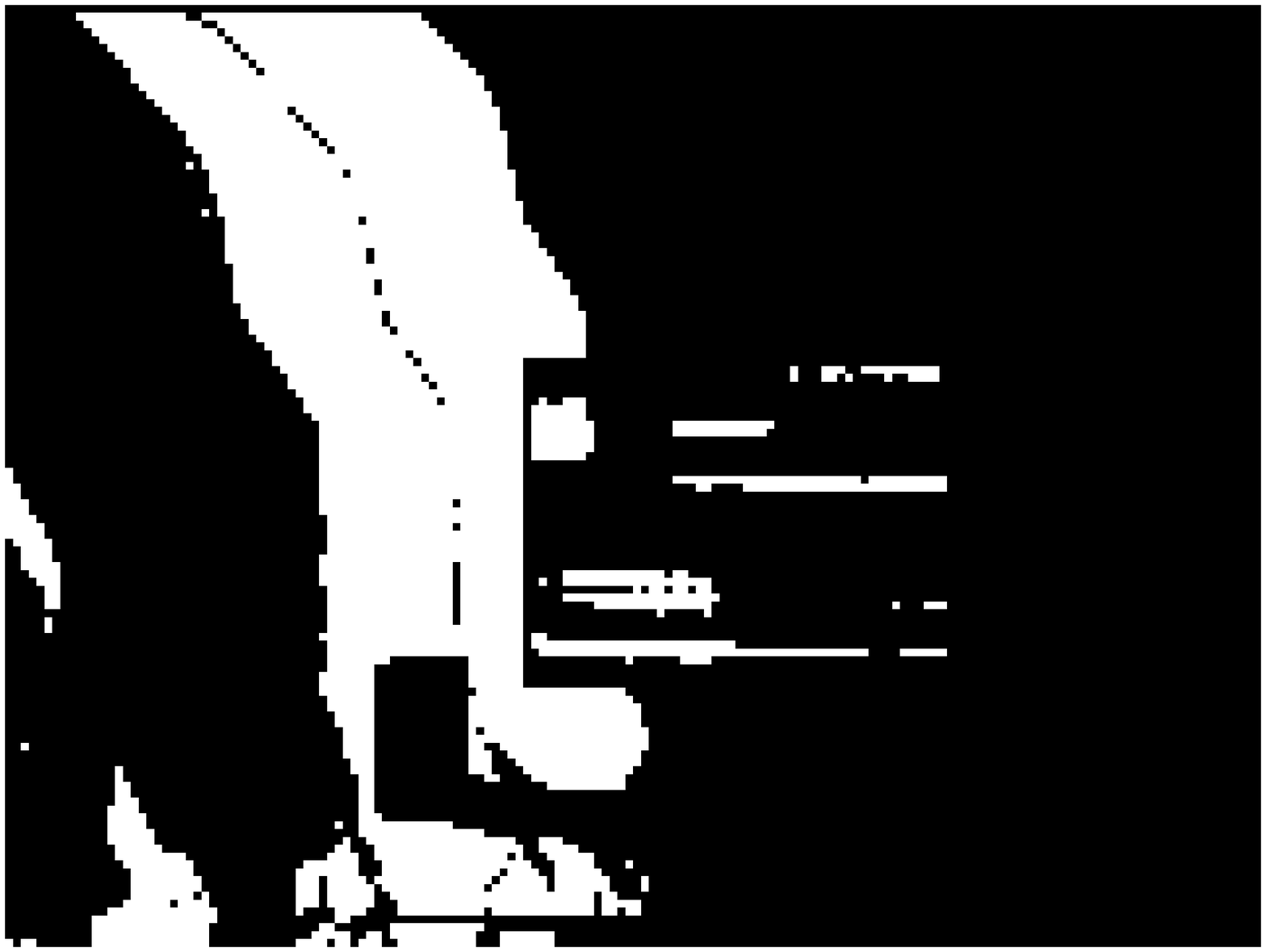}
		\includegraphics[width=15mm]{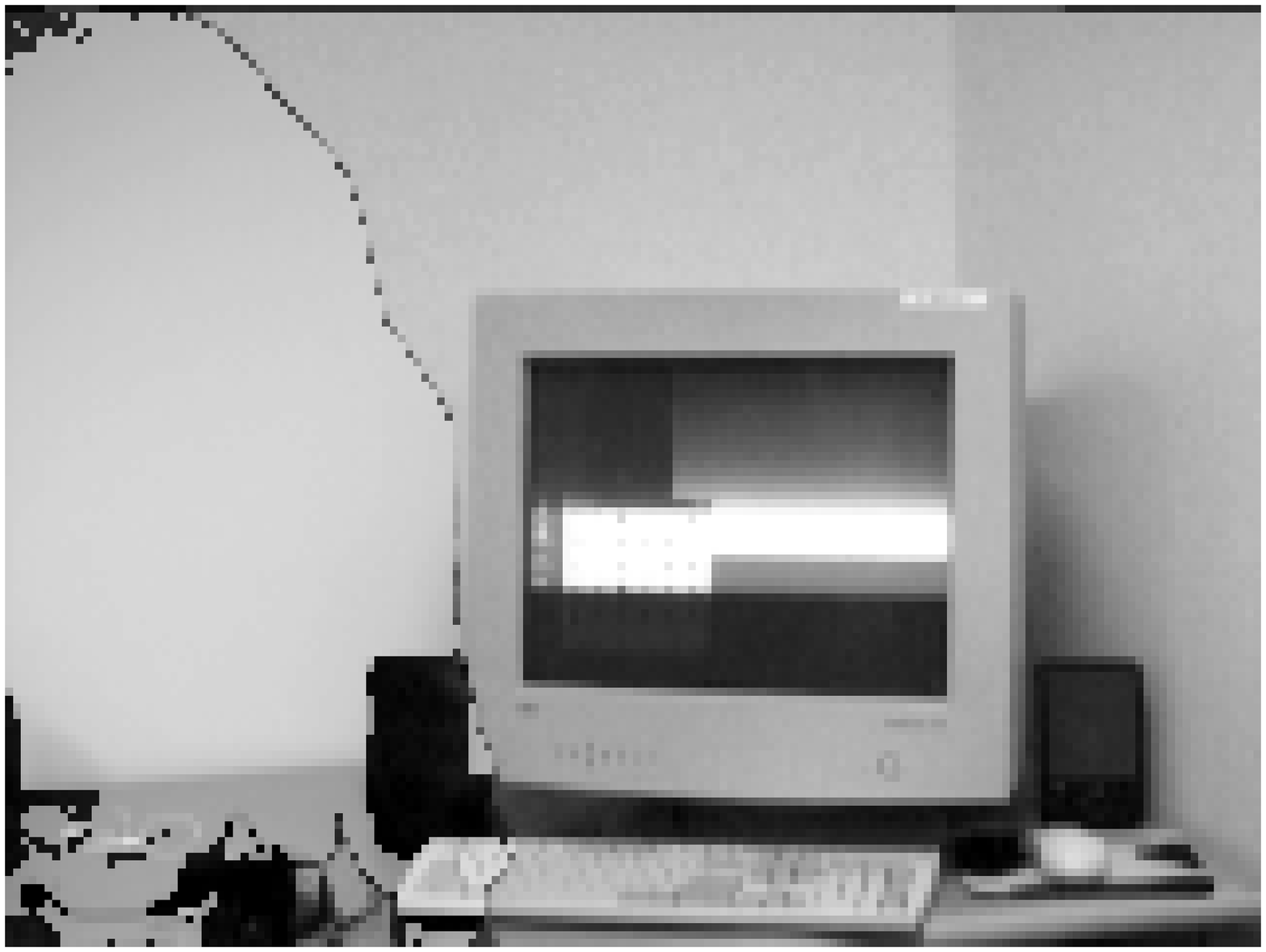}
		\includegraphics[width=15mm]{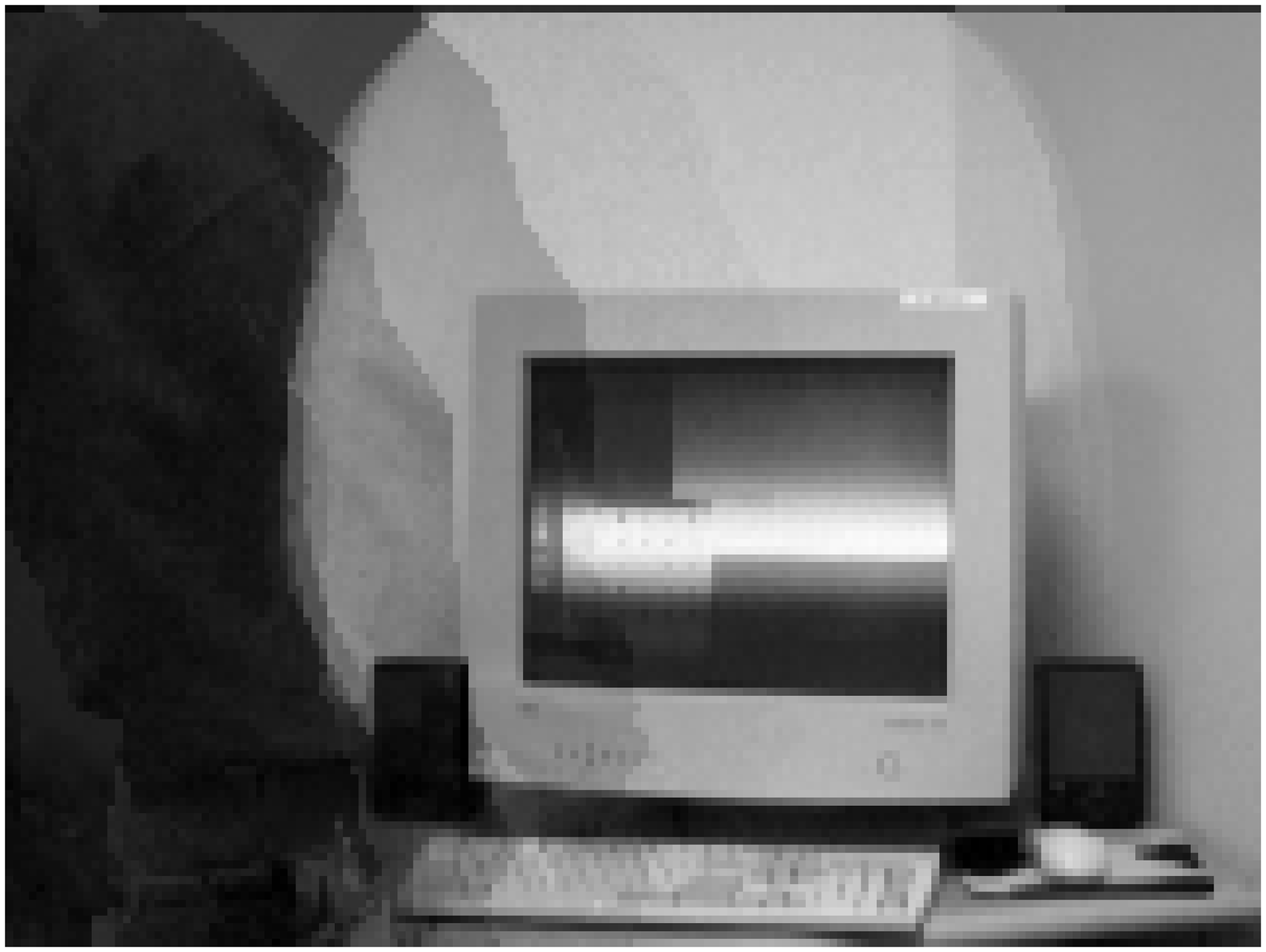}
		\includegraphics[width=15mm]{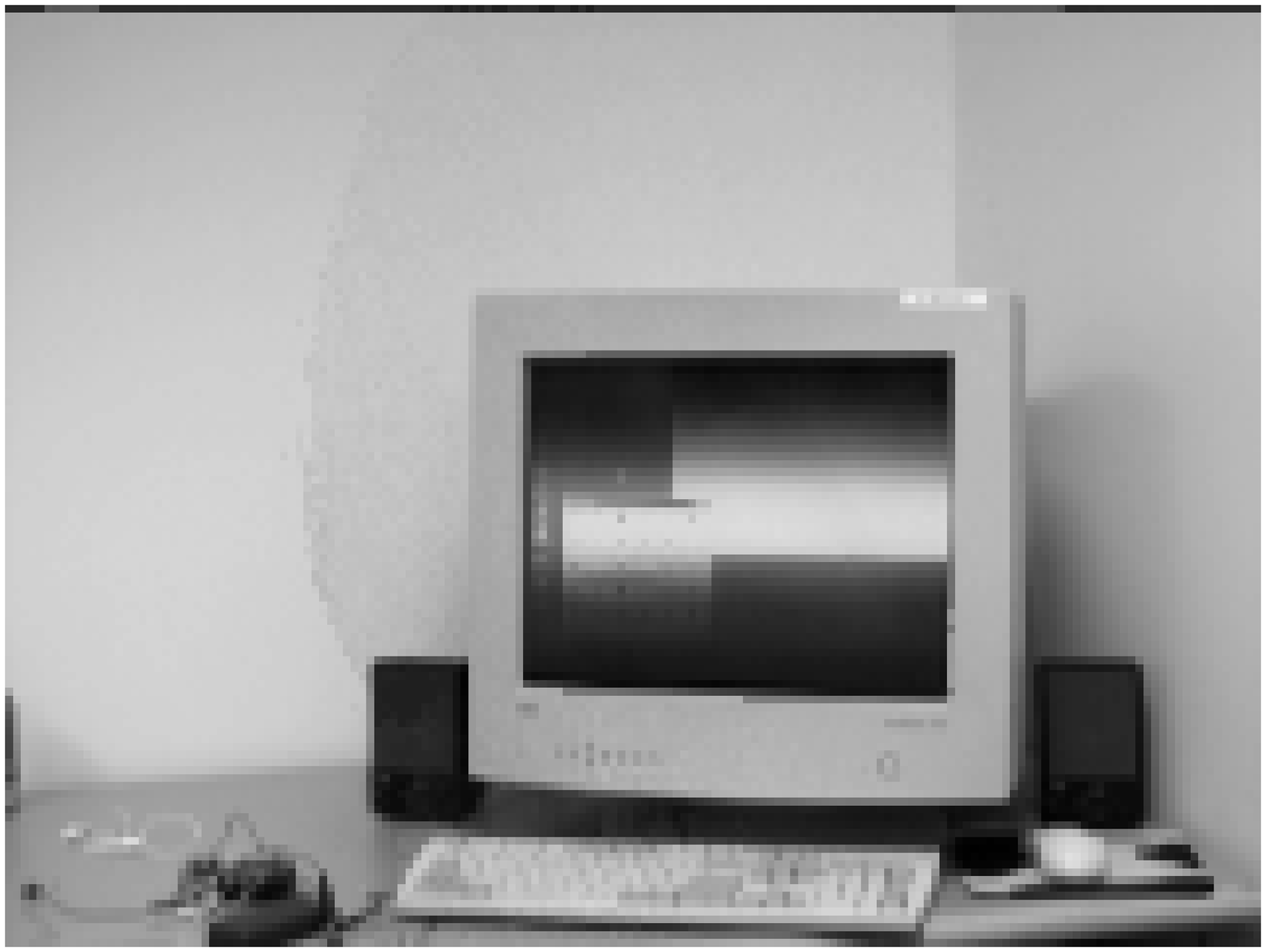}
		\includegraphics[width=15mm]{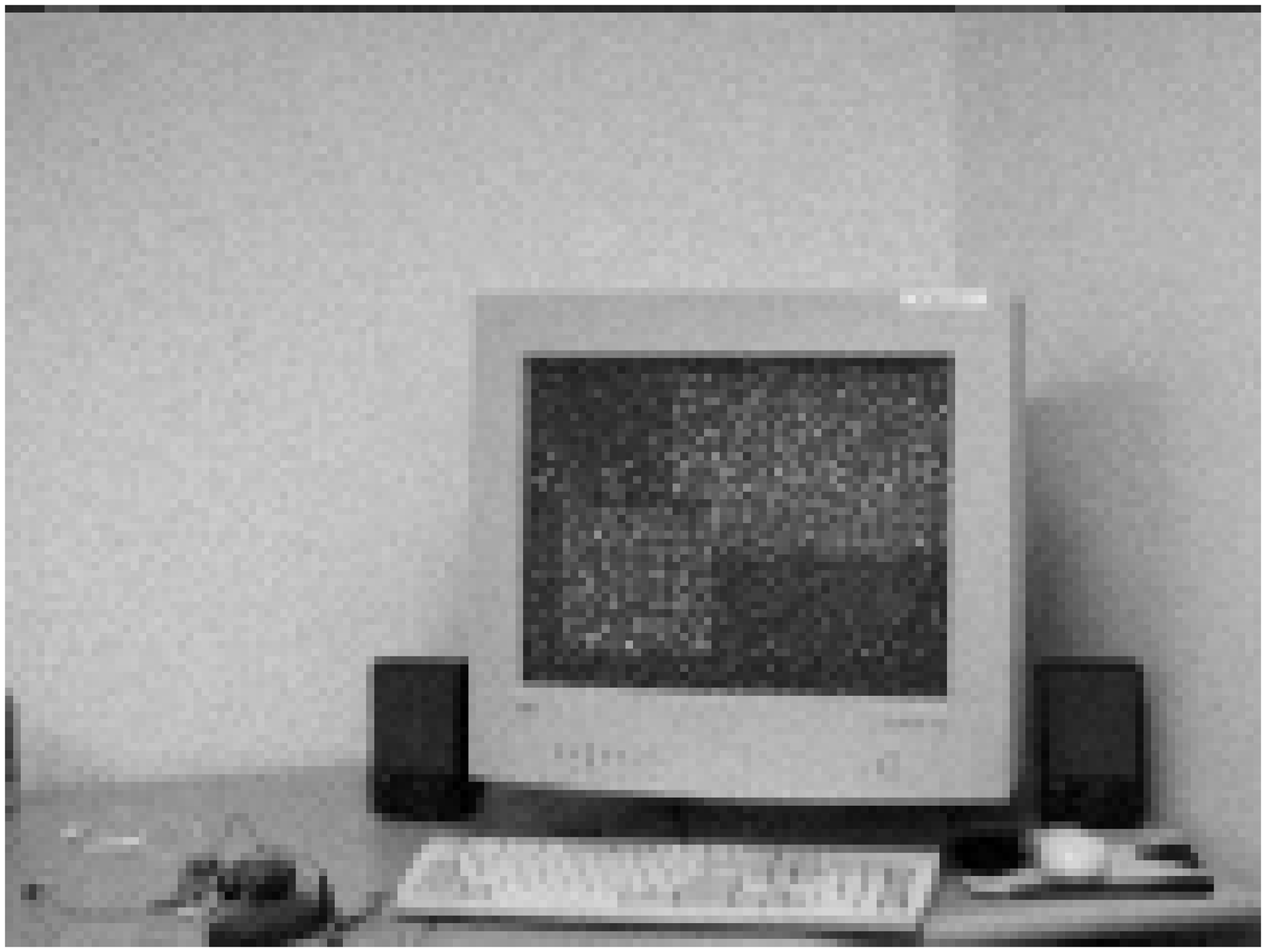}
		\includegraphics[width=15mm]{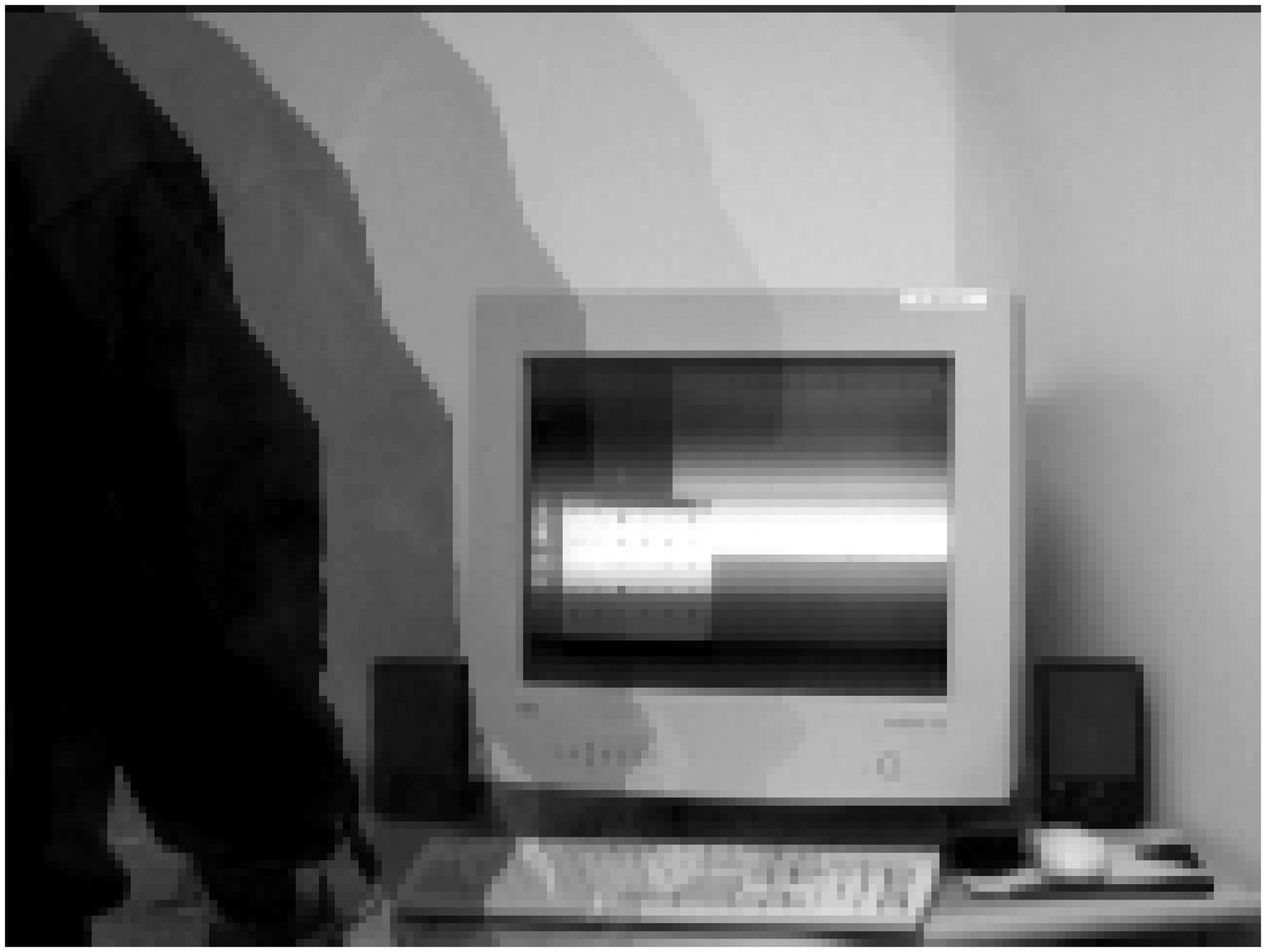}\\
		\includegraphics[width=15mm]{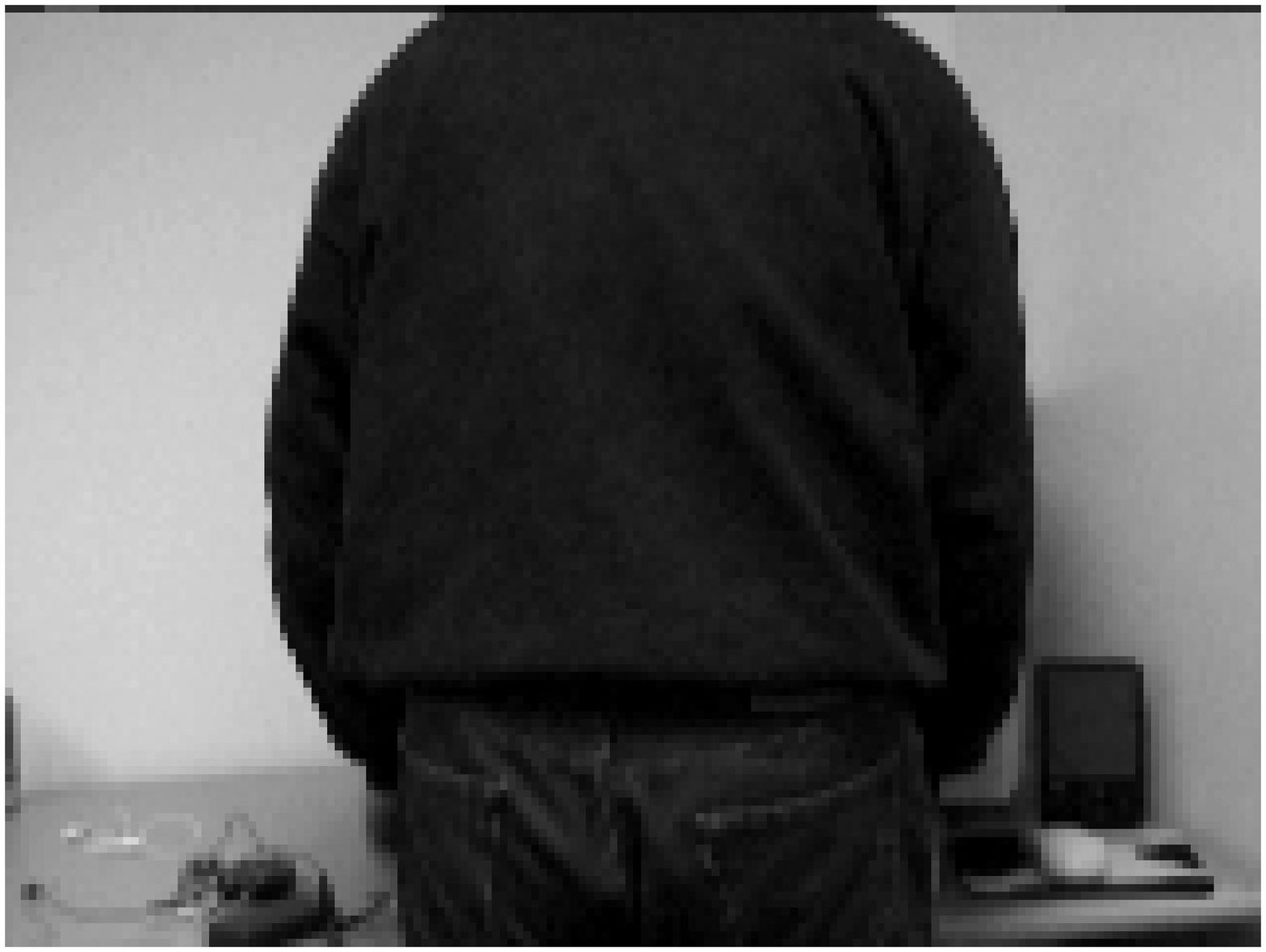}
		\includegraphics[width=15mm]{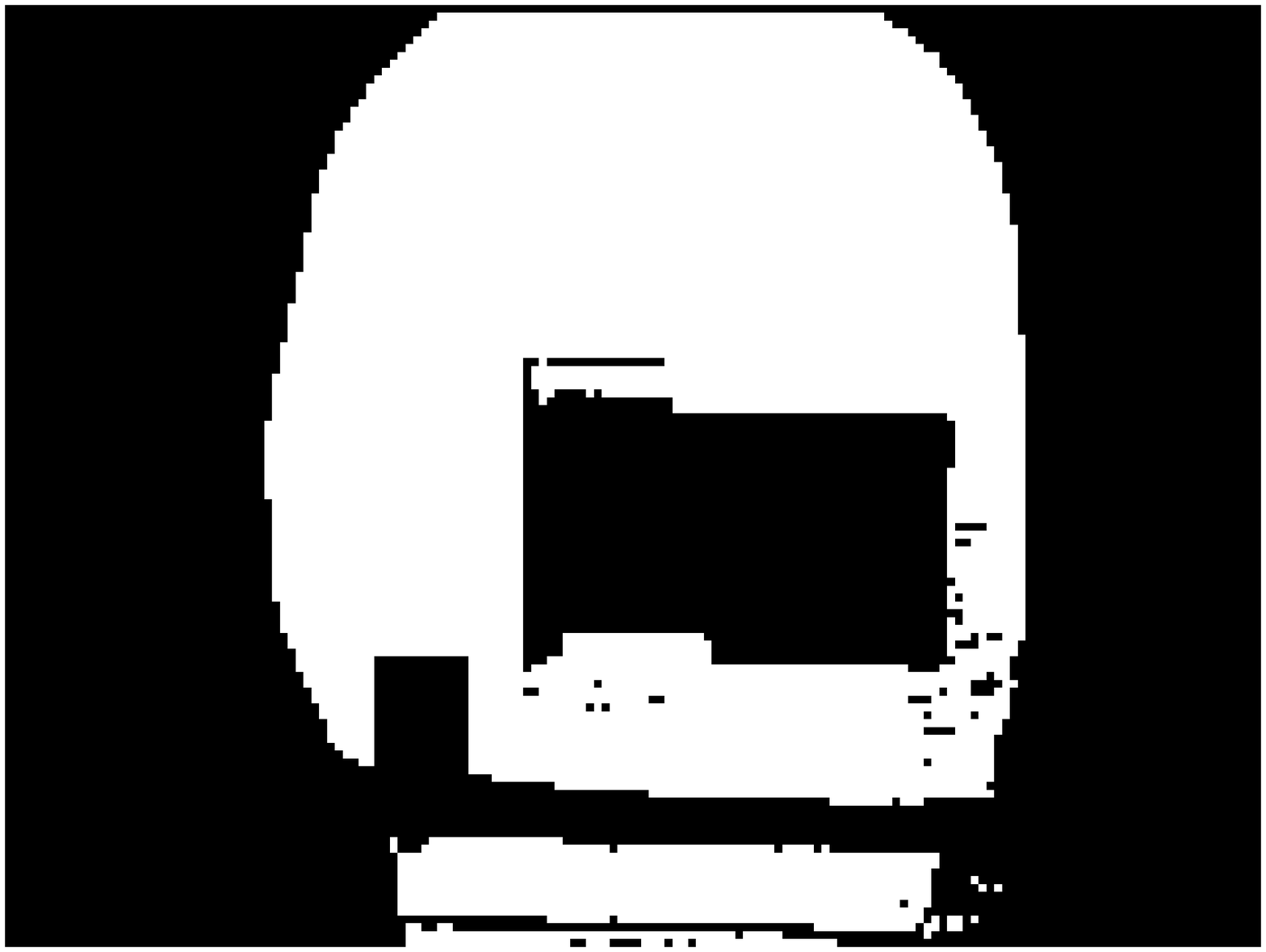}
		\includegraphics[width=15mm]{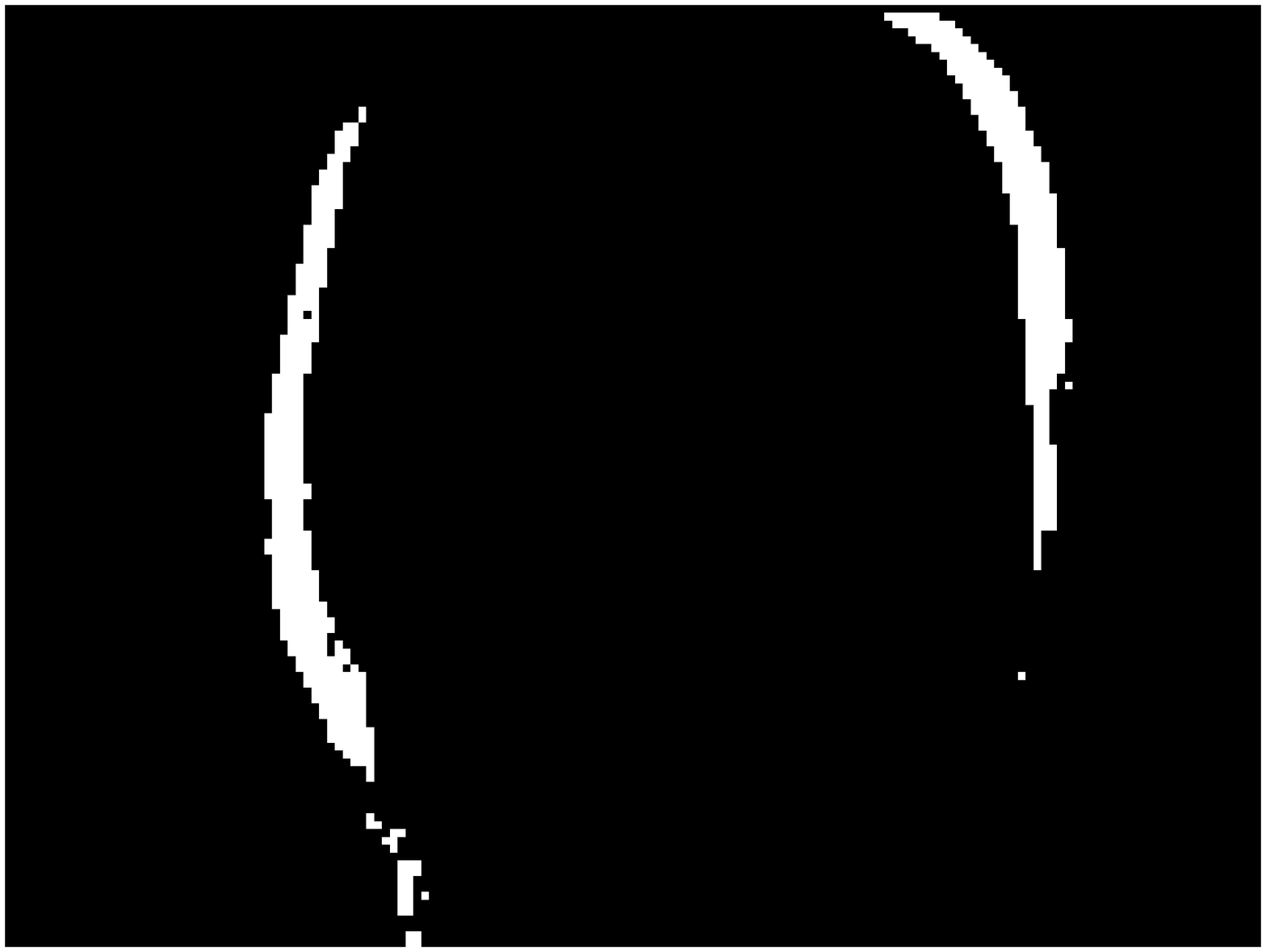}
		\includegraphics[width=15mm]{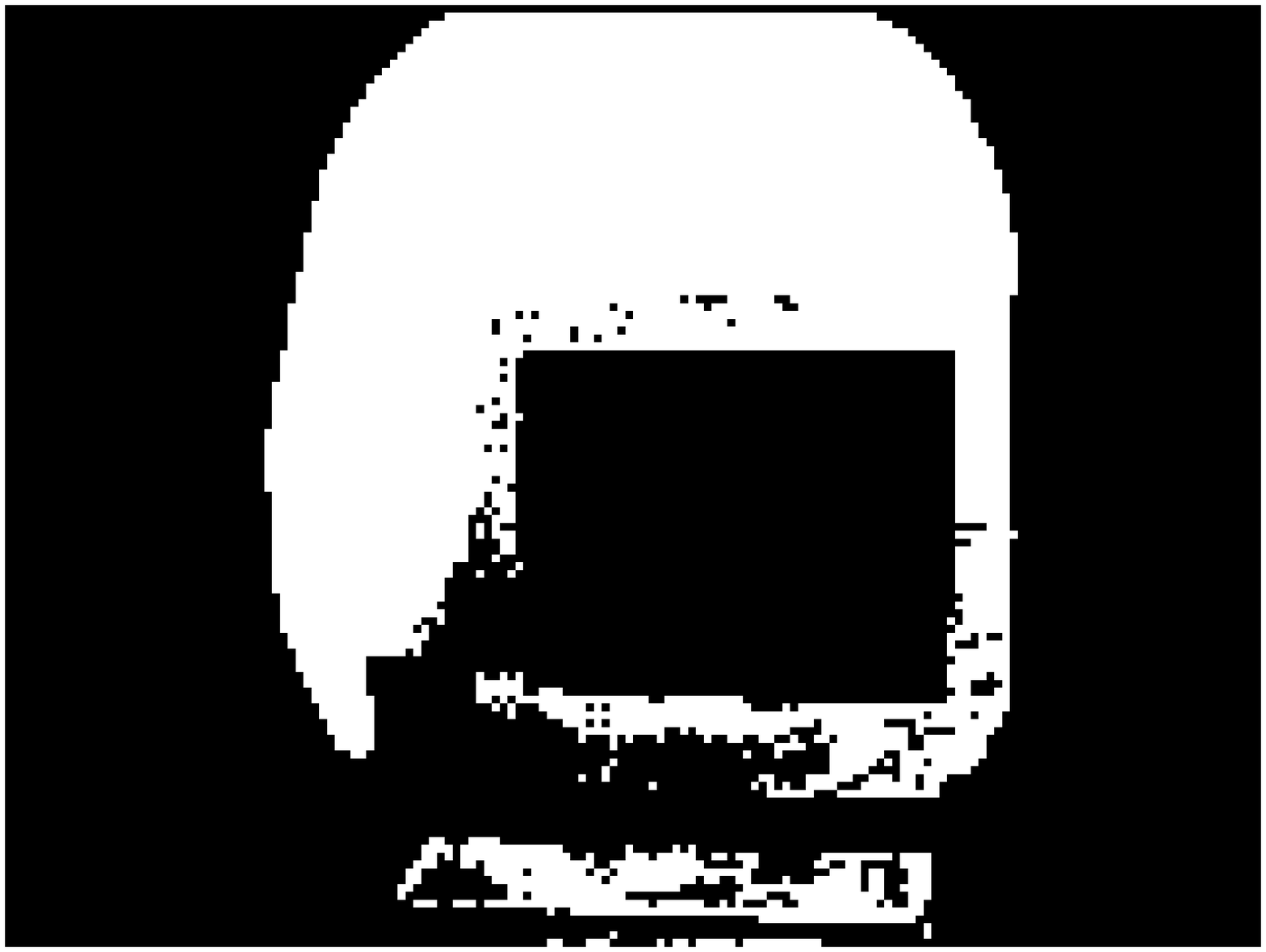}
		\includegraphics[width=15mm]{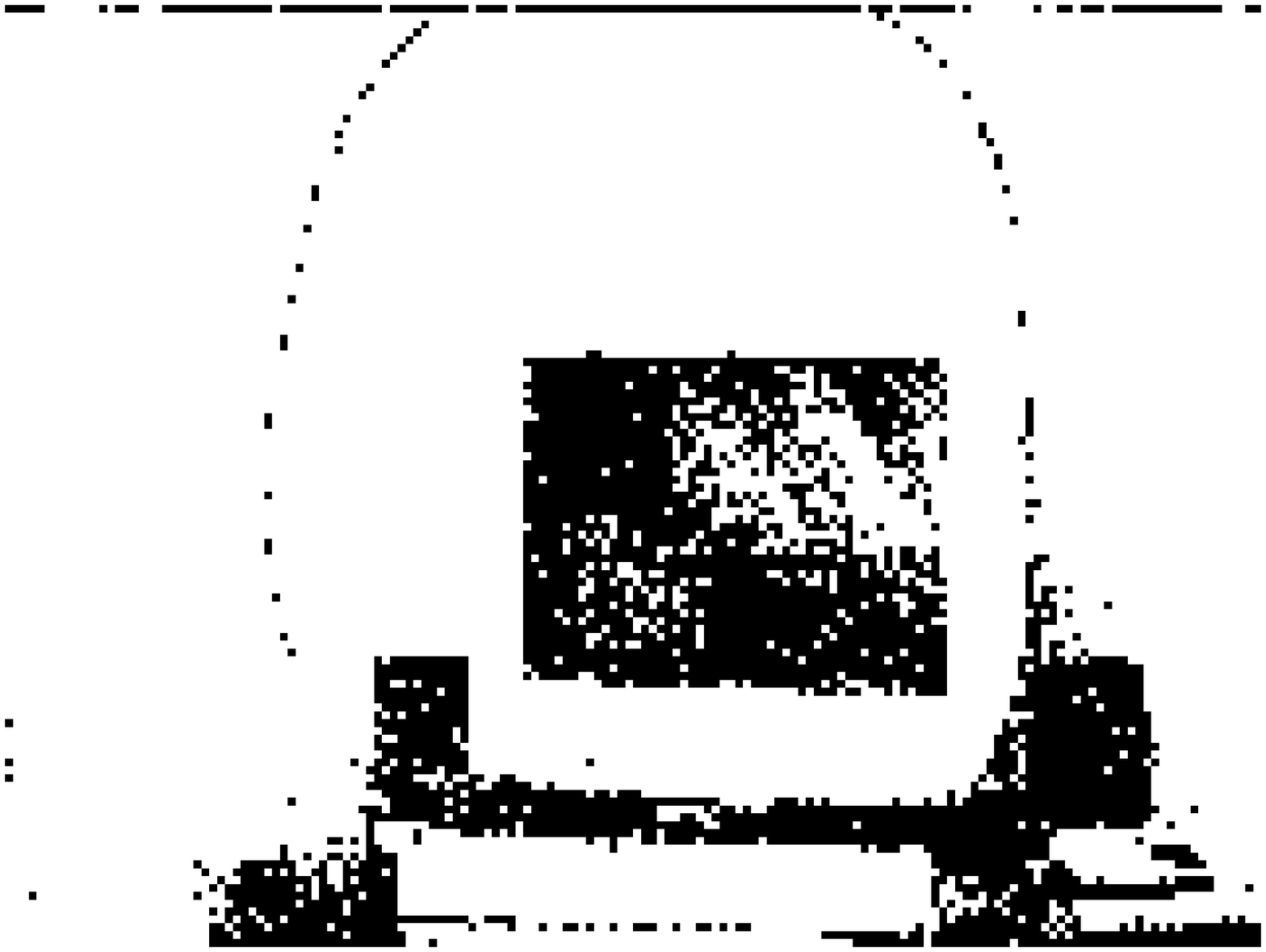}
		\includegraphics[width=15mm]{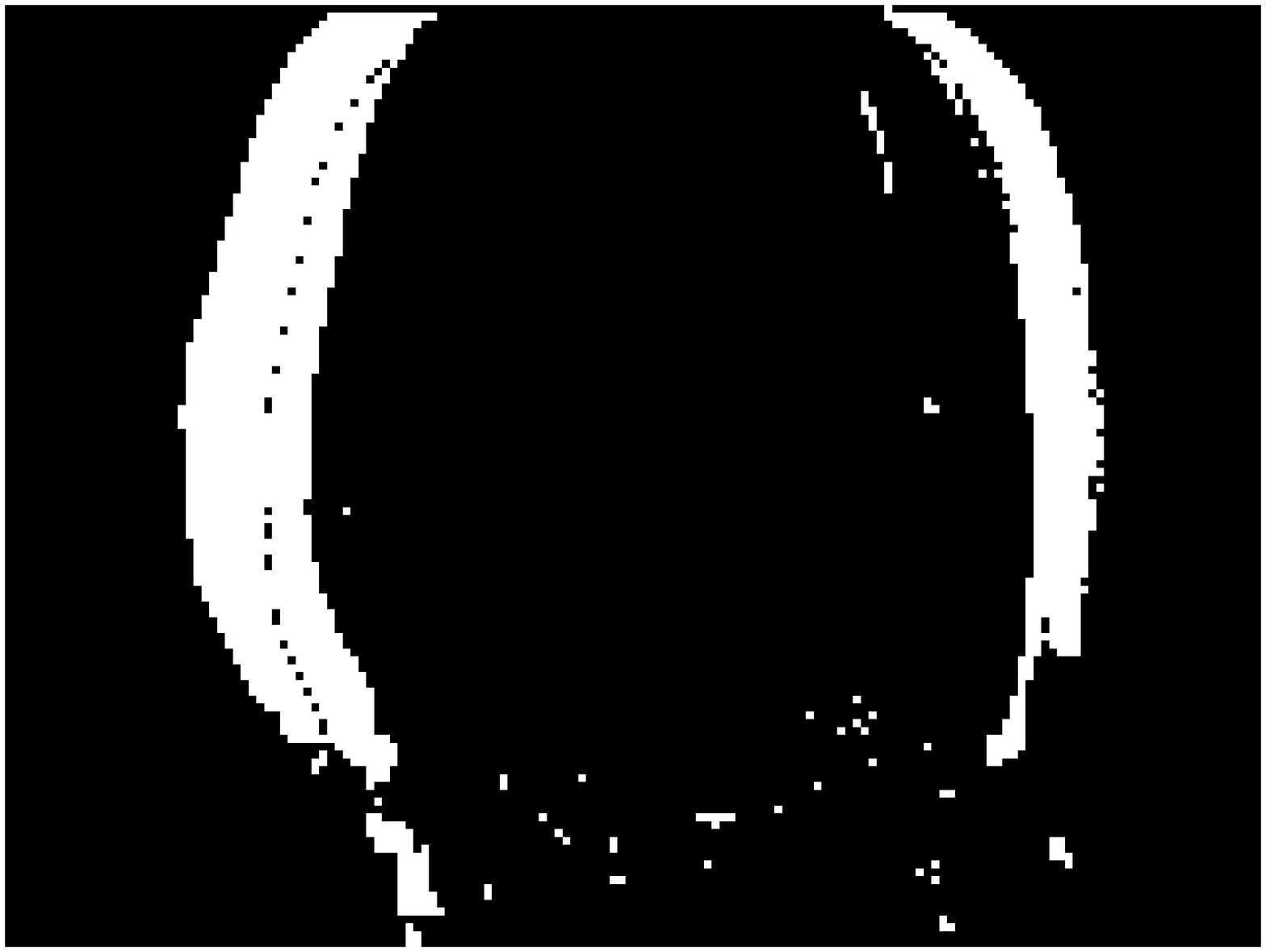}
		\includegraphics[width=15mm]{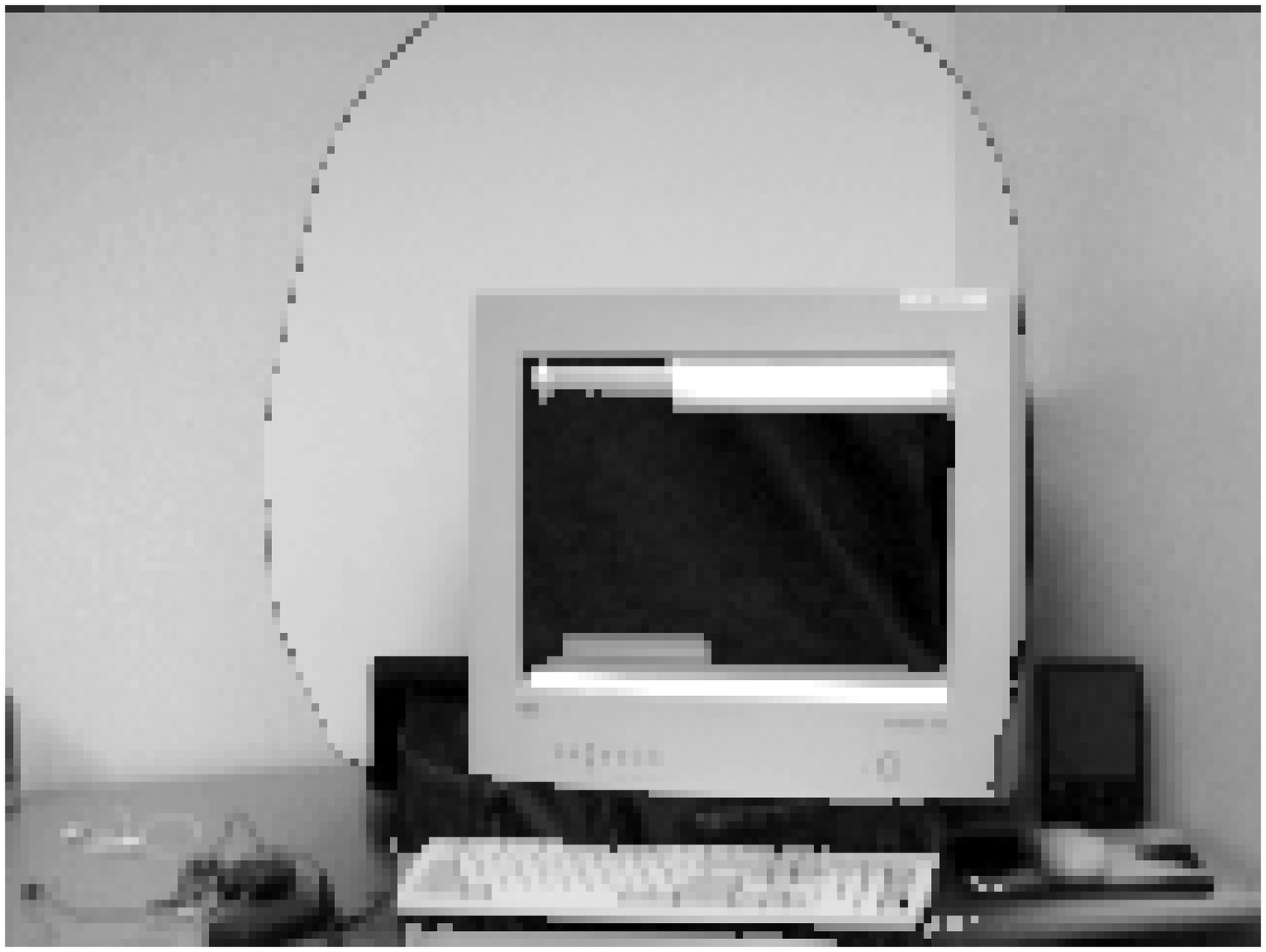}
		\includegraphics[width=15mm]{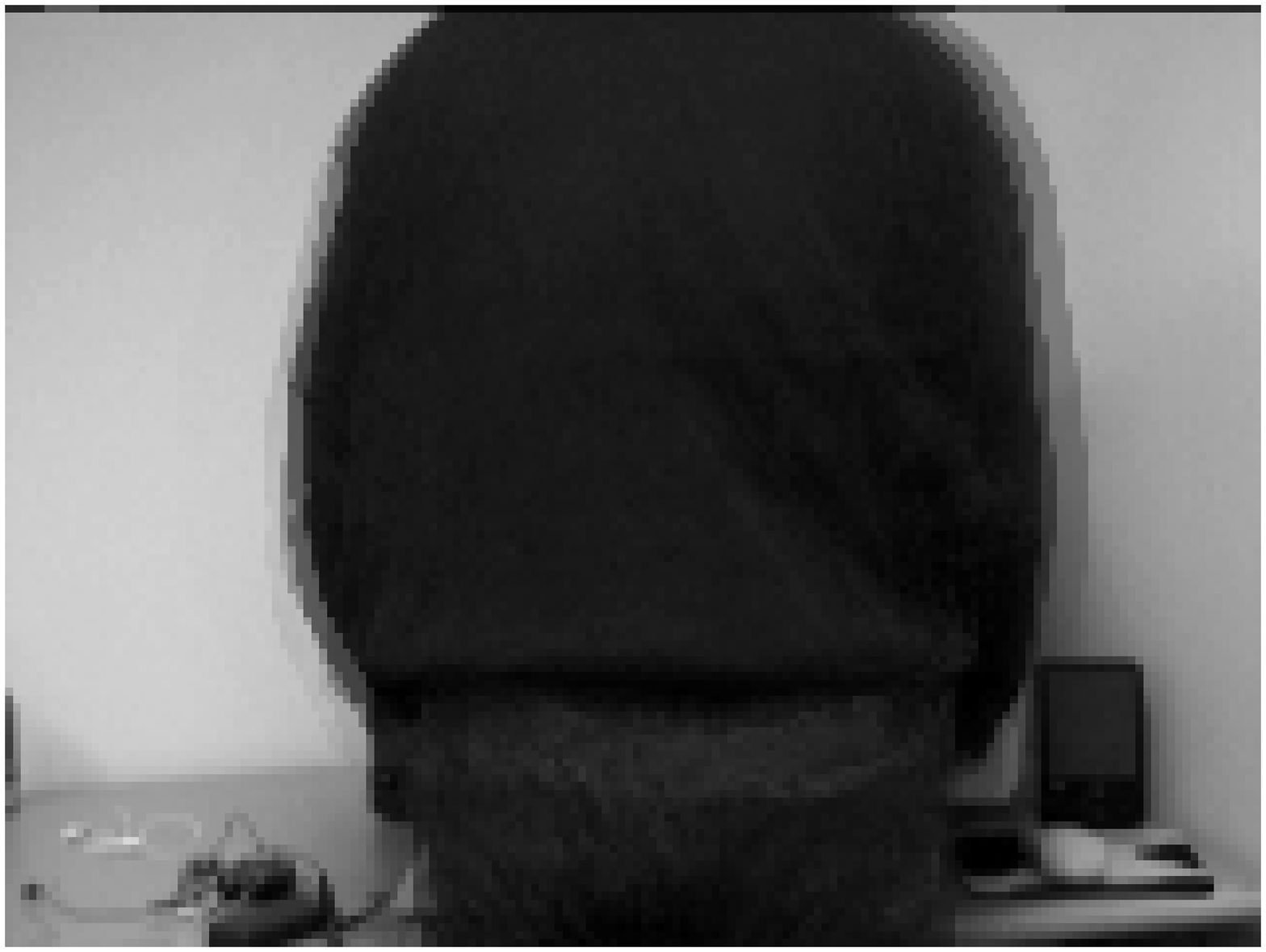}
		\includegraphics[width=15mm]{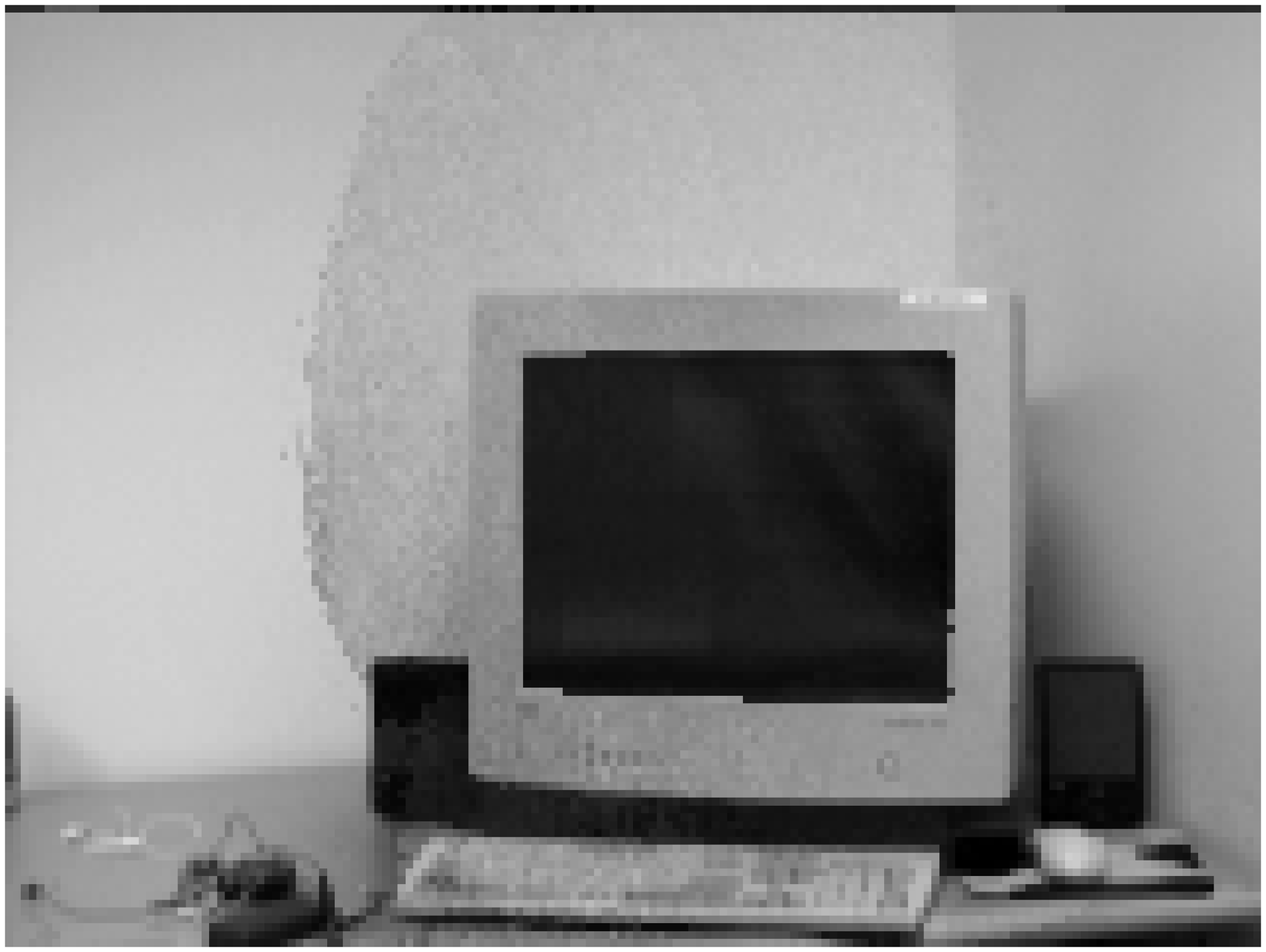}
		\includegraphics[width=15mm]{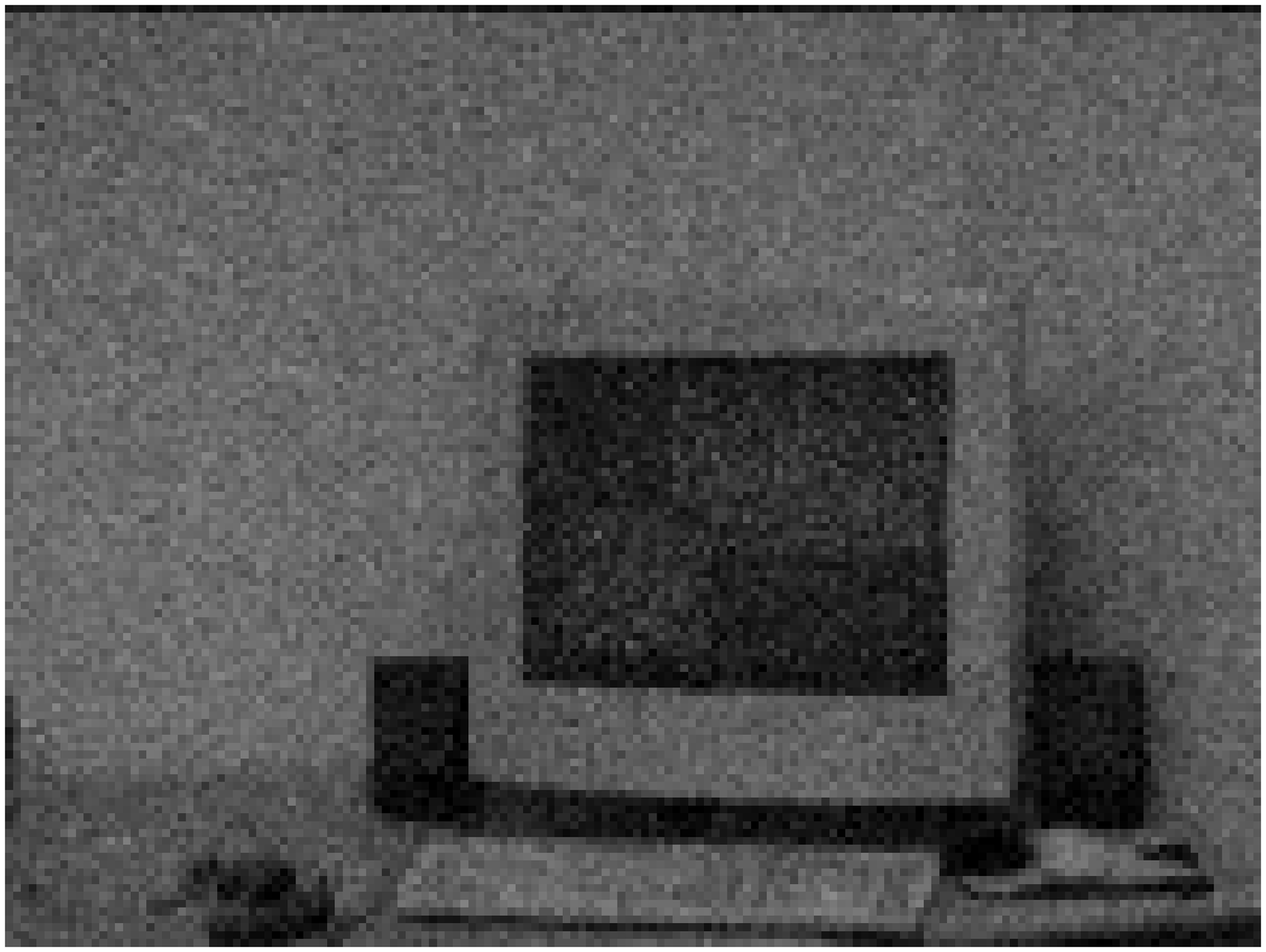}
		\includegraphics[width=15mm]{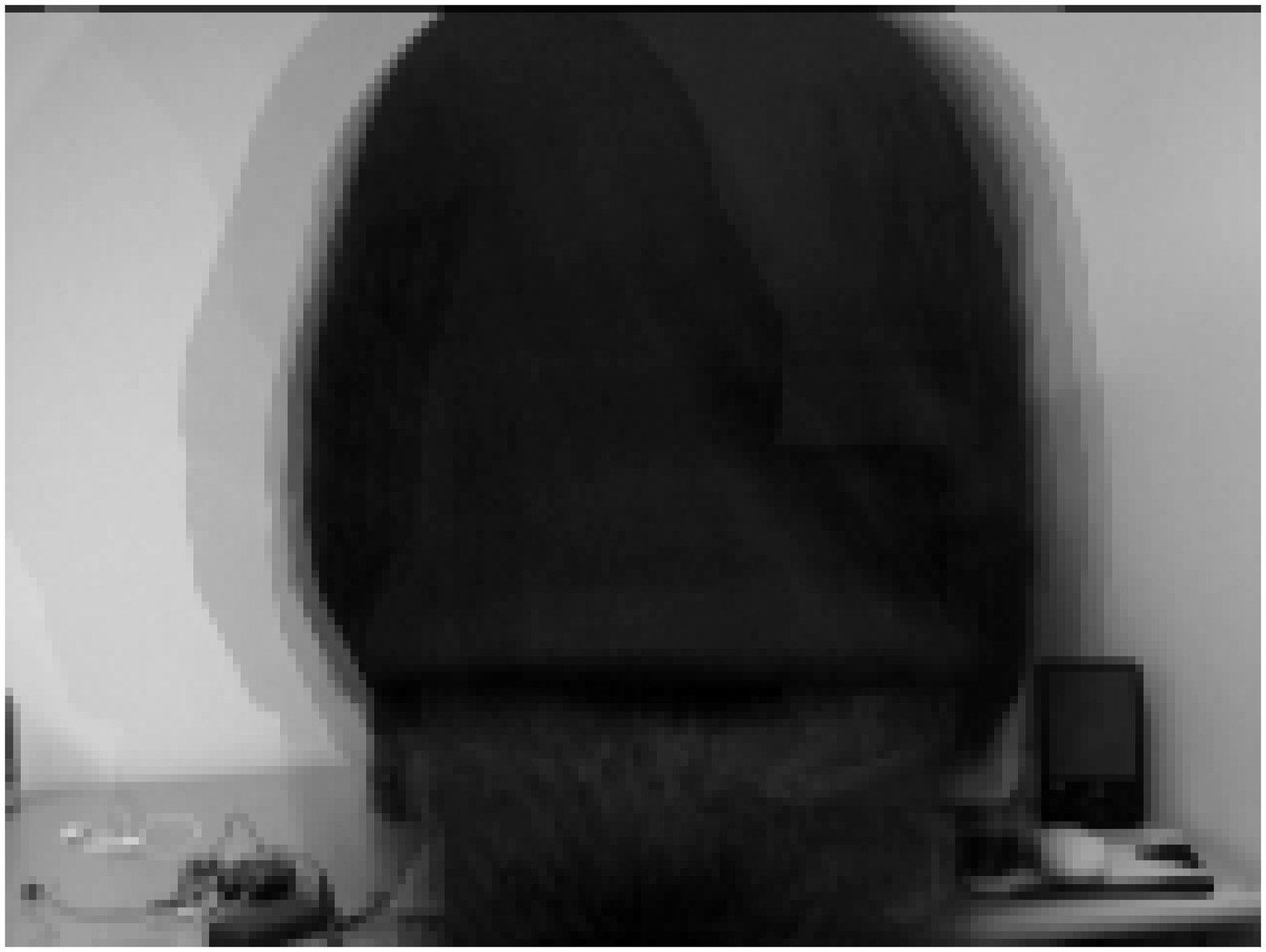}\\
	\hspace*{\fill}\makebox[0pt]{original }\hspace*{\fill}
		\hspace*{\fill}\makebox[0pt]{ReProCS}\hspace*{\fill}
		\hspace*{\fill}\makebox[0pt]{PCP}\hspace*{\fill}
		\hspace*{\fill}\makebox[0pt]{RSL}\hspace*{\fill}
		\hspace*{\fill}\makebox[0pt]{GRASTA}\hspace*{\fill}
		\hspace*{\fill}\makebox[0pt]{MG}\hspace*{\fill}
		\hspace*{\fill}\makebox[0pt]{ReProCS}\hspace*{\fill}
		\hspace*{\fill}\makebox[0pt]{PCP}\hspace*{\fill}
		\hspace*{\fill}\makebox[0pt]{RSL}\hspace*{\fill}
		\hspace*{\fill}\makebox[0pt]{GRASTA}\hspace*{\fill}
		\hspace*{\fill}\makebox[0pt]{MG}\hspace*{\fill}\\
		\hspace*{\fill}\makebox[0pt]{}\hspace*{\fill}
		\hspace*{\fill}\makebox[0pt]{(fg)}\hspace*{\fill}
		\hspace*{\fill}\makebox[0pt]{(fg)}\hspace*{\fill}
		\hspace*{\fill}\makebox[0pt]{(fg)}\hspace*{\fill}
		\hspace*{\fill}\makebox[0pt]{(fg)}\hspace*{\fill}
		\hspace*{\fill}\makebox[0pt]{(fg)}\hspace*{\fill}
		\hspace*{\fill}\makebox[0pt]{(bg)}\hspace*{\fill}
		\hspace*{\fill}\makebox[0pt]{(bg)}\hspace*{\fill}
		\hspace*{\fill}\makebox[0pt]{(bg)}\hspace*{\fill}
		\hspace*{\fill}\makebox[0pt]{(bg)}\hspace*{\fill}
		\hspace*{\fill}\makebox[0pt]{(bg)}\hspace*{\fill}
	
	\end{tabular}
\caption{\small{Original video sequence at $t=t_\train+42, 44, 52$ and its foreground (fg) and background (bg) layer recovery results using ReProCS (ReProCS-pCA) and other algorithms.
 For fg, we only show the fg support in white for ease of display.
}}
\label{PersonCompare}
\end{figure*}

\begin{figure*}
				\centering				
				\begin{subfigure}{0.3\textwidth}
\centering
					\begin{tabular}{cc}

		\includegraphics[width=16mm]{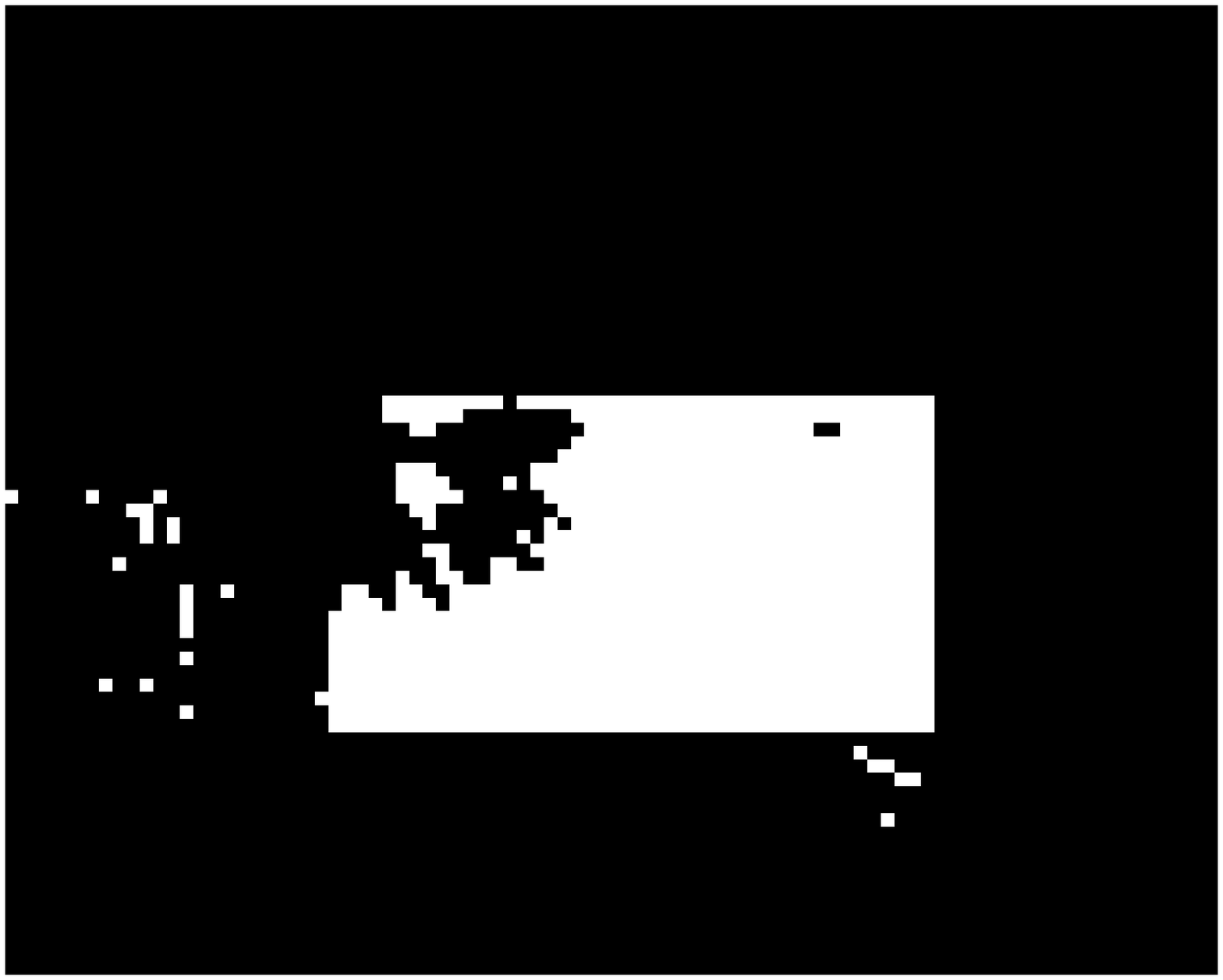}
		\includegraphics[width=16mm]{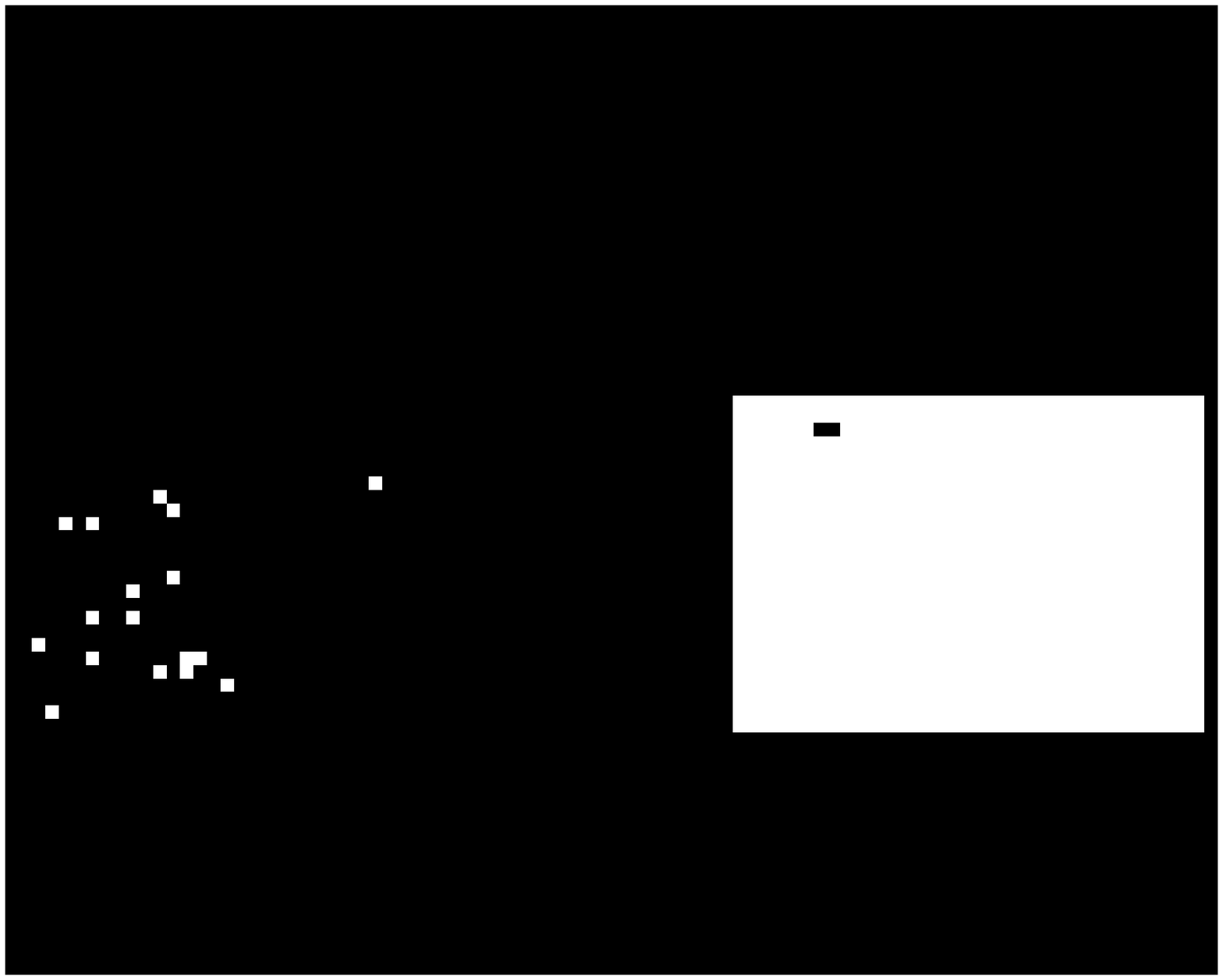}
		\includegraphics[width=16mm]{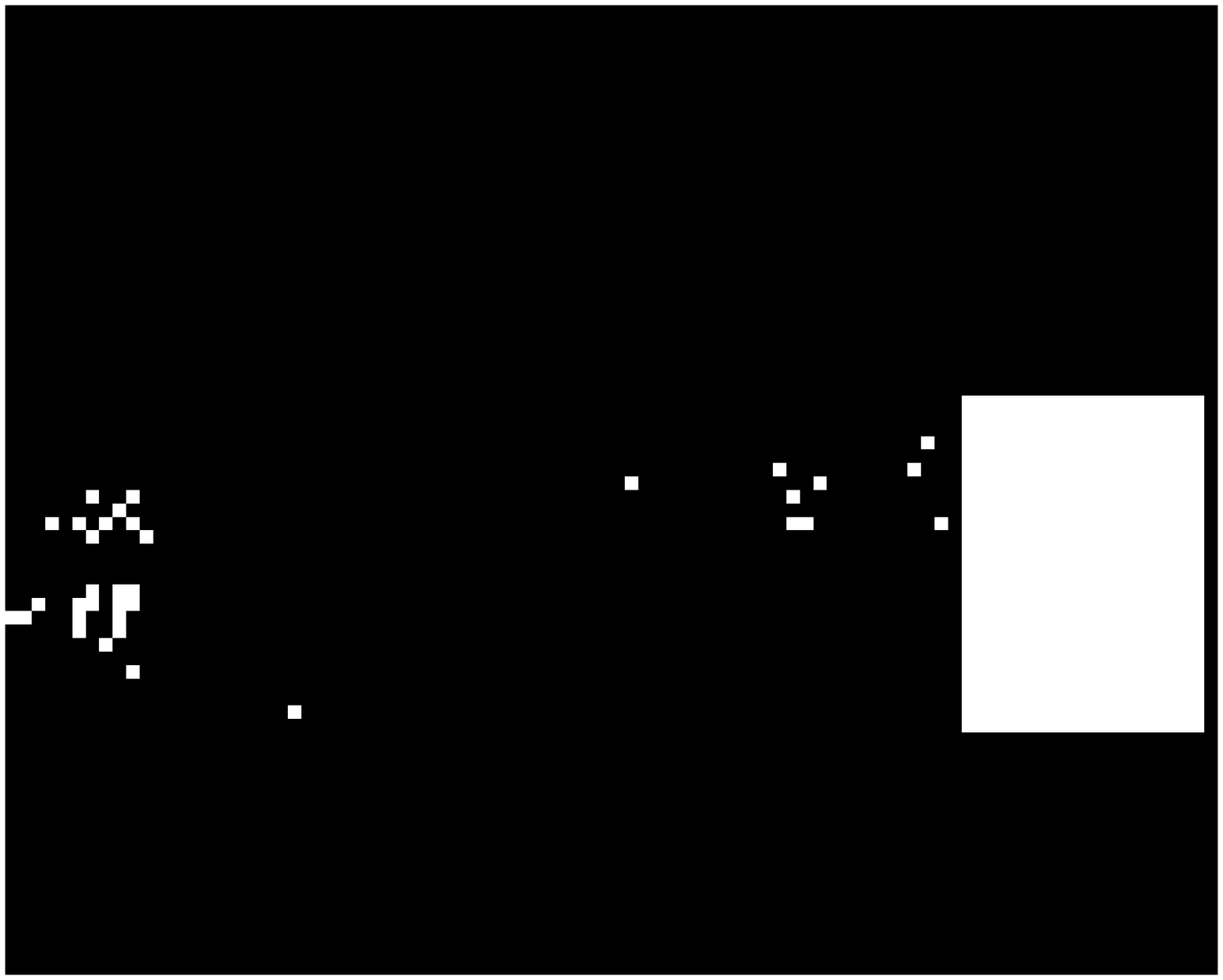}\\
					\end{tabular}
					\caption{$t=30, 60, 70$}
				\end{subfigure}%
				\hfill
				\begin{subfigure}{0.3\textwidth}
\centering
					\begin{tabular}{cc}

		\includegraphics[width=16mm]{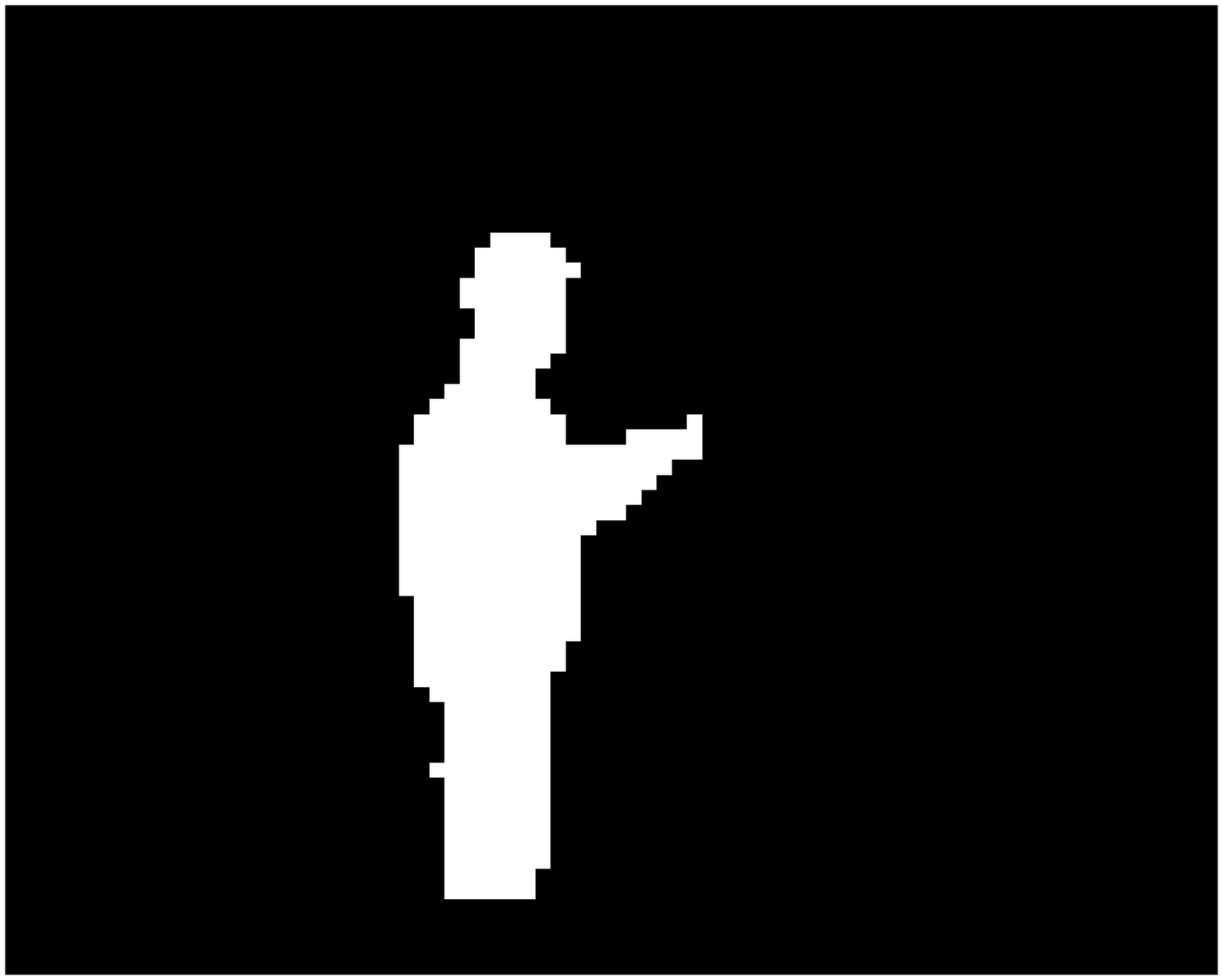}
		\includegraphics[width=16mm]{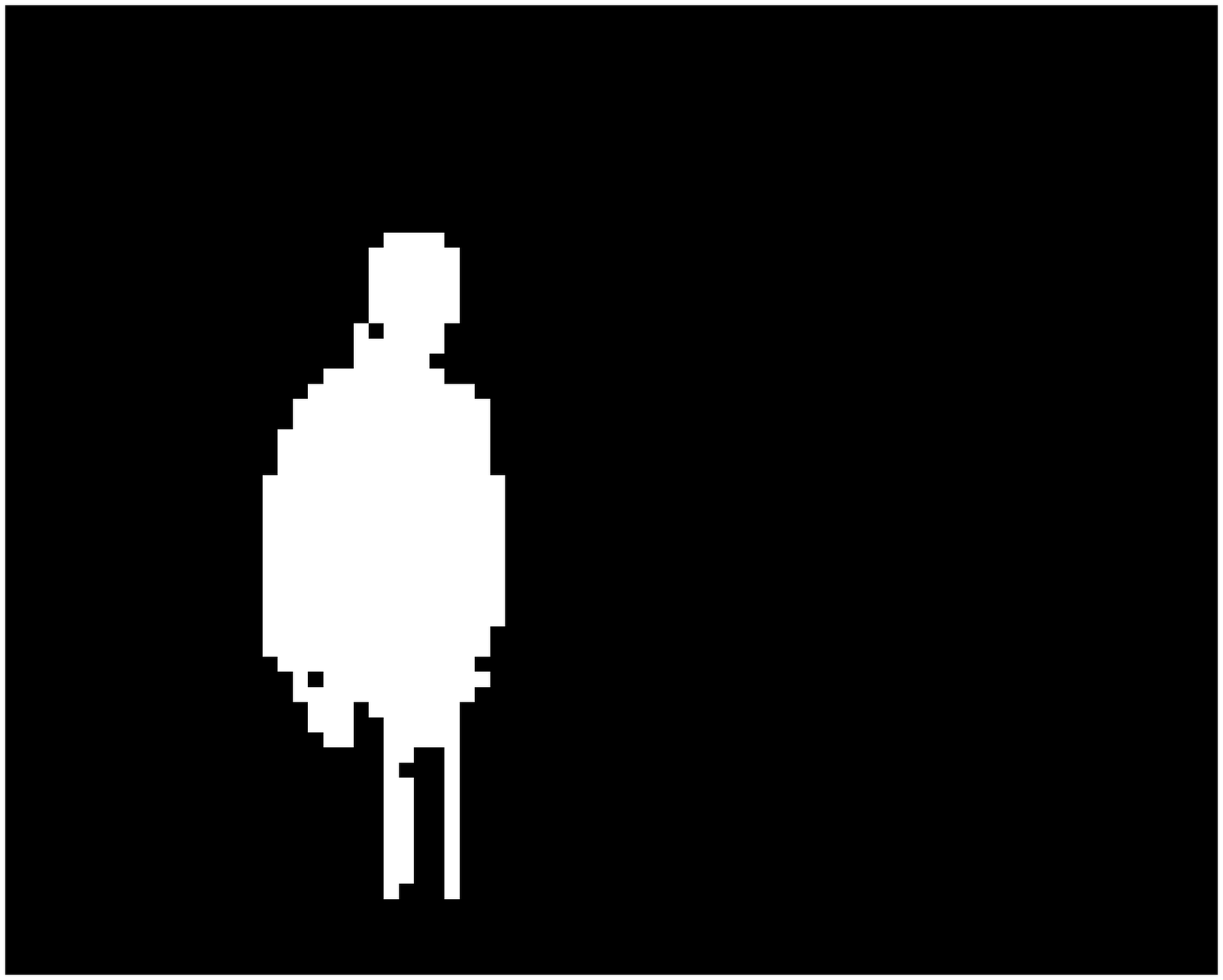}
		\includegraphics[width=16mm]{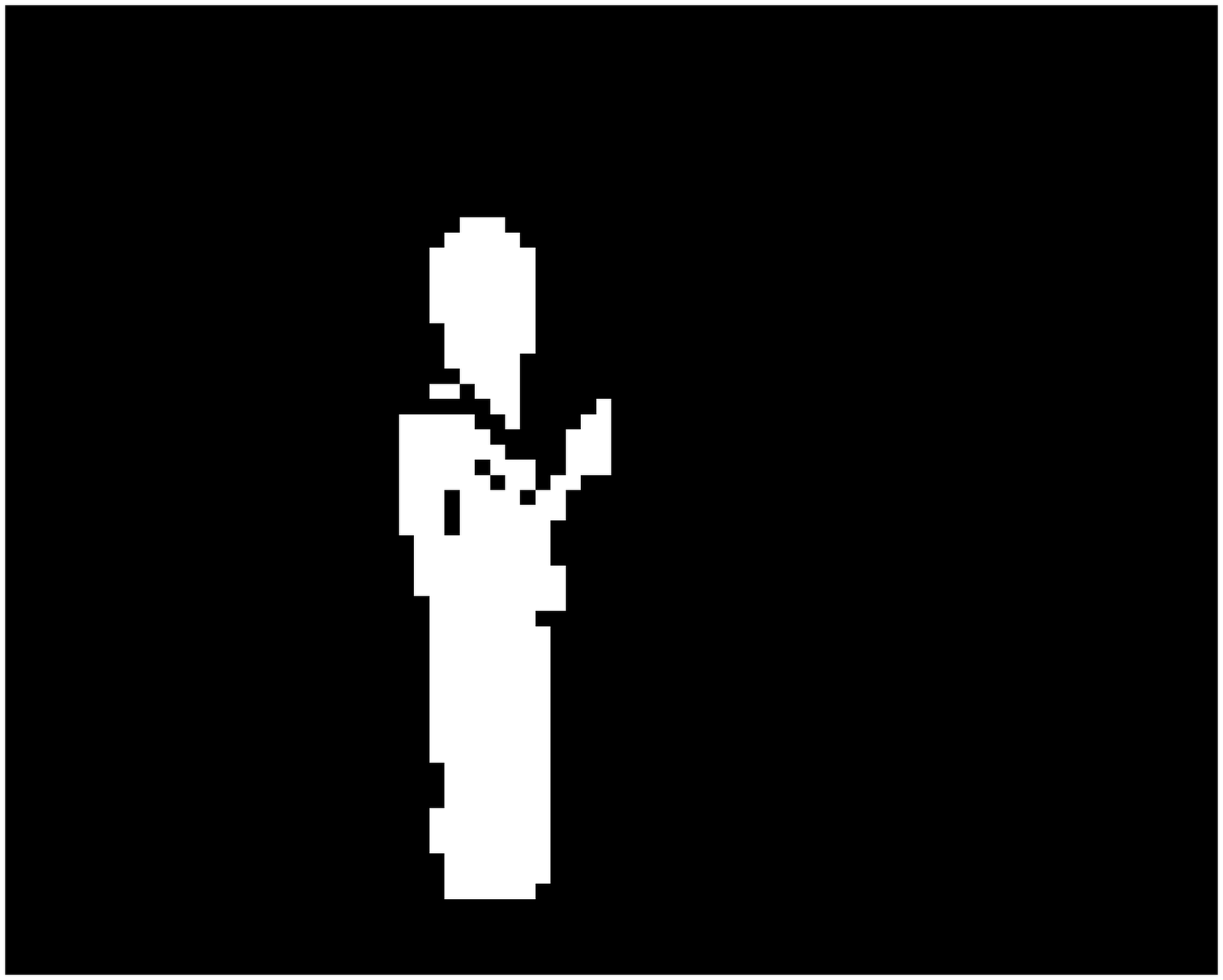}\\
					\end{tabular}
					\caption{$t=60, 120, 475$}
				\end{subfigure}%
			    \hfill
				\begin{subfigure}{0.3\textwidth}
\centering
					\begin{tabular}{cc}

		\includegraphics[width=16mm]{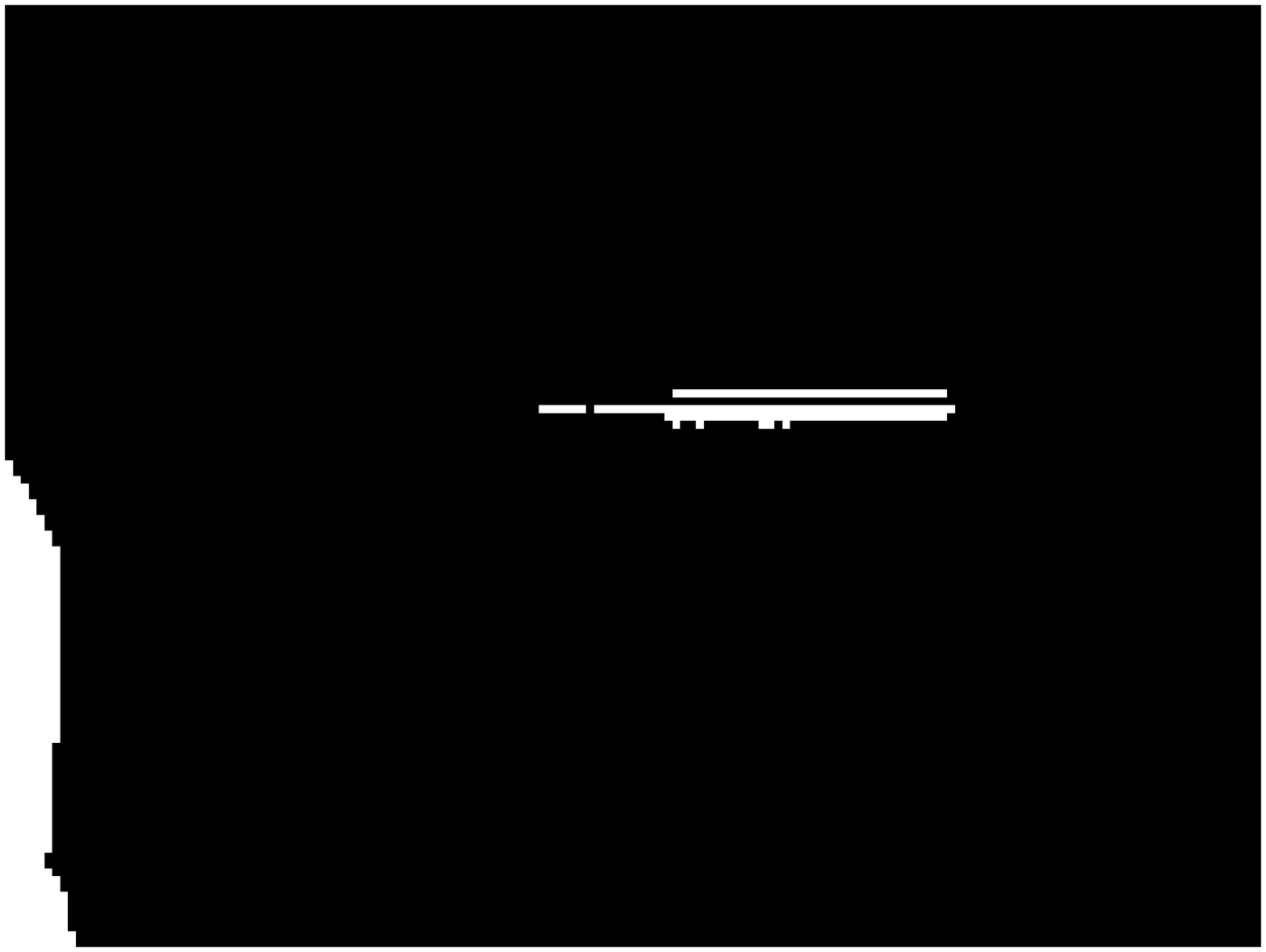}
		\includegraphics[width=16mm]{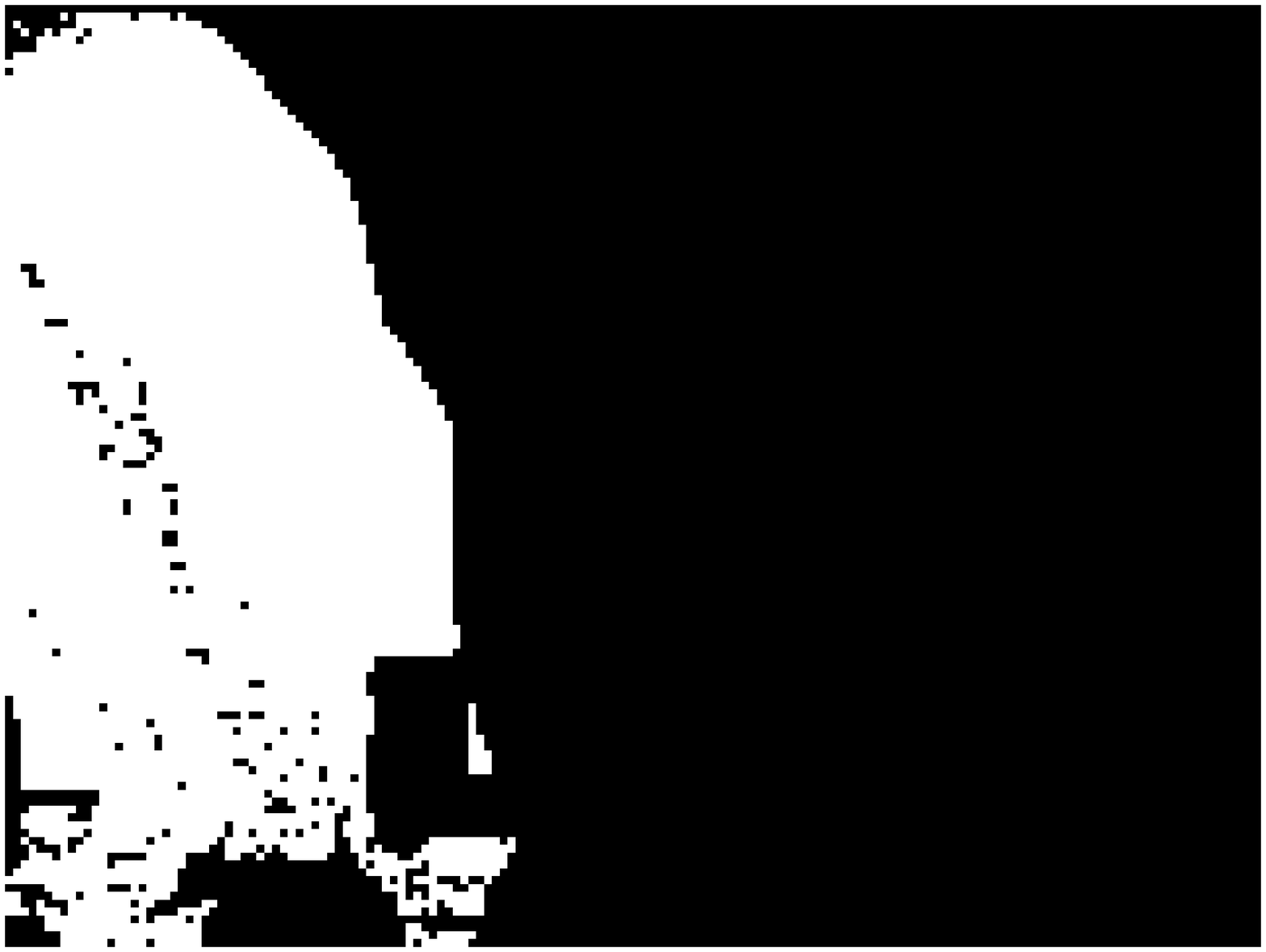}
		\includegraphics[width=16mm]{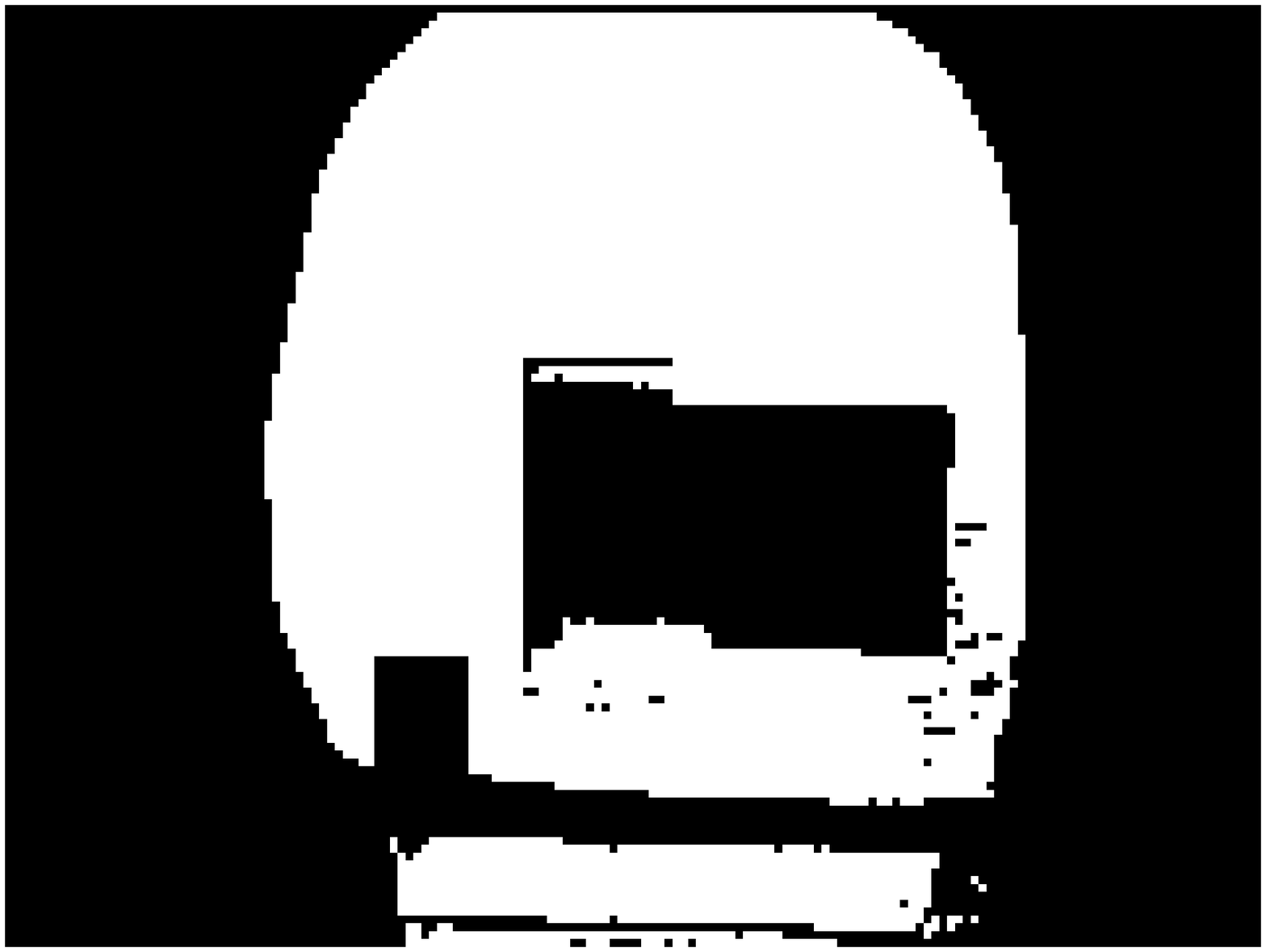}\\
					\end{tabular}
					\caption{$t=42, 44, 52$}
				\end{subfigure}				
\caption{\small{Foreground layer estimated by ReProCS-Recursive-PCA for the lake, curtain and person videos shown in Figs \ref{LakeCompare}, \ref{CurtainCompare} and \ref{PersonCompare}. As can be seen the recovery performance is very similar to that of ReProCS-pPCA (Algorithm \ref{PracReProCS}).
}}
\label{reprocs_rec_pca}
\end{figure*}

The second sequence consists of a person entering a room containing a computer monitor that contains a white moving region. Background changes due to lighting variations and due to the computer monitor. The person moving in the foreground occupies a very large part of the image, so this is an example of a sequence in which the use of weighted $\ell_1$ is essential (the support size is too large for simple $\ell_1$ to work). As can be seen from Fig. \ref{PersonCompare}, for most frames, ReProCS is able to recover the person correctly. However, for the last few frames which consist of the person in a white shirt in front of the white part of the screen, the resulting $S_t$ is too small even for ReProCS to correctly recover. The same is true for the other algorithms.  
Videos of all above experiments and of a few others are posted at \url{http://www.ece.iastate.edu/~hanguo/PracReProCS.html}. 

{\em Time Comparisons. }
The time comparisons are shown in Table \ref{speed}. In terms of speed, GRASTA is the fastest even though its performance is much worse. ReProCS is the second fastest. We expect that ReProCS can be speeded up by using mex files (C/C++ code) for the subspace update step. PCP and RSL are slower because they jointly process the entire image sequence. Moreover, ReProCS and GRASTA have the advantage of being recursive methods, i.e. the foreground/background recovery is available as soon as a new frame appears while PCP or RSL need to wait for the entire image sequence.

\begin{figure}
	\centering
	\begin{tabular}{cc}
		\includegraphics[width=16mm]{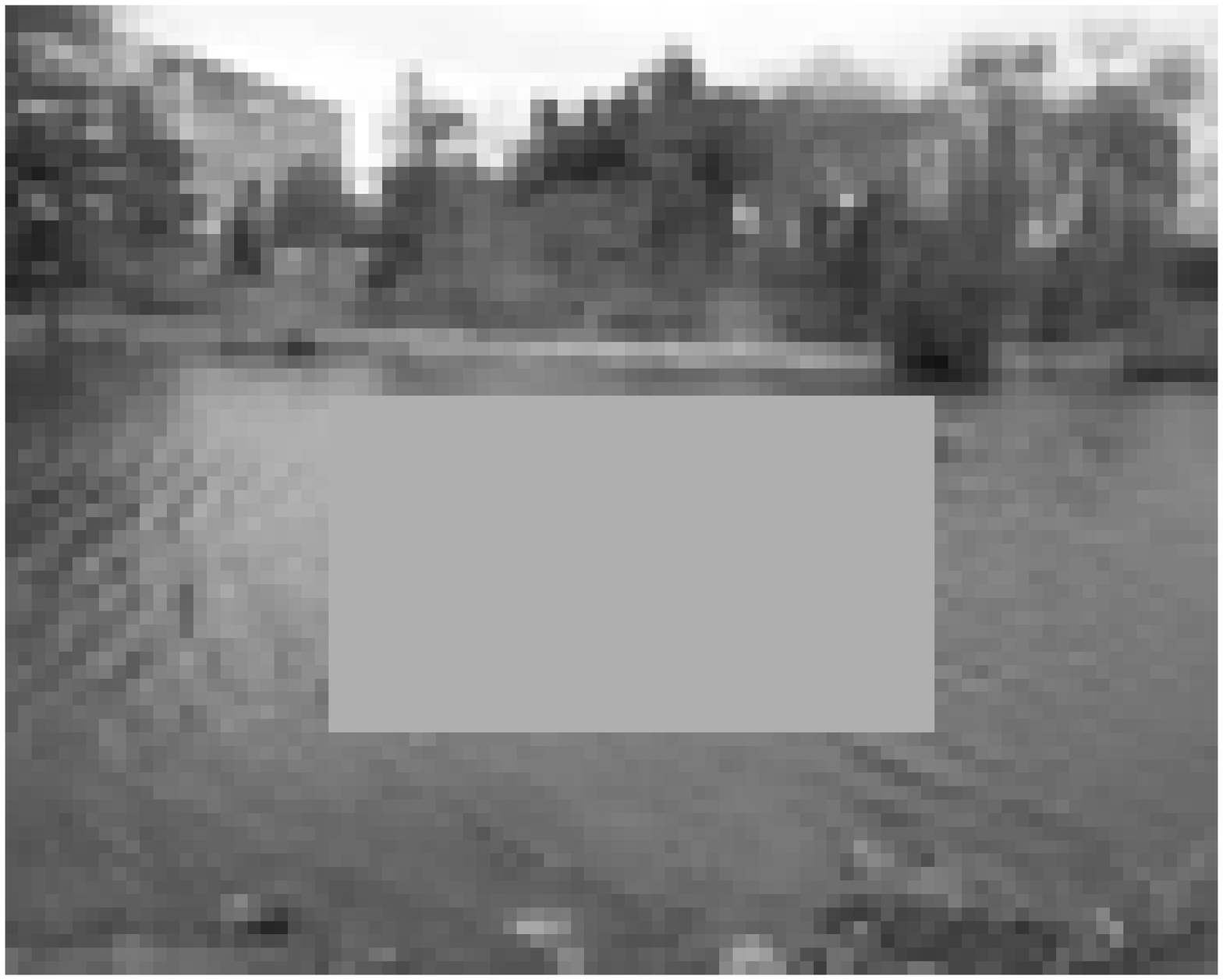}
		\includegraphics[width=16mm]{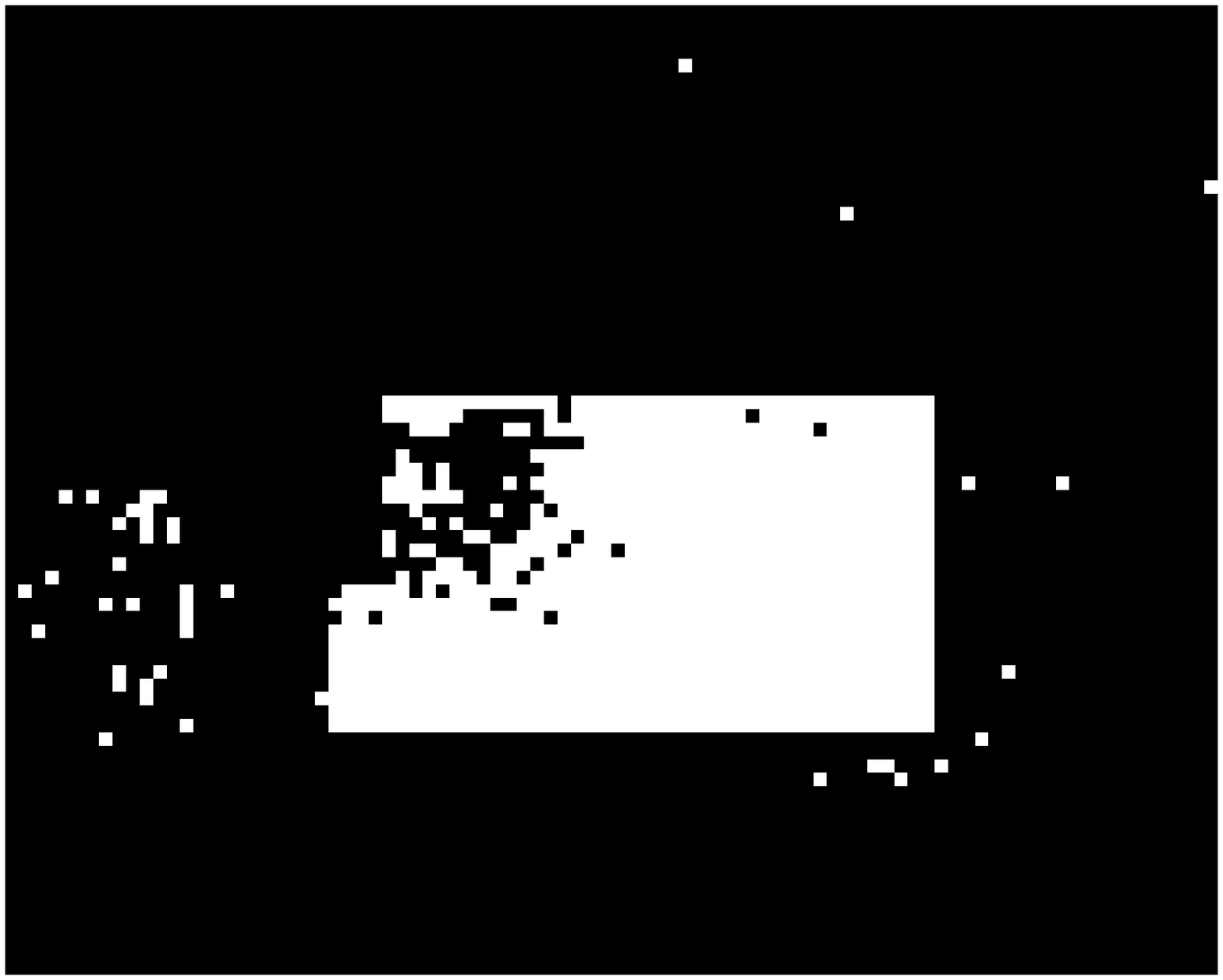}
		\includegraphics[width=16mm]{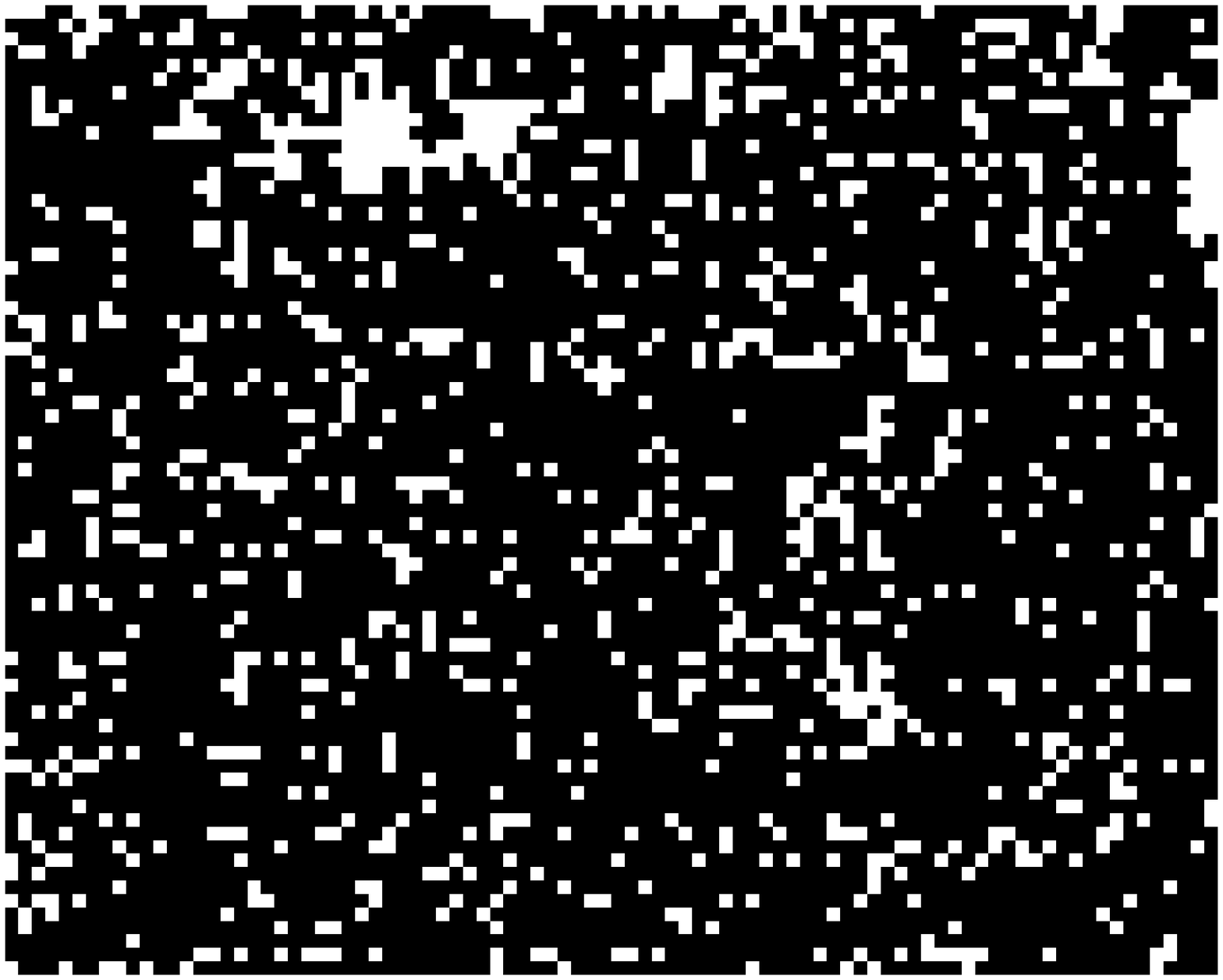}\\
		\includegraphics[width=16mm]{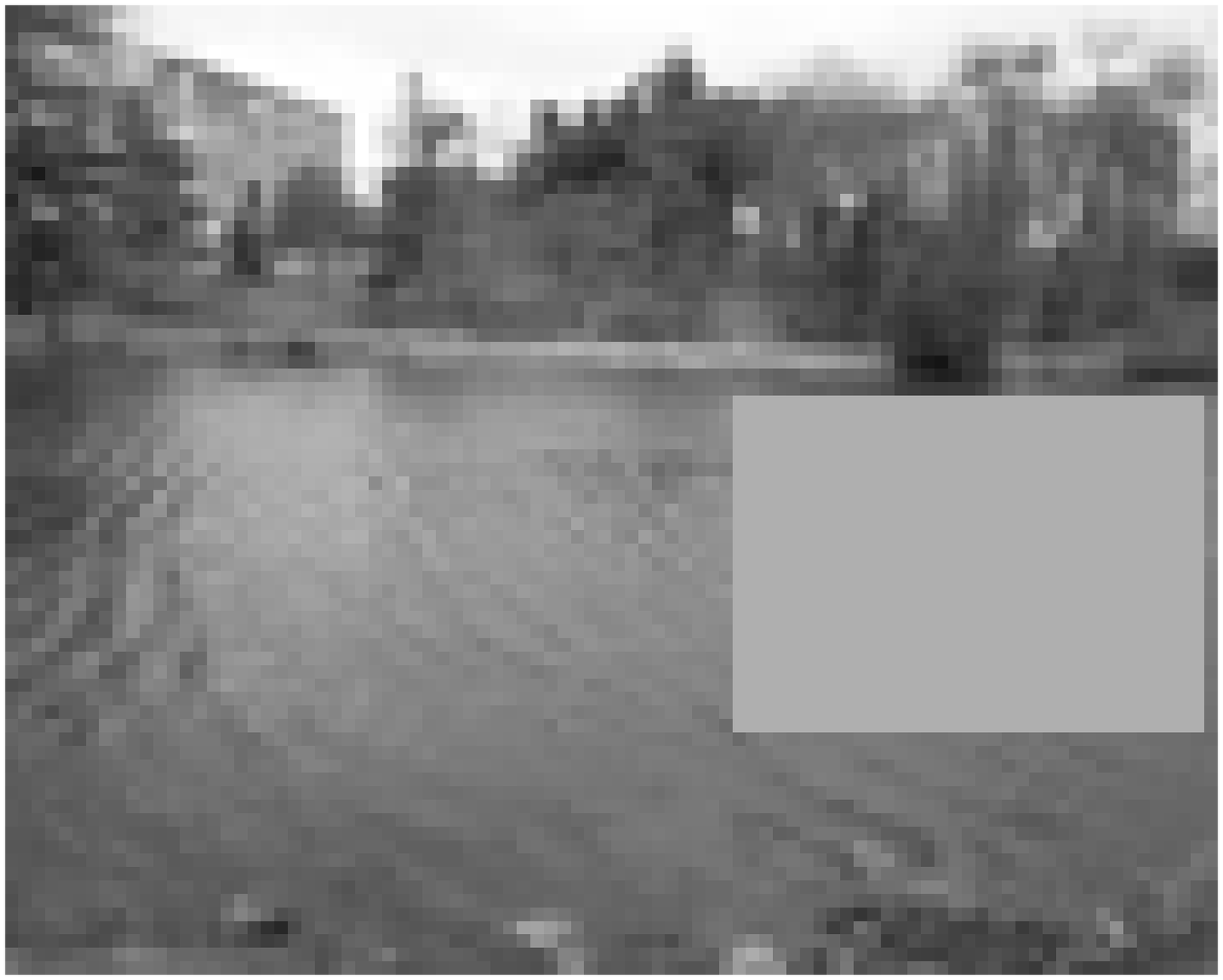}
		\includegraphics[width=16mm]{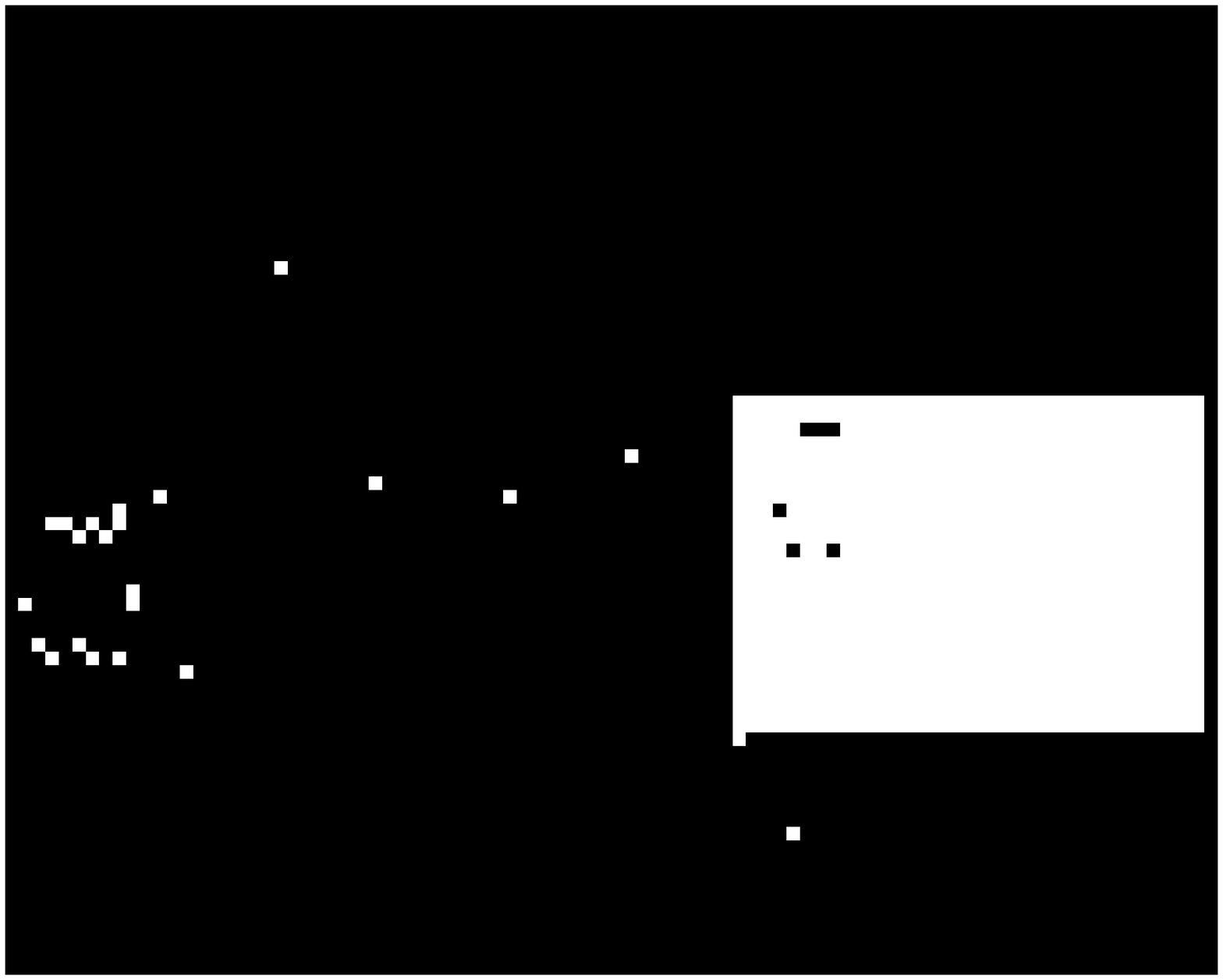}
		\includegraphics[width=16mm]{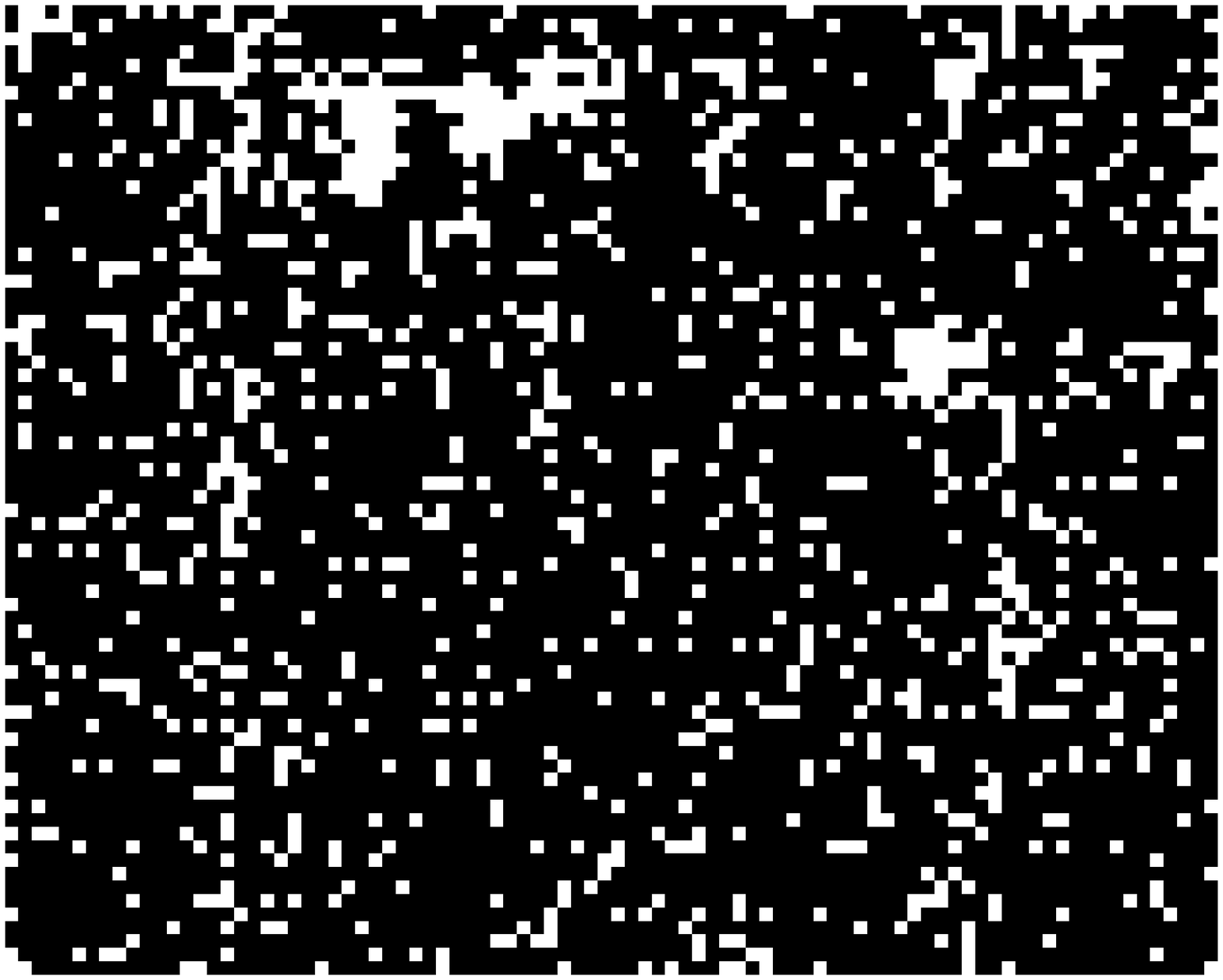}\\
		\includegraphics[width=16mm]{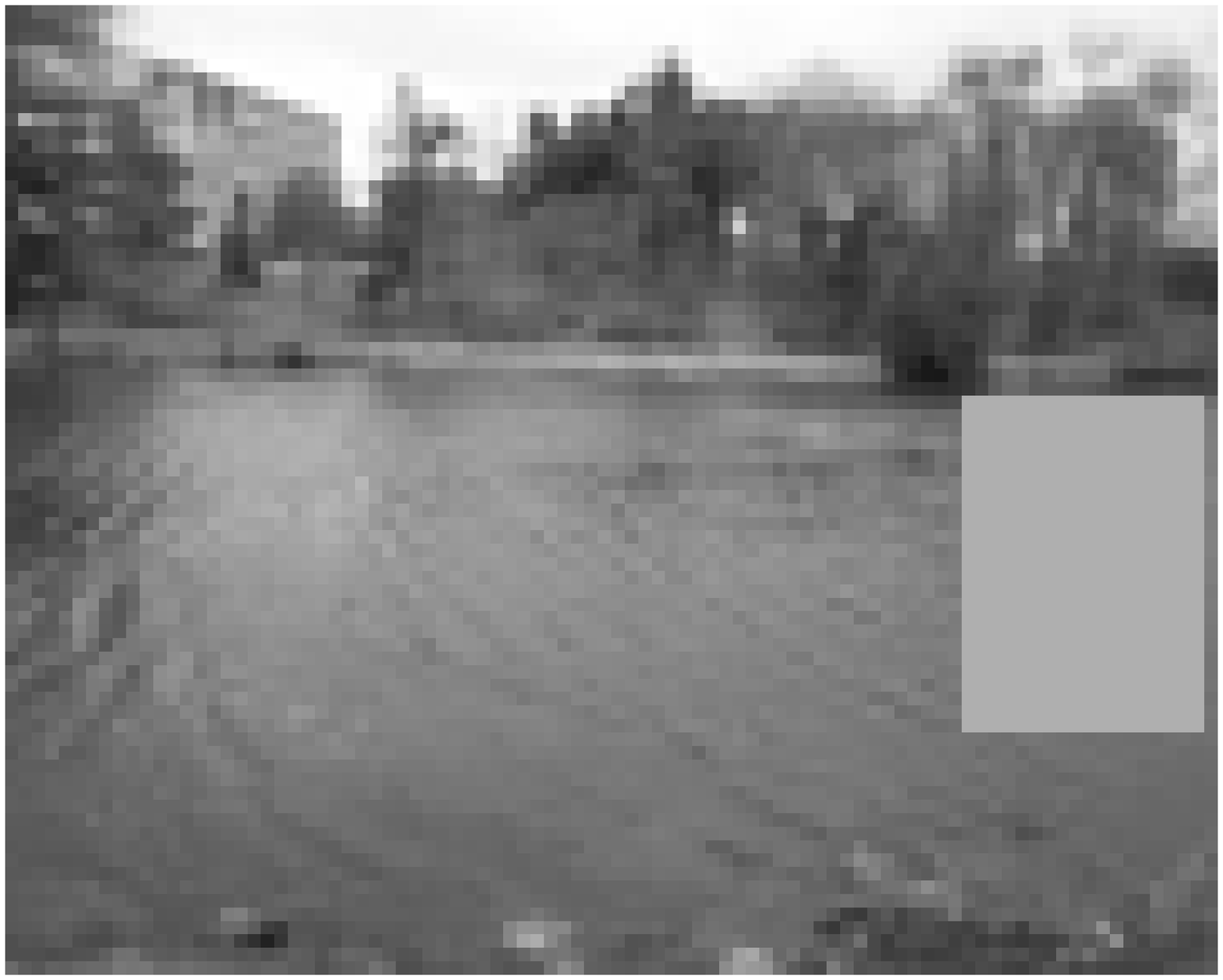}
		\includegraphics[width=16mm]{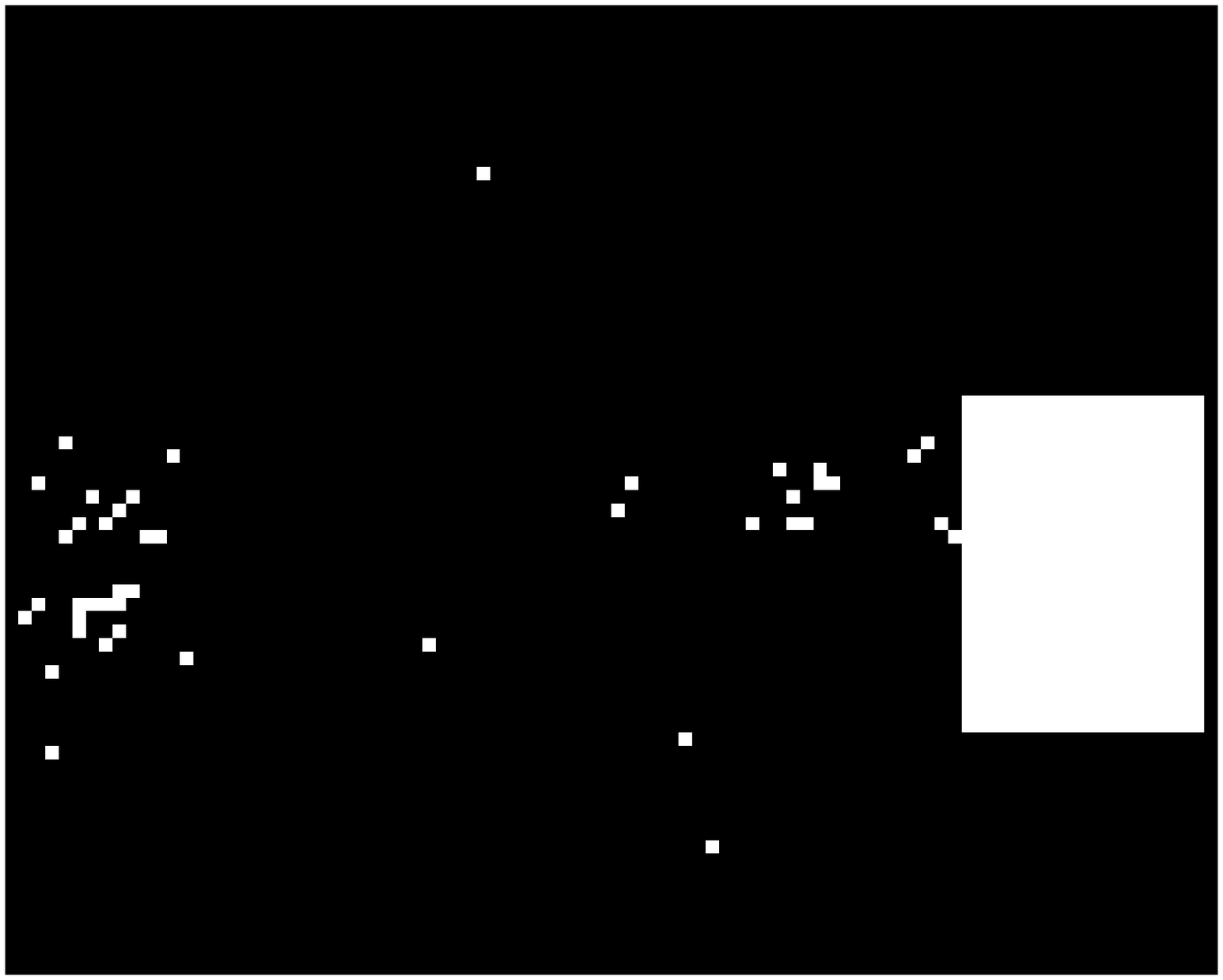}
		\includegraphics[width=16mm]{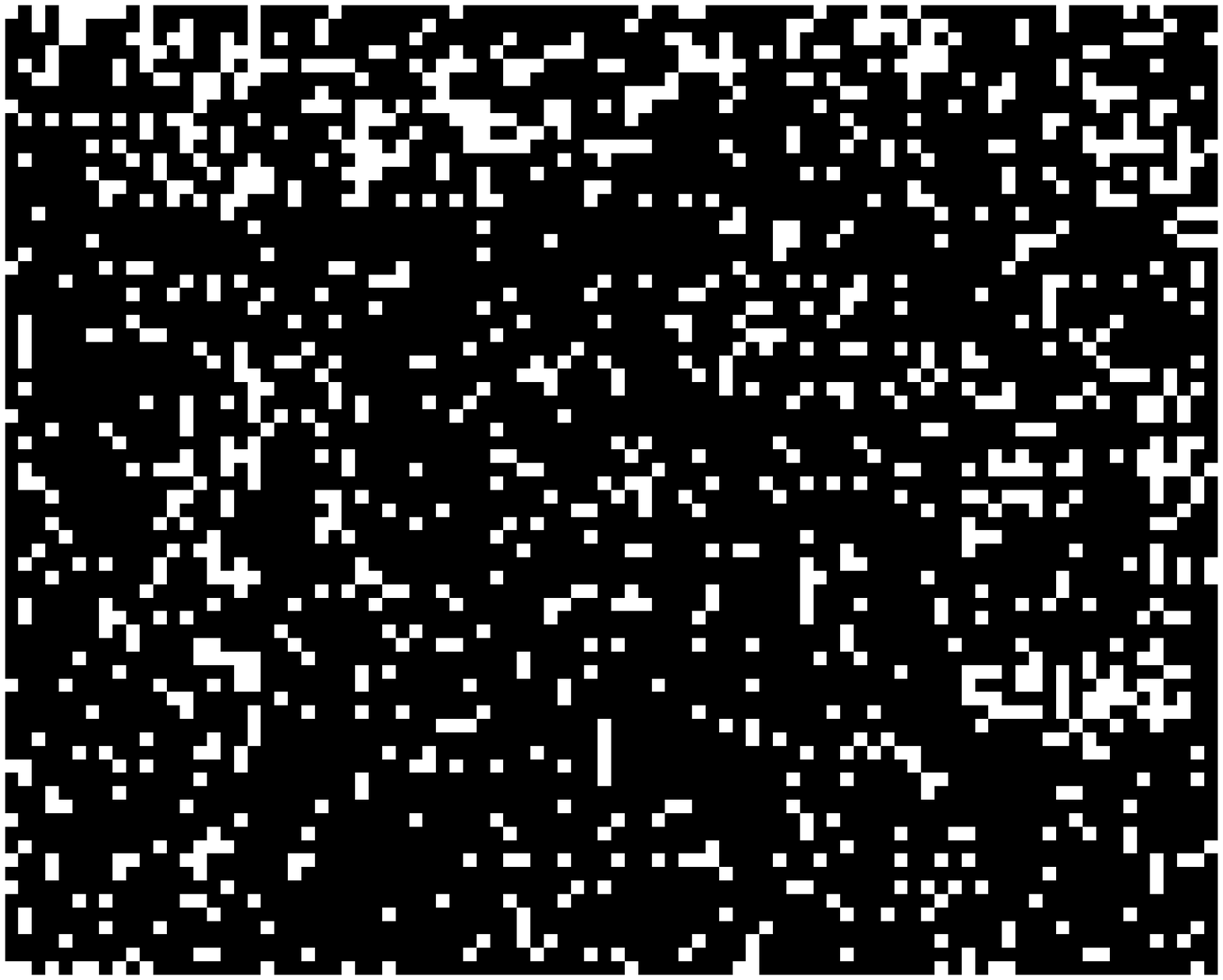}\\
		\hspace*{\fill}\makebox[0pt]{original }\hspace*{\fill}
		\hspace*{\fill}\makebox[0pt]{ReProCS}\hspace*{\fill}
		\hspace*{\fill}\makebox[0pt]{SparCS}\hspace*{\fill}
	\end{tabular}
	\caption{\small{Original video frames at $t=t_\train+30, 60, 70$ and foreground layer recovery by ReProCS and SparCS.}}\label{undersample}
\end{figure}

{\em Compressive ReProCS: comparisons for simulated video. }
We compare compressive ReProCS with SpaRCS \cite{SpaRCS} which is a batch algorithm for undersampled robust PCA / separation of sparse and low-dimensional parts(its code is downloaded from \url{http://www.ece.rice.edu/~aew2/sparcs.html}). No code is available for most of the other compressive batch robust PCA algorithms such as \cite{compressivePCP,rpca_reduced}.
SpaRCS is a greedy approach that combines ideas from CoSaMP \cite{cosamp} for sparse recovery and ADMiRA \cite{Bresler_matrix} for matrix completion. The comparison is done for compressive measurements of the lake sequence with foreground simulated as explained earlier. The matrix $B=A$ is $m \times n$ random Gaussian with $m = 0.7n$. Recall that $n = 6480$. The SpaRCS code required the background data rank and foreground sparsity as inputs. For rank, we used $\hat{r}$ returned by ReProCS, for sparsity we used the true size of the simulated foreground.
As can be seen from Fig. \ref{undersample}, SpaRCS does not work while compressive ReProCS is able to recover $S_t$ fairly accurately, though of course the errors are larger than in the full sampled case. All experiments shown in \cite{SpaRCS} are for very slow changing backgrounds and for foregrounds with very small support sizes, while neither is true for our data.

\begin{table*}
\centering
    \begin{tabular}{ | c || c | c | c | c || c | c |}
    \hline
    \multicolumn{7}{|c|}{(a) $(S_t)_i = 100$ for  $i \in T_t$ and $(S_t)_i = 0$ for $i \in T_t^c$}\\
    \hline
    \multicolumn{1}{|c|}{} & \multicolumn{4}{|c|}{$\mathbb{E} \| S-\hat{S} \|_F^2 / \mathbb{E}\| S \|^2_F$}& \multicolumn{2}{|c|}{$\mathbb{E} \| O-\hat{O} \|_F^2 / \mathbb{E}\| O \|^2_F$}\\
    \hline
    $|T_t|/n$   & ReProCS-pPCA  & RSL & PCP  & GRASTA  & adapted-iSVD & iRSL   \\
    \hline
    $9\%$  & $1.52\times 10^{-4}$ & $0.0580$ & $0.0021$ & $3.75\times 10^{-4}$ & $0.0283$ & $0.9105 $\\
    \hline
    $27\%$ & $1.90\times 10^{-4}$ & $0.0198$ & $0.6852$ & $0.1043$ & $0.0637$ &$0.9058 $\\
    \hline
    \hline
    \multicolumn{7}{|c|}{(b) $(S_t)_i = 10$ for  $i \in T_t$ and $(S_t)_i = 0$ for $i \in T_t^c$}\\
    \hline
    \multicolumn{1}{|c|}{} & \multicolumn{4}{|c|}{$\mathbb{E} \| S-\hat{S} \|_F^2 / \mathbb{E}\| S \|^2_F$}& \multicolumn{2}{|c|}{$\mathbb{E} \| O-\hat{O} \|_F^2 / \mathbb{E}\| O \|^2_F$}\\
    \hline
    $|T_t|/n$   & ReProCS-pPCA  & RSL & PCP  & GRASTA  & adapted-iSVD & iRSL   \\
    \hline
    $9\%$  & $0.0344$ & $8.7247$ & $0.2120$ & $0.1390$ & $0.2346$ & $0.9739 $\\
    \hline
    $27\%$ & $0.0668$ & $3.3166$ & $0.6456$ & $0.1275$ & $0.3509$ &$0.9778 $\\
    \hline
    \end{tabular}
\caption{\small{
{
Comparison of reconstruction errors of different algorithms for simulated data. Here, $|T_t|/n$ is the sparsity ratio of $S_t$, $\mathbb{E}[.]$ denotes the Monte Carlo average computed over 100 realizations and $\|.\|_F$ is the Frobenius norm of a matrix. Also, $S=[S_1,S_2,\dots S_{t_{\max}}]$  and $\Shat$ is its estimate; $(O_t)_i= (M_t)_i$ if $i \in T_t$ and $(O_t)_i=0$ otherwise and $\hat{O}_t$ is defined similarly with the estimates. $O$ and $\hat{O}$ are the corresponding matrices. We show error for $O$ for iRSL and adapted-iSVD since these algorithms can only return an estimate of the outlier support $T_t$; they do not return the background estimate.
}
}}
\label{tablesim}
\end{table*}

%
%

\begin{table*}
		\centering
		\begin{tabular}{c|c|c||c|c|c|c|c}
		\hline\hline
		DataSet & Image Size  & Sequence Length & ReProCS-pPCA & ReProCS-Recursive-PCA & PCP & RSL & GRASTA \\
		\hline
		Lake & $72\times 90$ & $1420+80$ &  $2.99+19.97 $ sec & $2.99+19.43$ sec  & 245.03 sec & 213.36 sec & $39.47+0.42$ sec \\
		Curtain & $64 \times 80 $ & $1755+1209$ & $4.37+159.02$ sec & $4.37+157.21$ sec & 1079.59 sec & 643.98 sec & $40.01+5.13$ sec \\
		Person & $120 \times 160$ & $200+52$ & $0.46+42.43$ sec & $0.46+41.91$ sec & 27.72 sec & 121.31 sec & $13.80+0.64$ sec \\
		\hline
		\end{tabular}
		\caption{\small{Comparison of speed of different algorithms. Experiments were done on a 64 bit Windows 8 laptop with 2.40GHz i7 CPU and 8G RAM. Sequence length refers to the length of sequence for training plus the length of sequence for separation. For ReProCS and GRASTA, the time is shown as training time + recovery time.
}}
\label{speed}
\end{table*}

%

\section{Conclusions and Future Work} \label{conc}
This work designed and evaluated Prac-ReProCS which is a practically usable modification of its theoretical counterpart that was studied in our earlier work \cite{rrpcp_perf, rrpcp_perf2,rrpcp_correct}. We showed that Prac-ReProCS has excellent performance for both simulated data and for a real application (foreground-background separation in videos) and the performance is better than many of the state-of-the-art algorithms from recent work. Moreover, most of the assumptions used to obtain its guarantees are valid for real videos. Finally we also proposed and evaluated a compressive prac-ReProCS algorithm.
In ongoing work, on one end, we are working on performance guarantees for compressive ReProCS \cite{rrpcp_globalsip} and on the other end, we are developing and evaluating a related approach for functional MRI. In fMRI, one is allowed to change the measurement matrix at each time. However if we replace $B=A$ by $A_t$ in Sec \ref{compressive} the compressive ReProCS algorithm does not apply because $A_t L_t$ is not low-dimensional \cite{zhan_icassp}. 

{ \appendix
\subsection{Detailed Discussion of Why ReProCS Works} \label{betasmall}

Define the subspace estimation error as
$$\SE(P,\Phat):=\|(I - \Phat \Phat') P\|_2$$
where both $P$ and $\Phat$ are {\em basis matrices.}
Notice that this quantity is always between zero and one. It is equal to zero when $\Span(\Phat)$ contains $\Span(P)$ and it is equal to one if $\Phat'P_i = 0$ for at least one column of $P$.

Recall that for $t \in [t_j, t_{j+1}-1]$, $P_{t} = [P_{(j-1)} R_j \setminus P_{(j),\old},  P_{(j),\new}]$. Thus, $L_t$ can be rewritten as
$$L_t = P_{(j-1)}a_{t,*} + P_{(j),\new} a_{t,\new} $$
where  $a_{t,\new}:=P_{(j),\new}' L_t$ and $a_{t,*}:= P_{(j-1)}' L_t$.


Let $c:=c_{\max}$ and $r_j:= r_0 + jc$.
We explain here the key idea of why ReProCS works \cite{rrpcp_perf,rrpcp_correct}. Assume the following hold besides the assumptions of Sec \ref{probdef}.
\ben
\item Subspace change is detected immediately, i.e. $\hat{t}_j = t_j$ and $c_{j,\new}$ is known. 

\item Pick a $\zeta \ll 1$. Assume that $\|L_t\|_2 \le \gamma_*$ for a $\gamma_*$ that satisfies $\gamma_* \le 1/\sqrt{r_{J} \zeta}$. Since $\zeta$ is very small, $\gamma_*$ can be very large.

\item Assume that $(t_{j+1} - t_j) \ge K \alpha$ for a $K$ as defined below. 

\item Assume the following model on the gradual increase of $a_{t,\new}$: for $t \in [t_j+ (k-1) \alpha, t_j+ k \alpha-1]$,  $\|a_{t,\new}\|_2 \le v^{k-1} \gamma_\new$ for a $1 < v \le 1.2$ and $\gamma_\new \ll \gamma_*$.

\item Assume that projection PCA ``works" i.e. its estimates satisfy $\SE(P_{(j),\new},\Phat_{(j),\new,k-1}) \le 0.6^{k-1} + 0.4 c \zeta $. The proof of this statement is long and complicated and is given in  \cite{rrpcp_perf,rrpcp_correct}. 

\item Assume that projection PCA is done $K$ times with $K$ chosen so that $0.6^{K-1} + 0.4 c\zeta \le c \zeta$.
\een


Assume that at $t=t_j-1$, $ \SE(P_{(j-1)},\Phat_{(j-1)}) \le r_{j-1} \zeta \ll 1$. We will argue below that $ \SE(P_{(j)},\Phat_{(j)}) \le r_{j} \zeta$. Since $r_j \le r_0 + J c$ is small, this error is always small and bounded.

First consider a $t \in [t_j, t_j+ \alpha)$. At this time, $\Phat_{t} = \Phat_{(j-1)}$. 
Thus,
\bea
\|\beta_t\|_2 \se \|(I - \Phat_{t-1} \Phat_{t-1}')L_t\|_2   \nn \\
              \sle \SE(P_{(j-1)},\Phat_{(j-1)}) \|a_{t,*}\|_2 + \|a_{t,\new}\|_2 \nn \\  
              \sle (r_{j-1} \zeta)  \gamma_* +  \gamma_\new \nn \\
              \sle \sqrt{\zeta} + \gamma_\new
\eea
By construction, $\sqrt{\zeta}$ is very small and hence the second term in the bound is the dominant one. By the slow subspace assumption $\gamma_\new \ll \|S_t\|_2$.  Recall that $\beta_t$ is the ``noise" seen by the sparse recovery step. The above shows that this noise is small compared to $\|S_t\|_2$.
Moreover, using Lemma \ref{delta_kappa} and simple arguments [see \cite[Lemma 6.6]{rrpcp_perf}], it can be shown that 
$$\delta_s(\Phi_{t}) \le \kappa_*^2 + r_{j-1} \zeta$$
is small. These two facts along with any RIP-based result for $\ell_1$ minimization, e.g. \cite{candes_rip}, ensure that $S_t$ is recovered accurately in this step. If the smallest nonzero entry of $S_t$ is large enough, it is possible get a support threshold $\omega$ that ensures exact support recovery. Then, the LS step gives a very accurate final estimate of $S_t$ and it allows us to get an exact expression for $e_t:=S_t - \Shat_t$. Since $\Lhat_t = M_t - \Shat_t$, this means that $L_t$ is also recovered accurately and $e_t = \Lhat_t-L_t$.
This is then used to argue that p-PCA at $t=t_j+\alpha-1$ ``works". 

Next consider $t \in [t_j+ (k-1) \alpha, t_j+ k \alpha-1]$. At this time, $\Phat_{t} = [\Phat_{(j-1)}, \Phat_{(j),\new,k-1}]$.
Then, it is easy to see that
\bea
\|\beta_t\|_2 \se \|(I - \Phat_{t-1} \Phat_{t-1}')L_t\|_2   \nn \\
              \sle \SE(P_{(j-1)},\Phat_{(j-1)}) \|a_{t,*}\|_2 + \SE(P_{(j),\new},\Phat_{(j),\new,k-1}) \|a_{t,\new}\|_2 \nn \\
              \sle (r_{j-1} \zeta) \ \gamma_* + (0.6^{k-1} + 0.4 c\zeta) \ v^{k-1} \gamma_\new  \nn \\
              \sle  \sqrt{\zeta} + 0.72^{k-1} \gamma_\new
\eea
Ignoring the first term, in this interval, $\|\beta_t\|_2 \le 0.72^{k-1} \gamma_\new$, i.e. the noise seen by the sparse recovery step decreases exponentially with every p-PCA step. This, along with a bound on $\delta_s(\Phi_{t})$ (this bound needs a more complicated argument than that for $k=1$, see  \cite[Lemma 6.6]{rrpcp_perf}), ensures that the recovery error of $S_t$, and hence also of $L_t=M_t-S_t$, decreases roughly exponentially with $k$. This is then used to argue that the p-PCA error also decays roughly exponentially with $k$.

Finally for $t \in [t_j + K \alpha, t_{j+1}-1]$, because of the choice of $K$, we have that $\SE(P_{(j),\new},\Phat_{(j),\new,K}) \le c \zeta$. At this time, we set $\Phat_{(j)} = [\Phat_{(j)-1}, \Phat_{(j),\new,K}]$. Thus, $\SE(P_{(j)}, \Phat_{(j)}) \le \SE(P_{(j-1)},\Phat_{(j)-1}) + \SE(P_{(j),\new},\Phat_{(j),\new,K}) \le r_{j-1}\zeta +c \zeta = r_j \zeta$.

}

\bibliographystyle{IEEEbib}
\bibliography{tipnewpfmt_kfcsfullpap}
\end{document}